\documentclass[
  english,
  font=cmss, 
  onecolumn, 
  11pt
]{article}

\usepackage[utf8]{inputenc}
\usepackage[english]{babel}

\usepackage{geometry}

\newlength{\marginparam}
\usepackage{titlesec} 
\usepackage{placeins}
\RequirePackage{pgfkeys}
\RequirePackage{pgfopts}

\usepackage{lmodern}

\usepackage{hyphenat}

\usepackage{caption}

\usepackage{subcaption}

\usepackage{booktabs}
\usepackage{tabularx}
\usepackage{multicol}
\usepackage{multirow}

\usepackage{graphicx}
\usepackage{colortbl}

\usepackage{xspace}

\usepackage{url}
\usepackage[backend=bibtex,
			natbib=true,
			bibstyle=ieee,
			citestyle=numeric-comp,
			sorting=none,
			giveninits=true,
			doi=true,
			urldate=comp,
			url=false, isbn=false, eprint=false]{biblatex}
\AtEveryBibitem{\clearfield{isbn}}
\AtEveryCitekey{\clearfield{url}}
\AtEveryBibitem{\clearfield{month}}
\AtEveryBibitem{\clearfield{day}}
\AtEveryBibitem{\clearfield{file}}
\AtEveryBibitem{\clearfield{language}}
\AtEveryBibitem{\clearfield{issn}}

\renewbibmacro*{doi+eprint+url}{%
  \printfield{doi}%
  \newunit\newblock%
  \iftoggle{bbx:eprint}{%
    \usebibmacro{eprint}%
  }{}%
  \newunit\newblock%
  \iffieldundef{doi}{%
    \usebibmacro{url+urldate}}%
    {}%
}

\usepackage[export]{adjustbox}
\usepackage[absolute, overlay]{textpos}

\usepackage{calc}
\usepackage{amsmath}
\usepackage{amsfonts}
\usepackage{amssymb}
\usepackage{siunitx}

\usepackage{tikzscale}
\usepackage[hidelinks,breaklinks]{hyperref}
\usepackage{tikz}
\usetikzlibrary{arrows, shapes, shapes.multipart, trees, positioning,
  backgrounds, fit, matrix, shapes.arrows, arrows.meta, pgfplots.groupplots}
\usepackage{pgfplots}
\pgfplotsset{compat=default}
\usepackage[mode=buildnew]{standalone}

\usepackage{fontawesome5}

\usepackage{defines}

\usepackage{ragged2e}
\newcolumntype{L}[1]{>{\raggedright\let\newline\\\arraybackslash\hspace{0pt}}m{#1}}
\newcolumntype{C}[1]{>{\centering\let\newline\\\arraybackslash\hspace{0pt}}m{#1}}
\newcolumntype{R}[1]{>{\raggedleft\let\newline\\\arraybackslash\hspace{0pt}}m{#1}}

\renewenvironment{abstract}{%
  \setlength{\parindent}{0mm}
  \vspace{15pt} 
  \noindent\rule{\textwidth}{0.4pt} 
  \par\vspace{5pt}
  {\textbf{\abstractname}}
  \normalfont\textcolor{black}{--}}{%
  \par\vspace{5pt}
  \noindent\rule{\textwidth}{0.4pt} 
}

\makeatletter
\renewcommand\subparagraph{\@startsection{subparagraph}{5}{\z@}{1.5ex \@plus1ex \@minus.2ex}%
{-1em}{\normalfont\normalsize\itshape}}
\makeatother

\usepackage{setspace}
\makeatletter
\renewcommand{\maketitle}{
    \noindent\parbox{0.99\textwidth}{
          \Large 
          \setstretch{1.05}
          \textbf{\@title}
        }\\[2em]
    \noindent{\large \@author}\\[0.5em]
    \textsuperscript{1}Institute for Computational Mechanics, TUM School of Engineering and Design, Technical University of Munich\\[0.1em]
    \textsuperscript{2}Department of Orthopedics and Sports Orthopedics, University Hospital rechts der Isar, TUM School of Medicine, Technical University of Munich\\[1em]
    \href{mailto:laura.engelhardt@tum.de}{laura.engelhardt@tum.de}, \href{mailto:renate.sachse@tum.de}{renate.sachse@tum.de}, \href{mailto:rainer.burgkart@tum.de}{rainer.burgkart@tum.de}, and \href{mailto:wolfgang.a.wall@tum.de}{wolfgang.a.wall@tum.de}
}
\makeatother

\title{Constitutive Models for Active Skeletal Muscle: Review, Comparison, and Application in a Novel Continuum Shoulder Model}
\author{\textbf{Laura Engelhardt}\textsuperscript{1} \href{https://orcid.org/0000-0003-2430-2196}{\faOrcid}, \textbf{Renate Sachse}\textsuperscript{1} \href{https://orcid.org/0000-0001-8895-9635}{\faOrcid}, \textbf{Rainer Burgkart}\textsuperscript{2}, and \textbf{Wolfgang A. Wall}\textsuperscript{1} \href{https://orcid.org/0000-0001-7419-3384}{\faOrcid}}

\begin{document}
\maketitle

\begin{abstract}
The shoulder joint is one of the functionally and anatomically most sophisticated articular systems in the human body. Both complex movement patterns and the stabilization of the highly mobile joint rely on intricate three-dimensional interactions among various components. Continuum-based finite element models can capture such complexity, and are thus particularly relevant in shoulder biomechanics. Considering their role as active joint stabilizers and force generators, skeletal muscles require special attention regarding their constitutive description.
In this contribution, we propose a constitutive description to model active skeletal muscle within complex musculoskeletal systems, focusing on a novel continuum shoulder model. We thoroughly review existing material models before analyzing three selected ones in detail: an active-stress, an active-strain, and a generalized active-strain approach.
To establish a basis for comparison, we identify the material parameters based on experimental stress-strain data obtained under various active and passive loading conditions. We discuss the concepts to incorporate active contractile behavior from a mathematical and physiological perspective, address analytical and numerical challenges arising from the mathematical formulations, and analyze the included biophysical principles of force generation in terms of physiological correctness and relevance for human shoulder modeling. Based on these insights, we present an improved constitutive model combining the studied models' most promising and relevant features. Using the example of a fusiform muscle, we investigate force generation, deformation, and kinematics during active isometric and free contractions. Eventually, we demonstrate the applicability of the suggested material model in a novel continuum mechanical model of the human shoulder.
\end{abstract}

\vspace{20pt}

\section{Introduction}
As one of the functionally and anatomically most complex articular systems in the human body, the shoulder joint combines mobility and stability in a unique musculoskeletal system.
The anatomical structure of the involved glenohumeral joint allows for an extensive range of motion, while passive and active soft tissues ensure the joint's integrity through static and dynamic mechanisms. Muscles, especially the rotator cuff and the deltoid, perform multiple essential functions. First, muscles actively stabilize the glenohumeral joint's bony structures through concavity compression and scapulohumeral balance. Second, muscles act as torque generators and enable complex movement patterns through their sophisticated interplay. 
Maintaining this delicate balance between mobility and stability is essential for proper shoulder function, yet it is easily disrupted by injury or pathological conditions. Despite the high incidence of shoulder disorders in clinical practice, understanding of the underlying biomechanics remains limited. Developing objective, reliable diagnostic procedures, and effective, monitorable treatments thus presents a major challenge for medical professionals and biomedical engineers.

Computational musculoskeletal models offer great potential to study the shoulder's biomechanics and physiology, investigate pathological conditions and (patient-specific) treatments, and accelerate developments of medical devices such as surgical tools, implants, or rehabilitation equipment for physical therapy.
While numerous reduced-dimensional multi-body models exist, research on comprehensive three-dimensional continuum mechanical models remains limited. 
Especially in a joint as complex as the shoulder, three-dimensional interactions between the geometrically complex components, sophisticated muscle fiber architectures, and directional material properties are, however, central to the shoulder's physiology. Here, continuum mechanical models can offer critical insights compared to a reduced-dimensional approach and help to further improve these highly efficient and desirable reduced-dimensional models. 

Considering their role as active joint stabilizers and force generators, skeletal muscles deserve special attention regarding their constitutive description. 
Current shoulder models either apply purely passive material models neglecting the muscle's active properties or use the active-stress material model from \cite{blemker_3d_2005} that generates internal forces and contractile deformations in response to a prescribed external stimulation.

Research on the constitutive modeling of active skeletal muscle is though fairly advanced. There exist various active constitutive models differing in the applied mathematical concept, rheological properties, modeled scales, and considered active force generation mechanisms.

Whether the active-stress muscle material model \cite{blemker_3d_2005} used in the existing shoulder models is the most suitable approach has not yet been investigated. The question of which material model best characterizes the shoulder's skeletal muscles at an appropriate level of detail while being computationally efficient and robust for such a large-scale application remains open.

In this paper, we aim to identify a suitable material model for modeling the active skeletal muscle components in a full three-dimensional continuum mechanical model of the human shoulder. To achieve this, we comprehensively review existing approaches, conduct a detailed study of three selected material models, and ultimately integrate the most promising and relevant properties into a modified material model suitable for our application scenario. 

In Chapter \ref{sec:review}, we provide an overview of current musculoskeletal models for the human shoulder and conduct a thorough review of existing constitutive descriptions for active skeletal muscle. 
We place particular focus on constitutive descriptions applicable to continuum mechanical musculoskeletal simulations, although our research extends beyond this scope. 
From the reviewed material models, we select three hyperelastic material models for further investigation in Chapter \ref{sec:materials}: the active-stress approach by Blemker et al. \cite{blemker_3d_2005}, which has already been applied to models of the human shoulder and knee, the microstructurally inspired generalized active-strain approach by Weickenmeier et al. \cite{weickenmeier_physically_2014}, and the mathematically well-posed active-strain approach by Giantesio et al. \cite{giantesio_strain-dependent_2017}.
We compare these considering physiological, mathematical, and computational aspects.
We discuss the concepts of modeling active material behavior from a mathematical and physiological perspective, address analytical and numerical problems arising from the mathematical formulations, and analyze the included biophysical principles of force generation in terms of physiological correctness and relevance considering the modeling of the human shoulder. Based on those insights, we present an improved constitutive model combining the studied models' most promising and relevant properties. 
To establish a basis for a numerical comparison, we fit the material parameters to a common set of experimentally obtained stress-strain data from the literature in Chapter \ref{sec:fitting}. Contrary to the original publications, we consider multiple active and passive loading conditions, as a single load case is generally insufficient to uniquely determine the material response.
By the example of a fusiform muscle geometry, we investigate force generation, deformation, and kinematics during active isometric and free contractions in Chapter \ref{sec:fusiform}. Eventually, we demonstrate the applicability of the suggested material model in simulations of a two-component muscle-bone model in Chapter \ref{sec:musclebone}, and a full continuum mechanical model of the human shoulder in Chapter \ref{sec:shoulder}.

\section{Literature review} \label{sec:review}

\subsection{Musculoskeletal models of the human shoulder} \label{sec:shoulder_models}
Computational models of the human shoulder can be primarily categorized into multi-body and continuum mechanical finite element models.

Reduced-dimensional multi-body models are based on rigid body dynamics and assume the body segments, i.e., the bones, as non-deforming rigid bodies. Muscles connect those rigid segments and are modeled as one-dimensional line actuators. For muscles with a broad attachment area, multiple actuators can be defined. Often, these approaches apply wrapping methods to geometrically constrain the muscle force path and prevent penetration between muscle and bone \cite{charlton_application_2001, garner_obstacle_2000}. 

Because muscles are assumed as simplified one-dimensional objects that deform independently of each other, multi-body models fail to capture a wide range of phenomena, such as contact or sliding interactions between the joint components, three-dimensional (non-uniform) deformations and stress distributions, or complex fiber arrangements and tendon morphologies.
Despite these disadvantages, multi-body models have been successfully applied in research and technology. Areas of application include investigations of movement actuation \cite{viceconti_biomechanics_2006}, muscle force and moment arm estimations \cite{pean_comprehensive_2019, webb_3d_2014}, and the simulation of neuromuscular control of prostheses \cite{blache_influence_2018} and surgical procedures \cite{holzbaur_model_2005}. A comprehensive overview of multi-body models of the shoulder and upper extremity can be found in \cite{bolsterlee_clinical_2013, dao_rigid_2016, prinold_musculoskeletal_2013, roupa_modeling_2022}.

In contrast, continuum mechanical models discretize muscles and other deformable structures in a full three-dimensional fashion.
These models can thus resolve internal stress and strain distributions and can account for contact and three-dimensional interactions between geometrically complex parts. Further, implementations involving sophisticated muscle fiber arrangements and tendon morphologies \cite{zheng_finite_2017, pean_comprehensive_2019, blemker_three-dimensional_2005}, complex (nonhomogeneous) constitutive behavior \cite{zheng_finite_2017, rohrle_two-muscle_2017}, and spatially varying muscle activation can be realized \cite{heidlauf_modeling_2013, heidlauf_multiscale_2014}.
Of course, such models come with additional challenges that for example include an increased computational cost and a higher complexity regarding the geometric design, discretization, methods of contact modeling, and solution techniques.  

While reduced-dimensional multi-body human shoulder models are common in literature, only few continuum-based models of the entire human shoulder exist. We conducted a throughout review of existing continuum mechanical shoulder models and in the following briefly summarize our findings.

To provide an overview, Table \ref{tab:shoulder_models} lists the reviewed models along with their distinctive features. 
The number of incorporated anatomical components varies, ranging from basic models incorporating only the most fundamental joint muscles to comprehensive models encompassing the entire upper limb musculature.
Bones are commonly considered rigid bodies or, in some cases, integrated into the finite element (FE) discretization and assigned a comparably high material stiffness.
Muscles are typically discretized using three-dimensional tetrahedral or hexahedral elements, except for the surface-based two-dimensional modeling approach in \cite{pean_surface-based_2020}.

The majority of reviewed models neglect the active contractile behavior of muscle tissue \cite{buchler_finite_2002, terrier_effect_2007, metan_fem_2014, duprey_numerical_2005, inoue_nonlinear_2013}. Instead, they solely account for the passive response and prescribe external forces or displacements to generate movement. Typically, those models employ hyperelastic, transversely isotropic, nonlinear material models to account for the passive muscle characteristics.
More recent publications assign active constitutive laws to the muscular components such that the prescribed activation controls the motion \cite{teran_creating_2005, webb_3d_2014, pean_comprehensive_2019, pean_computational_2022}. 
State of the art is the active-stress material model from \cite{blemker_3d_2005}, that is, to the best of our knowledge, so far the only model applied in the context of continuum mechanical modeling of the human shoulder.
\begingroup
\renewcommand{\arraystretch}{1.5}
\begin{table}[htpb]
\caption{Continuum mechanical models of the human shoulder in literature. Only models featuring at least one muscle actuating the glenohumeral (GH) joint are considered in this summary.}
\label{tab:shoulder_models}
\footnotesize
\begin{tabularx}{\textwidth}{p{0.9cm} p{3.45cm}p{3.55cm}p{2.95cm}p{1.77cm}p{2.1cm}}
\toprule
 & Model description \newline and objective & Model components & Discretization & Muscle \newline material & Contact \\ \midrule
Büchler \mbox{et al.} \cite{buchler_finite_2002} & FE model to quantify the impact of the humeral head shape on the stress distribution in the scapula & Humerus, scapula, sub\-scapularis, supra\-spi\-natus, infraspinatus, cartilage in GH joint &  \mbox{Bones: linear hexes}, rigid surface elements \newline \mbox{Muscles, cartilage}: \mbox{linear hexes}  & Passive \mbox{exponential} hyperelastic & Bone-muscle, GH joint \\
Terrier \mbox{et al.} \cite{terrier_effect_2007} & FE model to investigate the biomechanical influence of a supraspinatus deficiency & Humerus, scapula, rotator cuff, deltoid, cartilage in GH joint & Bones: rigid \newline \mbox{Muscles: linear hexes} with truss elements for fibers, cables \newline \mbox{Cartilage: linear hexes} & Passive fi\-ber-reinforced hyper\-elastic Neo-Hooke & Bone-muscle, GH joint \\
Metan \mbox{et al.} \cite{metan_fem_2014} & FE model to investigate stresses during adduction and abduction shoulder exercises & Humerus, scapula, clavi\-cu\-la, infra\-spinatus, sub\-scapularis, deltoid, triceps, ligaments & Bones: rigid, tets \newline \mbox{Muscles, ligaments:} hexes,  pents  & Passive \mbox{linear} elastic & GH joint \\
Duprey \mbox{et al.} \cite{duprey_numerical_2005} & FE model to predict injuries in impact scenarios based on the HUMOS full-body model & Humerus, scapula, clavi\-cu\-la, several muscles, \mbox{ligaments}, skin & Bones: shells, hexes \newline Muscles: \mbox{springs, shells, hexes} & Passive elastoplastic & Not defined \\
Inoue \mbox{et al.} \cite{inoue_nonlinear_2013} & FE model to investigate stress distribution in the rotator cuff tendons & Humerus, scapula, sub\-scapularis, su\-pra\-spi\-na\-tus, infra\-spinatus, acrom. deltoid, cartilage in GH joint & Hexes & Passive nonlinear elastic & \mbox{GH joint}, \mbox{subacromial} space \\
Teran \mbox{et al.} \cite{teran_creating_2005} & Finite volume model to simulate dynamic defor\-mation, inversely compute muscular activation & Bones and 30 muscles of the upper limb & Bones: rigid \newline Muscles: tets & Active stress material \cite{blemker_3d_2005} & Muscle-muscle \\
Webb \mbox{et al.} \cite{webb_3d_2014} & FE model to examine muscle fiber paths and moment arms & Humerus, scapula, clavi\-cula, \mbox{rotator} cuff muscles (incl. tendons), deltoid & Bones: \mbox{rigid surface elements} \newline \mbox{Muscles, tendons:} \mbox{linear hexes} & Active stress material \cite{blemker_3d_2005} & Muscle-muscle, muscle-tendon, bone-muscle \\
Pean \mbox{et al.} \cite{pean_comprehensive_2019} & Comprehensive three-dimensional FE model to investigate shoulder biomechanics & Humerus, scapula, clavi\-cula, rotator cuff, deltoid, eight additional shoulder muscles & Bones: \mbox{rigid surface elements} \newline Muscles: \mbox{linear hexes} & Active stress material \cite{blemker_3d_2005} & Bone-muscle \\
Pean \mbox{et al.} \cite{pean_surface-based_2020} & FE model with surface-based muscles to investigate shoulder biomechanics & Humerus, scapula, clavi\-cula, rotator cuff, deltoid, eight additional shoulder muscles  & Bones: \mbox{rigid surface elements} \newline Muscles: \mbox{membrane elements} & Active stress material \cite{blemker_3d_2005} & Bone-muscle \\
\bottomrule
\end{tabularx}
\end{table}
\endgroup

\subsection{Constitutive modeling of active skeletal muscle} \label{sec:muscle_models}
Research regarding the three-dimensional constitutive modeling of skeletal muscle tissue is fairly advanced, and there exist a variety of elaborate material models for both the passive characteristics and the active contractile behavior.
Typically, skeletal muscle is modeled with nonlinear, hyperelastic constitutive laws, e.g., \cite{ehret_continuum_2011, blemker_3d_2005, giantesio_strain-dependent_2017, calvo_passive_2010}. Some authors, such as \cite{pandolfi_coupled_2016, pandolfi_visco-electro-elastic_2017, latorre_strain-level_2017, khodaei_simulation_2013, ahamed_time-dependent_2016}, choose viscoelastic approaches to incorporate rate-dependent properties. Hypervisco-poroelastic constitutive approaches are presented, e.g., in \cite{wheatley_case_2017, wheatley_modeling_2018}. Due to the high water content, the tissue is mostly assumed as an incompressible or nearly incompressible material. Depending on the information incorporated, the constitutive models can be classified as purely phenomenological or multi-scale. A common approach is to consider the passive material response first and then incorporate the active material properties.

\paragraph{Passive constitutive models: Fiber and matrix contributions}
From a histological point of view, the muscle's passive behavior is governed by the extracellular matrix and the passive contribution of the embedded muscle fibers. As the fibers are arranged in parallel bundles, most material laws assume a transversely isotropic fiber orientation in an isotropic tissue matrix.

Purely phenomenological models fit the constitutive behavior through mathematical formulations reflecting the experimentally observed behavior. Typically, the modeling of hyperelastic behavior starts with the definition of a strain energy function $\Psi$. In accordance with the histological composition of muscle tissue, a common approach is to additively split the strain-energy function $\Psi^{\mathrm{p}}$ (where the index $\mathrm{p}$ point to the passive contribution) into the two respective parts, $\Psi_{\mathrm{fiber}}^{\mathrm{p}}$ and $\Psi_{\mathrm{matrix}}^{\mathrm{p}}$. 

The most popular choice for $\Psi_{\mathrm{matrix}}^{\mathrm{p}}$ is an isotropic Mooney-Rivlin constitutive law, as in \cite{rohrle_three-dimensional_2007, rohrle_physiologically_2012, rohrle_simulating_2010, rohrle_skeletal_2018, heidlauf_modeling_2013, meier_fem-simulation_2000, teran_finite_2003, teran_creating_2005, lemos_framework_2004, johansson_finite-element_2000, wheatley_validated_2017, wheatley_modeling_2018}. Other approaches apply rubber-like Ogden-type material models \cite{schmid_characterization_2019}, exponential Humphrey-type constitutive laws \cite{martins_comparative_2006, zhang_biomechanical_2016, marcucci_single_2017, pavan_effects_2019, pavan_investigation_2017, tang_3d_2009}, quadratic polynomial functions \cite{yamamura_effect_2014, kinugasa_influence_2016, chi_finite_2010}, or simpler Neo-Hooke \cite{oomens_finite_2003, hernandez-gascon_3d_2013, calvo_passive_2010, kuravi_3d_2021} and Saint-Venant-Kirchhoff relations \cite{lemos_realistic_2001}. In \cite{bol_micromechanical_2008, bol_finite_2006}, the extracellular matrix material is modeled by a rubber-like nonlinear stress-strain relation based on measurable physical muscle parameters. The work of Blemker et al. \cite{blemker_three-dimensional_2005, blemker_3d_2005, rehorn_effects_2010, sharafi_micromechanical_2010} proposes a transversely isotropic model, accounting explicitly for the extracellular matrix resistance to along-fiber shear and cross-fiber shear by two strain-energy components.
 Building on prior work \cite{kuravi_3d_2021}, a sophisticated model for the extracellular matrix featuring two preferred fiber directions for the included collagen fibers is presented in \cite{kuravi_predicting_2021}. 

The passive muscle fiber stress contribution $\Psi_{\mathrm{fiber}}^{\mathrm{p}}$ usually depends non-linearly on the current fiber stretch. Common choices include exponential functions, e.g., in \cite{johansson_finite-element_2000, oomens_finite_2003, martins_finite_2006, kuravi_3d_2021, calvo_passive_2010, grasa_3d_2011} or polynomial functions, e.g., in \cite{heidlauf_multi-scale_2016, rohrle_two-muscle_2017}. Another popular option is a piecewise-defined, experimentally-based function \cite{zajac_muscle_1989} as seen in \cite{blemker_3d_2005, rohrle_three-dimensional_2007}.
The authors in \cite{wheatley_validated_2017} assume fibers are oriented in an ellipsoidal distribution, which allows for a direction-dependent modulation of fiber stiffness.

Ehret et al. \cite{ehret_continuum_2011}, and others in succession \cite{weickenmeier_physically_2014,  giantesio_strain-dependent_2017, seydewitz_three-dimensional_2019}, circumvent an additive split into the matrix and fiber contributions by introducing a coupled exponential-type model. A similar concept is applied in \cite{walter_threedimensional_2022}.

In contrast to, what is called here, purely phenomenological models, multi-scale models exploit the hierarchical structure of skeletal muscle and incorporate micromechanical features. A common approach is to create representative volume elements for, e.g., the fiber muscle cells and the extracellular matrix. Through homogenization techniques, the microstructural information is projected to the macro scale and incorporated into a constitutive law on the continuum level. Such approaches are found, e.g., in \cite{spyrou_homogenization_2017, spyrou_multiscale_2019, bleiler_microstructurally-based_2019, konno_modelling_2021, he_multiscale_2022}.

A special concept is presented in \cite{yucesoy_specifically_2012}, where skeletal muscle is modeled as an elastically linked system of two independently meshed domains for the fiber and matrix constituents.

\paragraph{Active constitutive models: Active stress, active strain, and generalized active strain approaches}
To include the fibers' active contractile properties, two concepts -- the active stress and the active strain approach -- are commonly applied. For a detailed explanation, see, e.g., \cite{rohrle_skeletal_2018, rohrle_multiscale_2019, klotz_physiology-guided_2021}. Next to that, there exist generalized active strain approaches (also termed continuous approaches in literature) that follow neither the active stress nor the active strain approach. 

\begin{figure}[htb]
    \centering
    \begin{subfigure}[t]{0.32\textwidth}
    \centering
    \includegraphics{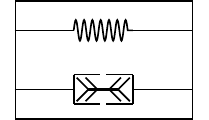}
    \subcaption{Active stress approach: Passive, elastic, and active element in parallel.}
    \label{fig:rheo_stress}
    \end{subfigure} \hfill
    \begin{subfigure}[t]{0.32\textwidth}
    \centering
    \includegraphics{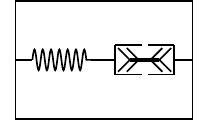}
    \subcaption{Active strain approach: Passive, elastic, and active element in series.}
    \label{fig:rheo_strain}
    \end{subfigure} \hfill
    \begin{subfigure}[t]{0.32\textwidth}
    \centering
    \includegraphics{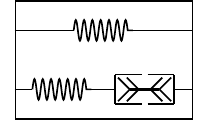}
    \subcaption{Generalized active strain approach: Active element in series and parallel to passive, elastic elements.}
    \label{fig:rheo_mixed}
    \end{subfigure}
    \caption{Rheological models representing different concepts of muscular activation as in \cite{klotz_physiology-guided_2021}.}
    \label{fig:rheology}
\end{figure}

The active stress approach adds an active stress term to the passive stress component such that the stress tensor (here given as the second Piola-Kirchhoff stress tensor) reads $\SecPK = \SecPK_{\mathrm{p}} + \SecPK_{\mathrm{a}}$. Often, the active fiber stress depends on an activation parameter that scales the maximal isometric active muscle force. In a rheological model, the active-stress approach is represented by a parallel arrangement of a passive, elastic spring and an active element, see Fig. \ref{fig:rheo_stress}. Examples of such hyperelastic, viscoelastic and poro-visco-hyperelastic material models are \cite{meier_fem-simulation_2000, oomens_finite_2003, teran_finite_2003, blemker_3d_2005, heidlauf_modeling_2013, heidlauf_multi-scale_2016, bol_micromechanical_2008, heidlauf_multiscale_2014, grasa_3d_2011, ramirez_active_2010, lamsfuss_skeletal_2021}, \cite{ahamed_time-dependent_2016, khodaei_simulation_2013, hernandez-gascon_3d_2013, spyrou_muscle_2011, sharifimajd_continuum_2013} and \cite{wheatley_case_2017, wheatley_modeling_2018}, respectively.
The main advantage of this concept is due to experimental practice and a straightforward interpretation of the active stress contribution \cite{ehret_continuum_2011, weickenmeier_physically_2014, giantesio_loss_2018}. In classical experiments on muscle tissue, both the muscle's force response in the passive resting state and the activated contractile state is tested. The characteristics of the resting state can then be attributed to the passive stress component, while the difference between the passive and the total activated stress-strain curve governs the \mbox{active stress term \cite{giantesio_loss_2018}}. Generally, the active stress tensor is considered a non-conservative contribution as it is not derived from the potential energy \cite{ambrosi_active_2012}. The active stress approach may thus violate the principle of energy conservation, possibly leading to numerical instabilities or non-physical predictions.

Opposed to that, the concept of active strain relies on a multiplicative decomposition of the deformation gradient into $\defgrad=\defgrada \defgrade$. While the active contribution $\defgrada$ maps the reference configuration onto a stress-free intermediate configuration, the elastic contribution $\defgrade$ maps from the intermediate configuration onto the current configuration. Since elastic energy is stored solely through $\defgrade$, the strain-energy function is expressed in terms of $\defgrade$ rather than $\defgrad$. Active contractile characteristics are commonly incorporated through an activation parameter in the formulation of  $\defgrada$. A representative rheological model consists of a passive, elastic spring in series with an active element, as shown in Fig. \ref{fig:rheo_strain}.
Hyperelastic and viscoelastic constitutive laws following the active strain approach can be found in \cite{giantesio_loss_2018, giantesio_strain-dependent_2017, hernandez-gascon_3d_2013} and \cite{pandolfi_coupled_2016, pandolfi_visco-electro-elastic_2017}, respectively. Due to the mathematical construction of the active strain approach, the strain-energy function inherits its mathematical properties from the underlying passive strain-energy function \cite{ambrosi_active_2012}. This includes properties such as frame invariance and rank-one ellipticity, which ensure that there is a guaranteed solution to the associated equilibrium equations \cite{ambrosi_active_2012}. These considerations do not apply for the active stress approach. In contrast to the active stress approach, the active contribution $\defgrada$ is not an experimentally observable quantity but rather more complex in its interpretation.

A generalized active strain concept was originally presented by Ehret et al. \cite{ehret_continuum_2011} and, in succession, adapted by Weickenmeier et al. \cite{weickenmeier_physically_2014} and Seydewitz et al. \cite{seydewitz_three-dimensional_2019}. Active properties are included by increasing the invariant accounting for the passive longitudinal fiber characteristics $I_\mathrm{p}$ by an active contribution $I_\mathrm{a}$, such that the combined invariant is $I = I_\mathrm{p} + I_\mathrm{a}$. According to \cite{klotz_physiology-guided_2021}, this is equal to applying the multiplicative decomposition of the deformation gradient to a single part of the given strain-energy function. A rheological representation is depicted in Fig. \ref{fig:rheo_mixed}. Advantage lies in the more physiological representation of the muscle tissue. On the cellular level, a sarcomere includes both an active component (actin-myosin complex) and a passive component (titin filaments) arranged in series. Modeling muscle as a parallel arrangement of the serially arranged sarcomere components and an elastic component representing the passive connective tissue provides a more accurate representation of tissue characteristics than a pure active stress or active strain approach. 

The approaches in \cite{martins_comparative_2006, martins_finite_2006, zhang_biomechanical_2016} and similarly in \cite{marcucci_single_2017, pavan_effects_2019, pavan_investigation_2017, tang_3d_2009} are expansions of the classic so-called Hill-type model to three dimensions. In this case, the total muscle force is -- equivalently to the generalized active strain approach -- estimated by adding the forces from a passive spring and the serial arrangement of a passive spring and a contractile active element. 

\paragraph{Activation characteristics: Influences on muscular activation and force generation}
A muscle's potential for force production is governed by various factors, such as its geometry, histological composition, neural activity, its current state of motion and deformation, and its contraction history. While geometric factors, such as size and fiber architecture, are considered in the geometric representation of the finite element model, histology-, activity-, and motion-related factors are commonly included in the material description. In any of the concepts presented above, the active contribution, be it $\SecPK_\mathrm{a}$, $\defgrada$ or $I_\mathrm{a}$, involves the computation of an activation quantity accounting for a varying number of those effects.

Experimentally observable force-stretch-, force-velocity- and force-stimulation-frequency-dependencies are commonly included in a phenomenological fashion.

Thereof, the force-stretch-dependency is considered in most publications. Popular choices for its mathematical description include (piecewise-defined) exponential \cite{rohrle_two-muscle_2017, yamamura_effect_2014}, linear \cite{meier_fem-simulation_2000, heidlauf_multi-scale_2016}, or parabolic \cite{heidlauf_modeling_2013, rohrle_three-dimensional_2007, blemker_3d_2005, lemos_realistic_2001, chi_finite_2010} formulations. Besides that, sigmoid functions \cite{grasa_3d_2011, ramirez_active_2010, hernandez-gascon_3d_2013}, and a normalized Weibull distribution \cite{ehret_continuum_2011, pavan_effects_2019} were proposed in the literature.

Less common is the additional inclusion of a force-velocity-dependency. Often, a hyperbolic relation based on the work in \cite{van_leeuwen_muscle_1992} and \cite{hill_heat_1938} is chosen \cite{bol_new_2007, bol_micromechanical_2008, ehret_continuum_2011, ramirez_active_2010, meier_fem-simulation_2000, johansson_finite-element_2000}. Other authors present exponential and arcus-tangent functions, see \cite{lemos_framework_2004, yamamura_effect_2014, tang_3d_2009, pavan_effects_2019} and \cite{martins_finite_2006}.

The simplest approach to account for the neural activity (or, in other words, the stimulation frequency) is to linearly scale the active contribution with an activation factor, see, e.g., \cite{blemker_3d_2005, teran_finite_2003, kinugasa_influence_2016, chi_finite_2010, lemos_realistic_2001}.
To simulate temporal variations of muscular activation, e.g., the successive build-up of a fused tetanic contraction state, some authors include a time-dependent activation function \cite{marcucci_single_2017, pavan_effects_2019, pavan_investigation_2017, johansson_finite-element_2000, yamamura_effect_2014, martins_finite_2006, tang_3d_2009, zhang_biomechanical_2016}.
More sophisticated formulations such as \cite{meier_fem-simulation_2000, bol_new_2007, bol_micromechanical_2008, ramirez_active_2010, hernandez-gascon_3d_2013} resolve the time-dependent activation level on the scale of milliseconds by a superposition of single muscle twitches. An additional composition into different fiber types, as proposed by Ehret et al. \cite{ehret_continuum_2011}, featuring different twitch force amplitudes and frequencies, is accounted for by a weighted sum of the contributions. 

Despite experimental evidence (see \cite{seydewitz_three-dimensional_2019} for a summary), history-dependent effects such as force depression and force enhancement are less commonly included in the constitutive description. To include these effects, the authors in \cite{seydewitz_three-dimensional_2019} propose an extension of the constitutive law in \cite{ehret_continuum_2011} by a so-called dynamic function. This function accounts for a dynamic force-velocity-dependency and the effects of force depression and force enhancement by evaluating a differential equation.
Opposed to this phenomenological approach, the work in \cite{heidlauf_multi-scale_2016} accounts for force enhancement effects on the micro-scale. At the sarcomere level, force enhancement is primarily governed by actin-titin interactions. To incorporate these interactions, they combine their multi-scale chemo-electro-mechanical model with a 'sticky-spring' mechanism for actin-titin interactions \cite{rode_titin-induced_2009}.

Other multi-scale approaches link the macroscopic constitutive model to detailed mathematical descriptions of electrical, biophysical, and chemical processes at the microscopic level.

Monodomain or bidomain equations are frequently used to model the action potential propagation along a muscle fiber, as seen in, e.g., \cite{fernandez_modelling_2005, heidlauf_modeling_2013, heidlauf_multiscale_2014, schmid_characterization_2019, maier_mesh_2022} and \cite{rohrle_bridging_2008, rohrle_multiscale_2013}, respectively.
An approach integrating the mechanism of electromechanical delay, i.e., the time difference between the muscle's stimulation and a measurable produced force, is further proposed in \cite{schmid_characterization_2019}. 
The authors of \cite{bol_coupled_2011} incorporate an electric field that triggers mechanical activation once the electric potential exceeds a certain threshold.
A phenomenological model of motor-unit recruitment driven by neural activity is coupled to the continuum level in \cite{rohrle_physiologically_2012, rohrle_multiscale_2013}.

Chemical processes such as calcium-concentration-driven muscle activation and calcium activation dynamics, i.e., the release of calcium from the sarcoplasmic reticulum, are added, e.g., in \cite{fernandez_modelling_2005, oomens_finite_2003, heidlauf_modeling_2013, heidlauf_multiscale_2014} and \cite{oomens_finite_2003, heidlauf_modeling_2013, heidlauf_multiscale_2014}.
To describe the de- and attachment of cross-bridges during muscle contraction on a molecular level, partial differential equations based on the Huxley sliding filament theory are applied in \cite{oomens_finite_2003, stojanovic_multi-scale_2020}.
One of the most detailed descriptions of the electrophysiological behavior of a half-sarcomere on the cellular level is presented in \cite{shorten_mathematical_2007}, and is coupled to continuum mechanical constitutive laws in \cite{heidlauf_modeling_2013, heidlauf_multiscale_2014, rohrle_physiologically_2012, rohrle_multiscale_2013}. It models the entire pathway from electrical excitation to muscle cell contraction through differential equations, thereby including electrochemical models of the membrane electrophysiology, calcium (activation) dynamics, cross-bridge dynamics, and fatigue.

Besides the basic modeling of physiologically realistic behavior of healthy skeletal muscle, the study of specific (pathological) biological processes is an ongoing research topic. Examples include models for damage \cite{lamsfuss_computational_2022}, fatigue \cite{heidlauf_modeling_2013, heidlauf_multiscale_2014} and age-related loss of activation \cite{giantesio_continuum_2017}.

\section{Material models for active skeletal muscle} \label{sec:materials}
The choice of an appropriate material model is essential to obtain physiologically realistic kinematics and stress results. Based on the literature review in the previous section, we select three material models for a detailed investigation. 
Following a very brief introduction to the basic continuum mechanical quantities in Section \ref{sec:continuum_basis}, Section \ref{sec:ASA_GASA_ASE_material} summarizes the three chosen models and highlights the modifications made in our work. In Section \ref{sec:compare_materials}, we compare the approaches and critically discuss the respective advantages and disadvantages. Drawing from our findings, we propose a modified material model combining the essential properties for an application to human shoulder modeling in Section \ref{sec:modified_material}. 

\subsection{Continuum mechanical basics} \label{sec:continuum_basis}
In nonlinear continuum mechanics, the deformation gradient $\defgrad=\frac{\mathrm{d}\mathbf{x}}{\mathrm{d}\mathbf{X}}$, with the Jacobi determinant \mbox{$J=\det\defgrad$}, serves as the primary measure of deformation. $\mathbf{x}$ and $\mathbf{X}$ denote the coordinates of a material point in current and reference configuration, respectively.
The right Cauchy-Green tensor $\C$ is an important quantity to calculate the strains with regard to the reference configuration and is defined as 
\begin{equation}
\C = \defgrad^\top \defgrad.
\label{eq:C}
\end{equation}
Following a multiplicative decomposition of the deformation gradient into isochoric and volumetric parts, the modified right Cauchy-Green tensor $\modC = J^{-2/3} \C$, which describes the isochoric contribution, is introduced. All modified, i.e., isochoric, quantities are indicated by $(\bar{\bullet})$ in this work. 

Hyperelastic material laws postulate the existence of a strain-energy function $\Psi(\C)$.
To account for the fiber direction in a transversely isotropic material model, a structural tensor $\M$ can be incorporated into the strain-energy function, such that $\Psi(\C, \M)$. Assuming the fiber direction in reference configuration as the unit vector $\mathbf{m}$, the structural tensor is computed to $\M = \mathbf{m} \otimes \mathbf{m}$. The stretch in fiber direction is
\begin{equation}
\lambda = \sqrt{\C : \M}.
\label{eq:fiber_stretch}
\end{equation}

The second Piola-Kirchhoff stress tensor $\SecPK$ is derived from the strain-energy function as
\begin{equation}
\SecPK = 2 \frac{\partial \Psi}{\partial \C}
\end{equation}
while the first Piola-Kirchhoff stress tensor $\FirstPK$ results from the push-forward operation
\begin{equation}
\FirstPK = \defgrad \SecPK.
\end{equation}

Solving a continuum mechanical problem with the finite element method, usually requires the linearization of the constitutive equation. Therefore, the forth-order elasticity tensor $\CC$ is computed to 
\begin{equation}
\CC = 4 \frac{\partial^2 \Psi}{\partial \C^2}.
\end{equation}

\subsection{Selected material models} \label{sec:ASA_GASA_ASE_material}
\begin{figure}[htb]
\centering
\includegraphics[width=\textwidth]{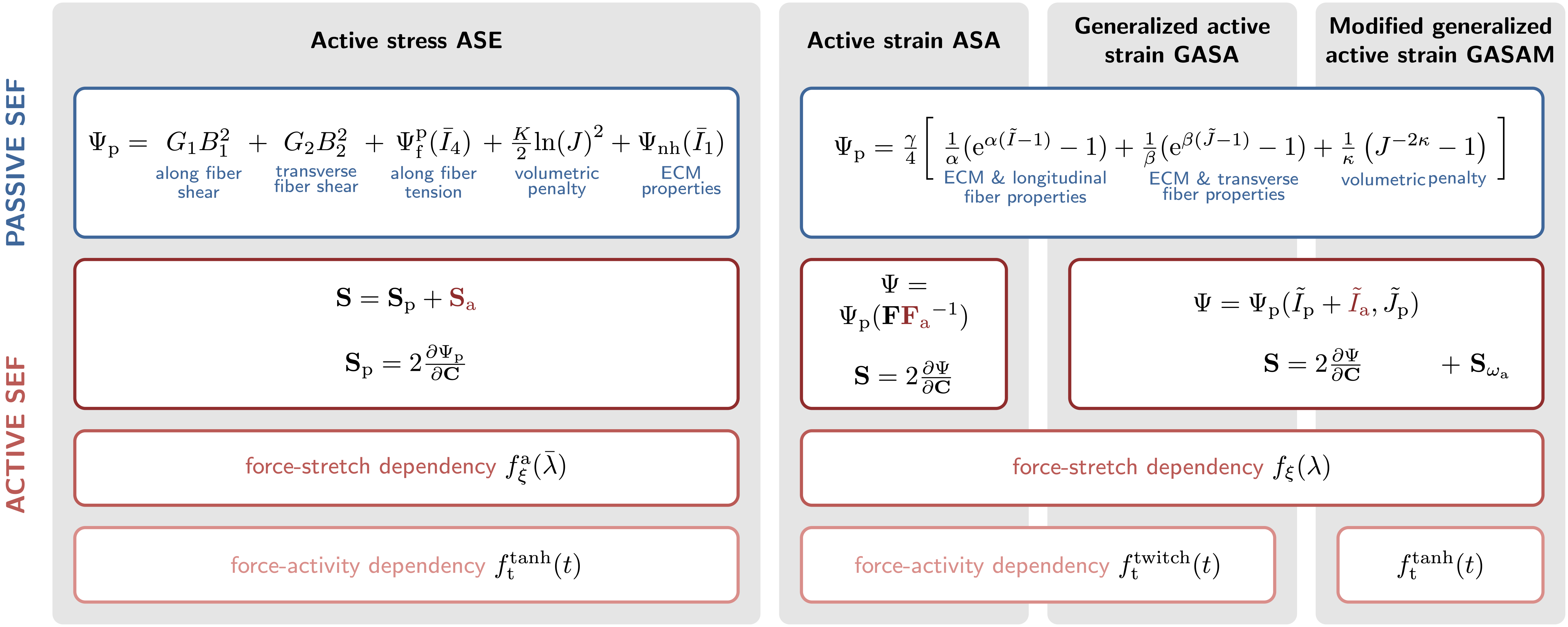}
\caption{Schematic overview of the investigated constitutive laws for active skeletal muscle tissue.} 
\label{fig:overview_materials}
\end{figure}

We evaluate three hyperelastic and nearly incompressible material models from the literature based on either the active strain, active stress, or the generalized active strain approach. In accordance with the anatomical predominant unidirectional fiber alignment on the local scale, all of the selected material models assume a transversely isotropic fiber distribution with respect to this preferred fiber direction.
Blemker et al.'s active stress model \cite{blemker_3d_2005}, here named \BLE, is chosen due to its successful application to several single muscles \cite{blemker_three-dimensional_2005, knaus_3d_2022, fiorentino_musculotendon_2014, rehorn_effects_2010} and muscle tissue parts \cite{sharafi_micromechanical_2010} but also comprehensive models of the human shoulder \cite{pean_comprehensive_2019, webb_3d_2014} and the human knee \cite{kiapour_finite_2014}. Weickenmeier et al.'s model \cite{weickenmeier_physically_2014}, termed \WKM, is selected because it incorporates muscle activation through a novel generalized active strain approach and allows for seamless integration of micromechanical data. Giantesio et al.'s model \cite{giantesio_strain-dependent_2017}, abbreviated \GIANT, is a variant of the aforementioned \WKM-model but uses an active strain approach to include activation in a mathematically well-posed manner. Fig. \ref{fig:overview_materials} provides a schematic overview of the constitutive laws. Table  \ref{tab:material_parameters_naming_summary} summarizes material parameters and abbreviations used in the following.

\begin{table}[htb]
\caption{Overview of the material parameters for the \BLE-, \GIANT-, and \WKM-models in Blemker et al. \cite{blemker_3d_2005} (used in this work in a variant, i.e., in combination with the Neo-Hooke material model \cite[p.238]{holzapfel_nonlinear_2000}), Giantesio et al. \cite{giantesio_strain-dependent_2017} and Weickenmeier et al. \cite{weickenmeier_physically_2014}, respectively.}
\begin{tabularx}{\textwidth}{lp{6.5cm}lX}
\toprule
\multicolumn{2}{l}{\BLE} & \multicolumn{2}{l}{\GIANT and \WKM} \\ \midrule
$\Gone$                 & along fiber shear modulus & $\alpha$ & parameter related to along fiber properties \\
$\Gtwo$                 & transverse fiber shear modulus & $\beta$ & parameter related to transverse fiber properties  \\
$\Pone$                 & magnitude of passive along fiber tension & $\gamma$ & stiffness parameter \\
$\Ptwo$                 & exponential growth rate of passive along fiber tension  & $\omegazero$ & weighting factor for isotropic tissue constituent  \\
$\kappaBle$                & bulk modulus  & $\kappa$ & incompressibility parameter \\
$\sigmamax$             & maximum isometric stress  & $\Na$ & number of activated muscle units per reference cross-section area \\
$\lambdaOptBle$            & optimal fiber stretch  & $\lambdaOpt$ &  optimal fiber stretch  \\
$\lambdaStar$           & minimum linear fiber stretch & $\lambdaMin$ & minimum fiber stretch  \\
$\alphaa$   & amplitude of time-dependent activation & $F_i$ & twitch force of motor unit type $i$ \\
$c$                     & frequency of time-dependent activation & $T_i$ & twitch contraction time of motor unit type $i$ \\
                        &   & $I_i$ & interstimulus interval of motor unit \mbox {type $i$} \\
                        &   & $\rho_i$ & fraction of motor unit type $i$  \\
\multicolumn{2}{l}{Neo-Hooke} &  &  \\ \midrule
$\mu$                 & shear modulus  &  &  \\ \bottomrule
\end{tabularx}
\label{tab:material_parameters_naming_summary}
\end{table}

\subsubsection{Active stress approach (\BLE)} \label{sec:ASE}
Blemker et al. \cite{blemker_3d_2005} present a purely phenomenological material model with a fiber-stretch-dependent activation, named \BLE in this work. Following the concept of active stress, activation is modeled by adding an active stress contribution to the passive stress. Near-incompressibility is achieved through a decoupled strain-energy function involving a purely isochoric part $\PsiIso$ and a purely volumetric part $\PsiVol$. 

The isochoric part is formulated with respect to the modified invariants
\begin{equation}
	\modIone = \trace \modC, \quad \quad
	\modIfour = \modC : \M = \modlambda^2, \quad \quad 
	\modIfive = \modC^2 : \M,
	\label{eq:modIs}
\end{equation}
and the strain invariants
\begin{equation}
	\Bone =\sqrt{\frac{\modIfive}{\modIfour^2}-1} \quad \text{and} \quad \Btwo = \acosh \left(\frac{\modIone \modIfour - \modIfive}{2 \sqrt{\modIfour}} \right).
	\label{eq:strainInvariants}
\end{equation}
Considering the bulk modulus $\kappaBle$,  the along fiber shear modulus $\Gone$, and the transverse fiber shear modulus $\Gtwo$, the proposed strain-energy function reads
\begin{equation}
	\Psi = \PsiIso + \PsiVol = {\underbrace{\Gone \Bone^2}_{\Psi_\mathrm{lfs}}} + {\underbrace{\Gtwo \Btwo^2}_{\Psi_\mathrm{tfs}}} + {\PsiTotFib} + {\underbrace{\tfrac{\kappaBle}{2} {\ln(J)}^2}_{\PsiVol}}.
	\label{eq:Psi_Ble}
\end{equation}
It involves the contributions $\Psi_\mathrm{lfs}$ and $\Psi_\mathrm{tfs}$ accounting distinctively for shear along and transverse to the fiber direction. The term $\PsiTotFib$ can be attributed to active and passive tension and compression along the fiber direction ($\PsiAFib$ and $\PsiPFib$, respectively). 
With the total Cauchy fiber stress $\sigmafibtot$, $\PsiTotFib$ is implicitly given by the equation
\begin{equation}
	\frac{\partial \PsiTotFib}{\partial \modlambda} = \frac{\sigmafibtot}{\modlambda}.
\end{equation}

Accounting for the active stress, $\sigmafibtot$ comprises an active part $\sigmafibact$ and a passive part $\sigmafibact$. Considering the maximal isometric fiber stress $\sigmamax$, we compute $\sigmafibtot$ to
\begin{equation}
     \sigmafibtot = \underbrace{\sigmamax \tfrac{\modlambda}{\lambdaOptBle} \alphaa \falphat \factive}_{\sigmafibact} + \underbrace{\sigmamax \modlambda \fpassive}_{\sigmafibpas}.
\label{eq:sigmafibtot_blemker}
\end{equation}
The amplitude $\alphaa$ scales the active contribution. In contrast to the original formulation in \cite{blemker_3d_2005}, we introduce an additional time-dependent function $\falphat$. By this integration, we can establish a time-dependent activation profile comparable to one in the \WKM-model introduced in the upcoming section \ref{sec:GASA}. To mimic the successive build-up of twitch forces up to a fused tetanized level, we choose the $\tanh$-function
\begin{equation}
\falphat = \falphat(t) = \tanh \left(c \left(t - t_0\right)\right),
\label{eq:ft_blemker}
\end{equation} with the frequency $c$ and activation start time $t_0$. Setting $\falphat = 1$ results in the original material model in \cite{blemker_3d_2005}. 
The functions $\factive$ and $\fpassive$ in Eq. \eqref{eq:sigmafibtot_blemker} account for the experimentally observed active and passive force-stretch-dependencies, respectively. Assuming the maximal isometric fiber stress $\sigmamax$ occurs at the optimal fiber stretch $\lambdaOptBle$, the active stretch-dependency is given as in the original publication as
\begin{equation}
    \factive = \factive(\modlambda) = \left\{
  \begin{array}{ll}
  9 \left(\tfrac{\modlambda}{\lambdaOptBle} - 0.4\right)^2 &\text{if } \modlambda \leq 0.6 \lambdaOptBle \\
  1 - 4 \left(1 - \tfrac{\modlambda}{\lambdaOptBle} \right)^2&\text{if } 0.6 \lambdaOptBle < \modlambda < 1.4 \lambdaOptBle \\ 
  9  \left(\tfrac{\modlambda}{\lambdaOptBle} - 1.6\right)^2 &\text{if } \modlambda \geq 1.4 \lambdaOptBle.
  \end{array} \right.
  \label{eq:fxia_blemker}
\end{equation}
Different from the original formulation, we assume that passive fibers solely produce a stress response in all tensile states, i.e., when $\modlambda > 1$ and not as originally when $\modlambda > \lambdaOptBle$.
Considering the minimum linear fiber stretch $\lambdaStar$ and the parameters $\Pone$ and $\Ptwo$, the passive stretch-dependency reads
\begin{equation}
  \fpassive = \fpassive(\modlambda) = \left\{
  \begin{array}{ll}
    \Pthree \modlambda + \Pfour  & \text{if } \modlambda \geq \lambdaStar  \\
    \Pone \left(\eexp{\Ptwo  \left(\modlambda - 1\right)} - 1\right)  & \text{if } \lambdaStar > \modlambda > 1  \\
    0 & \text{if } 1 \geq \modlambda 
  \end{array} \right.
\label{eq:fxip_blemker}
\end{equation}
with 
\begin{equation}
    \Pthree = \Pone \Ptwo \eexp{\Ptwo \left({\lambdaStar} - 1\right)} \quad \text{and} \quad
    \Pfour = \Pone \left(\eexp{\Ptwo  \left(\lambdaStar - 1\right)} - 1\right) -  \lambdaStar\Pthree.
\end{equation}
Since $\lambdaOptBle$ is now only involved in the computation of $\factive$, the active and passive behavior is decoupled and the material parameters can be fitted to the two scenarios independently. 

We provide the derivation of the second Piola-Kirchhoff stress tensor in the Appendix for the reader's convenience, as these equations have not been published so far. The presented equations reflect the additive composition of a passive and an active stress component, as it is characteristic of the active stress concept. 
Details about the elasticity tensor derivation are given in the supplementary materials.

\paragraph{Remark 1}
For passive compression along the fiber direction (i.e., $\modlambda<1$), $\fpassive$ and in succession $\sigmafibtot$ become zero. If shear contributions vanish as well, the entire stress response is zero (see also the \mbox{remark in \cite{walter_threedimensional_2022}}). 
From a modeling perspective, this can be attributed to the fact that the material neglects the compressive stiffness of the fiber surrounding tissue. Instead, it solely incorporates components directly associated with the muscle fibers. To account for the influence of the fiber surrounding tissue and circumvent numerical difficulties arising from the lack of stiffness in plain compressive states, this work pairs the material model with the isometric Neo-Hookean material model in \cite[p.238]{holzapfel_nonlinear_2000} with the strain-energy function $\PsiNeoHooke(\modIone)$.

\paragraph{Remark 2}
The stress computation exhibits singularities in case the \mbox{argument $\theta$} of $\acosh$ in the invariant $\Btwo$ in Eq. \eqref{eq:strainInvariants} becomes $\theta = 1$. To calculate the stress, the derivative of $\Btwo$ with respect to $\modIfour$ is formed and accounted for in the auxiliary variable $A_2 = \modIfour^{-1/2} \frac{\acosh (\theta)}{\sqrt{\theta^2 - 1}}$ (see Eq. \eqref{eq:helper_As}). The zero in the denominator thus leads to a singularity for the case that $\theta = 1$. 
In an analytical setting, we can compute the limit, such that $A_2 = \lim \limits_{\theta \to 1} \modIfour^{-1/2} \frac{\acosh (\theta)}{\sqrt{\theta^2 - 1}} = \modIfour^{-1/2}$. In a numerical evaluation, the singularity can be circumvented by adding a very small contribution $\epsilon$ to $\theta$ such that $\theta = 1 + \epsilon$.
This behavior can be attributed to the chosen invariants, initially published by Criscione et al. \cite{c_criscione_physically_2001}. As previously noted by Bleiler et al. \cite{bleiler_microstructurally-based_2019}, the derivative of the invariant $\Btwo$ becomes singular in case of vanishing shears.

\subsubsection{Generalized active strain approach (\WKM)} \label{sec:GASA}
The generalized active strain approach by Weickenmeier et al. \cite{weickenmeier_physically_2014}, here named \WKM, is based on the fully incompressible model for passive and active muscle presented by Ehret et al. in \cite{bol_micromechanical_2008, ehret_continuum_2011}. On this basis, Weickenmeier et al. \cite{weickenmeier_physically_2014} propose two compressible constitutive descriptions, allowing to model the muscle tissue as nearly incompressible. The so-called coupled approach circumvents the commonly applied additive volumetric-isochoric split of the strain-energy function. Since it has been proven to be advantageous in maintaining incompressible behavior, we employ this coupled approach in our forthcoming studies. In contrast to the active stress and active strain concept, activation is achieved through the modification of an invariant.

The proposed strain-energy function incorporates the material parameters $\alpha$, $\beta$ and $\gamma$, and the incompressibility parameter $\kappa$. The weighting parameters $\omegazero$ and $\omegap$, related by $\omegazero+\omegap=1$, describe the percentage contribution of the extracellular matrix and the muscle fibers, respectively. 
While the structural tensor $\M$ accounts for the muscle fiber alignment, the isotropic matrix contribution is included in the structural tensor $\Ltilde=\frac{\omegazero}{3}\mathbf{I} + \omegap \M$.
Considering the activation parameter $\omegaa$, the two general invariants $\Itilde$ (with its passive and active parts $\Iptilde$ and $\Itilde_\mathrm{a}$, respectively) and $\Jtilde$ are introduced as
\begin{equation}
  \Itilde= \Iptilde + \Itilde_\mathrm{a} \quad \text{with} \quad  \Iptilde = \C : \Ltilde \quad \text{and} \quad \Itilde_\mathrm{a} = \C : (\omegaa \M),
  \quad \quad \text{and} \quad  \quad
  \Jtilde = \cof(\C) : \Ltilde.
  \label{eq:I,J}
\end{equation}
Based on those quantities, the strain-energy function is defined as 
\begin{equation}
  \Psi = \tfrac{\gamma}{4}
  \left[
  \tfrac{1}{\alpha} \left(\eexp{\alpha(\Itilde-1)}-1\right) +
  \tfrac{1}{\beta} \left(\eexp{\beta(\Jtilde-1)}-1\right) +
  \tfrac{1}{\kappa} \left(\det(\C)^{-\kappa} - 1 \right) 
  \right].
  \label{eq:Psi_WKM}
\end{equation}

For the computation of the activation parameter $\omegaa$, two assumptions are made: first, the model nominal stress response to a uniaxial deformation along the fiber direction matches the experimentally measured total nominal stress, and second, this measured total nominal stress can be additively decomposed into passive and active contributions. Based on these considerations, $\omegaa$ can be explicitly expressed in terms of the active nominal stress $\Pact$. 
Assuming $W_0(\chi^*)$ is the principal branch of the Lambert $W$ function, given as the solution of the inverse function $\chi = W(\chi^*) \eexp{W(\chi^*)}$, the activation parameter is obtained as
\begin{equation}
  \omegaa =
  \left\{ \begin{array}{ll}
    0 & \text{if } \Pact=0 \\
    \frac{W_0(\chi^*)}{\alpha \lambda^2} - \frac{1}{2 \lambda} \IptildePrime & \text{else} 
  \end{array} \right.
\quad \text{with} \quad
  \chi^* = \Pact \tfrac{2\alpha\lambda}{\gamma} 
  \eexp{\tfrac{\alpha}{2} (2-2\Iptilde + \lambda \IptildePrime)}
  + \tfrac{\alpha}{2} \lambda \IptildePrime 
  \eexp{\tfrac{\alpha}{2} \lambda \IptildePrime} .
  \label{eq:wa}   
\end{equation}
$\Iptilde$ (and its derivative with respect to $\lambda$, $\Iptilde'$), denotes the passive part of the first generalized invariant $\Itilde$ for uniaxial tension and are given in Eq. \eqref{eq:Iptilde} in the Appendix.
 
The active nominal stress $\Pact$ accounts for the force-stretch-dependency through $\fxi$ and for the force-velocity-dependency through $\fv$. It further incorporates the term $\Popt \ft$ in which $\Popt$ is the peak level of the active nominal stress, and $\ft$ is a dimensionless, normalized, time-dependent function, such that
\begin{equation}
  \Pact = \Popt \ft \fxi \fv.
  \label{eq:P_act}
\end{equation}
The total active force created by $n_{\mathrm{MU}}$ muscle motor units of type $i$ is calculated as the sum of the force responses $\Ft^i$ at time $t$ weighted by the corresponding fraction in the muscle $\rho_i$. $\Popt\ft$ then results from multiplication with the number of activated muscle units per unit reference cross-section area $\Na$ according to
\begin{equation}
  \Popt\ft  = \Na \sum_{i=1}^{n_{\mathrm{MU}}} \rho_i \Ft^i.
  \label{eq:Poptft}
\end{equation}
$\Ft^i$ results from superposition of single twitches characterized by the experimentally observed microstructural quantities $T_i$, ${F}_i$, and $I_i$. The twitch contraction time $T_i$ defines the time until the peak twitch force ${F}_i$ in the ascending phase of a single twitch response is reached \cite{gorelick_mechanomyographic_2007, fuglevand_models_1993}. $I_i$ denotes the interstimulus interval. For a detailed explanation of the computation of $\Ft^i$, we refer to the original publication \cite{ehret_continuum_2011}.

The stretch-dependency $\fxi$ is chosen as a function representing experimentally observed behavior.
It depends on $\lambdaMin$, the minimal fiber stretch at which myofilaments still overlap and $\lambdaOpt$, the fiber stretch associated to the maximal twitch force. Its mathematical description reads 
\begin{equation}
  \fxi = \fxi(\lambda) = \left\{
  \begin{array}{ll}
    \frac{\lambda -\lambdaMin}{\lambdaOpt-\lambdaMin} \exp \frac{(2 \lambdaMin- \lambda-\lambdaOpt) (\lambda-\lambdaOpt) }{2 (\lambdaMin-\lambdaOpt)^2}  & \text{if } \lambda > \lambdaMin \\
    0 & \text{if } \lambda \leqslant \lambdaMin. \\
  \end{array} \right.
  \label{eq:fxi}
\end{equation}
For comparative reasons, the velocity-dependency $\fv$ is neglected in this work and set to $\fv=1$.

\subsubsection{Active strain approach (\GIANT)}\label{sec:ASA}
Based on the same incompressible model \cite{ehret_continuum_2011} as the compressible \WKM-approach \cite{weickenmeier_physically_2014} introduced in the previous section, Giantesio et al. \cite{giantesio_strain-dependent_2017} propose an active strain approach, here termed \GIANT. A common approach to enforce the incompressibility condition, i.e., $J=1$, is to add an additional contribution to the strain energy function that penalizes deviations from $J=1$. To this end, we incorporate a volumetric penalty term similar to the coupled formulation in \cite{weickenmeier_physically_2014}.

The active strain approach relies on a multiplicative decomposition of the deformation gradient $\defgrad$ into an elastic part $\defgrade$, associated with the elastic deformation and an active part $\defgrada$, resulting from an internal active deformation, such that $\defgrad = \defgrade \defgrada$. 
Considering the activation parameter $\omegaa$, the active deformation gradient is defined as
\begin{equation}
\defgrada = (1-\omegaa) \M + \frac{1}{\sqrt{1-\omegaa}} (\I - \M) \quad \text{with} \quad \det(\defgrada) = 1.
\label{eq:Fa}
\end{equation}
The strain-energy function is expressed in terms of the elastic Cauchy-Green strain tensor $\Ce = \defgrade^T \defgrade$ instead of $\C$. With the elastic general invariants
\begin{gather}
	\Ietilde= \Ce : \Ltilde \quad \text{and} \quad \Jetilde = \cof(\Ce) : \Ltilde,
	\label{eq:Ie,Je}
\end{gather}
the strain energy function thus reads
\begin{equation}
	\Psi = \Psie + \PsiVol = \tfrac{\gamma}{4}
	\left[
	\tfrac{1}{\alpha} \left( \eexp{\alpha(\Ietilde-1)}-1 \right) +
	\tfrac{1}{\beta} \left( \eexp{\beta(\Jetilde-1)}-1\right)	\right] +
	\tfrac{\gamma}{4 \kappa} \left[\det(\C)^{-\kappa} - 1 \right].
	\label{eq:Psi_Giant}
\end{equation}

The computation of the activation parameter $\omegaa$ relies on the same two assumptions as mentioned for the \WKM- model. Consequently, $\omegaa$ is implicitly given as the solution of the equation
\begin{equation}
\tfrac{1}{\alpha} \eexp{\alpha\left(\Ietilde (\lambda, \omegaa)-1\right)}+\tfrac{1}{\beta} \eexp{\beta\left(\Jetilde (\lambda, \omegaa)-1\right)}=\tfrac{1}{\alpha} \eexp{\alpha\left(\Iptilde(\lambda)-1\right)}+\tfrac{1}{\beta} \eexp{\beta\left(\Jptilde(\lambda)-1\right)}+\tfrac{4}{\gamma} 
\Popt\ft \int_{\lambdaMin}^{\lambda} \fxi(\tilde{\lambda})  d\tilde{\lambda},
\label{eq:waimplicit}
\end{equation}
with the generalized elastic invariants for uniaxial tension, $\Ietilde$ and $\Jetilde$, and their passive counterparts, $\Iptilde$ and $\Jptilde$, given in Eq. \eqref{eq:Iptilde} in the Appendix.
The stretch- and time-dependencies included in the above equation are formulated in the same fashion as for the \WKM-model, and are given in Eq. \eqref{eq:fxi} and \eqref{eq:Poptft}, respectively. 
We apply a standard Newton-Raphson algorithm to determine $\omegaa$ from Eq. \eqref{eq:waimplicit}.

Again, as a service to the reader, we present the derivation of the second Piola-Kirchhoff stress tensor in the Appendix. Further, the derivation of the elasticity tensor is provided in the supplementary materials.

\paragraph{Remark} 
Since in the passive case $\defgrada = \mathbf{I}$, and thus $\defgrade = \defgrad$ and $\Ce = \C$, the \WKM- and \GIANT-model coincide in absence of any activation. In the active case, the nominal stress in the fiber direction due to uniaxial loading along the fibers $P_\mathrm{tot}$ is identical. We recall, that both models determine the activation parameter $\omegaa$ such that the equation $P_\mathrm{tot}(\omegaa) = P_\mathrm{act} + P_\mathrm{pas}$ is fulfilled. Since $P_\mathrm{pas}$ and $P_\mathrm{act}$ coincide, $P_\mathrm{tot}$ must also be equivalent.

\subsection{Comparison of the selected approaches} \label{sec:compare_materials}
In the following, we analyze the presented material models and compare them considering the models' ability to represent physiological reality, their mathematical properties, resulting numerical challenges, and aspects of computational efficiency.

\paragraph{Activation concept: Physiological representation and mathematical properties}
In Section \ref{sec:muscle_models}, we discussed the different activation concepts and assessed how well the models reflect physiological reality (consider the rheological representations in Fig. \ref{fig:rheology}). Comparing the three material models against this background, the generalized active strain model (\WKM) stands out as the physiologically most realistic. It comprehensively represents the tissue structure and its mechanical properties, incorporating both serial elastic properties of the sarcomeres (titin filaments) and the parallel elastic properties of the connective tissue. The active stress model (\BLE) accounts for the connective tissue’s elasticity but neglects the sarcomeres’ serial elasticity, while the active strain approach (\GIANT) captures the sarcomeres’ active and passive elastic characteristics but disregards the connective tissue’s parallel contribution.

We have further outlined the mathematical properties associated with the different activation concepts in Section \ref{sec:muscle_models}. As typical for the active strain approach, the \GIANT-model's active strain-energy function, preserves the elliptic properties of the underlying passive strain-energy function, thereby ensuring well-posedness of the associated balance equations (see \cite{giantesio_strain-dependent_2017} for a full discussion). For the \BLE-model, the active stress is not derived from a potential, as it becomes evident through the implicit definition of $\PsiTotFib$. Although we did not examine the model's elliptic properties in detail, we emphasize that the well-posedness of the equilibrium problem is not given by construction, but depends on the specific active stress tensor.

\paragraph{Passive material model}
Examining the passive material models, we find differences in the construction of the model equations and the parametric control of model properties.
The \BLE-model separates the contributions for different loading modes, with the parameters $\Gone$, $\Gtwo$ and $\sigmamax$ distinctively addressing along fiber shear, transverse fiber shear and along fiber tension. In our variant, the Neo-Hookean contribution accounts for the isotropic ECM stiffness through the parameter $\mu$. Conversely, the \WKM- and \GIANT-model exhibit a more convoluted structure, where the parameters $\alpha$, $\beta$, and $\gamma$ describe the combined properties of anisotropic fibers and isotropic ECM. 
For the \WKM- and \GIANT-model, the degree of anisotropy can be easily controlled through $\omegazero$. The \BLE-model links the anisotropic invariant $\modIfour$ with several parameters, making it more challenging to control the level of anisotropy. 

It could be argued that splitting the stress response into components associated with distinct loading modes (\BLE-model) simplifies fitting the model stress to experimental measurements. However, as further discussed in Section \ref{sec:fitting}, fitting the combined stress response (\WKM- and \GIANT-model) has also proven to be straightforward. Against this background, we do not prefer one material model over the other. 

\paragraph{Activation level: Implicit or explicit computation}
In the \GIANT-model, computing the activation parameter $\omegaa$ involves solving an implicit equation, introducing additional numerical challenges associated with the iterative solver, including the selection of step size and initial guess, as well as possible convergence problems. While we haven't conducted specific tests to precisely determine its impact on the computation effort, we expect and experienced this to perform worse than an explicit computation. As $\omegaa$ is implicitly defined, its derivatives are approximated using central differences. Selecting an appropriate step size for the central differences scheme thus presents a manageable, yet additional challenge. 

For the \WKM-model, the activation parameter $\omegaa$ is explicitly given. However, its computation involves the principal branch of the Lambert W function, $W_\mathrm{0}$, which is defined implicitly. Here, the same considerations as above apply. Notably, the authors of \cite{giantesio_strain-dependent_2017} find that incorporating the additional stress contribution $\dPsidomegaa\DomegaaDC$ leads to a fully explicit expression for $\omegaa$. 

In contrast to those two, the \BLE-model formula contain no additional implicit equations.

\begin{figure}[t]
    \centering
    \tikzstyle{every node}=[font=\small]
    \begin{minipage}[t]{0.5\textwidth}
    \centering
    	\includegraphics[height=4.81cm]{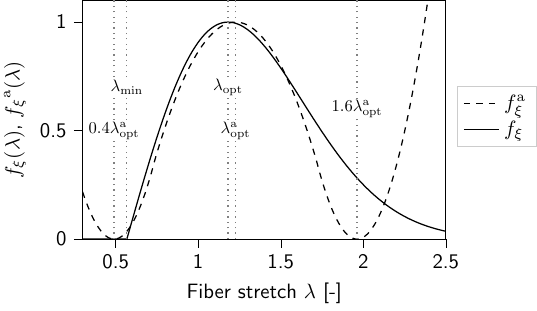}
    	\captionof{figure}{Active force-stretch-dependencies $\fxi$ in Eq. \eqref{eq:fxi} and $\factive$ in Eq. \eqref{eq:fxia_blemker}.}
    \label{fig:fxi_normalized}
    \end{minipage} \hfill
    \begin{minipage}[t]{0.47\textwidth}
    \centering
    	\includegraphics[height=4.81cm]{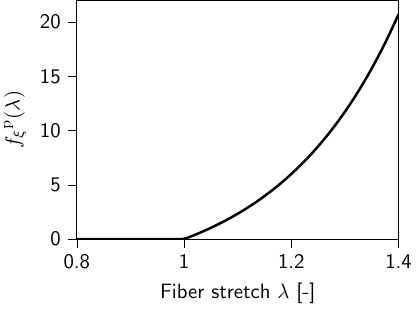}
    	\caption{Adapted passive force-stretch-dependency $\fpassive$ in Eq. \eqref{eq:fxip_blemker}. The original $\fpassive$ in \cite{blemker_three-dimensional_2005} is zero for $\lambda \leq \lambdaOptBle$.}
        \label{fig:fxip_blemker}
     \end{minipage} 
\end{figure}

\paragraph{Force-stretch-dependencies}
A closer look at the active force-stretch-dependencies $\fxi$ and $\factive$ in Fig. \ref{fig:fxi_normalized} reveals some numerically problematic and physiologically unrealistic features. Due to the non-smooth definition of the \GIANT- and \WKM-model's stretch-dependency $\fxi$ in Eq. \eqref{eq:fxi}, $\fxi$ and, in conclusion, also the stress response is not continuously differentiable at $\lambda=\lambdaMin$. This could lead to numerical difficulties. In contrast, the \BLE-model's stretch-dependency $\factive$ in Eq. \eqref{eq:fxia_blemker} is continuously differentiable in the entire stretch regime. For values $\lambda<0.4\lambdaOptBle$ and $\lambda>1.6\lambdaOptBle$, the stretch-dependency $\factive$, however, shows an unphysiological rise. The convergence to zero values for large fiber stretches and the absence of an active contribution for values below a certain minimal stretch is thus better captured by the \GIANT- and \WKM-model's $\fxi$. We further note that $\factive$ is symmetric with respect to $\lambdaOpt$, whereas $\fxi$ can represent non-symmetric force-stretch relations. 

For both $\fxi$ and $\factive$, the parameter $\lambdaOpt$ represents the fiber stretch related to the maximal isometric active stress. 
In $\fxi$, the parameter $\lambdaMin$ describes the minimal fiber stretch at which muscle activity is observed - an experimentally measurable and interpretable quantity. In contrast, with $\factive$, this value is preset to $0.4\lambdaOpt$, which limits the options for adjusting the minimal actively contracting fiber length.

Apart from the active stretch-dependency $\factive$, the \BLE-model considers the passive stretch-dependency $\fpassive$ (see Eq. \eqref{eq:fxip_blemker}) depicted in Fig. \ref{fig:fxip_blemker}. Similarly, this function is not continuously differentiable at $\lambda=1$. We further note that the parameter $\lambdaStar$ is a pure phenomenological quantity with no physiological meaning.

\begin{figure}[t]
    \centering
    \includegraphics[height=4.81cm]{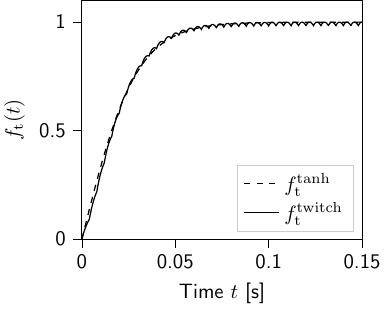}
    \captionof{figure}{Time-dependent activation functions $\ft$ in Eq. \eqref{eq:Poptft} and $\falphat$ in Eq. \eqref{eq:ft_blemker}.}
    \label{fig:ft_normalized}
\end{figure}

\paragraph{Time-dependent activation functions}
Fig. \ref{fig:ft_normalized} compares the time-dependent activation functions $\ft$ and $\falphat$. While the superposition of individual twitch forces in the computation of $\ft$ for the \GIANT- and \WKM-model is crucial for observing the time-dependent evolution of active forces at a millisecond scale, it can be disregarded for our application. Still, we acknowledge the use of physical, experimentally measurable, and well-interpretable microstructural parameters in $\ft$. Due to its lower computational expense, we opt for the $\tanh$ time-dependency $\falphat$ proposed for the \BLE-model for this application. 

\paragraph{Consistency of stress tensor derivation}
Comparing the stress terms of the \GIANT- and \WKM-model in Eq. \eqref{eq:S_Giant} and \eqref{eq:S_WKM}, we notice that the \GIANT-model's stress response considers the dependence of the activation parameter $\omegaa$ on the stretch $\lambda$ in the stress derivation through the term $\SecPK_2 =  2 \frac{\partial \Psie}{\partial \omegaa}$. As already pointed out in \cite{giantesio_strain-dependent_2017}, the \WKM-model neglects this dependency. For a correct stress derivation in the context of hyperelasticity, this contribution would also have to be included in the \WKM-model, as described in detail in \cite{giantesio_strain-dependent_2017}.

 \paragraph{Numerical treatment of singularities}
Circumventing the singularities in the computation of the \BLE-model invariant's derivative, mentioned in Section \ref{sec:ASE}, involves adding a small numerical contribution. While this is considered to not affect the solution to a considerable extent, it is not particularly elegant.

\subsection{A modified constitutive description of active muscle designed for complex musculoskeletal models} \label{sec:modified_material}
Aiming to combine the optimal properties of the three material models for our application, we propose a fourth material model, named \COMBI-model in the following. The \WKM-model from \cite{weickenmeier_physically_2014} serves as a basis since the generalized active strain approach most accurately models the structure and physiology of muscle tissue and there is no clear preference considering the underlying passive material models. We perform two modifications:
\begin{itemize}
    \item We add the additional term $\SecPK_{\omegaa}$, which takes into account the derivative $\dPsidomegaa \DomegaaDC$, to $\SecPK$ in \mbox{Eq. \eqref{eq:S_WKM}}. A positive side effect of this modification is that $\omegaa$ can now be given by an explicit and computationally less expensive equation. For a detailed explanation, we refer to \cite{giantesio_strain-dependent_2017}. 
    \item Instead of the elaborate calculation of $\Popt \ft$ via the superposition of the twitch forces, we use the smooth function $\Popt \falphat$. $\Popt$ is now prescribed as a material parameter and specifies the amplitude of the $\tanh$-function.
\end{itemize}
The term $\SecPK_{\omegaa}$ is computed from the strain-energy function in Eq. \eqref{eq:Psi_WKM} to 
\begin{equation}
    \SecPK_{\omegaa} = 2 \dPsidomegaa\DomegaaDC = \tfrac{\gamma}{4}\mathrm{e}^{\alpha(\Itilde-1)} \lambda \DomegaaDlambdaM \M. 
    \label{eq:combi_secpk_omegaa}
\end{equation}
With the explicit formulation of the activation level in \cite{giantesio_strain-dependent_2017},
\begin{equation}
\omegaa = \frac{1}{\alpha \lambda^2} \ln\phi 
\quad \text{with} \quad 
    \phi = 1 + \tfrac{4\alpha}{\gamma} {\mathrm{e}}^{\alpha \left(1- \Iptilde \right)} \Popt \falphat  \int_{\lambdaMin}^{\lambda} \fxi(\tilde{\lambda})  d\tilde{\lambda},
\end{equation}
the derivative reads
\begin{equation}
\DomegaaDlambdaM =  \frac{1}{\alpha\lambda^2}  \left( \frac{1}{\phi} \phi' - \frac{2}{\lambda} \ln\phi \right)
\quad \text{with} \quad 
    \phi' = \tfrac{4\alpha}{\gamma} {\mathrm{e}}^{\alpha \left(1 - \Iptilde \right)} \Popt \falphat  \left(\fxi - \alpha \Iptilde' \int_{\lambdaMin}^{\lambda} \fxi(\tilde{\lambda})  d\tilde{\lambda} \right).
\end{equation}
The additional contribution $\CC_{\omegaa}$ to the elasticity tensor is provided in the supplementary materials. For a visual comparison between our modified model and the three models selected from the literature, we refer to Fig. \ref{fig:overview_materials}.
\section{Material parameter identification} \label{sec:fitting}
In order to establish a basis for comparison between the four materials, we fit their parameters to a common set of experimental stress-strain data.
One load case is generally not enough to uniquely determine the material response. Unlike the original publications, we thus consider multiple active and passive load conditions to determine a unique set of parameters representing the experimentally observed data.
To this end, we compute the analytical stress as a function of a given deformation and use this function to fit the material parameters to the experimental stress-strain curves.
The material models are implemented into the solid finite element framework of our comprehensive and well tested in-house research simulation code 4C \mbox {(implemented in \Cplus) \cite{baci}}. For verification purposes, we compare the numerically calculated stress responses with the analytical solutions.

\subsection{Experimental data and associated load cases}
The experimental data serving as a basis for the subsequent fitting of the material parameters was selected according to the following criteria. If available, we preferably chose human specimen data. Since we are interested in the continuum mechanical characteristics rather than the behavior of isolated fibers, we only consider muscle tissue sample data for the fitting. To ensure the comparability of experimental results across different load cases, we use data obtained at comparable quasi-static strain rates (less than $\SI{0.05}{\s^{-1}}$). 

In total, data corresponding to six different load cases is incorporated into the fitting. Table \ref{tab:overview_loading_abbreviatons} gives an overview of the load cases and their abbreviations, the respective literature reference, and whether the data was obtained in the active or passive muscle state.
While the passive muscle material behavior is fitted to data representing all six load cases, the active response is fitted solely to data obtained from uniaxial tension along the fiber direction. To the best of our knowledge, unfortunately, there is no published data testing the active muscle response in load cases different from uniaxial tension. 


\begin{table}[b]
\centering
\caption{Experimental data used in the parameter fitting: Load case, muscle state, abbreviation and reference.}
\begin{tabularx}{\textwidth}{l p{7.0cm}X p{5.0cm}}
\toprule
Abbreviation & Load case & State & Reference \\ \midrule
\multirow{3}{*}{\parbox{1.0cm}{UTCAF\newline}} &
\multirow{3}{*}{\parbox{7.0cm}{Uniaxial tension and compression along fiber direction}}
&  active &  \cite{hawkins_comprehensive_1994} \\
 & & passive & \cite{einarsson_muscle_2010}, mean of supraspinatus \newline and deltoid measurements \\ \arrayrulecolor{gray}\midrule
UTCTF & Uniaxial tension and compression transversal \newline to fiber direction & \multirow{3}{*}{passive} &  \multirow{3}{*}{\cite{morrow_transversely_2010}} \\
SAF & Simple shear along fiber direction & &  \\ \arrayrulecolor{gray}\midrule
PSAF & Pure shear along fiber direction & \multirow{3}{*}{passive} &  \multirow{3}{*}{\smash{\parbox[t]{5.0cm}{\cite{bol_anisotropy_2014}, mean of measurements \newline of differently sized samples}}} \\
PSTF & Pure shear transversal to fiber direction & &  \\
PSTIF & Pure shear transversal to isometrically \newline constrained fibers & &  \\ \arrayrulecolor{black}\bottomrule
\end{tabularx}
\label{tab:overview_loading_abbreviatons}
\end{table}

With the muscle fibers aligned in the $\mathbf{e}_3$-direction, the deformation gradients corresponding to the aforementioned load cases are listed in Table \ref{tab:defgrad_and_loading_sketches}.

\begin{table}[tb]
\centering
\caption{Load cases and associated deformation gradients $\defgrad$ for which the stress response is computed analytically and numerically during the material parameter fitting. The cubes’ dimensions are $d \times d \times d$ with $d=1$, and the geometry is discretized using one linear hexahedral element. Muscle fibers are aligned in $\mathbf{\mathbf{e}}_3$-direction and indicated in red. The deformation is expressed in terms of the stretch $\lambdaBC$ and the shear $\nuBC$ (only for SAF). The Dirichlet boundary conditions enforcing the respective deformation in the simulation are detailed in Table \ref{tab:constraints_cube} in the Appendix.}
\begin{tabularx}{\textwidth}{Xcccccc}
\toprule
    & UTCAF & UTCTF & SAF & PSAF & PSTF & PSTIF \\ \midrule
    \begin{minipage}{.06\textwidth}
      \includegraphics[width=\linewidth]{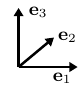}
    \end{minipage}
    &\begin{minipage}{.125\textwidth}\centering
      \includegraphics[width=\linewidth, height=\linewidth]{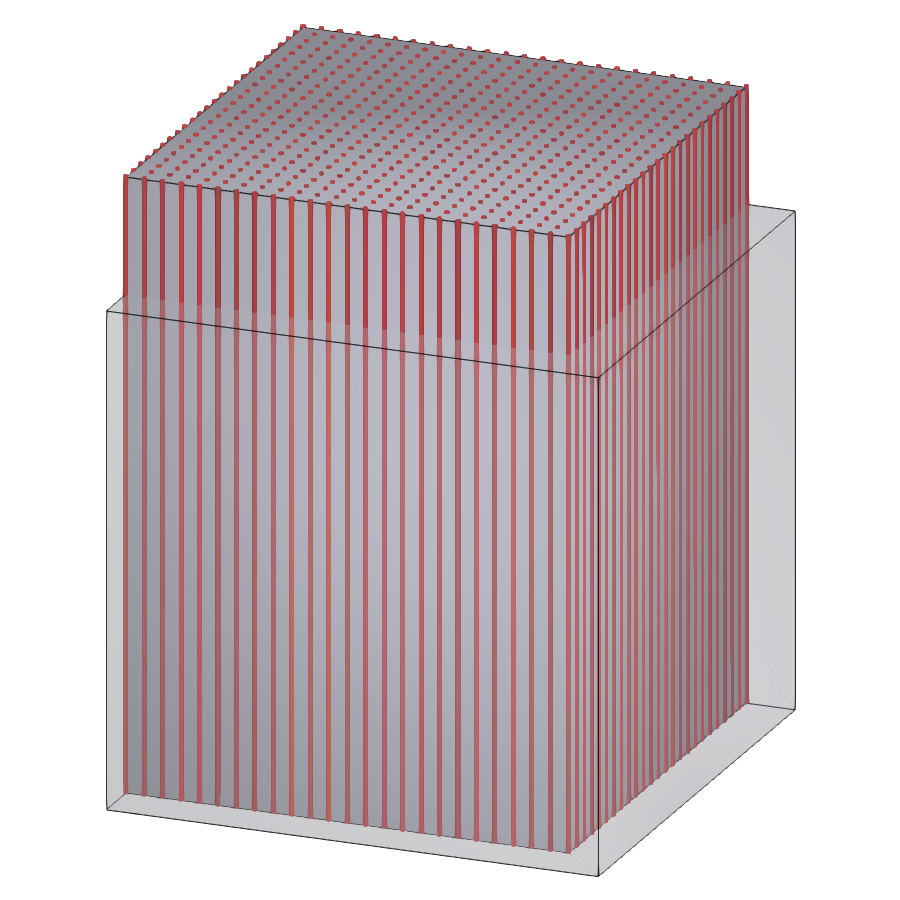}
    \end{minipage} 
    &\begin{minipage}{.125\textwidth}\centering
      \includegraphics[width=\linewidth, height=\linewidth]{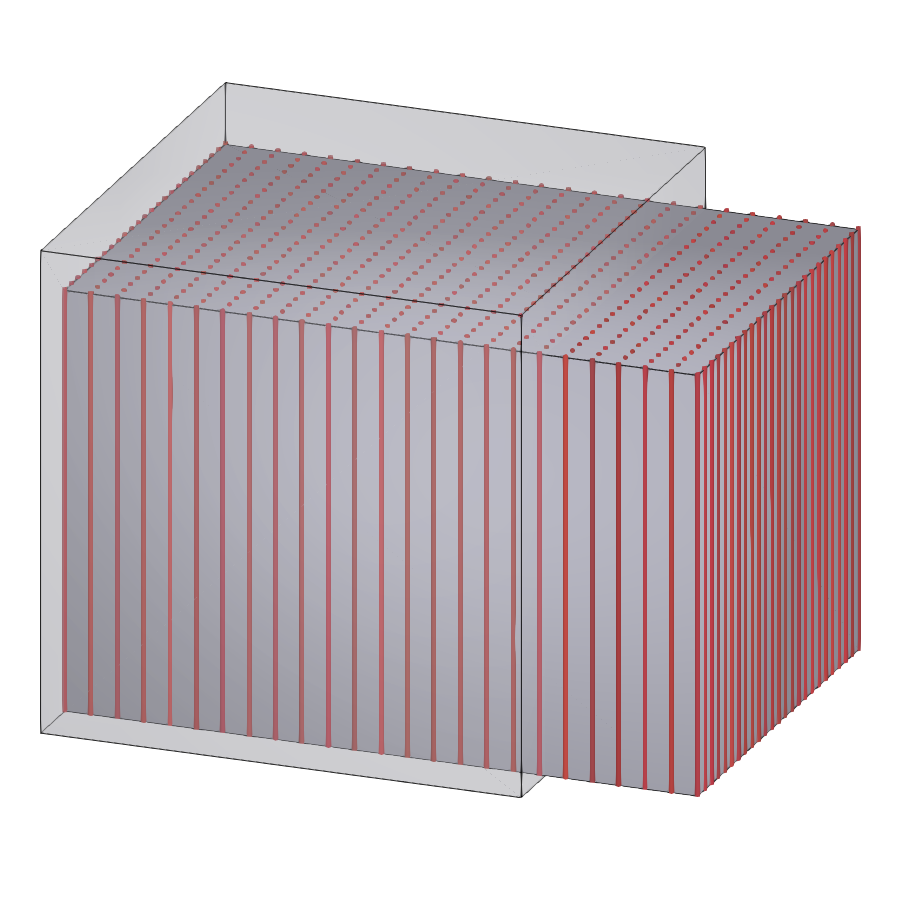}
    \end{minipage}
    &\begin{minipage}{.125\textwidth}\centering
      \includegraphics[width=\linewidth, height=\linewidth]{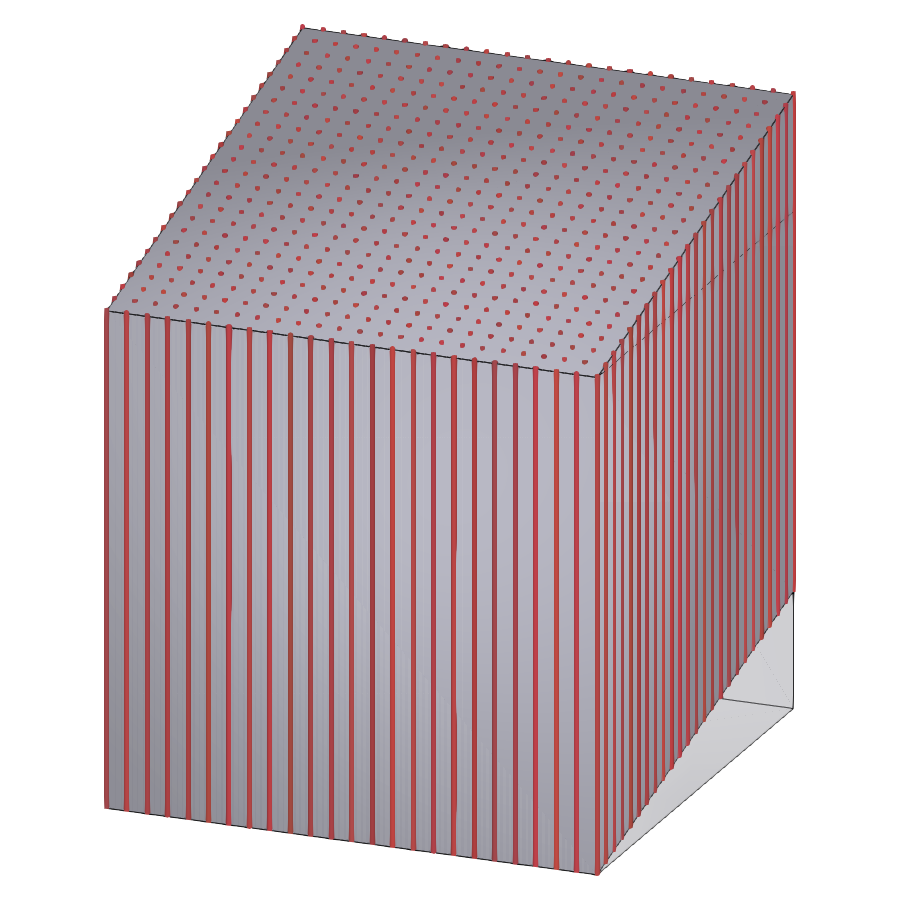}
    \end{minipage}
    &\begin{minipage}{.125\textwidth}\centering
      \includegraphics[width=\linewidth, height=\linewidth]{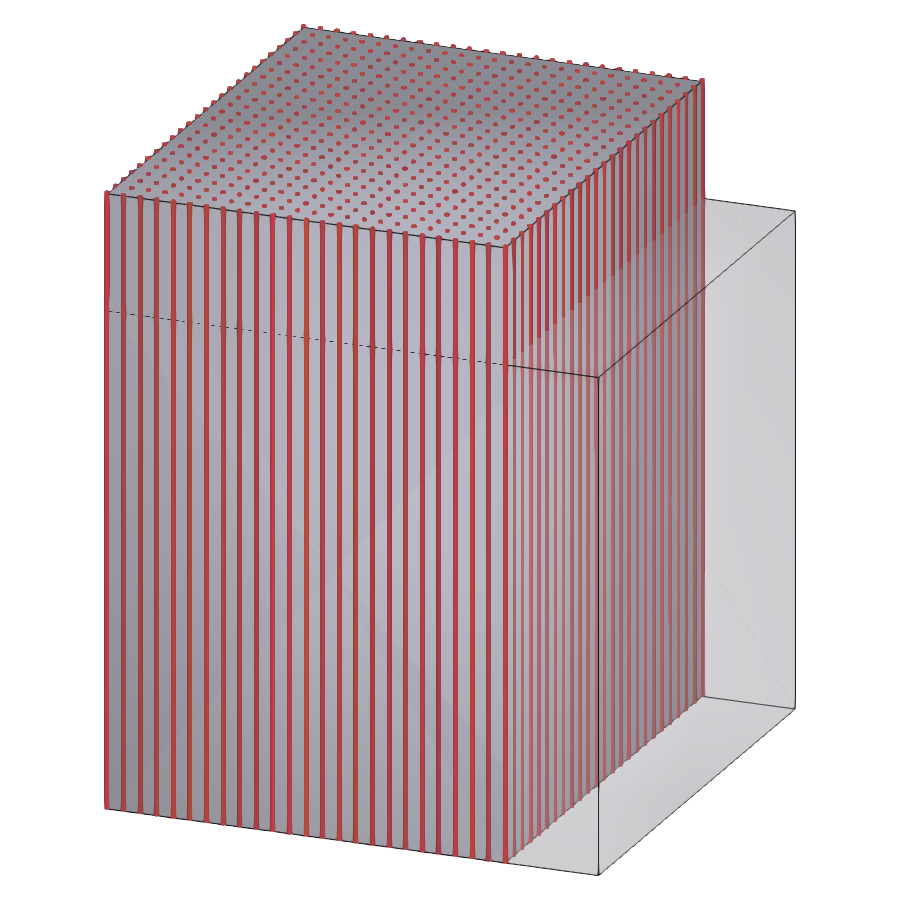}
    \end{minipage}
    &\begin{minipage}{.125\textwidth}\centering
      \includegraphics[width=\linewidth, height=\linewidth]{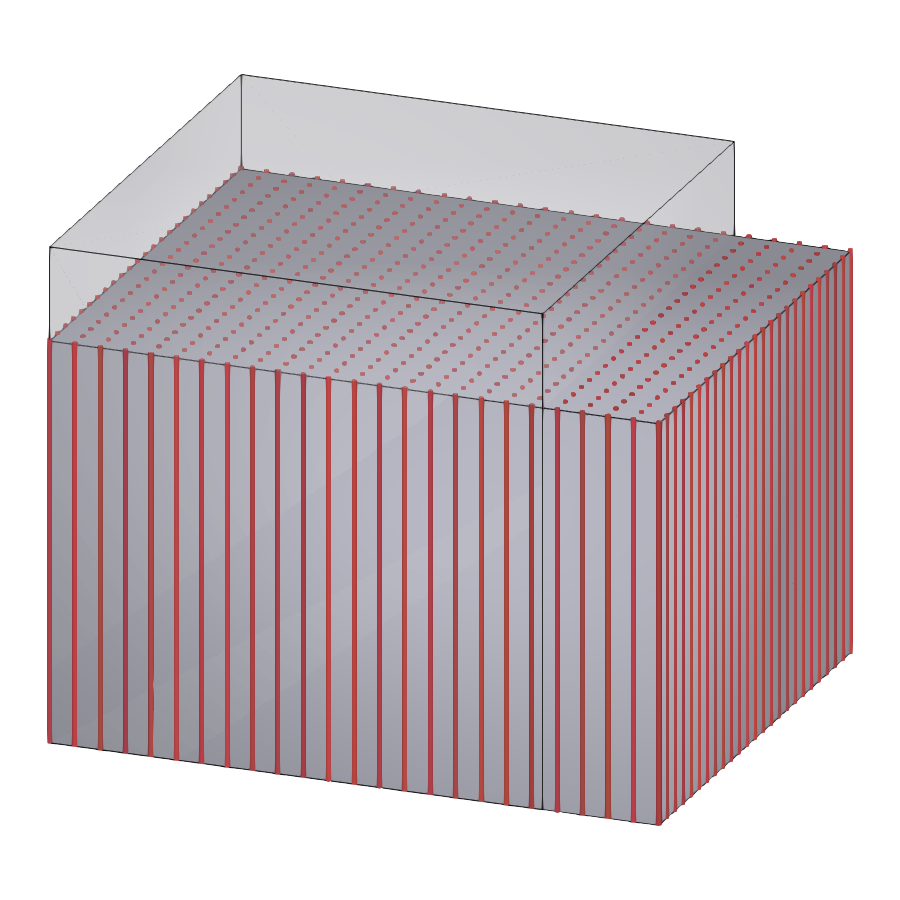}
    \end{minipage}
    &\begin{minipage}{.125\textwidth}\centering
      \includegraphics[width=\linewidth, height=\linewidth]{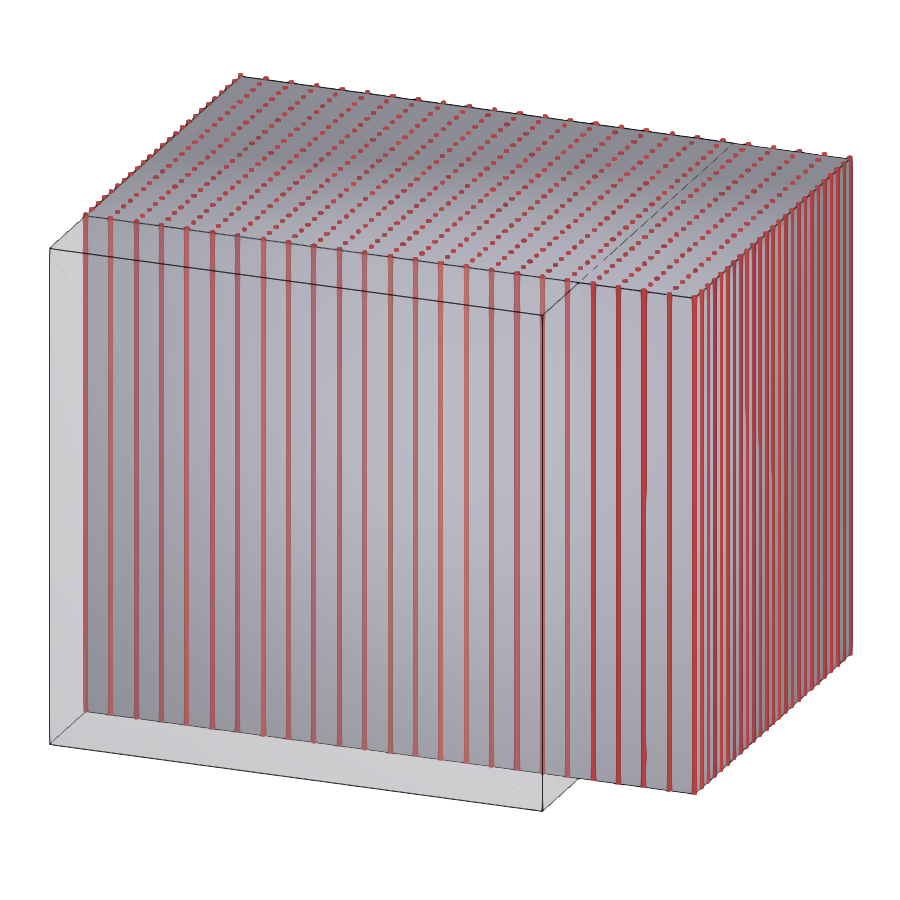}
    \end{minipage} \\
    & & & & & & \\
$\defgrad
$&\begingroup \setlength{\arraycolsep}{1.5pt}$
	\begin{bmatrix}
		\frac{1}{\sqrt{\lambdaBC}} & 0& 0\\
		0& \frac{1}{\sqrt{\lambdaBC}} & 0\\
		0& 0& \lambdaBC 
	\end{bmatrix}
$\endgroup & \begingroup \setlength{\arraycolsep}{1.5pt}$
	\begin{bmatrix}
		\lambdaBC & 0& 0\\
		0& \frac{1}{\sqrt{\lambdaBC}} & 0\\
		0& 0& \frac{1}{\sqrt{\lambdaBC}} 
	\end{bmatrix}
$\endgroup & $
	\begin{bmatrix}
		1   & 0 &    0\\
		0  & 1 &  0\\
		0  & \nuBC & 1
	\end{bmatrix}
$&$
    \begin{bmatrix}
		\frac{1}{\lambdaBC} & 0 & 0\\
		0  & 1 & 0\\
		0  & 0 & \lambdaBC 
	\end{bmatrix}
$&$
	\begin{bmatrix}
		\lambdaBC   & 0 &    0\\
		0  & 1 &    0\\
		0  & 0 &    \frac{1}{\lambdaBC} 
	\end{bmatrix}
$&$
	\begin{bmatrix}
		\lambdaBC   & 0 &    0\\
		0  & \frac{1}{\lambdaBC}  &    0\\
		0  & 0 &    1
	\end{bmatrix}
$\\
\bottomrule
\end{tabularx}
\label{tab:defgrad_and_loading_sketches}
\end{table}

\subsection{Analytical stress-strain responses}
As noted in \cite{weickenmeier_physically_2014} and \cite{blemker_3d_2005}, the compressible and incompressible formulation of the material models described in Section \ref{sec:materials} coincide for the case that the incompressibility parameters $\kappa$ and $K$, respectively, approach infinity.
In contrast to the nearly incompressible formulations presented in Section \ref{sec:materials}, for an analytical interpretation, we consider the fully incompressible formulations as also given in \cite{giantesio_strain-dependent_2017} for the \GIANT-model and in \cite{ehret_continuum_2011} for the \WKM-model. The fully incompressible formulation of the \BLE-model is obtained as the isochoric contribution with the unmodified strain measures and invariants. 
In the simulation, we then apply appropriate incompressibility parameters $\kappaBle$ and $\kappa$ to recover the close-to incompressible state. 

With the deformation gradients $\defgrad$ for the six load cases in Table \ref{tab:defgrad_and_loading_sketches}, we derive the first Piola-Kirchhoff stress $\mathbf{P}$ in the respective load direction. The analytical expressions are provided in Table \ref{tab:analytical_responses} in the Appendix. Analyzing the equations highlights the importance of using multiple load modes for the fitting of the passive material parameters. As an example, fitting the \BLE-model to experimental data solely obtained from UTCAF would result in arbitrary values of the parameters $\Gone$ and $\Gtwo$, as those do not appear in the corresponding equation in Table \ref{tab:analytical_responses}.

\subsection{Parameter identification through a least-squares fit}
We fit the material parameters of these analytical stress-strain responses to the experimental data in Table \ref{tab:overview_loading_abbreviatons} by solving a least-squares minimization problem. For this purpose, we employ the Trust Region Reflective algorithm \cite{branch_subspace_1999} implemented in the 
\texttt{scipy.optimize.least\_squares} method from the Python SciPy library (version 1.7.2) \cite{2020SciPy-Optimize}. 

Since the experimental data for the active load case UTCAF$^\mathrm{act}$ was obtained under isometric conditions at a tetanic activation level \cite{hawkins_comprehensive_1994}, the time-dependent activation functions $\ft$ and $\falphat$ are set to $1$. This also means that the parameters involved in the computation of $\ft$ and $\falphat$ cannot be determined from the experimentally determined stress-strain curves.
Still, the active parameters $I_i$, $F_i$, $T_i$, and $\rho_i$ in $\ft$ are physically measurable micromechanical quantities whose values we adopt from \cite{ehret_continuum_2011}. To create a comparable time-dependent activation function $\falphat$, its parameter $c$, governing the time-dependent rise of the activation, is set to match the slope of $\ft$. Fig. \ref{fig:ft_normalized} shows the normalized functions $\ft$ and $\falphat$.

To quantify the deviation of the fitted analytical stress-strain response from the experimental data, we compute three error measures. Based on the \(L_\infty\) norm, we first define the relative error $\varepsilon_\mathrm{\infty}$ (see Eq. \eqref{eq:varepsiloninf}) to assess the maximum absolute deviation. With the \(L_1\) norm providing a metric for the total absolute deviation, we second evaluate the relative error $\varepsilon_\mathrm{1}$ (see Eq. \eqref{eq:varepsilon1}). Since we minimized the sum of squared distances between experimental and fitted value, we third compute the \(L_2\) norm-based relative error \(\varepsilon_\mathrm{2}\).

\subsection{Results}
\begin{figure}[h]
\begin{minipage}{0.75\linewidth}
    \centering 
    \tikzstyle{every node}=[font=\small]
    \begin{subfigure}[t]{0.32\textwidth}
        \includegraphics[trim=10.2 0 0 0, clip, height=0.7\linewidth]{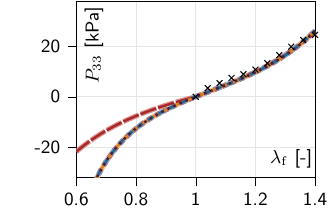}
        \caption{UTCAF}
      \label{fig:fitting_passive_UTCAF}
    \end{subfigure}\hfill
    \begin{subfigure}[t]{0.32\textwidth}
        \includegraphics[trim=10.2 0 0 0, clip, height=0.7\linewidth]{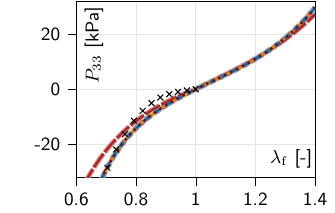}
        \caption{PSAF}
      \label{fig:1}
    \end{subfigure}\hfill
    \begin{subfigure}[t]{0.32\textwidth}
    	\includegraphics[trim=1.9 0 0 0, clip, height=0.7\linewidth]{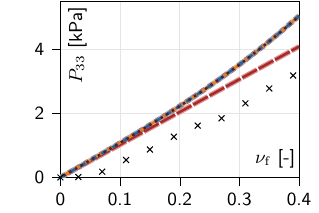}
    	\caption{SAF}
      \label{fig:1}
    \end{subfigure} \\
    \par\bigskip 
    \begin{subfigure}[t]{0.32\textwidth}
    	\includegraphics[trim=10 0 0 0, clip, height=0.7\linewidth]{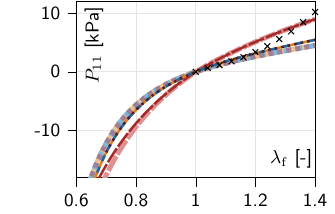}
    	\caption{UTCTF}
      \label{fig:1}
    \end{subfigure}\hfill
    \begin{subfigure}[t]{0.32\textwidth}
    	\includegraphics[trim=10 0 0 0, clip, height=0.7\linewidth]{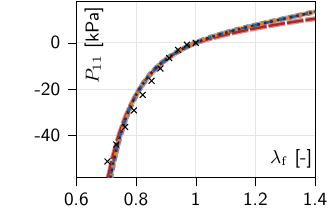}
    	\caption{PSTF}
      \label{fig:1}
    \end{subfigure}\hfill
    \begin{subfigure}[t]{0.32\textwidth}
    	\includegraphics[trim=10 0 0 0, clip, height=0.7\linewidth]{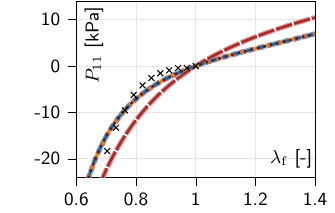}
    	\caption{PSTIF}
      \label{fig:1}
    \end{subfigure}
\end{minipage}\hfill
\begin{minipage}{0.2\linewidth}
     \includegraphics[width=\linewidth]{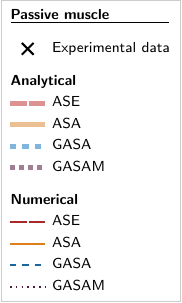}
\end{minipage}
\caption{Passive stress-stretch responses with fitted parameters for six different load cases alongside the experimental data listed in Table \ref{tab:overview_loading_abbreviatons}. Results of the analytical equations are plotted alongside the numerical results. The obtained curves overlap because the results are nearly identical.}
\label{fig:fitting_passive}
\end{figure}

\begin{figure}[h]
\begin{minipage}{0.75\linewidth}
    \centering 
    \tikzstyle{every node}=[font=\small]
    \begin{subfigure}[t]{0.32\textwidth}
        \includegraphics[trim=5 0 0 0, clip, height=0.7\linewidth]{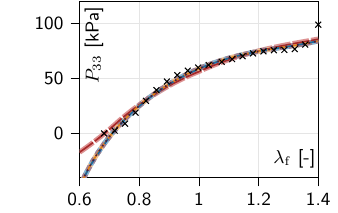}
        \caption{UTCAF}
      \label{fig:fitting_active}
    \end{subfigure}\hfill
    \begin{subfigure}[t]{0.32\textwidth}
        \includegraphics[trim=5 0 0 0, clip, height=0.7\linewidth]{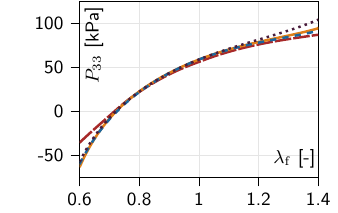}
        \caption{PSAF}
      \label{fig:1}
    \end{subfigure}\hfill
    \begin{subfigure}[t]{0.32\textwidth}
    	\includegraphics[trim=1.9 0 0 0, clip, height=0.7\linewidth]{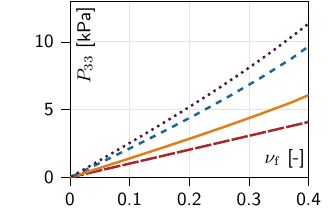}
    	\caption{SAF}
      \label{fig:1}
    \end{subfigure} \\
    \par\bigskip 
    \begin{subfigure}[t]{0.32\textwidth}
    	\includegraphics[trim=10 0 0 0, clip, height=0.7\linewidth]{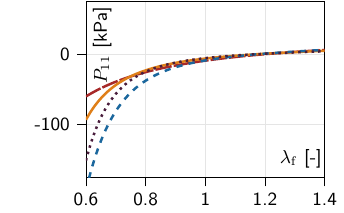}
    	\caption{UTCTF}
      \label{fig:1}
    \end{subfigure}\hfill
    \begin{subfigure}[t]{0.32\textwidth}
    	\includegraphics[trim=10 0 0 0, clip, height=0.7\linewidth]{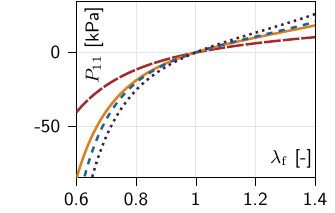}
    	\caption{PSTF}
      \label{fig:1}
    \end{subfigure}\hfill
    \begin{subfigure}[t]{0.32\textwidth}
    	\includegraphics[trim=5 0 0 0, clip, height=0.7\linewidth]{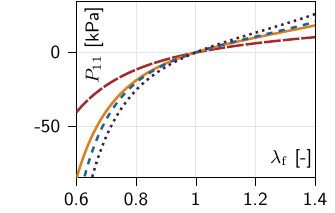}
    	\caption{PSTIF}
      \label{fig:1}
    \end{subfigure}
\end{minipage}\hfill
\begin{minipage}{0.2\linewidth}
     \includegraphics[width=\linewidth]{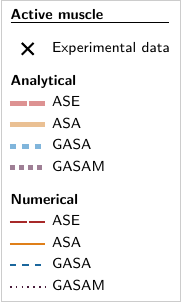}
\end{minipage}
\caption{Active stress-stretch responses with fitted parameters for six different load cases. For the sake of a better visibility, the plotting resolution is chosen such that only the twitch maxima are depicted for the \GIANT- and \WKM-model. Since in the active case only experimental data from uniaxial tension in the fiber direction was used for the parameter identification, the remaining load cases show solely the numerically obtained stress-strain response.}
\label{fig:fitting_all_active}
\end{figure}

Table \ref{tab:fitting_all} lists the parameter values obtained from the fitting and the literature. The experimental data and the analytical and computational results are shown in Fig. \ref{fig:fitting_passive} and \ref{fig:fitting_all_active} for the passive and active load cases, respectively. Table \ref{tab:relative_errors} lists the computed error measures.

As expected, the nearly-incompressible formulations used in the simulation coincide with the analytical incompressible responses for the chosen incompressibility parameters. Further, as designed, the passive responses for the \WKM-, \GIANT-, and \COMBI-model coincide.

\begin{table}[t]
\centering
\caption{Material model parameters fitted to the experimental data listed in Table \ref{tab:overview_loading_abbreviatons}. $I_i$, $F_i$, $T_i$, and $\rho_i$ are considered fixed and adopted from \cite{ehret_continuum_2011}. $c$ is set to match the slope of $\ft$.}
\begin{tabular}{lrl p{0.05 \textwidth} lrl p{0.05 \textwidth} lrl}
\toprule
\multicolumn{3}{l}{\BLE} & &  \multicolumn{3}{l}{\WKM and \GIANT} & & \multicolumn{3}{l}{\COMBI} \\ \midrule
$\Gone$ & 0.1000 & $\si{\kilo \pascal}$ &  & $\alpha$ & 2.3796 & -  &  & $\alpha$ & 2.3796 & -\\
$\Gtwo$ & 0.0500 & $\si{\kilo \pascal}$ &  & $\beta$ & 0.5161 & -  &  & $\beta$ & 0.5161 & - \\
$\Pone$ & 3.6055 & - &  & $\gamma$ & 27.1072 & $\si{\kilo \pascal}$ &  & $\gamma$ & 27.1072 & $\si{\kilo \pascal}$ \\
$\Ptwo$ & 4.4883 & - &  & $\omegazero$ & 0.6388 & -  &  & $\omegazero$ & 0.6388 & - \\
$\kappaBle$ & 10000 & $\si{\kilo \pascal}$ &  & $\kappa$ & 1000 & - &  & $\kappa$ & 1000 & -\\
$\lambdaOpt$ & 1.2264 & - &  & $\lambdaOpt$ & 1.1806 & - &  & $\lambdaOpt$ & 1.1806 & -\\
$\lambdaStar$ & 1.4000 & - &  & $\lambdaMin$& 0.5680 & - &  & $\lambdaMin$& 0.5680 & -   \\
$\sigmamax$ & 1.1450 & $\si{\kilo \pascal}$ &  & $\Na$  & 0.4619 & $\si{\mm^{-1}}$ &  & $\Popt$  & 64.6809 & $\si{\kilo \pascal}$\\
$\alphaa$ & 69.5471 & - &  & $F_i$ & 2.5, 4.4, 76.8 & $\SI{0.001}{\newton}$ &  & $c$ & 34.4017 & -   \\
$c$& 34.4017 & - &  & $T_i$ & 0.02, 0.011, 0.011 & $\si{s}$ & & & &\\
&  &  &  & $I_i$ & 0.004, 0.004, 0.004 & $\si{s}$ & & & &\\
&  &  &  & $\rho_i$ & 0.05, 0.29, 0.66 & - & & & & \\
\multicolumn{2}{l}{Neo-Hooke} &  &  \\ \midrule
$\mu$ & 10 & $\si{\kilo \pascal}$  &  \\ \bottomrule
\end{tabular}
\label{tab:fitting_all}
\end{table}

\paragraph{Goodness of fit}
Qualitatively, all material models approximate the experimental data well. An exception is the UTCTF load case in tension. Contrary to the experimental data, suggesting a stiffness increase for rising stretches, the fitted stress responses flatten.

Quantitatively, the quality of the fit differs both between the material models and among the load cases. 

The experimental observations for the passive load cases UTCAF, PSAF, and PSTF are well approximated with all material models, with $\varepsilon_\mathrm{1} \leq 0.30$. For UTCAF, the fit is exceptionally good, with $\varepsilon_\mathrm{1} \leq 0.12$ in the passive case and $\varepsilon_\mathrm{1} \leq 0.06$ in the active case. 
For those three load cases, we observed no considerable differences when comparing the material models considering the error measures. A minor exception is the $\varepsilon_\mathrm{\infty}$ error: Compared to the \WKM-, \GIANT-, and \COMBI-models, the \BLE-model exhibits a larger $\varepsilon_\mathrm{\infty}$ error for PSAF and a smaller $\varepsilon_\mathrm{\infty}$ error for PSTF. 

For UTCTF, SAF, and PSTIF, the deviations between the experimental data and the fitted responses are more pronounced.  
For UTCTF, the errors obtained for the \BLE-model (e.g., $\varepsilon_\mathrm{1} = 0.26$) suggest a slightly better fit compared to the other material models ($\varepsilon_\mathrm{1} = 0.39$). While the \WKM/\GIANT/\COMBI-model better approximates stresses due to small stretches, the \BLE-model matches stresses in the high stretch regime more closely. 
The fitted responses for SAF deviate the strongest from the experimental observations. Although the \BLE-model provides a better overall approximation than the other three models, errors are still moderately high (e.g., $\varepsilon_\mathrm{1} = 0.50$).
For PSTIF, the \WKM/\GIANT/\COMBI-model delivers a good fit with errors below $0.26$. In contrast, errors of the \BLE-model response are considerably higher (e.g., $\varepsilon_\mathrm{1} = 0.77$).

As mentioned before, for the load cases PSAF, PSTF, and PSTIF, we averaged the experimental measurements from differently sized samples and used these averages in the fitting procedure. We note that compared to the original data in \cite{bol_anisotropy_2014} and the associated standard deviations, our fitted material response remains within the experimentally determined range.

Although the experimental data is not perfectly matched, we rate the fitting as satisfactory. It has to be considered that the experimental data originate from different trials, and test conditions are not guaranteed to be the same. Potential measurement errors must also be taken into account. Depending on the load scenario, one or the other material provides the more accurate result. From this perspective, none of the material models is preferred over the other.

\begin{table}[t]
\centering
\caption{Relative errors \(\varepsilon_\mathrm{1}\) and \(\varepsilon_\mathrm{\infty}\) between the experimental data and the fitted analytical model stress responses for different load cases.}
\begin{tabular}{p{3cm}cccccccc}
\toprule
\multirow{2}{*}{Load case} & \multicolumn{2}{c}{\(\varepsilon_\mathrm{\infty}\)} & & \multicolumn{2}{c}{\(\varepsilon_\mathrm{1}\)} & & \multicolumn{2}{c}{\(\varepsilon_\mathrm{2}\)} \\ \cmidrule{2-3} \cmidrule{5-6}  \cmidrule{8-9}
 & \BLE & \GIANT/GASA(M) && \BLE & \GIANT/GASA(M) && \BLE & \GIANT/GASA(M) \\
\midrule
UTCAF & 0.10 & 0.09 && 0.11 & 0.12 && 0.11 & 0.11\\
UTTF & 0.17 & 0.56 && 0.26 & 0.39 && 0.24 & 0.47\\
PSAF & 0.24 & 0.15 && 0.26 & 0.29 && 0.22 & 0.24\\
PSTF & 0.16 & 0.28 && 0.13 & 0.15 && 0.13 & 0.17\\
PSTIF & 0.33 & 0.19 && 0.77 & 0.26 && 0.58 & 0.21\\
SAF & 0.27 & 0.47 && 0.50 & 0.72 && 0.41 & 0.62\\
UTCAF$^\mathrm{act}$ & 0.14 & 0.15 && 0.06 & 0.05 && 0.07 & 0.06 \\    
\bottomrule
\end{tabular}
\label{tab:relative_errors}
\end{table}

\paragraph{Differences between the model stress responses}
Where no experimental data was available, the stress responses of the different material models are - to no surprise - different. Although, qualitatively, the model responses are comparable, quantitatively, there are major differences. 

First, we evaluate the results for passive muscle. For compressive states of UTCTF and tensile states of PSTIF, the \BLE-model behaves slightly stiffer than the other models. The opposite is true for compressive states of UTCAF. Differences between the material models for tensile states of PSAF and PSTF are minor. To determine which model better represents reality, further experimental evidence is needed.

Considering the active muscle behavior, we observe significant differences between the computational material model responses for PSTIF, SAF, and compressive states of UTCTF. In all three cases, the \BLE-model behaves the least stiff, followed by the \GIANT-model. Contrarily, deviations between the material model responses are comparably small for UTCTF in tension, PSAF, and PSTF.  
Differences between the \BLE-model response and the other three are partly explained by the non-matching passive material model responses. Deviations between the \COMBI- and \WKM-model are caused by the additional stress term $\SecPK_{\omegaa}$. Since all active material parameters can be uniquely fitted through the UTCAF load case, we rule out the possibility that the differences are due to random, undetermined parameters. Hence, the remaining differences are attributed to the use of different activation concepts. Without any additional experimental data, it is challenging to determine the quantitative superiority of either of the material models. Against this background, the physiologically realistic modeling of muscle activation becomes particularly relevant, and our decision for the generalized active strain approach seems justified.

\FloatBarrier
\section{Numerical examples} \label{sec:numerics}
To demonstrate the applicability of the material models to biomechanically relevant scenarios -- in particular human shoulder biomechanics -- and investigate the material behavior using the fitted parameters, we consider three numerical examples: A simple fusiform muscle, a two-component model consisting of one bone and one muscle, and a full human shoulder model. Simulations are again conducted using 4C \cite{baci}.

\subsection{Fusiform muscle contraction} \label{sec:fusiform}
\begin{figure}[htb]
\centering
\begin{minipage}[b]{0.5\textwidth}
\centering
\begin{subfigure}[t]{\textwidth} \centering
    \includegraphics[width=0.75\textwidth]{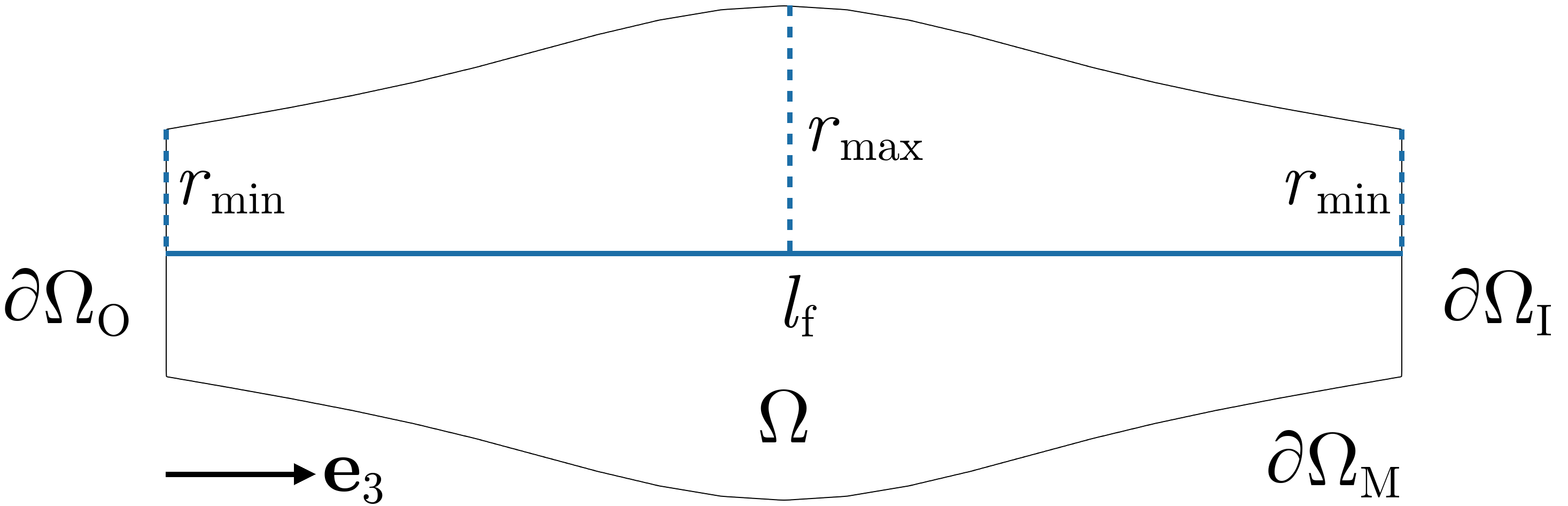}
    \caption{Geometric dimensions.}
    \label{fig:fusiform_geo}
\end{subfigure}\\ \vspace{0.25cm}
\begin{subfigure}[b]{\textwidth} \centering
    \includegraphics[width=0.8\textwidth]{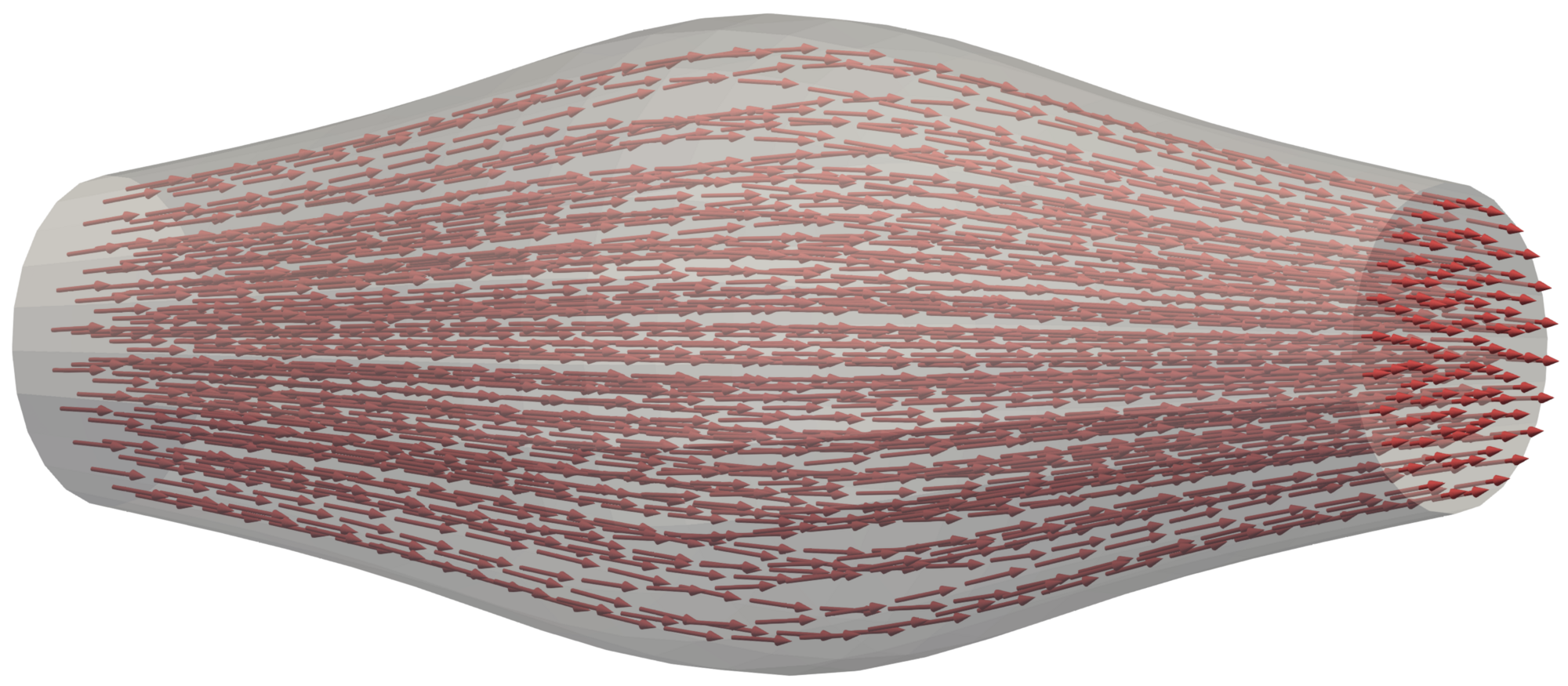}
    \caption{Fiber directions $\mathrm{\mathbf{m}}$ visualized in red.}
    \label{fig:fusiform_fiber}
\end{subfigure}
\end{minipage}
\hfill
\begin{subfigure}[b]{0.45\textwidth} \centering
    \includegraphics[width=0.9\textwidth]{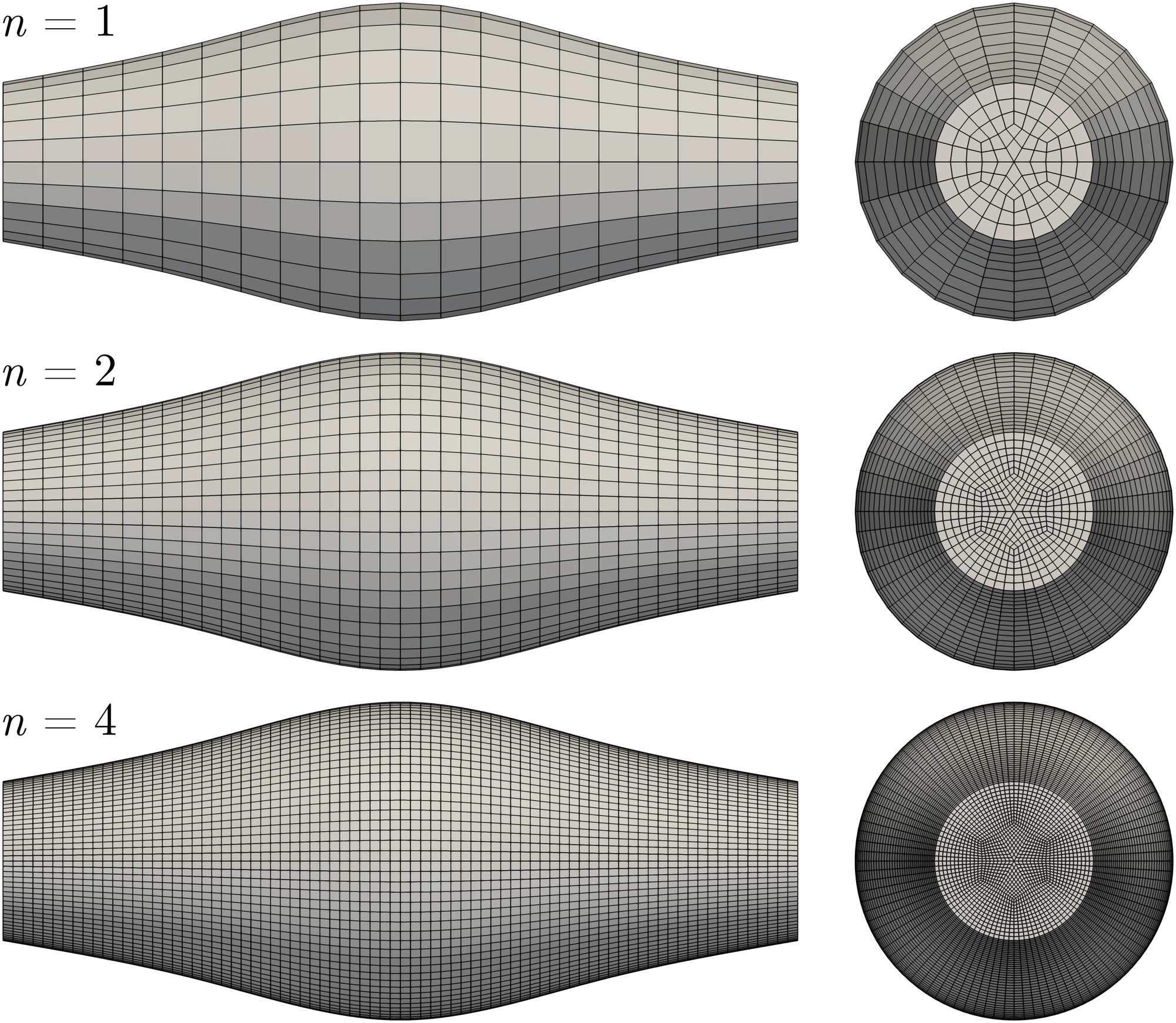}
    \caption{Hexahedral meshes for the refinement levels $n$.}
    \label{fig:fusiform_geo_meshes}
\end{subfigure}
\caption{Fusiform muscle geometry, mesh, and fibers.}
\label{fig:fusiform_geo_fiber_mesh}
\end{figure}

\paragraph{Geometry and mesh}
In the first step, we consider the geometry of a fusiform muscle with length $l_\mathrm{f}$ in the $\mathbf{e}_3$-direction and a circular cross-section with varying radius $r$ as depicted in Fig. \ref{fig:fusiform_geo}. The outer contour along the $\mathbf{e}_3$-axis is described by spline curves through the points $(r,l) = (r_\mathrm{min},0)$, $(\frac{1}{2} (r_\mathrm{max}-r_\mathrm{min}), \tfrac{1}{4} l_\mathrm{f})$ and $(r_\mathrm{max}, \tfrac{1}{2} l_\mathrm{f})$ such that the muscle's radius increases from $r_\mathrm{min}$ at the ends to $r_\mathrm{max}$ at its center. We choose $l_\mathrm{f}=\SI{100}{\mm}$, $r_\mathrm{min}=\SI{10}{\mm}$, and $r_\mathrm{max}=\SI{20}{\mm}$. With reported mean cross-section areas of $\SI{438}{\milli \meter^2}$ \cite{bouaicha_cross-sectional_2016}, $\SI{370}{\milli \meter^2}$ \cite{holzbaur_upper_2007}, and $\SIrange[range-phrase=-]{294}{360}{\milli \meter^2}$ \cite{yanagisawa_appropriate_2009}, and lengths of $\SI{115}{\milli \meter^2}$ \cite{langenderfer_musculoskeletal_2004}, those measures approximately represent an average-sized teres minor muscle. 

We use Cubit 13.2. \cite{cubit_version_2012} to create linear hexahedral element meshes with three different refinement levels $n=1,2,4$ as specified in Table \ref{tab:mesh_refinement_data} and shown in Fig. \ref{fig:fusiform_geo_meshes}. To prevent the occurrence of locking phenomena, we apply the F-bar element technology \cite{de_souza_neto_design_1996}.

Similar to \cite{choi_skeletal_2013}, we compute the normalized elementwise fiber direction $\mathrm{\mathbf{m}}$ as the solution of the Laplacian problem $\Delta \Phi = 0$ on the muscle domain $\Omega$ with Dirichlet boundary conditions $\Phi = \hat{\Phi}$ prescribed on the outer muscle boundary surface $\partial \Omega_\mathrm{M}$ (excluding the origin and insertion surfaces $\partial \Omega_\mathrm{O}$ and $\partial \Omega_\mathrm{I}$).
$\hat{\Phi}$ is determined using a rule-based approach, according to which the fiber vectors describe a continuous path from $\partial \Omega_\mathrm{O}$ to $\partial \Omega_\mathrm{I}$ and are tangential to $\partial \Omega_\mathrm{M}$. Fig. \ref{fig:fusiform_fiber} shows the resulting fiber directions $\mathrm{\mathbf{m}}$.

\begin{table}[htb]
    \centering
    \caption{Mesh quantities of the fusiform muscle geometry for different refinement levels. The number of elements in the circumferential direction is counted along the intersection $\partial \Omega_\mathrm{O} \cap \partial \Omega_\mathrm{M}$.}
    \begin{tabular}{L{2cm}ccccc}
    \toprule
        \multirow{2}{1.8cm}{Refinement level $n$} & \multicolumn{3}{c}{Number of elements} & & \multicolumn{1}{c}{Number of nodes} \\ \cmidrule{2-4} \cmidrule{6-6}
        & circumferential & longitudinal & total & &  total\\ \midrule
        1 & 24 & 20 & 1920 & & 2289\\
        2 & 48 & 40 & 15360 & & 16769\\
        4 & 96 & 80 & 122880 & & 128385\\
        \bottomrule
    \end{tabular}
    \label{tab:mesh_refinement_data}
\end{table}

\paragraph{Simulation scenarios}
We simulate two physiologically relevant scenarios: an isometric contraction and a free concentric contraction. In an isometric contraction, activation leads to a change in tension while the muscle length remains constant. Isometric contractions are responsible for holding tasks and support in the musculoskeletal system and therefore play a crucial role in stabilizing the shoulder joint. In a free concentric contraction, the produced active forces cause a muscle shortening since no external forces act against the contraction direction. Concentric contractions thus generate motion. 

In the isometric contraction scenario, we apply Dirichlet boundary conditions to fix both lateral ends to zero displacement in all three coordinate directions. 
For the free contraction, we fix the origin surface to zero displacement in all three directions. To mimic the attachment to bone, we ensure no relative displacement occurs between the insertion surface nodes.
Hence, on the insertion surface, we prescribe zero Dirichlet conditions in the $\mathbf{e}_1$- and $\mathbf{e}_2$-directions and multipoint constraints in the $\mathbf{e}_3$-direction. The zero Dirichlet conditions ensure that the insertion surface nodes remain fixed in-plane, while the multipoint constraints ensure that the insertion surface nodes displace uniformly in contraction direction.

We restrict the analysis to the quasi-static case and neglect inertia effects. The simulations are repeated for the three mesh refinement levels in Table \ref{tab:mesh_refinement_data} and the four material models introduced in Section \ref{sec:materials} with the parameters in Table \ref{tab:fitting_all}.

\paragraph{Simulation results}
As discussed later in more detail, the results for the three mesh refinements exhibit no significant qualitative differences and only slight variations in quantity. We thus focus our evaluation on the results obtained with mesh refinement $n=4$. 

\begin{figure}[b]
\centering
\includegraphics[height=5cm]{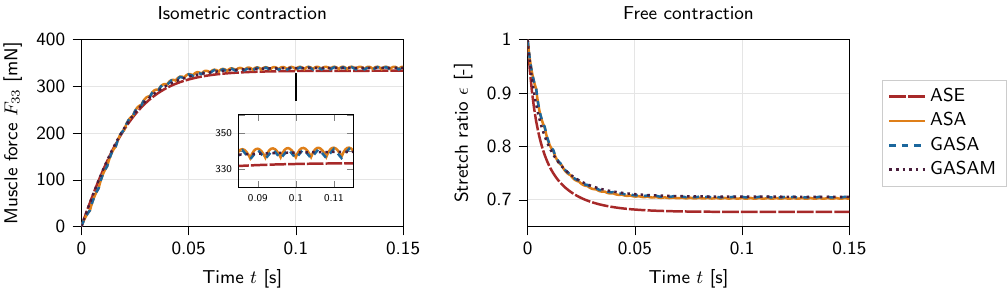}
\captionof{figure}{Simulated muscle force $F_{33}$ and stretch ratio $\epsilon$ for an isometric and a free contraction of the fusiform muscle ($n=4$).}
\label{fig:fusiform_plot_iso_free}
\end{figure}

First, we compare quantities on the global level. For the isometric contraction, the muscle force $F_{33}$ is computed as the surface integral of the Cauchy stress $\sigma_{33}$ over the central cross-section area at $l=\SI{5}{cm}$ in the current configuration. For the free contraction, the stretch ratio $\epsilon$ serves as a measure of the percentage change in length and is evaluated as $\epsilon = 1-\frac{\Delta l_f}{l_f}$. Results are displayed in Fig. \ref{fig:fusiform_plot_iso_free} over time $t$.

\begin{figure}[h]
    \centering 
    \begin{minipage}[t]{0.8\textwidth}
    \begin{subfigure}[t]{0.18\textwidth}
    \centering \begingroup
    \sbox0{\includegraphics{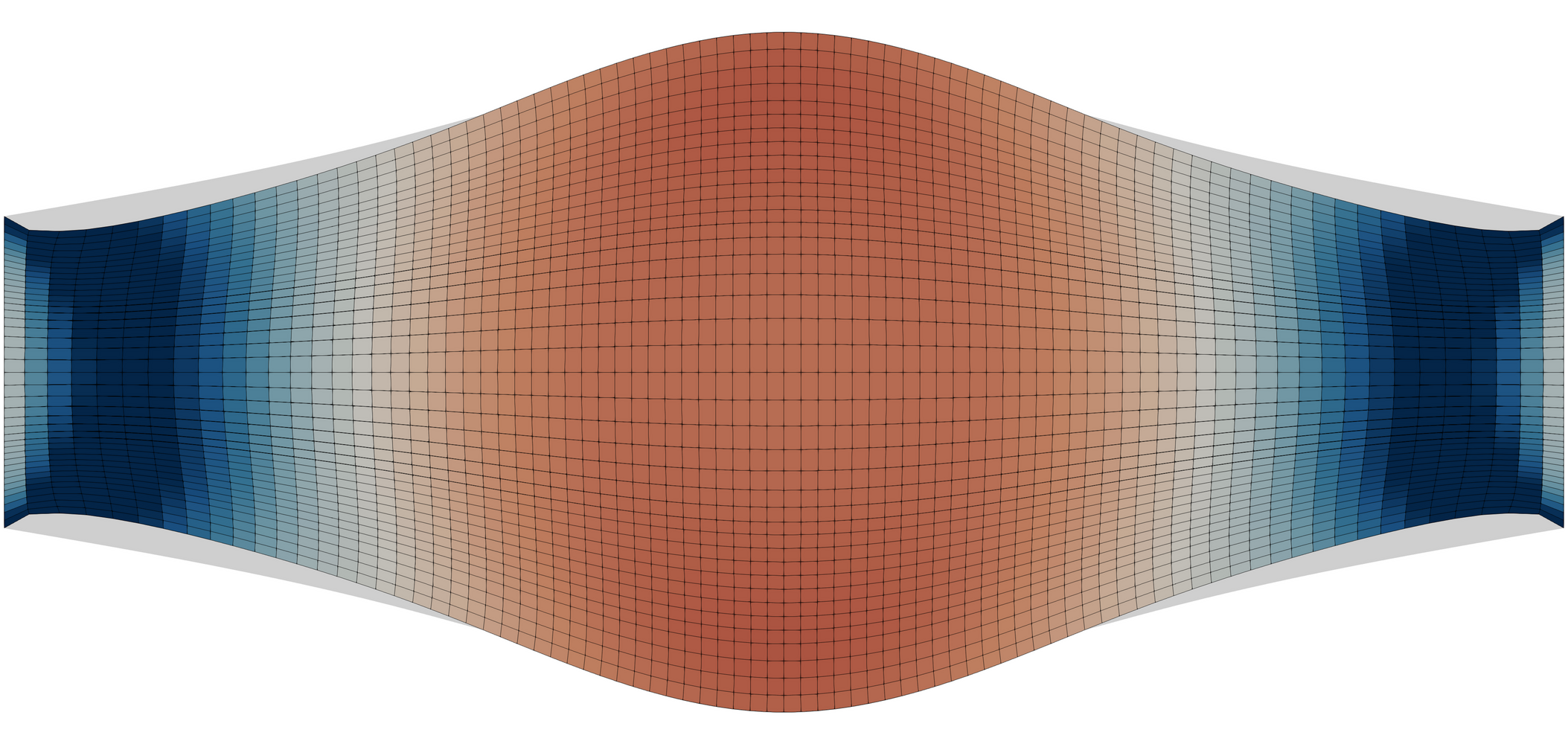}}
    \includegraphics[clip,trim={.5\wd0} 0 0 0, height=3cm]{figures/fig_10_slice_zx_fiber_stretch_blemker_n4_ic_element.png}
    \endgroup \caption{\BLE} \label{fig:unknown}
    \end{subfigure}\hfill
    \begin{subfigure}[t]{0.18\textwidth}
    \centering \begingroup
    \sbox0{\includegraphics{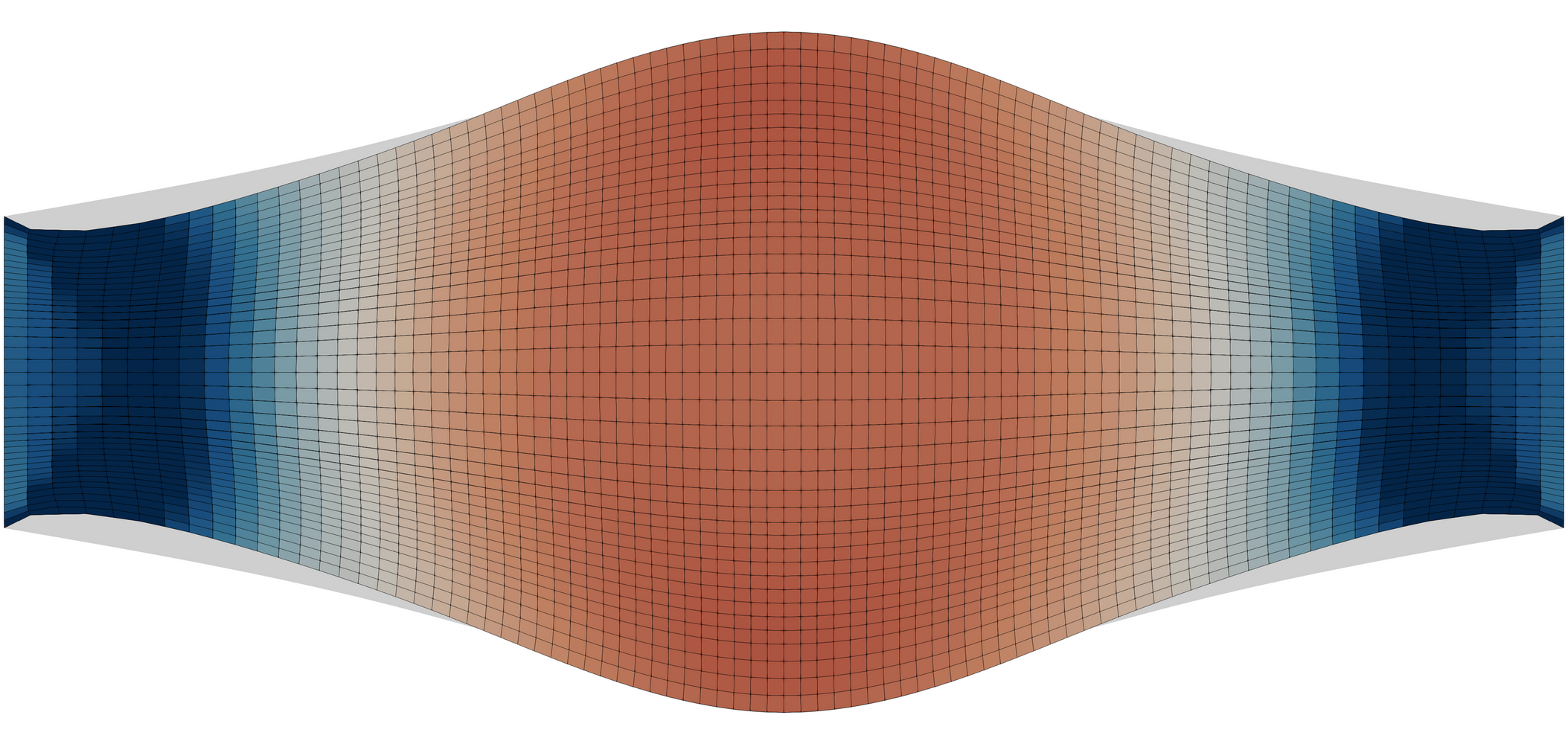}}
    \includegraphics[clip,trim={.5\wd0} 0 0 0, height=3cm]{figures/fig_10_slice_zx_fiber_stretch_giantesio_n4_ic_element.png}
    \endgroup \caption{\GIANT} \label{fig:unknown}
    \end{subfigure}\hfill
    \begin{subfigure}[t]{0.18\textwidth}
    \centering \begingroup
    \sbox0{\includegraphics{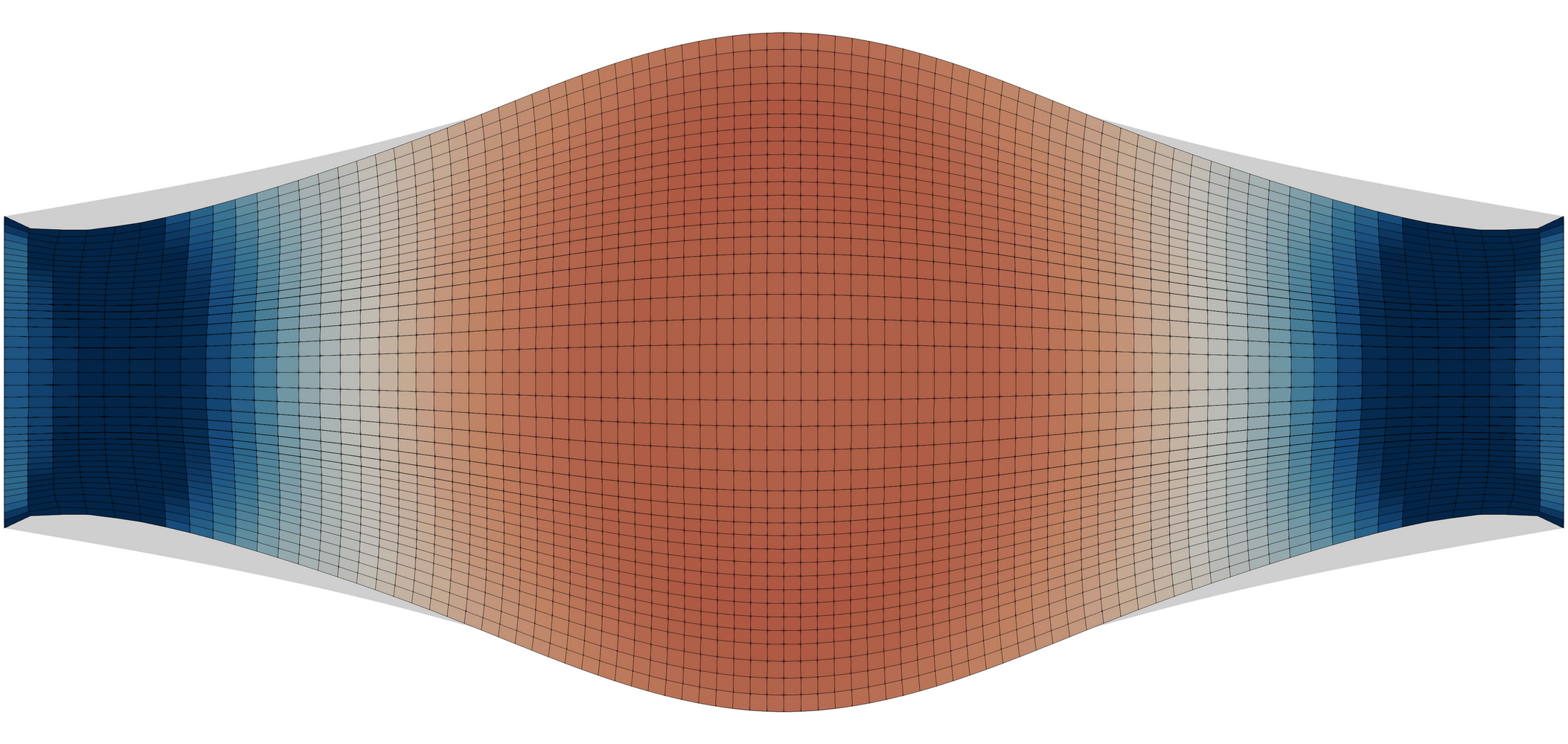}}
    \includegraphics[clip,trim={.5\wd0} 0 0 0, height=3cm]{figures/fig_10_slice_zx_fiber_stretch_weickenmeier_n4_ic_element.png}
    \endgroup \caption{\WKM} \label{fig:unknown}
    \end{subfigure}\hfill
    \begin{subfigure}[t]{0.18\textwidth}
    \centering \begingroup
    \sbox0{\includegraphics{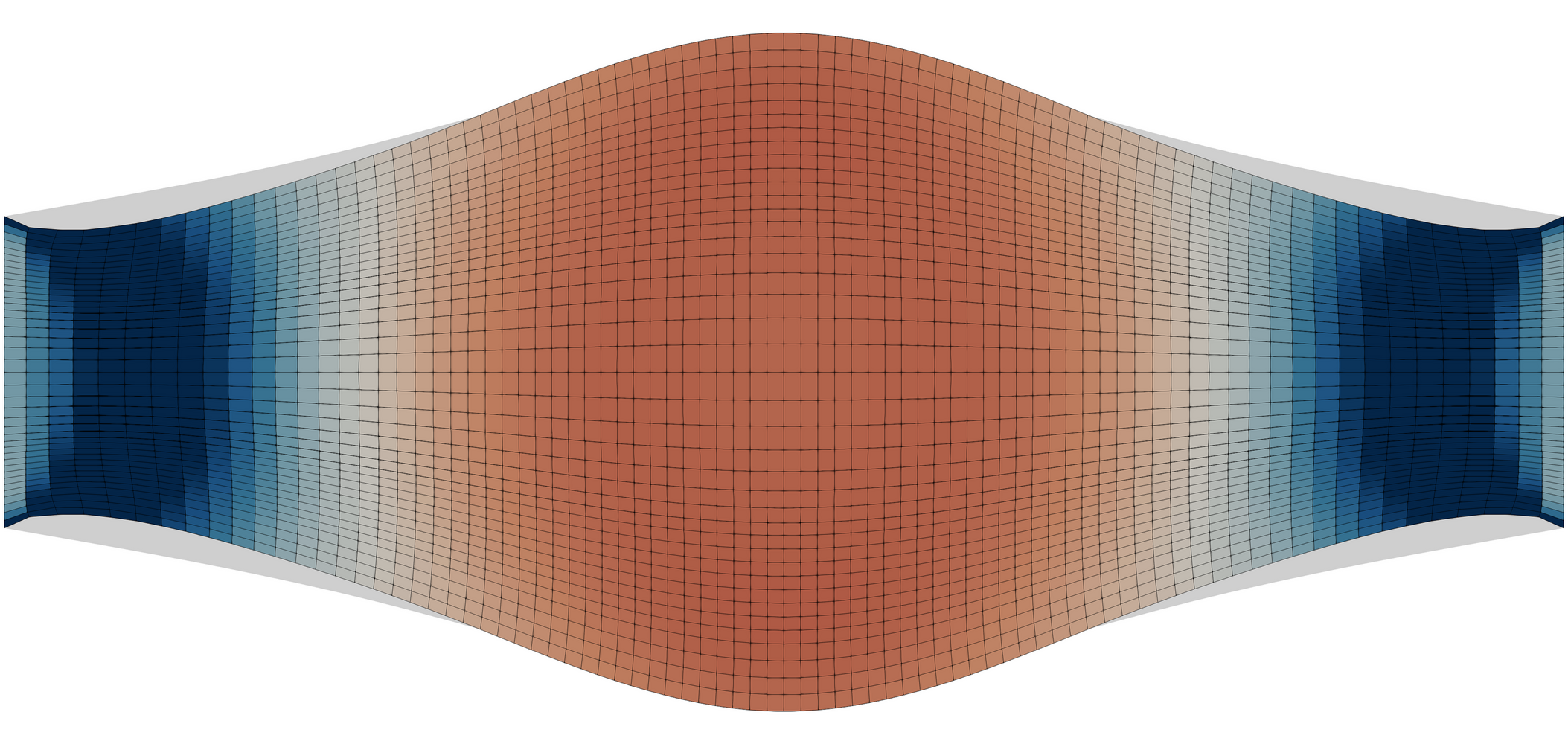}}
    \includegraphics[clip,trim={.5\wd0} 0 0 0, height=3cm]{figures/fig_10_slice_zx_fiber_stretch_combi_n4_ic_element.png}
    \endgroup \caption{\COMBI} \label{fig:unknown}
    \end{subfigure}
    \end{minipage}
    \hfill
    \begin{minipage}[t]{1.29cm}
        \raggedleft
    	\includegraphics[height=3cm]{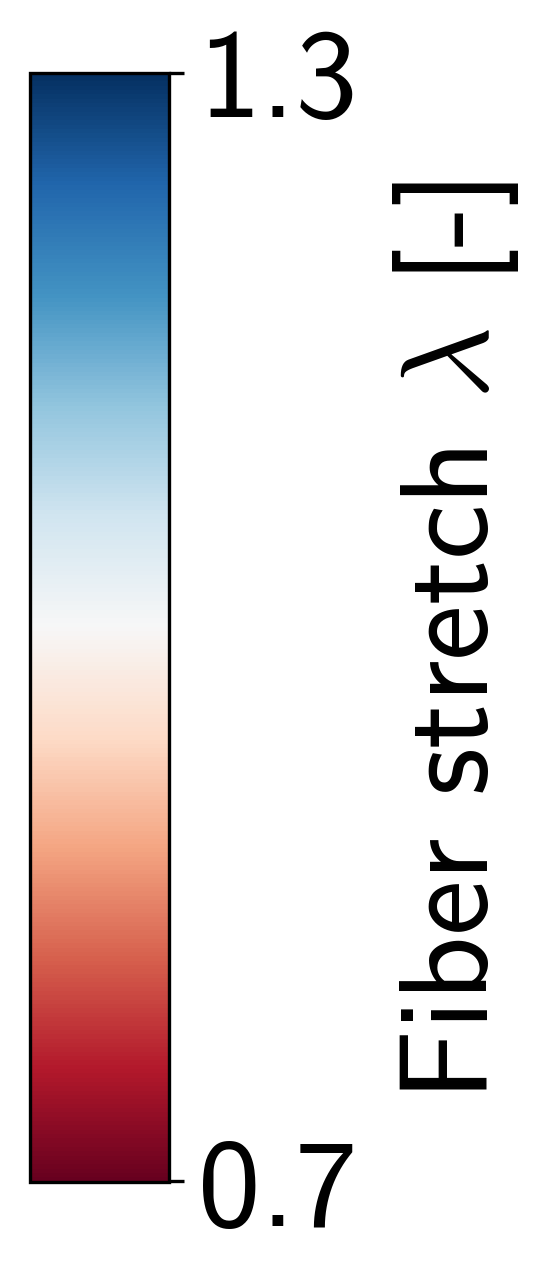}
    \end{minipage}
    \caption{Fiber stretch $\lambda$ in the axial cross-section of the fusiform muscle ($n=4$) for an isometric contraction in the tetanized state at $t=\SI{0.15}{\s}$. The initial configuration is displayed in grey. Only half the symmetric muscle is visualized.}
	\label{fig:fusi_iso_stretch_vis}
\end{figure}

\begin{figure}[bt]
\centering
\begin{minipage}[c]{0.86\textwidth}
\raggedright
    \parbox{\LWCap}{\centering \phantom{phantom text}} \hspace*{-10pt}
    \parbox{\LW}{\centering $t=\SI{0.005}{s}$ \\ 
    \vspace*{10pt}}
   \hspace*{6pt}
    \parbox{\LW}{\centering $t=\SI{0.02}{s}$ \\ 
    \vspace*{10pt}}
   \hspace*{6pt}
    \parbox{\LW}{\centering $t=\SI{0.15}{s}$ \\ 
    \vspace*{10pt}}
    \\ 
    \vspace*{6pt}
    \parbox{\LWCap}{\subcaption{\raggedright \BLE}} \hspace*{-10pt}
    \parbox{\LW}{ \begingroup\sbox0{\includegraphics{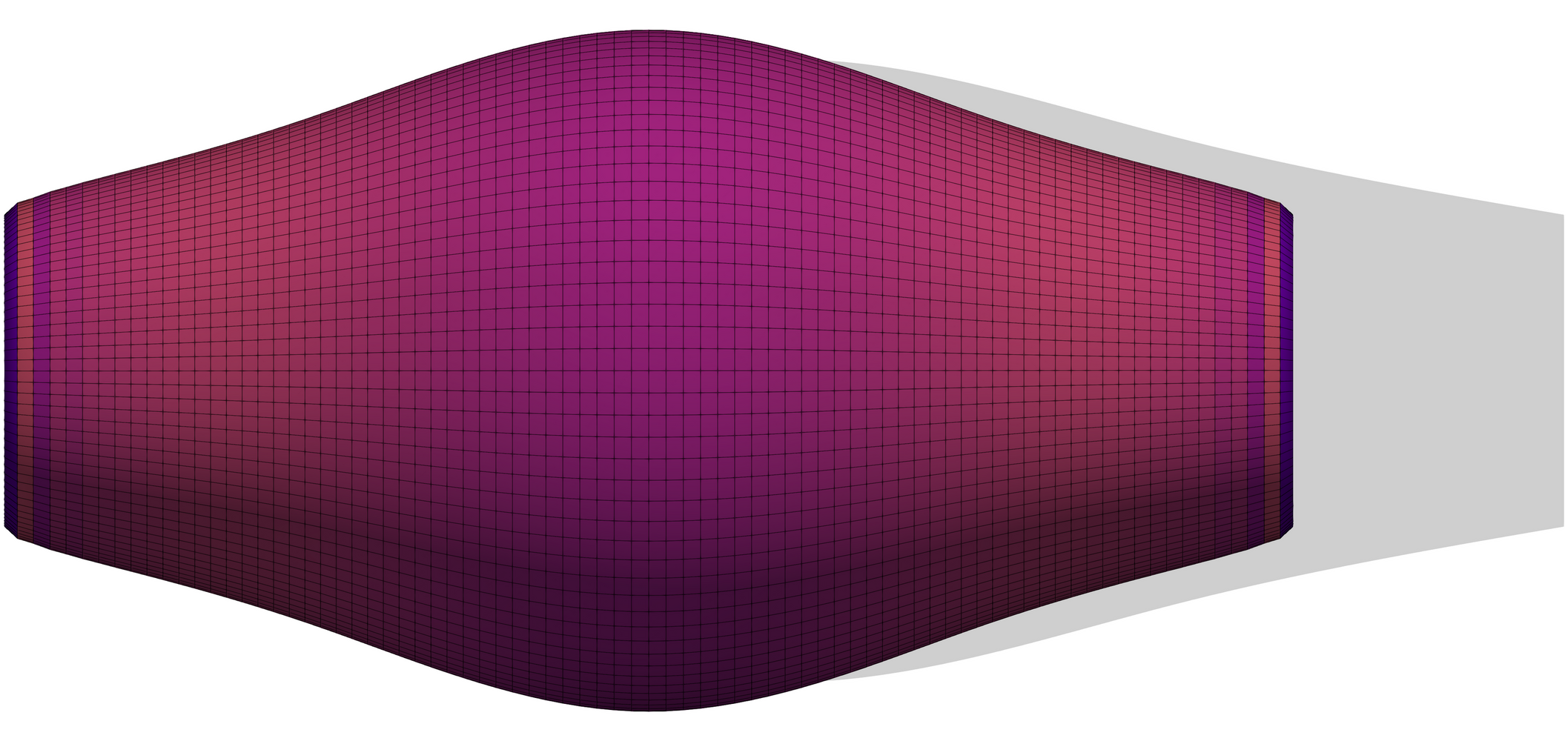}}
    \includegraphics[clip,trim=0 {.5\ht0} 0 0, width=\linewidth,left]{figures/fig_12_solid_fiber_stretch_blemker_n4_fc_element_0005.png}
    \endgroup 
    \\
    \begingroup\sbox0{\includegraphics{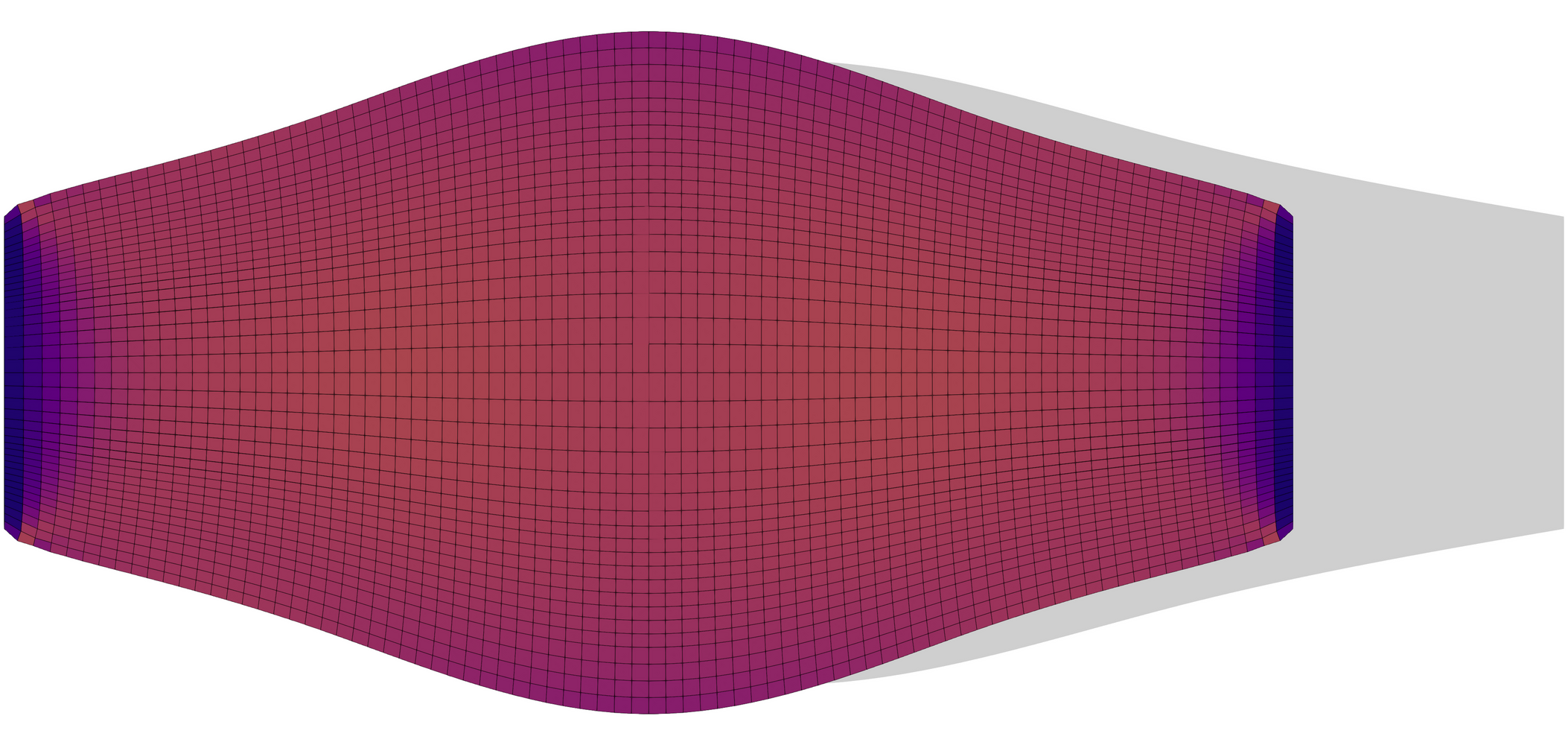}}
    \includegraphics[clip,trim=0 0 0 {.5\ht0}, width=\linewidth,left]{figures/fig_12_slice_zx_fiber_stretch_blemker_n4_fc_element_0005.png}
    \endgroup}\hspace*{6pt}
    \parbox{\LW}{ \begingroup\sbox0{\includegraphics{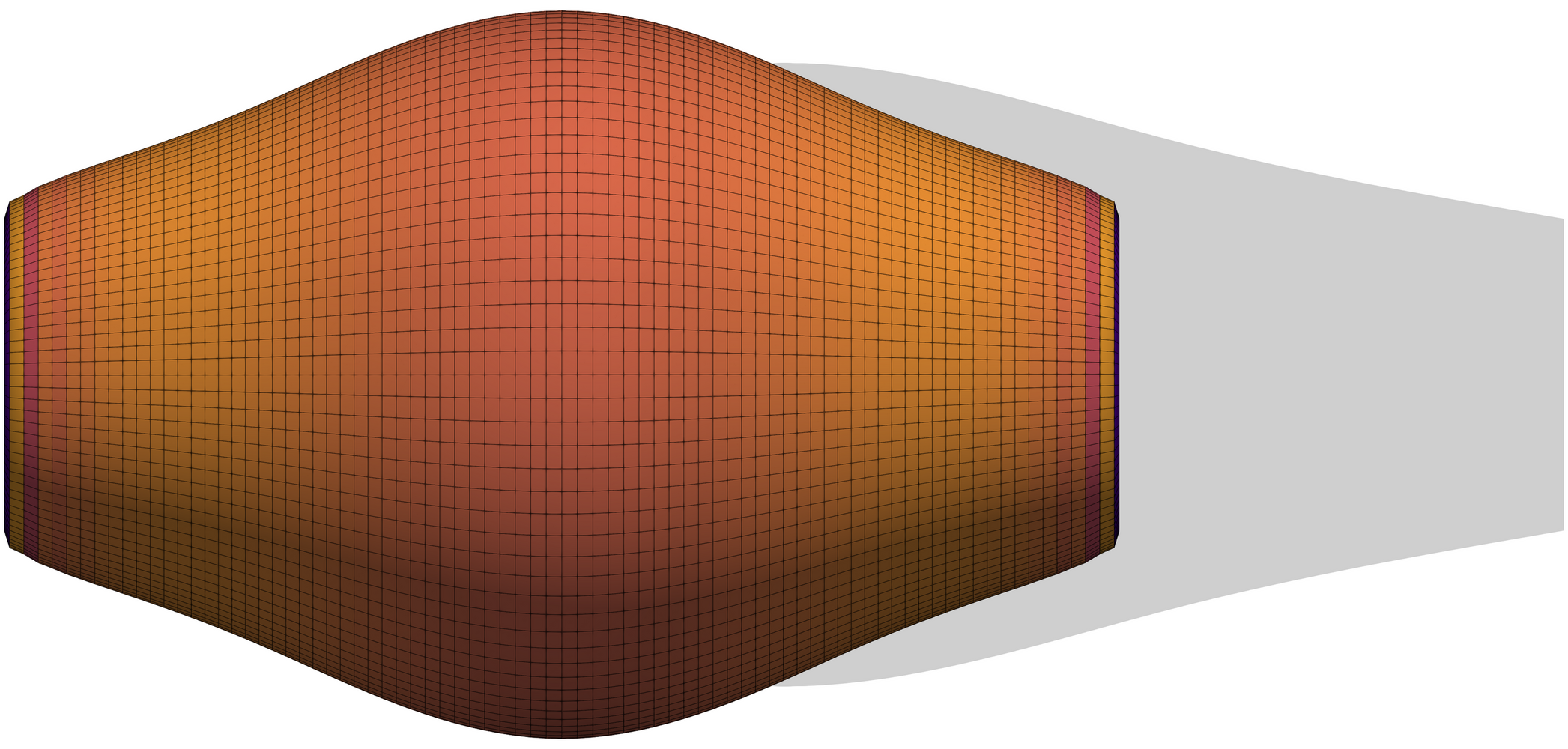}}
    \includegraphics[clip,trim=0 {.5\ht0} 0 0, width=\linewidth,left]{figures/fig_12_solid_fiber_stretch_blemker_n4_fc_element_0020.png}
    \endgroup
    \\
    \begingroup\sbox0{\includegraphics{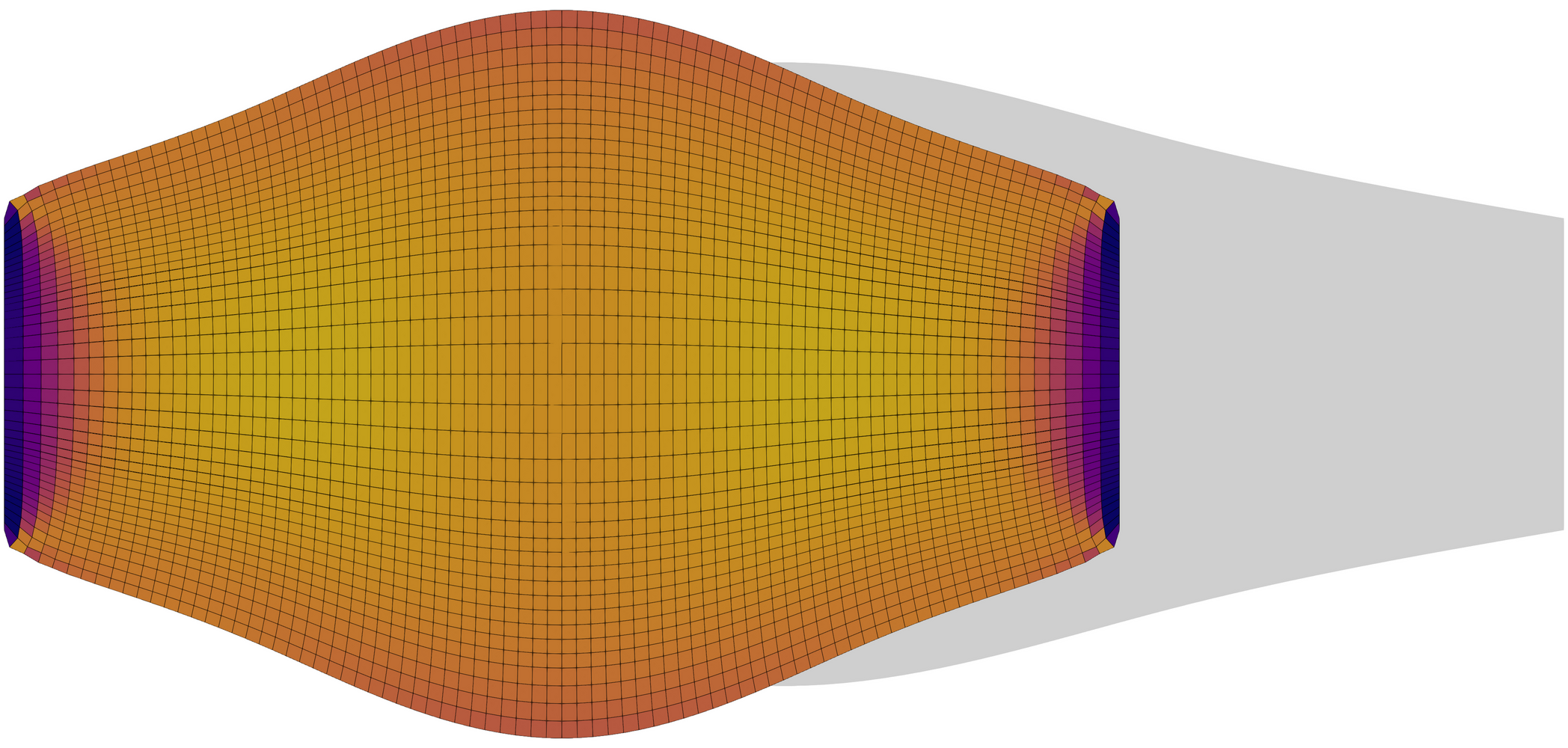}}
    \includegraphics[clip,trim=0 0 0 {.5\ht0}, width=\linewidth,left]{figures/fig_12_slice_zx_fiber_stretch_blemker_n4_fc_element_0020.png}
    \endgroup}\hspace*{6pt}
    \parbox{\LW}{ \begingroup\sbox0{\includegraphics{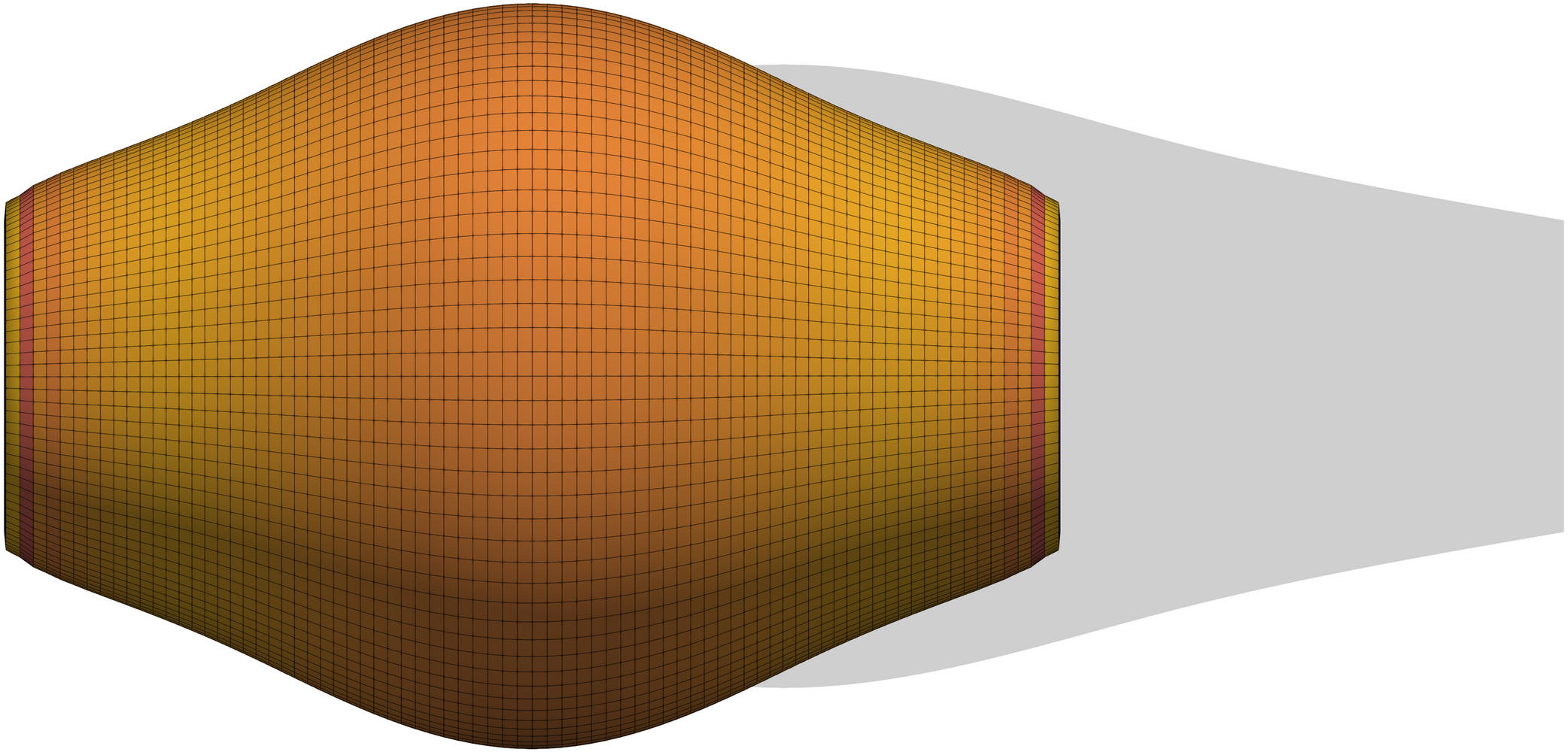}}
    \includegraphics[clip,trim=0 {.5\ht0} 0 0, width=\linewidth,left]{figures/fig_12_solid_fiber_stretch_blemker_n4_fc_element_0150.png}
    \endgroup
    \\
    \begingroup\sbox0{\includegraphics{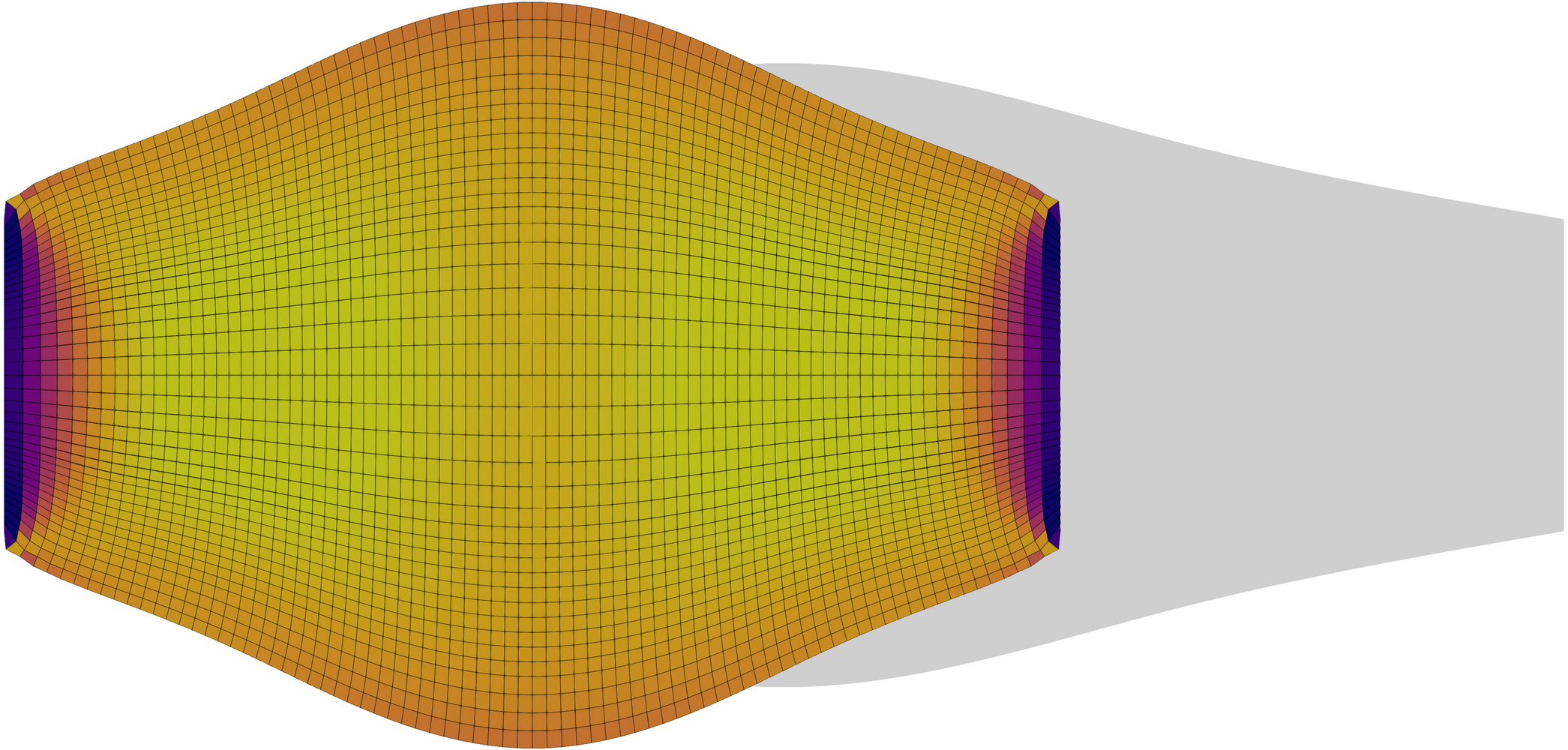}}
    \includegraphics[clip,trim=0 0 0 {.5\ht0}, width=\linewidth,left]{figures/fig_12_slice_zx_fiber_stretch_blemker_n4_fc_element_0150.png}
    \endgroup} \\ 
    \vspace*{6pt}
    \parbox{\LWCap}{\subcaption{\raggedright \GIANT}} \hspace*{-10pt}
    \parbox{\LW}{ \begingroup\sbox0{\includegraphics{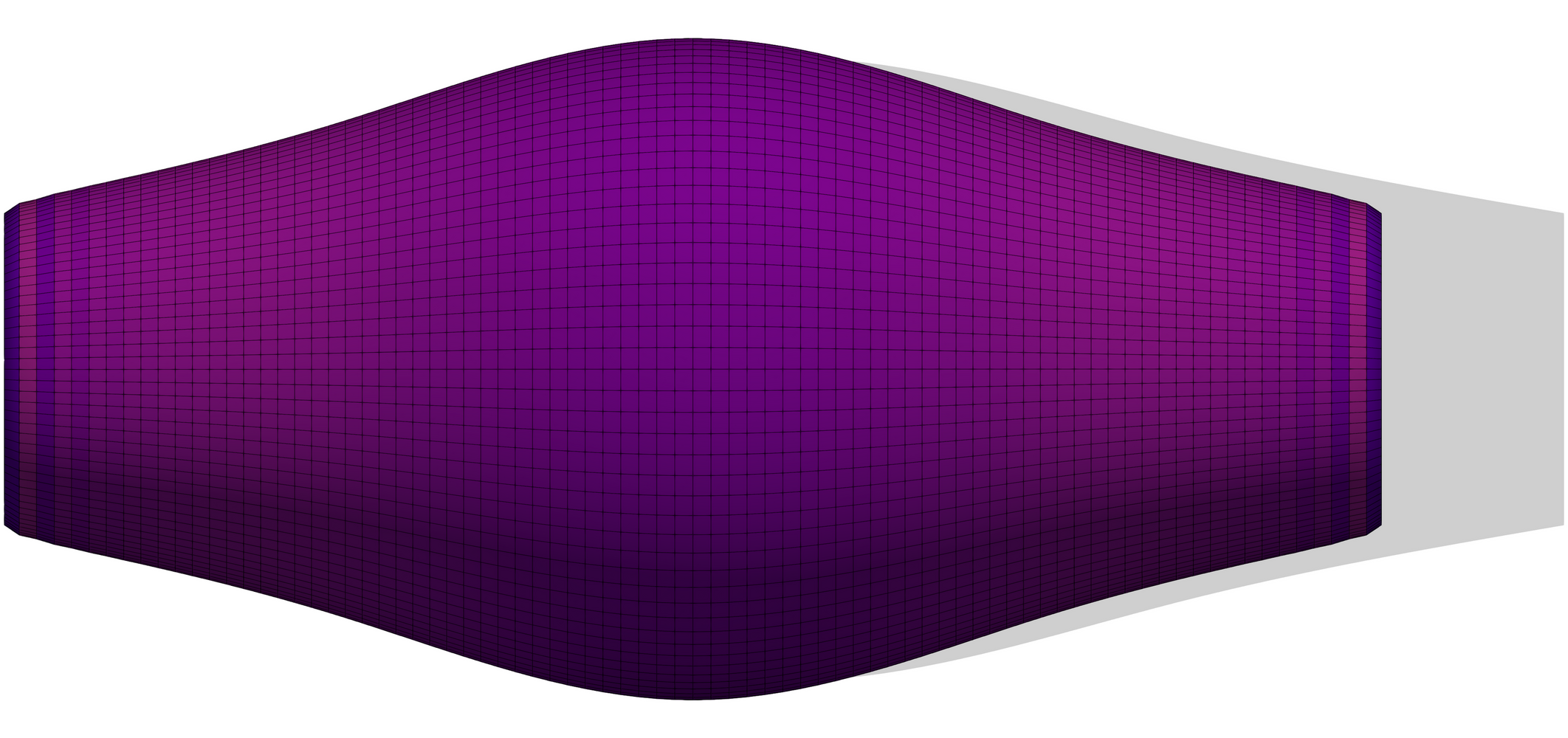}}
    \includegraphics[clip,trim=0 {.5\ht0} 0 0, width=\linewidth,left]{figures/fig_12_solid_fiber_stretch_giantesio_n4_fc_element_0005.png}
    \endgroup 
    \\ 
    \begingroup\sbox0{\includegraphics{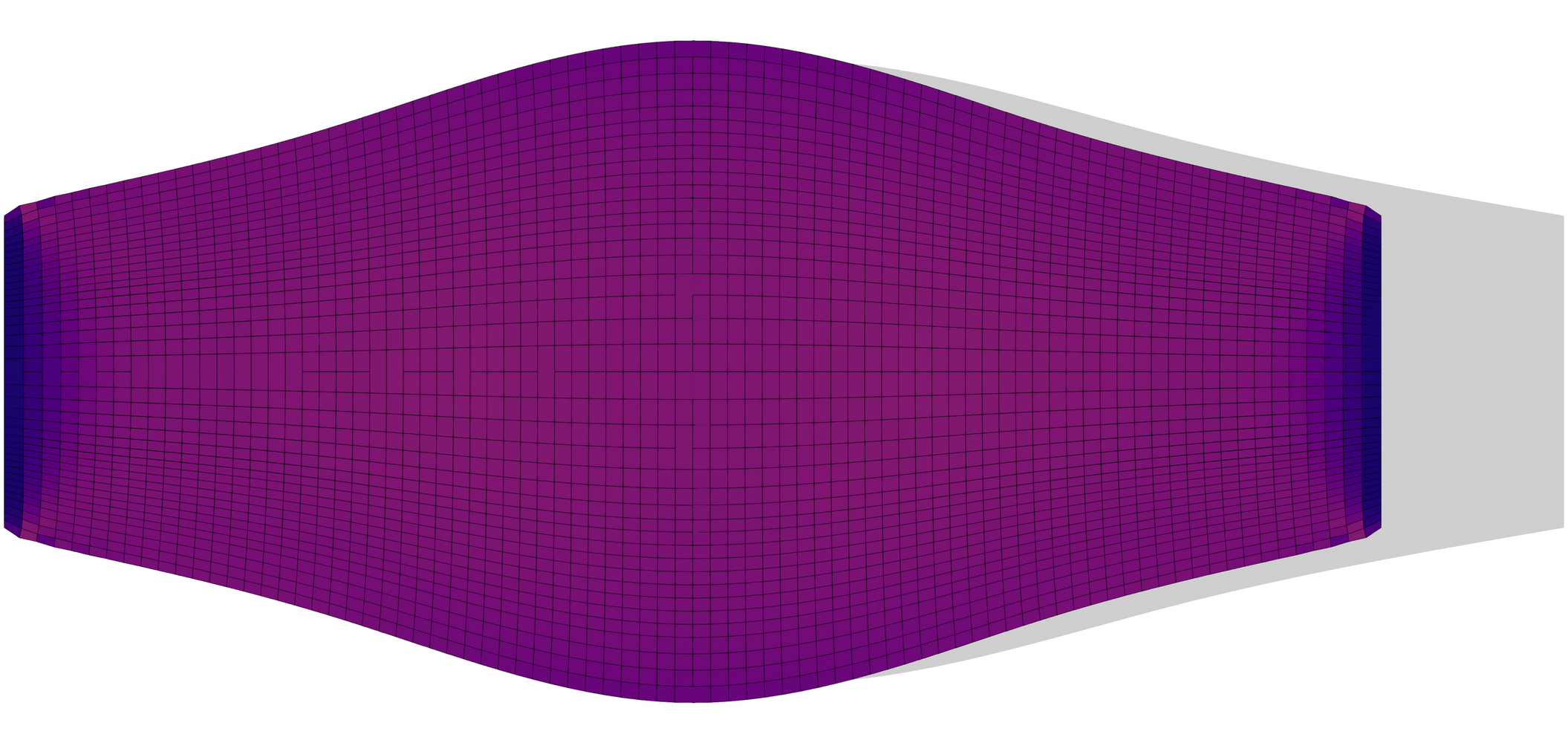}}
    \includegraphics[clip,trim=0 0 0 {.5\ht0}, width=\linewidth,left]{figures/fig_12_slice_zx_fiber_stretch_giantesio_n4_fc_element_0005.png}
    \endgroup}\hspace*{6pt}
    \parbox{\LW}{ \begingroup\sbox0{\includegraphics{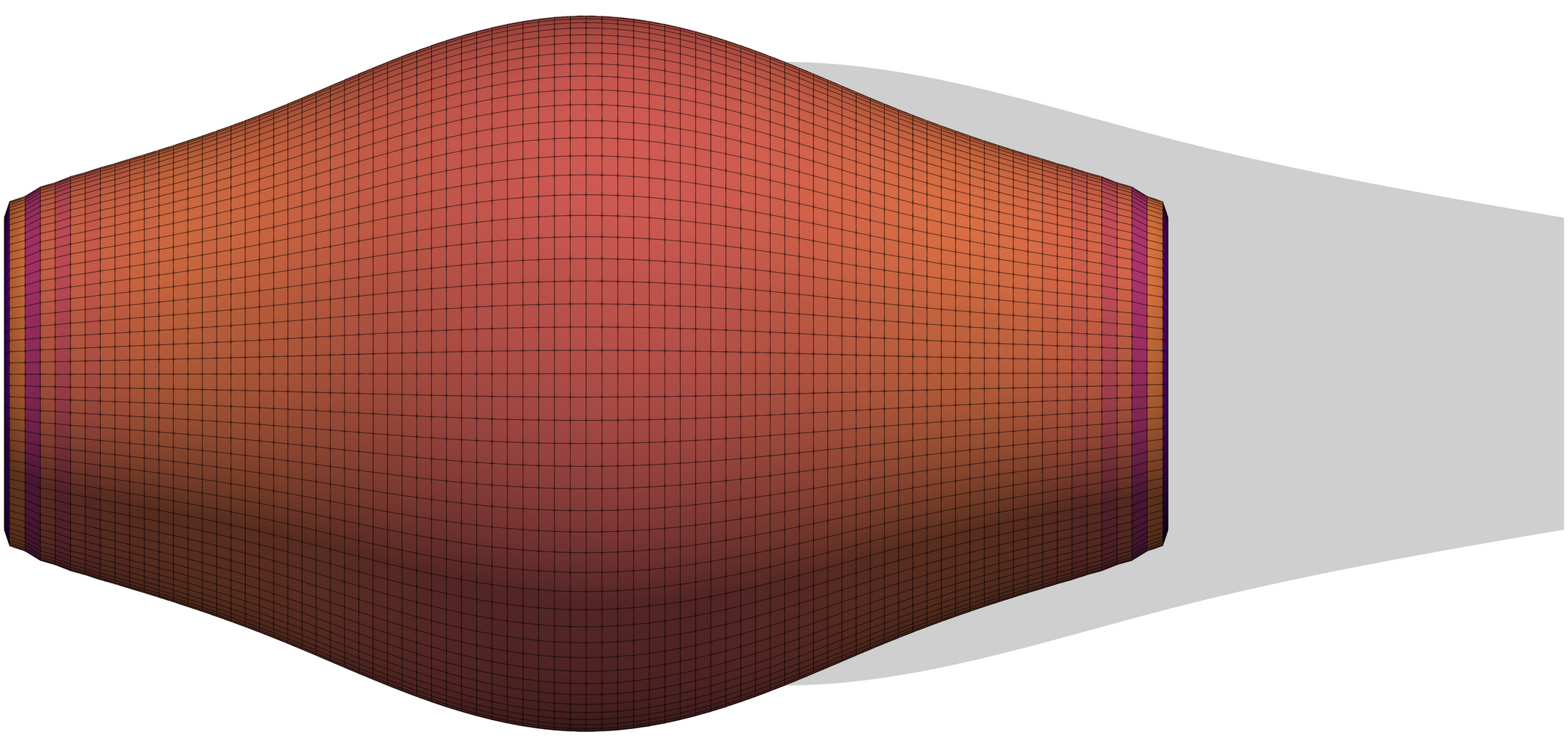}}
    \includegraphics[clip,trim=0 {.5\ht0} 0 0, width=\linewidth,left]{figures/fig_12_solid_fiber_stretch_giantesio_n4_fc_element_0020.png}
    \endgroup
    \\ 
    \begingroup\sbox0{\includegraphics{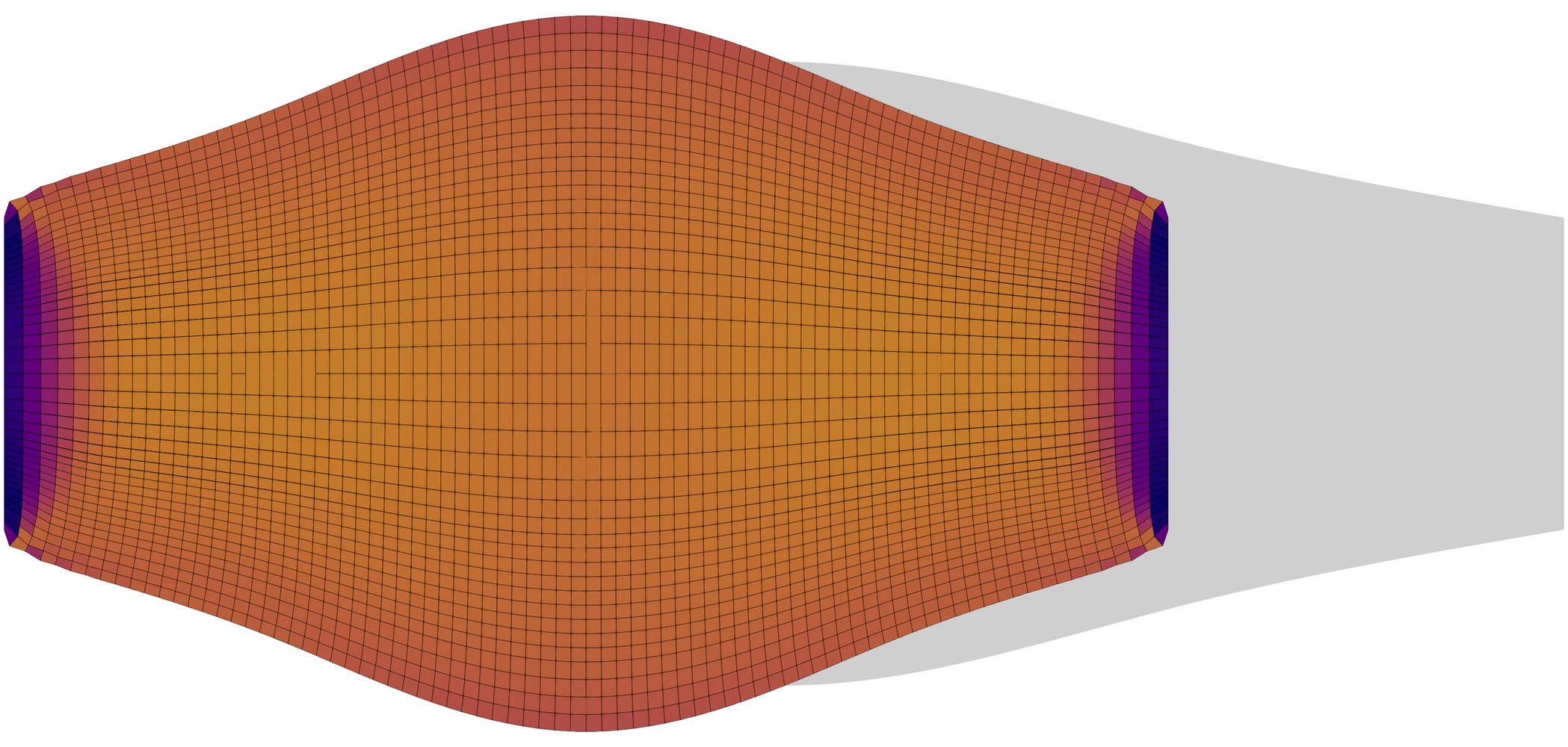}}
    \includegraphics[clip,trim=0 0 0 {.5\ht0}, width=\linewidth,left]{figures/fig_12_slice_zx_fiber_stretch_giantesio_n4_fc_element_0020.png}
    \endgroup}\hspace*{6pt}
    \parbox{\LW}{ \begingroup\sbox0{\includegraphics{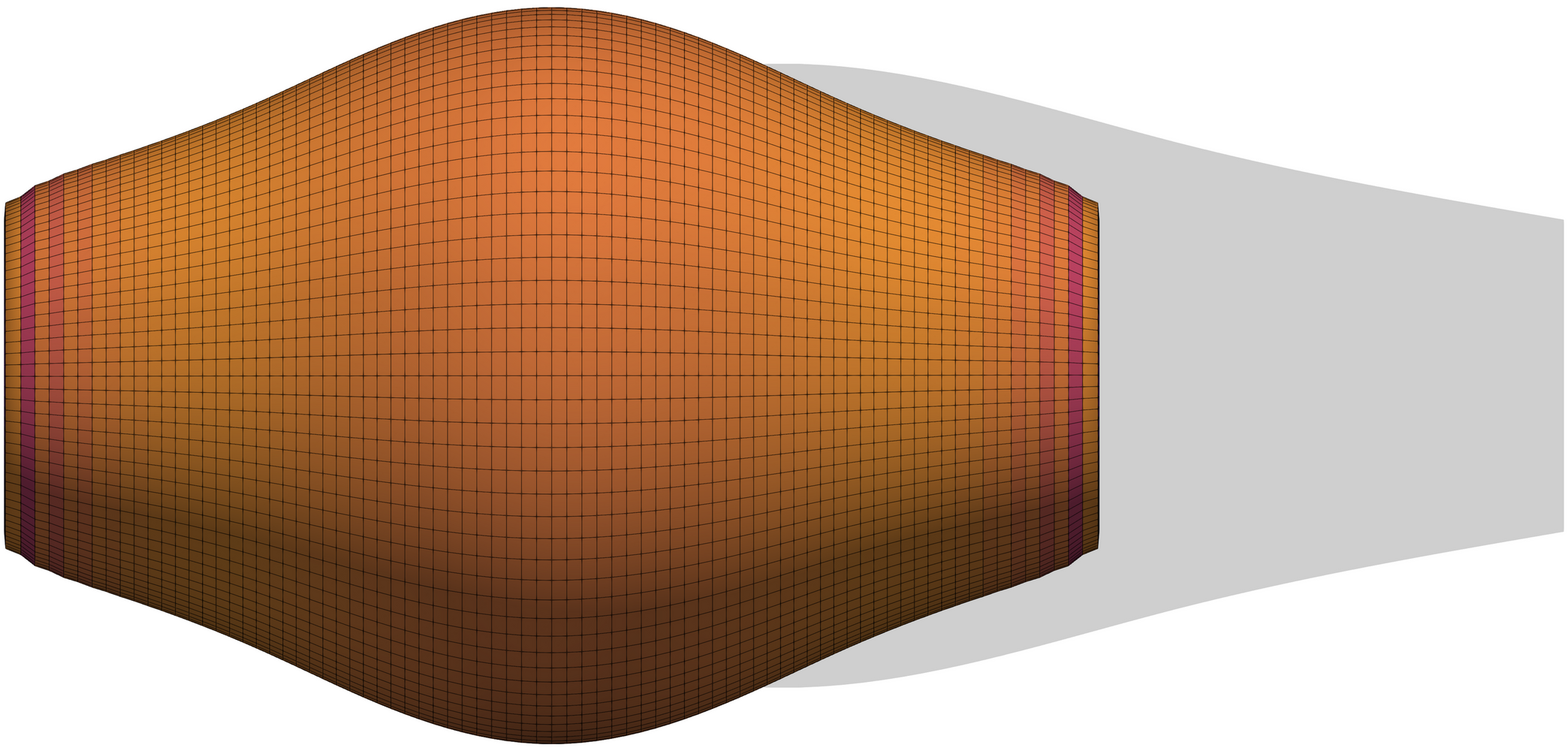}}
    \includegraphics[clip,trim=0 {.5\ht0} 0 0, width=\linewidth,left]{figures/fig_12_solid_fiber_stretch_giantesio_n4_fc_element_0150.png}
    \endgroup
    \\ 
    \begingroup\sbox0{\includegraphics{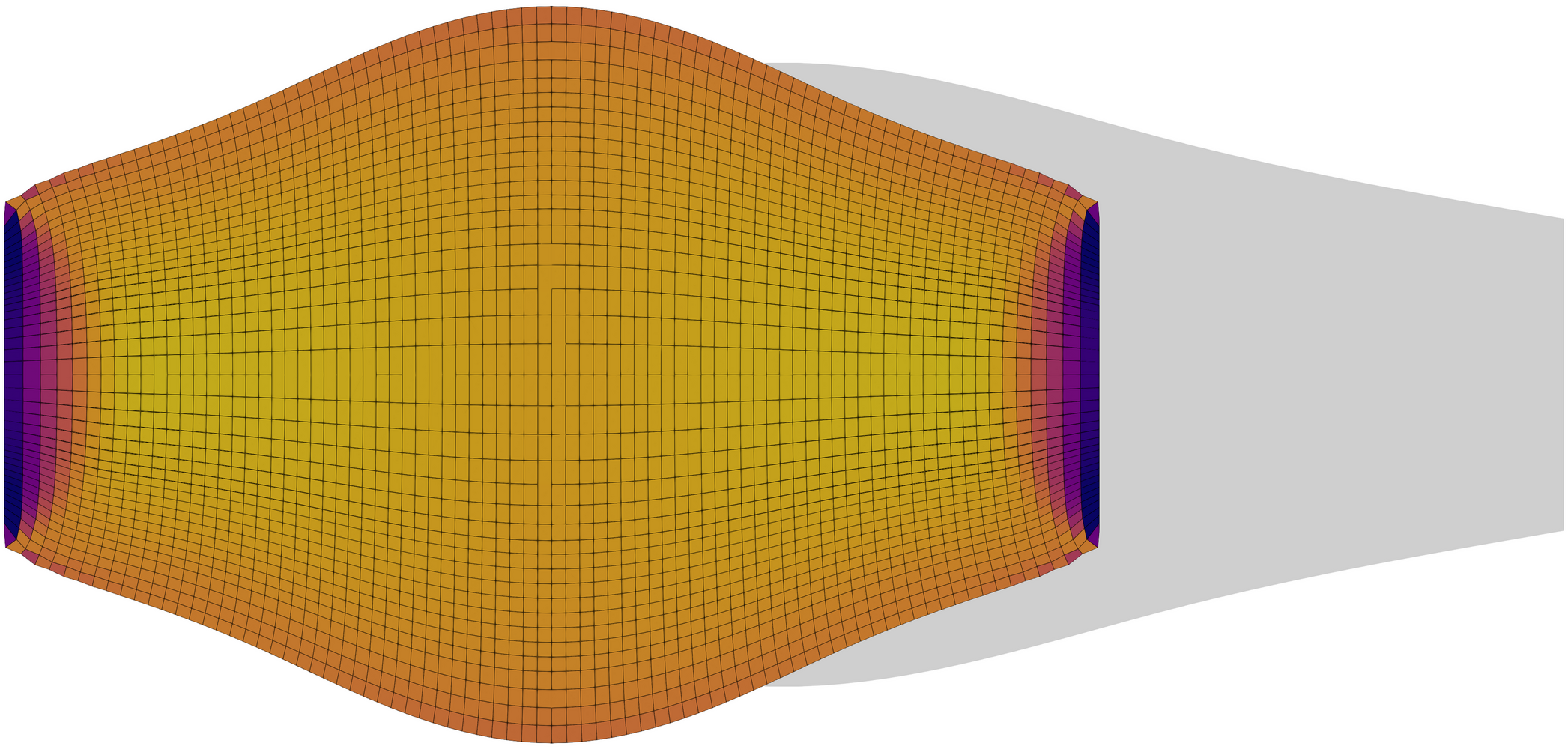}}
    \includegraphics[clip,trim=0 0 0 {.5\ht0}, width=\linewidth,left]{figures/fig_12_slice_zx_fiber_stretch_giantesio_n4_fc_element_0150.png}
    \endgroup} \\
    \vspace*{6pt}
    \parbox{\LWCap}{\subcaption{\raggedright \WKM}} \hspace*{-10pt}
    \parbox{\LW}{ \begingroup\sbox0{\includegraphics{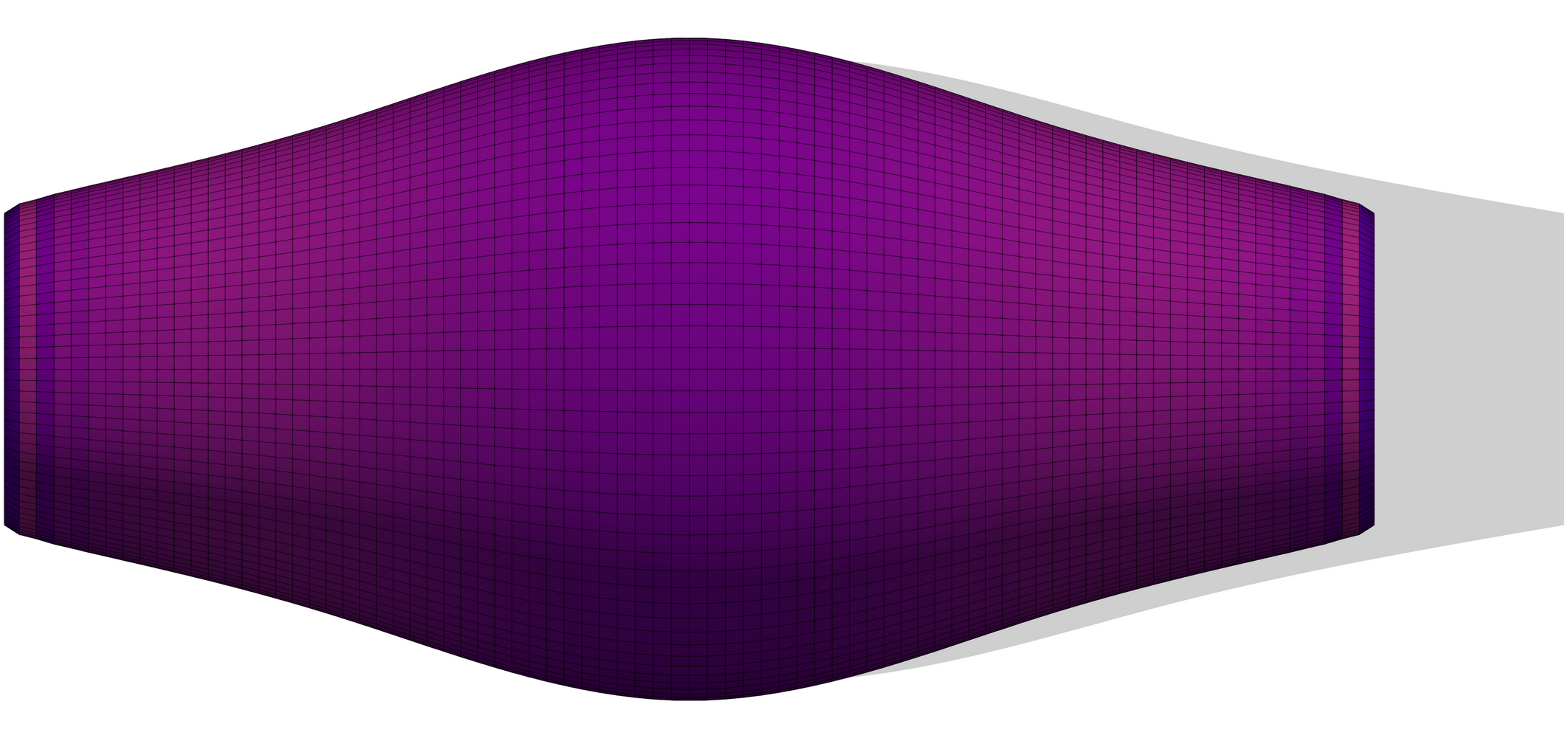}}
    \includegraphics[clip,trim=0 {.5\ht0} 0 0, width=\linewidth,left]{figures/fig_12_solid_fiber_stretch_weickenmeier_n4_fc_element_0005.png}
    \endgroup 
    \\
    \begingroup\sbox0{\includegraphics{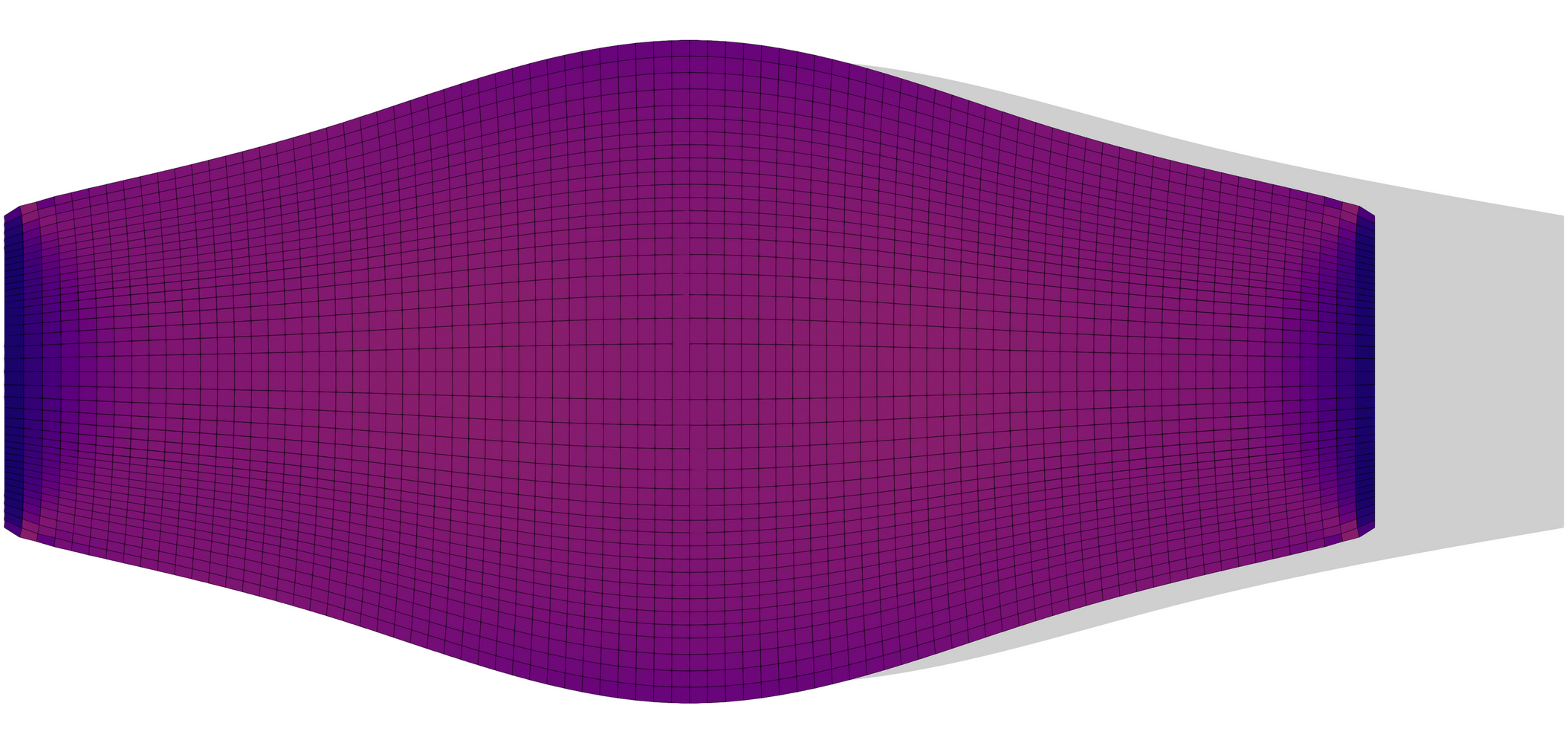}}
    \includegraphics[clip,trim=0 0 0 {.5\ht0}, width=\linewidth,left]{figures/fig_12_slice_zx_fiber_stretch_weickenmeier_n4_fc_element_0005.png}
    \endgroup}\hspace*{6pt}
    \parbox{\LW}{ \begingroup\sbox0{\includegraphics{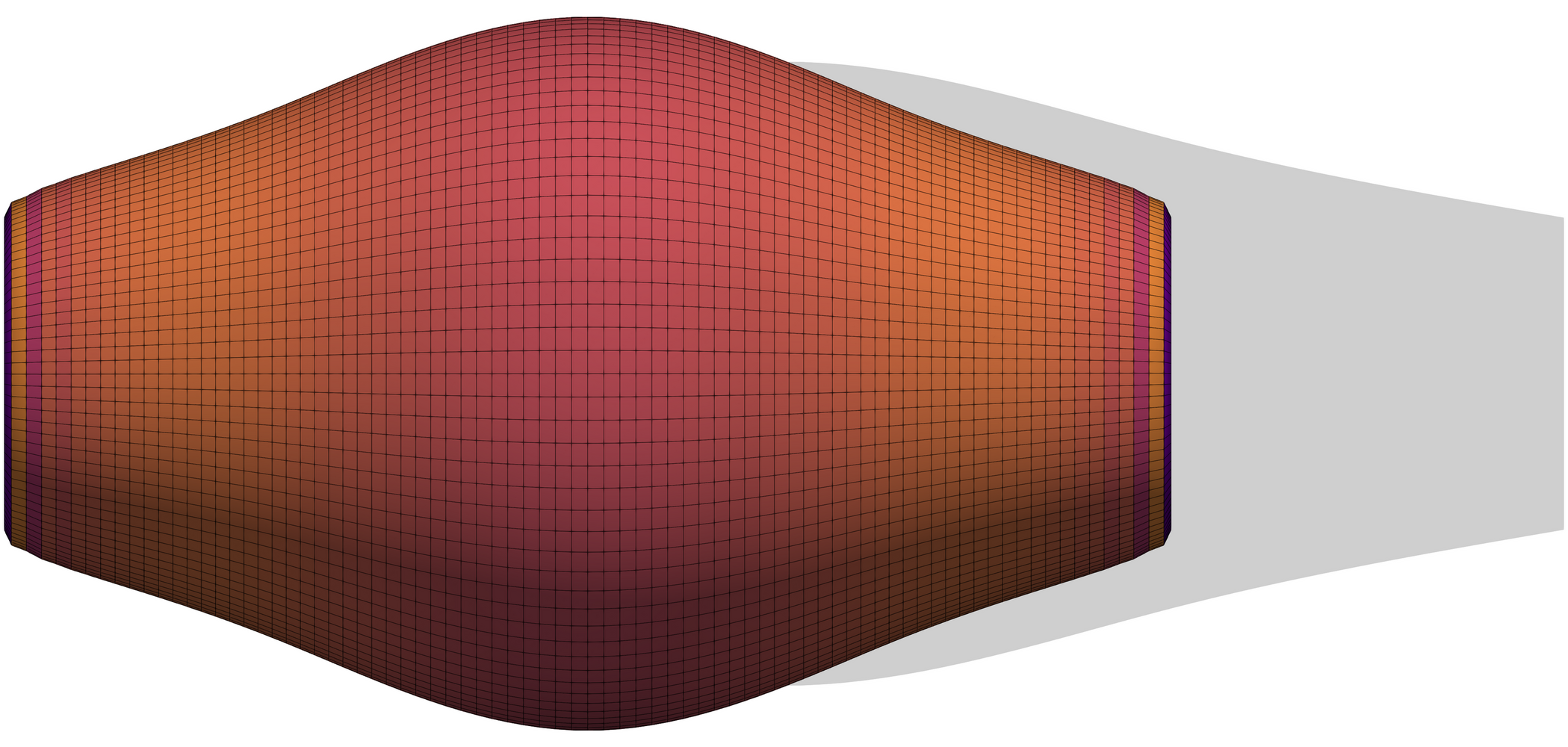}}
    \includegraphics[clip,trim=0 {.5\ht0} 0 0, width=\linewidth,left]{figures/fig_12_solid_fiber_stretch_weickenmeier_n4_fc_element_0020.png}
    \endgroup
    \\
    \begingroup\sbox0{\includegraphics{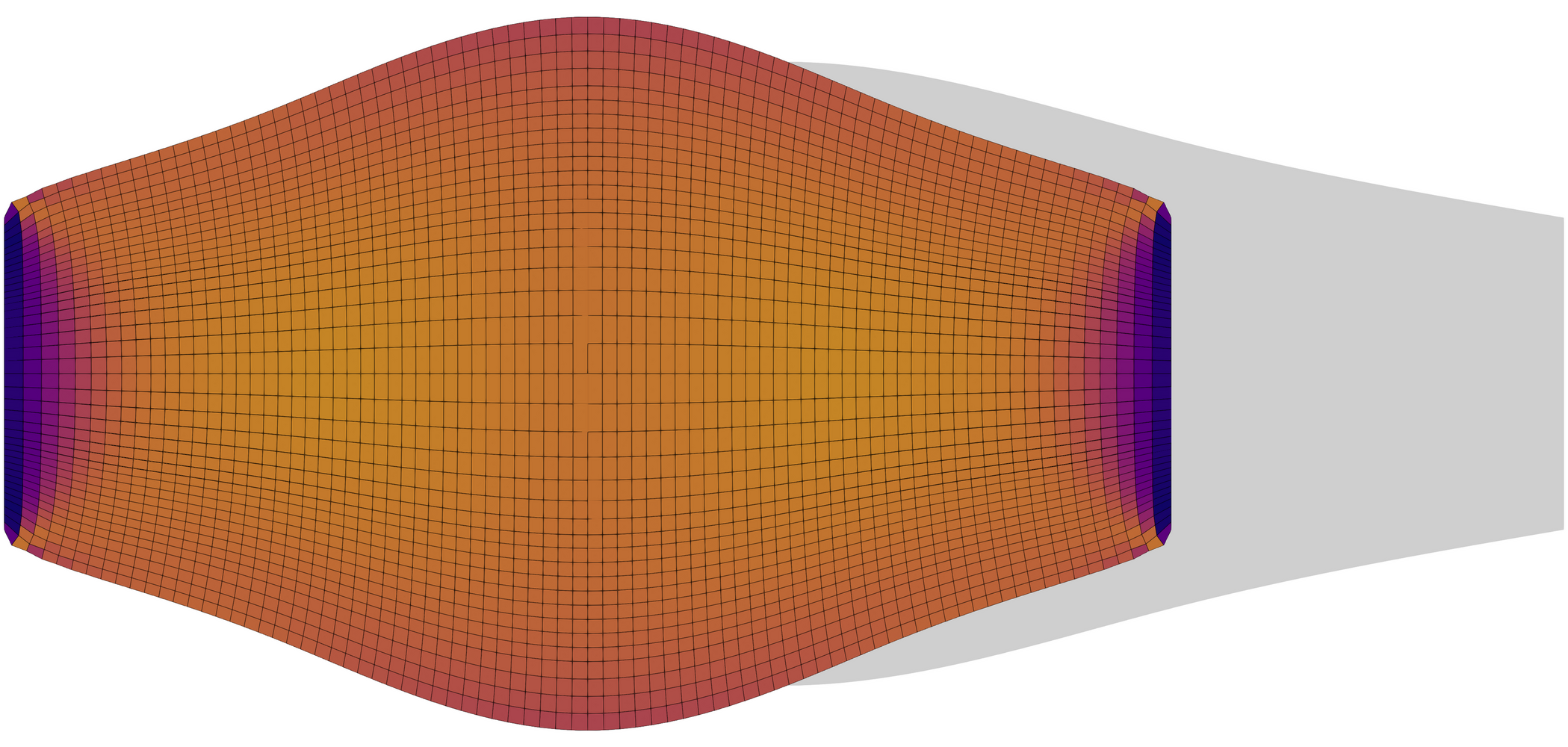}}
    \includegraphics[clip,trim=0 0 0 {.5\ht0}, width=\linewidth,left]{figures/fig_12_slice_zx_fiber_stretch_weickenmeier_n4_fc_element_0020.png}
    \endgroup}\hspace*{6pt}
    \parbox{\LW}{ \begingroup\sbox0{\includegraphics{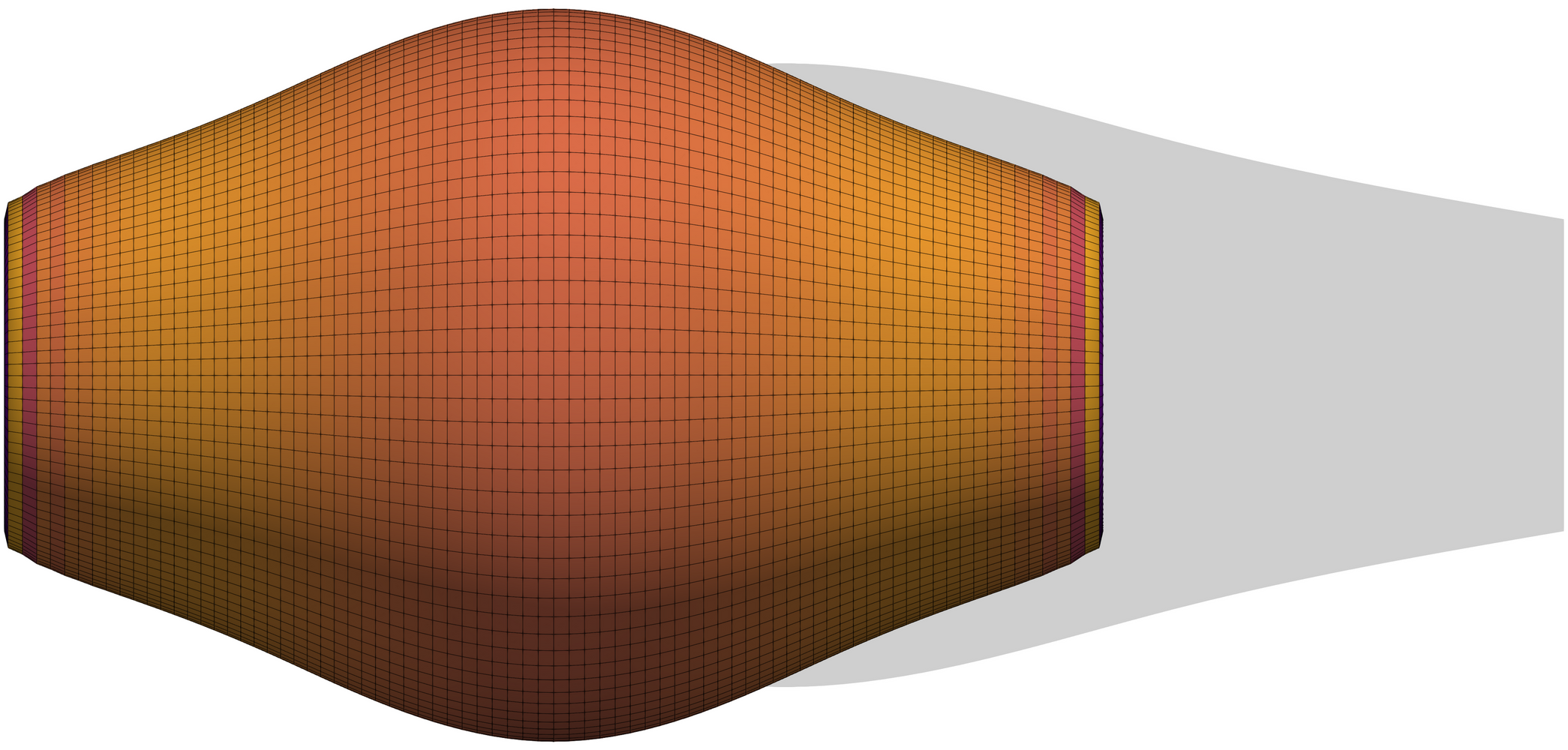}}
    \includegraphics[clip,trim=0 {.5\ht0} 0 0, width=\linewidth,left]{figures/fig_12_solid_fiber_stretch_weickenmeier_n4_fc_element_0150.png}
    \endgroup
    \\
    \begingroup\sbox0{\includegraphics{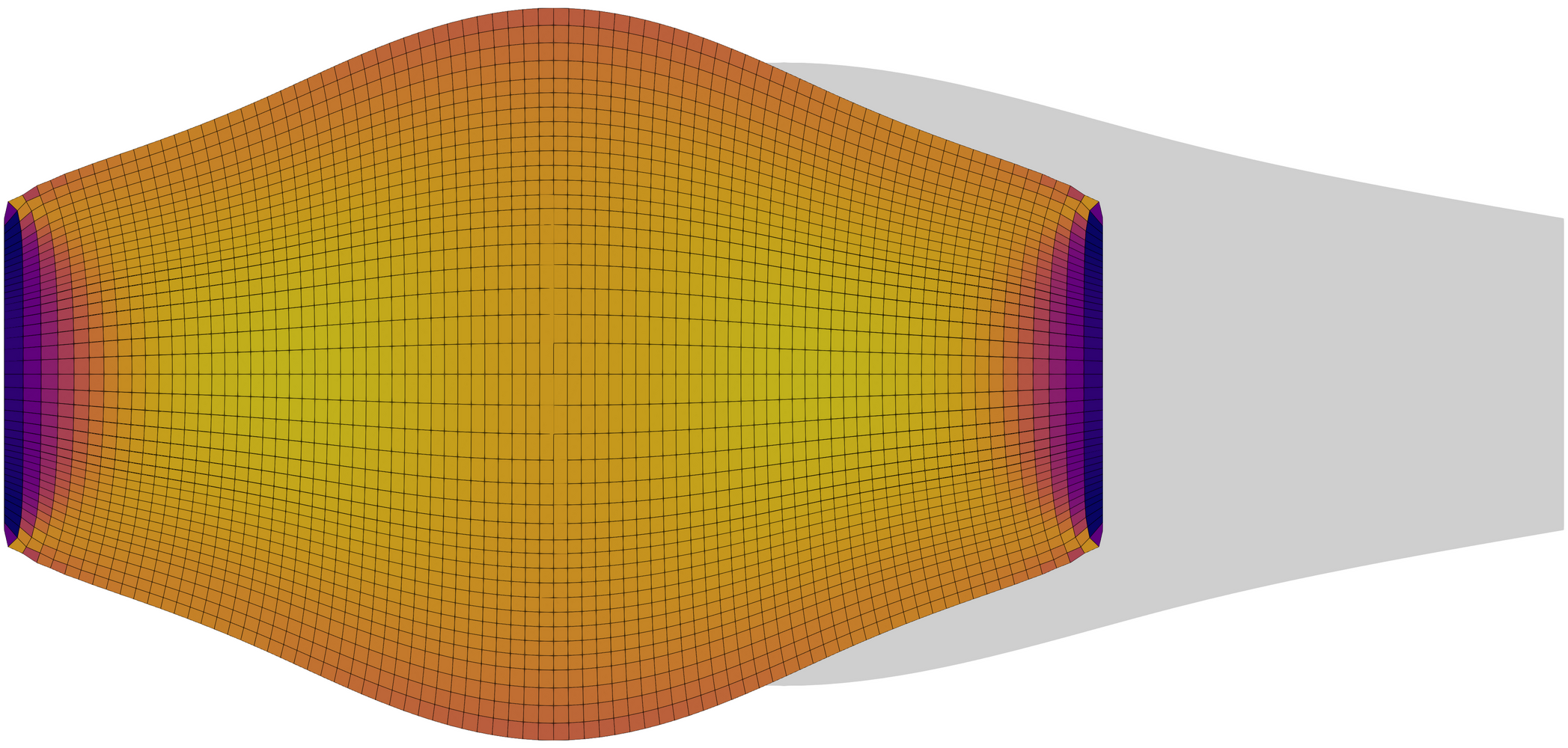}}
    \includegraphics[clip,trim=0 0 0 {.5\ht0}, width=\linewidth,left]{figures/fig_12_slice_zx_fiber_stretch_weickenmeier_n4_fc_element_0150.png}
    \endgroup}\\
    \vspace*{6pt}
    \parbox{\LWCap}{\subcaption{\raggedright \COMBI}} \hspace*{-10pt}
    \parbox{\LW}{ \begingroup\sbox0{\includegraphics{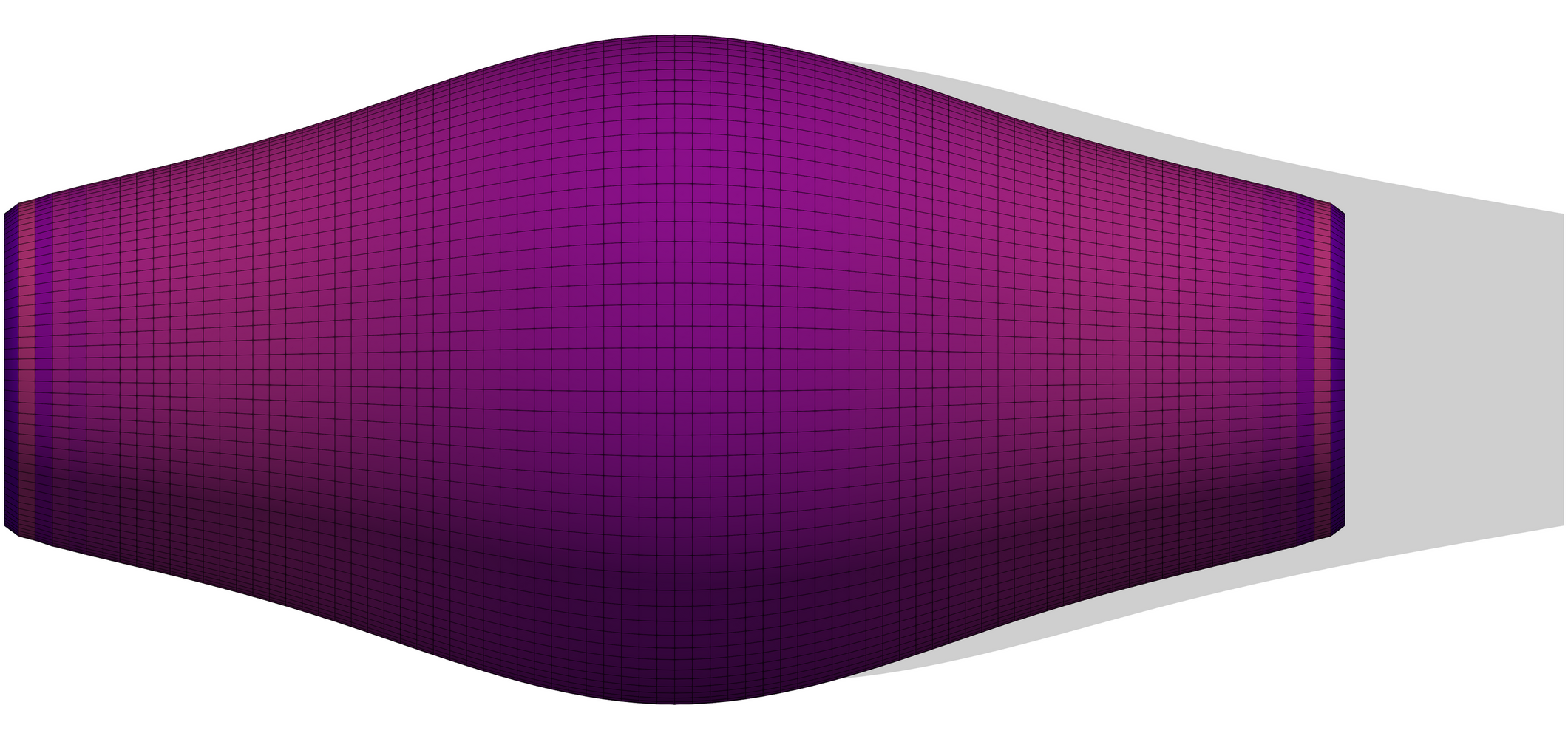}}
    \includegraphics[clip,trim=0 {.5\ht0} 0 0, width=\linewidth,left]{figures/fig_12_solid_fiber_stretch_combi_n4_fc_element_0005.png}
    \endgroup 
    \\
    \begingroup\sbox0{\includegraphics{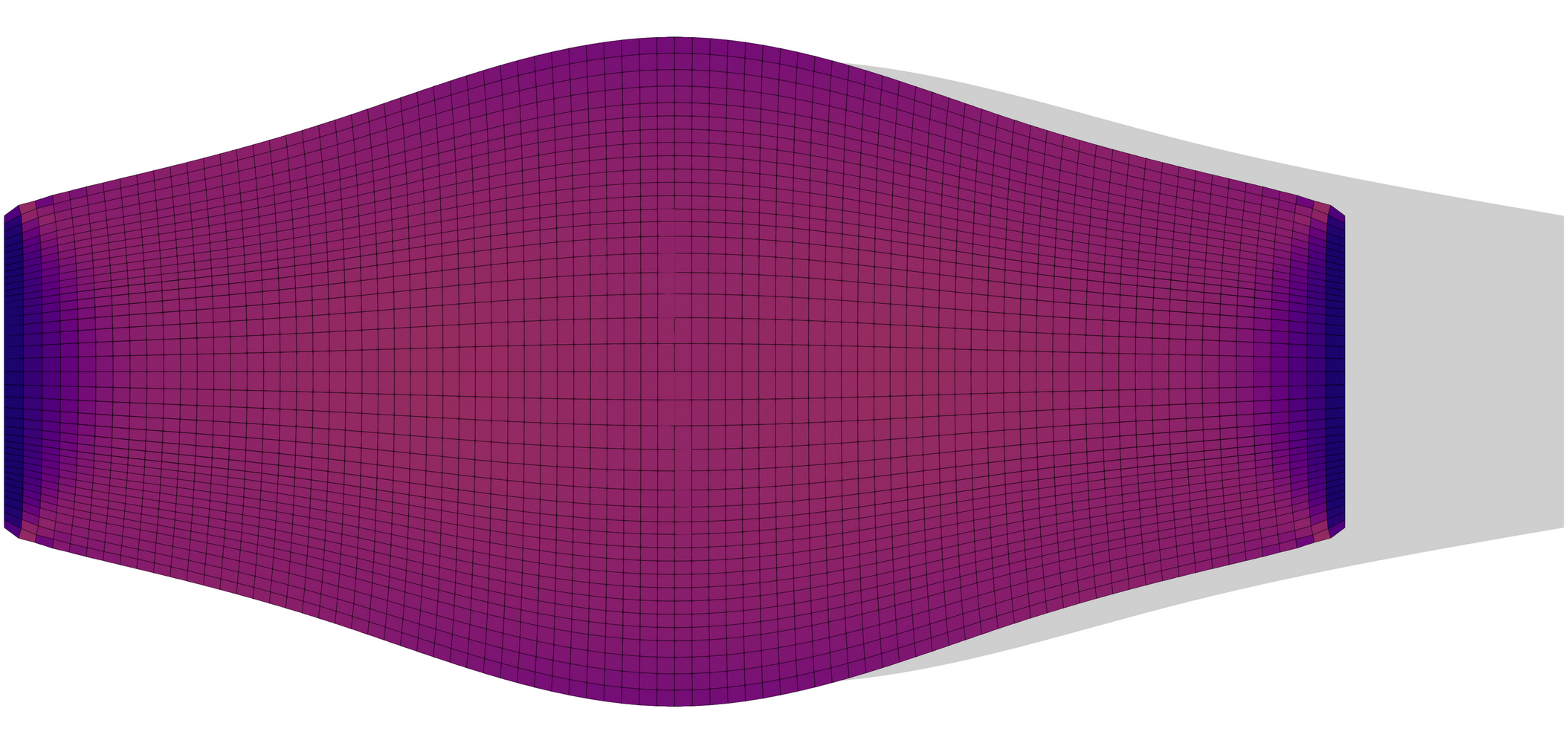}}
    \includegraphics[clip,trim=0 0 0 {.5\ht0}, width=\linewidth,left]{figures/fig_12_slice_zx_fiber_stretch_combi_n4_fc_element_0005.png}
    \endgroup}\hspace*{6pt}
    \parbox{\LW}{ \begingroup\sbox0{\includegraphics{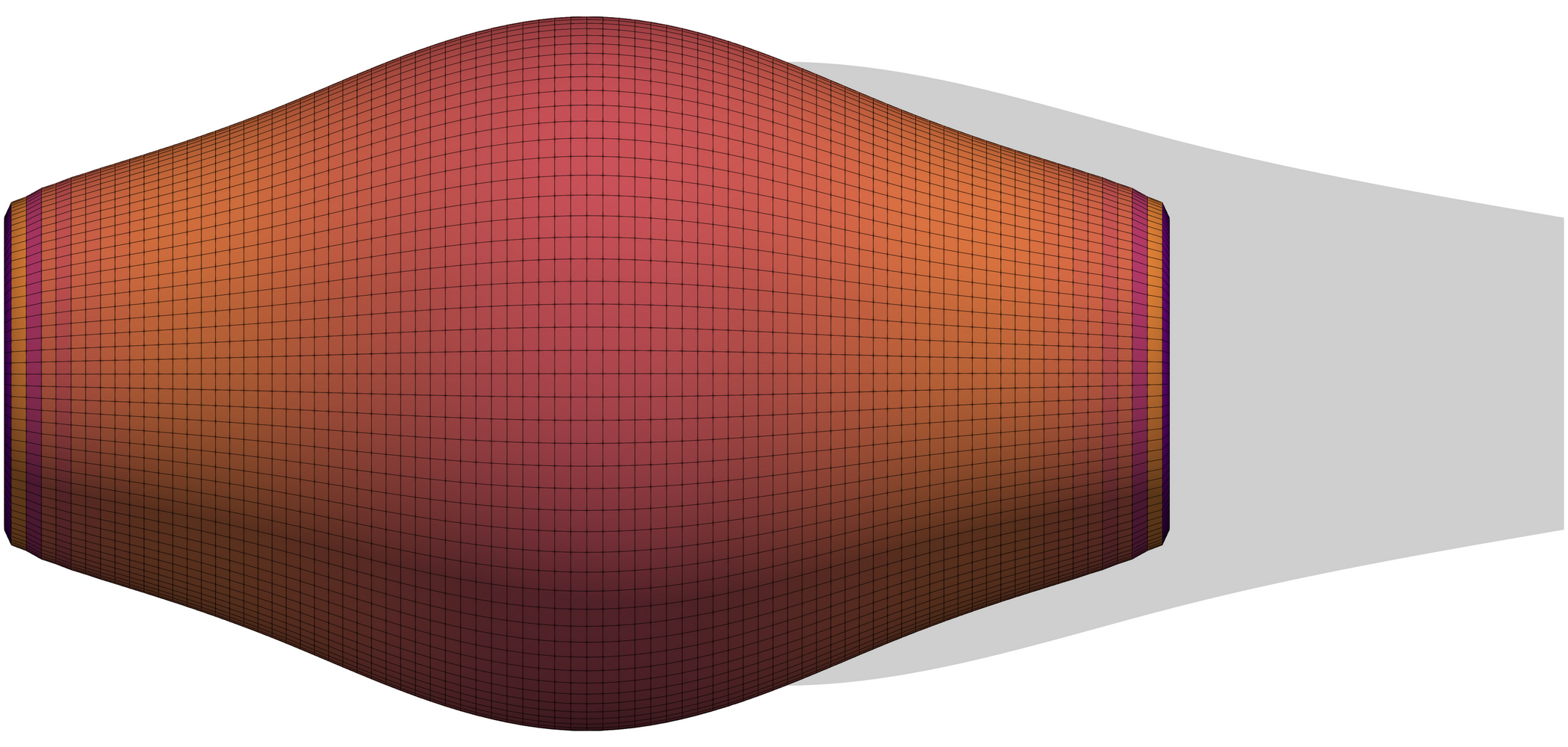}}
    \includegraphics[clip,trim=0 {.5\ht0} 0 0, width=\linewidth,left]{figures/fig_12_solid_fiber_stretch_combi_n4_fc_element_0020.png}
    \endgroup
    \\
    \begingroup\sbox0{\includegraphics{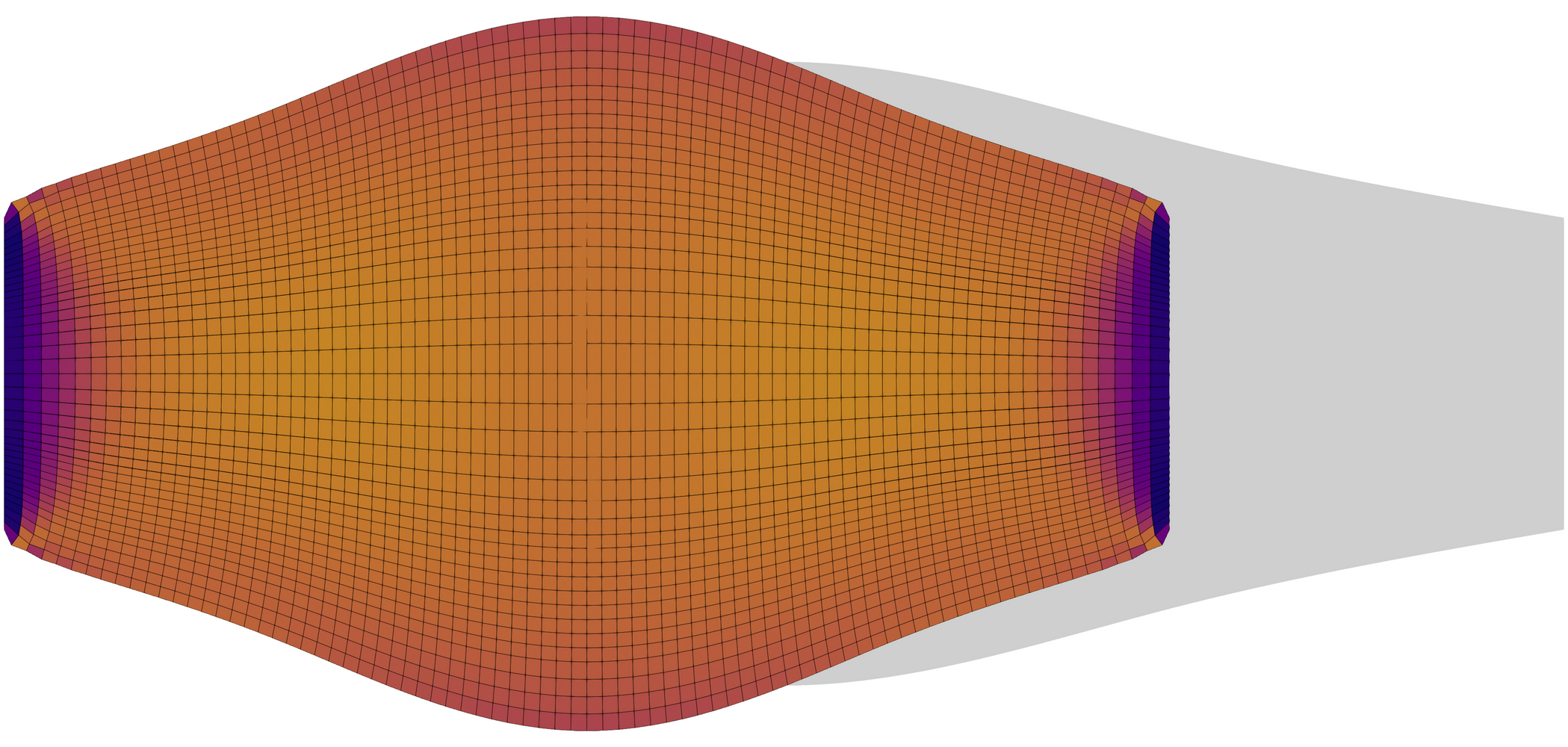}}
    \includegraphics[clip,trim=0 0 0 {.5\ht0}, width=\linewidth,left]{figures/fig_12_slice_zx_fiber_stretch_combi_n4_fc_element_0020.png}
    \endgroup}\hspace*{6pt}
    \parbox{\LW}{ \begingroup\sbox0{\includegraphics{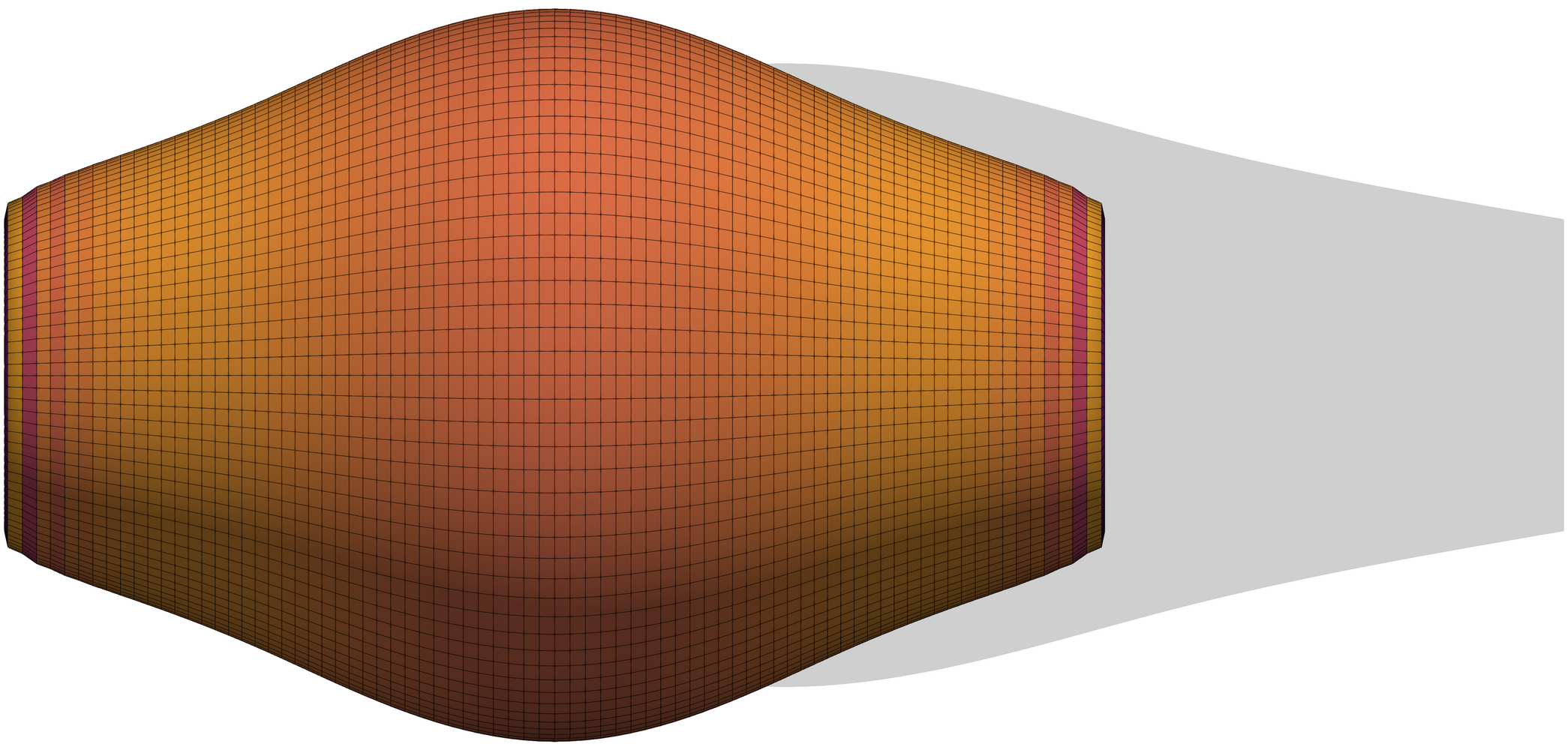}}
    \includegraphics[clip,trim=0 {.5\ht0} 0 0, width=\linewidth,left]{figures/fig_12_solid_fiber_stretch_combi_n4_fc_element_0150.png}
    \endgroup
    \\
    \begingroup\sbox0{\includegraphics{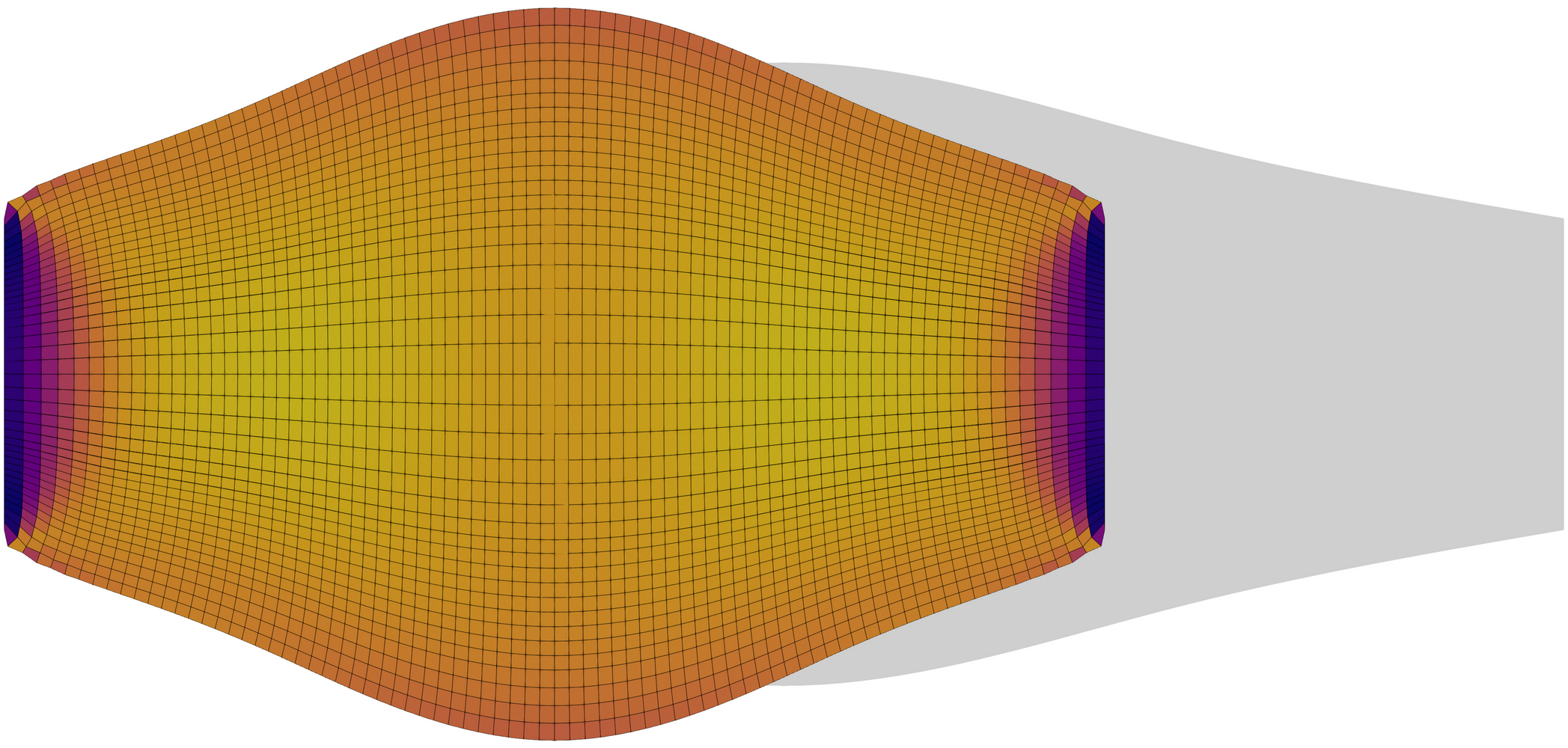}}
    \includegraphics[clip,trim=0 0 0 {.5\ht0}, width=\linewidth,left]{figures/fig_12_slice_zx_fiber_stretch_combi_n4_fc_element_0150.png}
    \endgroup}
\end{minipage}
\hfill
\begin{minipage}[c]{1.29cm}
\raggedleft
\includegraphics[height=3cm]{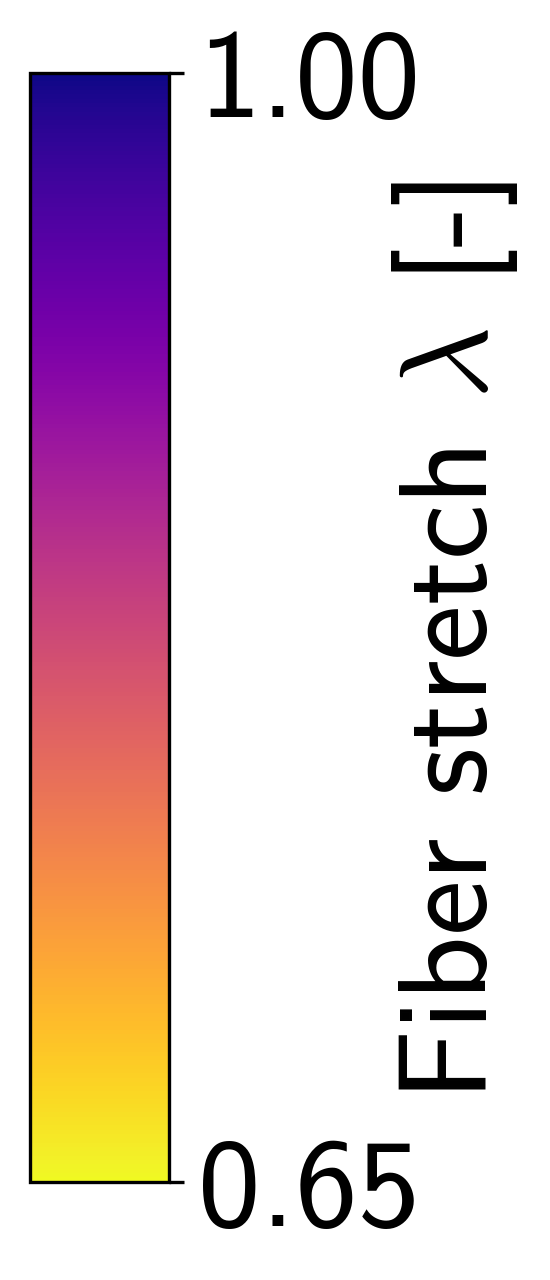}
\end{minipage}
\caption{Fiber stretch $\lambda$ for a free contraction of the fusiform muscle ($n=4$) at selected times. Results are visualized on the surface (top) and in the axial cross-section (bottom) in comparison to the initial configuration (grey).}
\label{fig:fusi_free_stretch_vis}
\end{figure}

\begin{figure}[htb]
\begin{minipage}[c]{0.86\textwidth}
\raggedright
    \parbox{\LWCap}{\centering \phantom{phantom text}} \hspace*{-10pt}
    \parbox{\LW}{\centering $t=\SI{0.005}{s}$ \\ 
    \vspace*{10pt}}
    \hspace*{6pt}
    \parbox{\LW}{\centering $t=\SI{0.02}{s}$ \\ 
    \vspace*{10pt}}
    \hspace*{6pt}
    \parbox{\LW}{\centering $t=\SI{0.15}{s}$ \\ 
    \vspace*{10pt}}
    \\ 
    \vspace*{6pt}
    \parbox{\LWCap}{\subcaption{\raggedright \BLE}} \hspace*{-10pt}
    \parbox{\LW}{ \begingroup\sbox0{\includegraphics{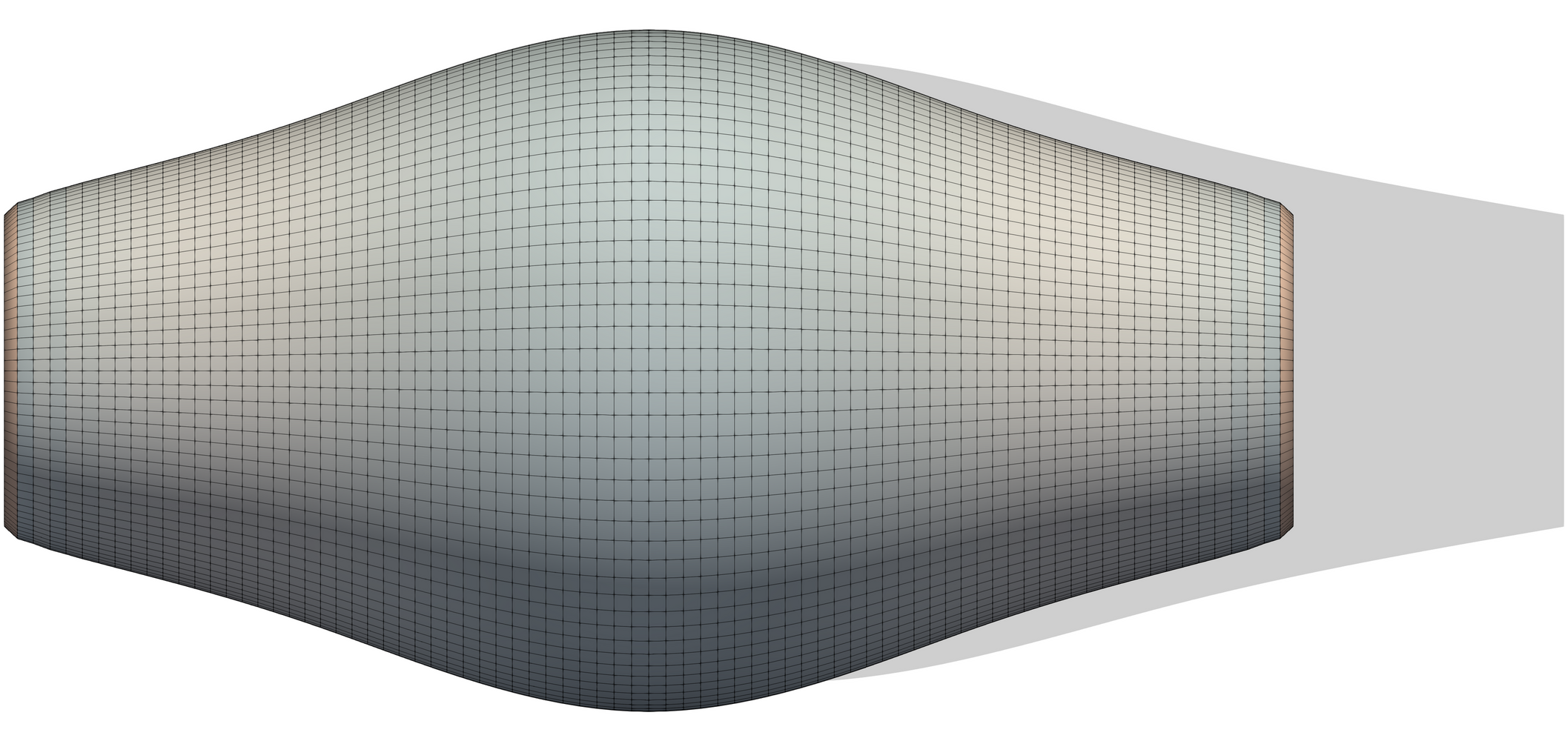}}
    \includegraphics[clip,trim=0 {.5\ht0} 0 0, width=\linewidth,left]{figures/fig_13_solid_cauchy_fib_dir_blemker_n4_fc_element_0005.png}
    \endgroup 
    \\
    \begingroup\sbox0{\includegraphics{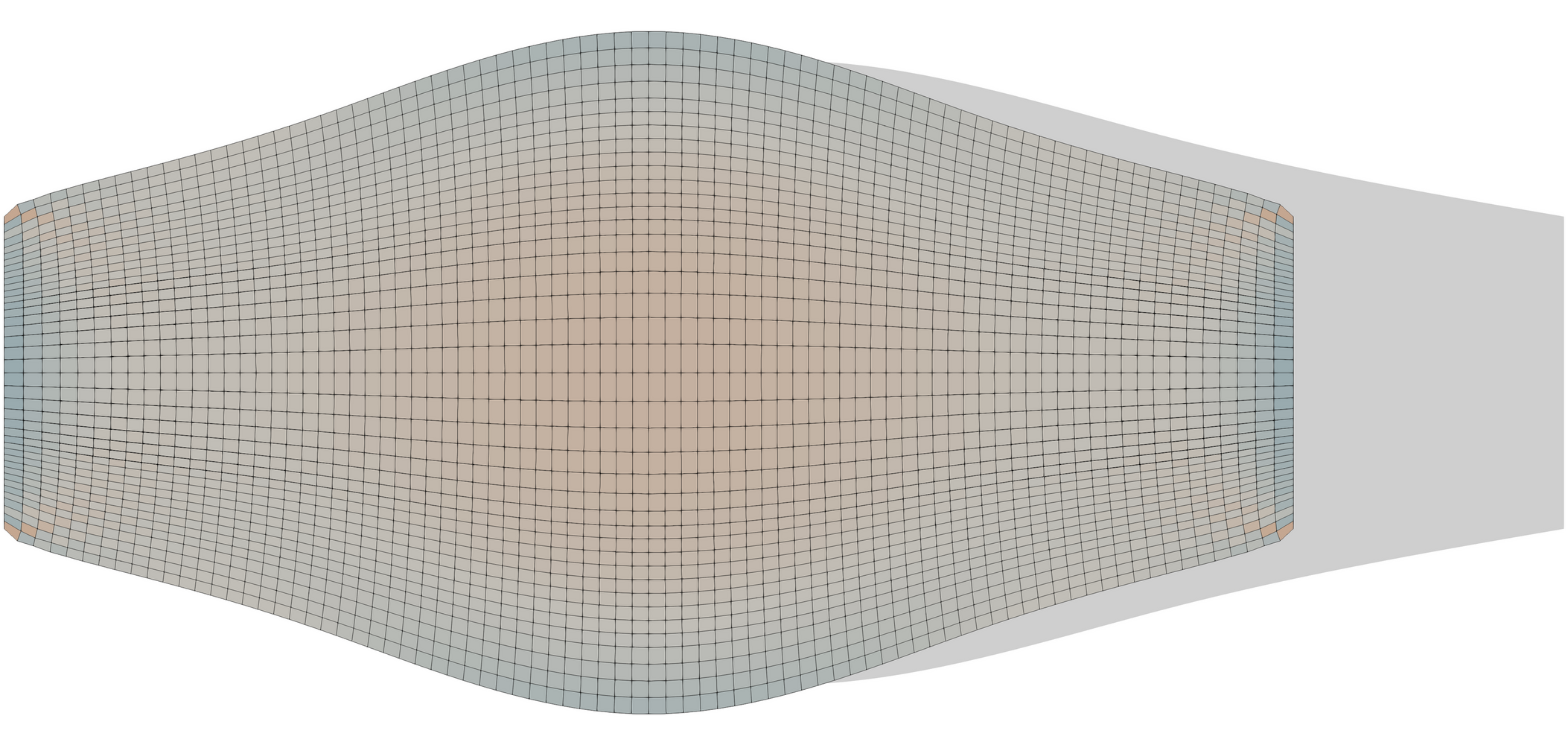}}
    \includegraphics[clip,trim=0 0 0 {.5\ht0}, width=\linewidth,left]{figures/fig_13_slice_zx_cauchy_fib_dir_blemker_n4_fc_element_0005.png}
    \endgroup} \hspace*{6pt}
    \parbox{\LW}{ \begingroup\sbox0{\includegraphics{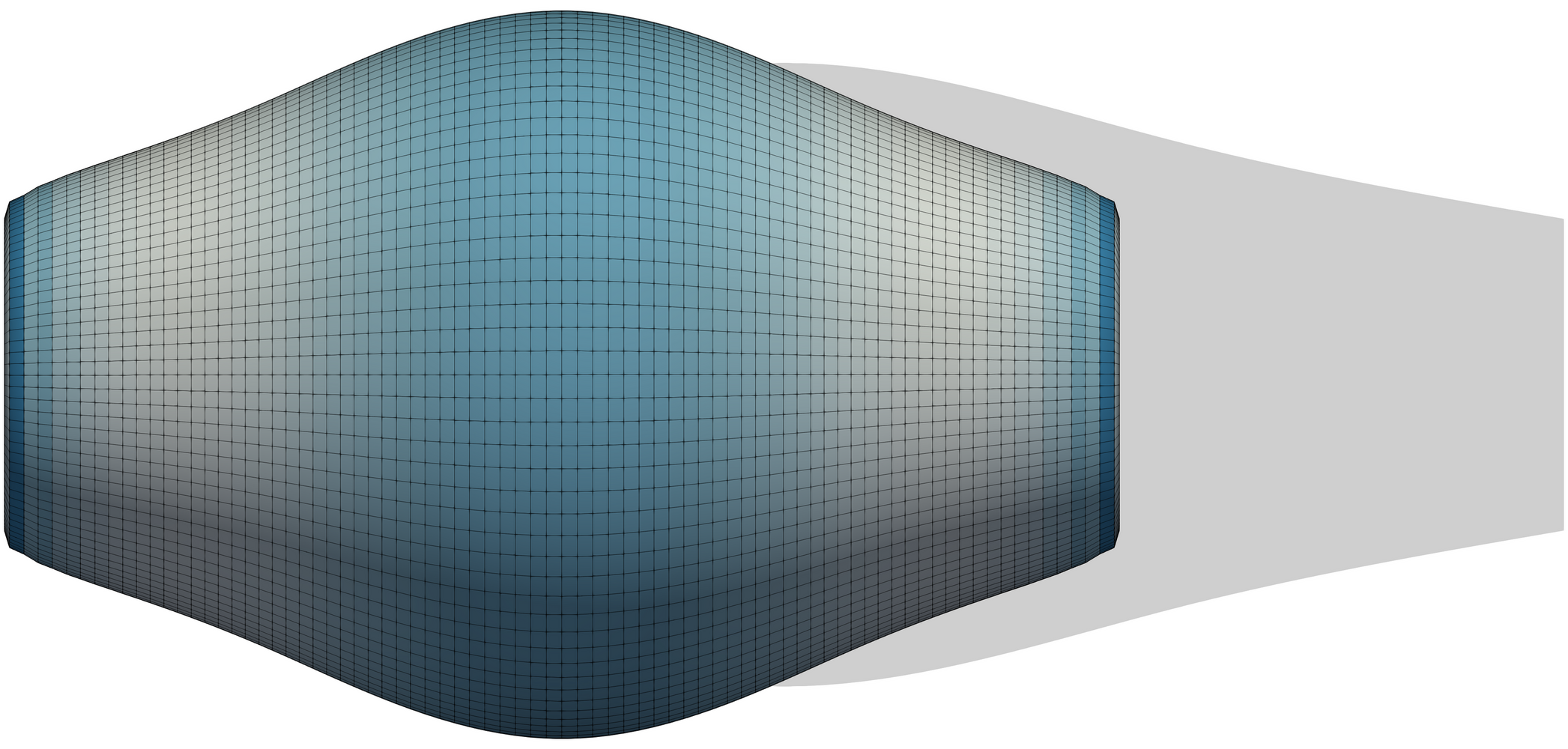}}
    \includegraphics[clip,trim=0 {.5\ht0} 0 0, width=\linewidth,left]{figures/fig_13_solid_cauchy_fib_dir_blemker_n4_fc_element_0020.png}
    \endgroup
    \\
    \begingroup\sbox0{\includegraphics{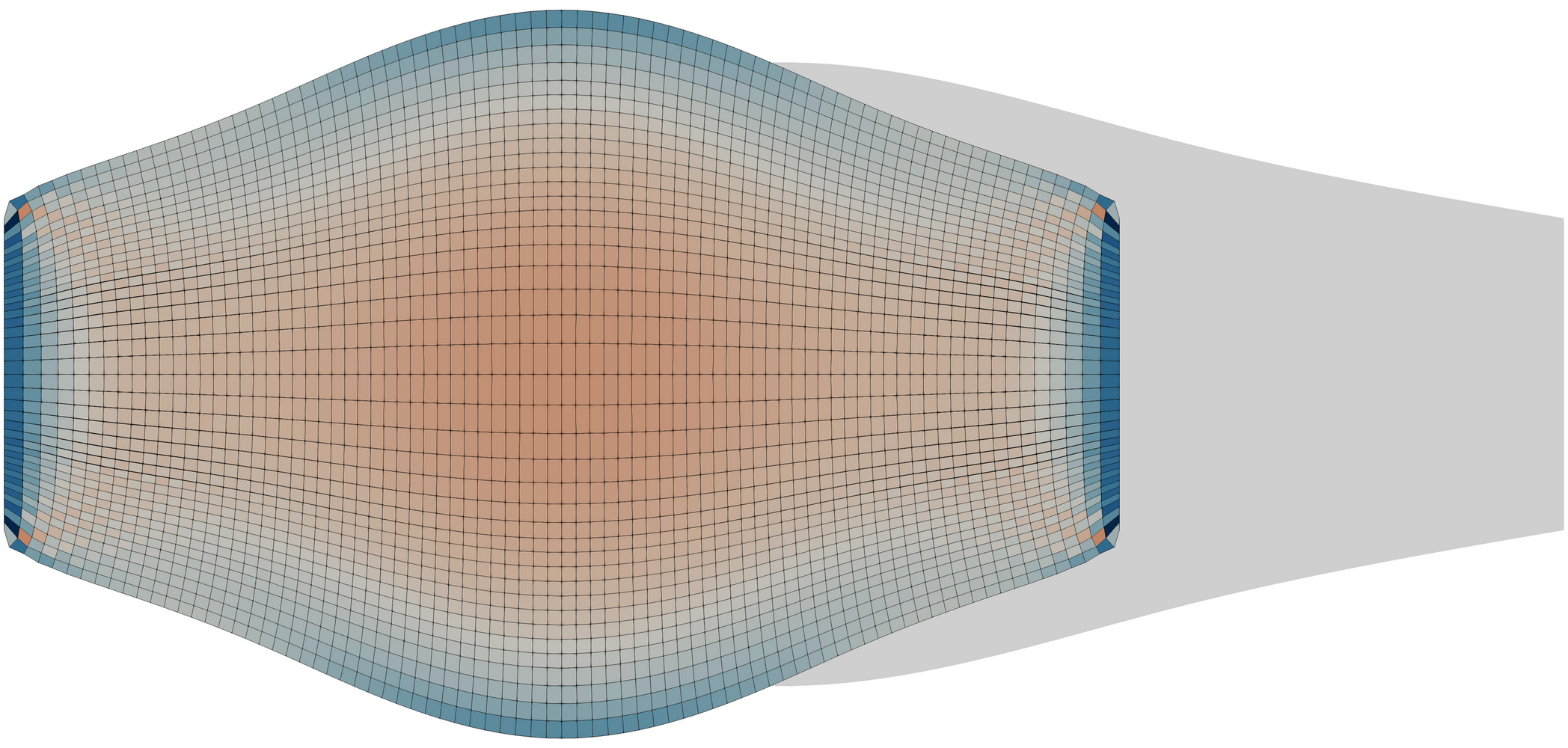}}
    \includegraphics[clip,trim=0 0 0 {.5\ht0}, width=\linewidth,left]{figures/fig_13_slice_zx_cauchy_fib_dir_blemker_n4_fc_element_0020.png}
    \endgroup} \hspace*{6pt}
    \parbox{\LW}{ \begingroup\sbox0{\includegraphics{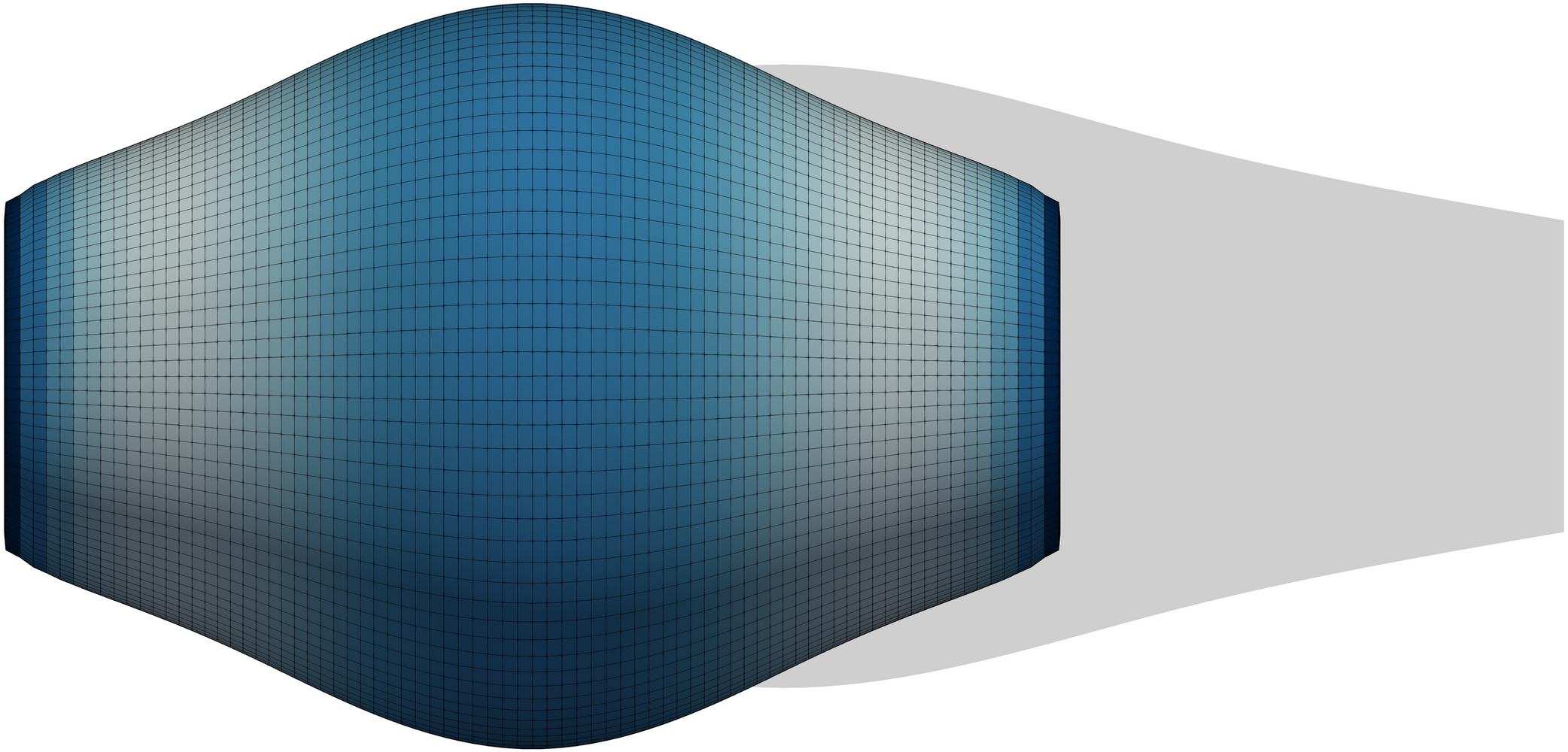}}
    \includegraphics[clip,trim=0 {.5\ht0} 0 0, width=\linewidth,left]{figures/fig_13_solid_cauchy_fib_dir_blemker_n4_fc_element_0150.png}
    \endgroup
    \\
    \begingroup\sbox0{\includegraphics{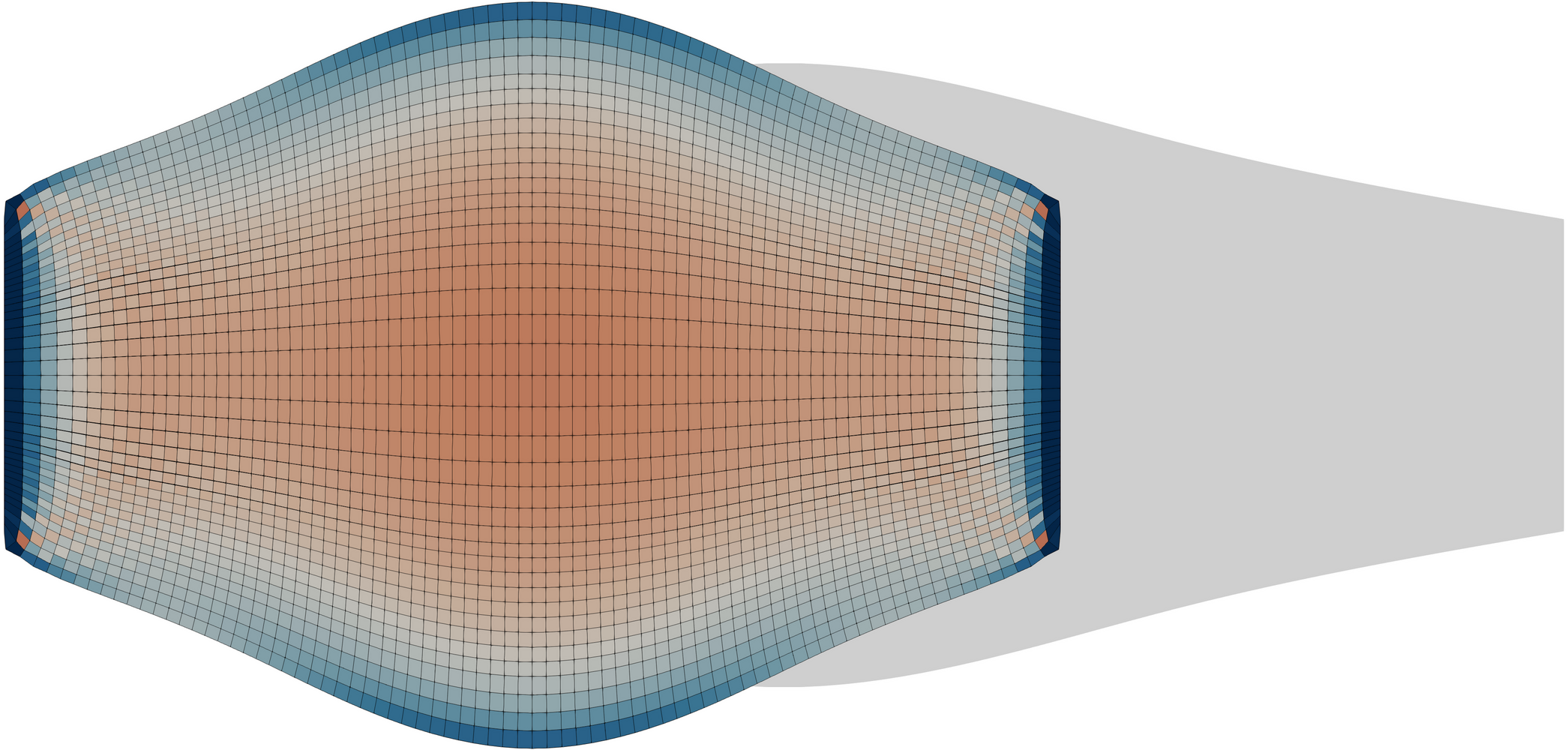}}
    \includegraphics[clip,trim=0 0 0 {.5\ht0}, width=\linewidth,left]{figures/fig_13_slice_zx_cauchy_fib_dir_blemker_n4_fc_element_0150.png}
    \endgroup} \\ 
    \vspace*{6pt}
    \parbox{\LWCap}{\subcaption{\raggedright \GIANT}} \hspace*{-10pt}
    \parbox{\LW}{ \begingroup\sbox0{\includegraphics{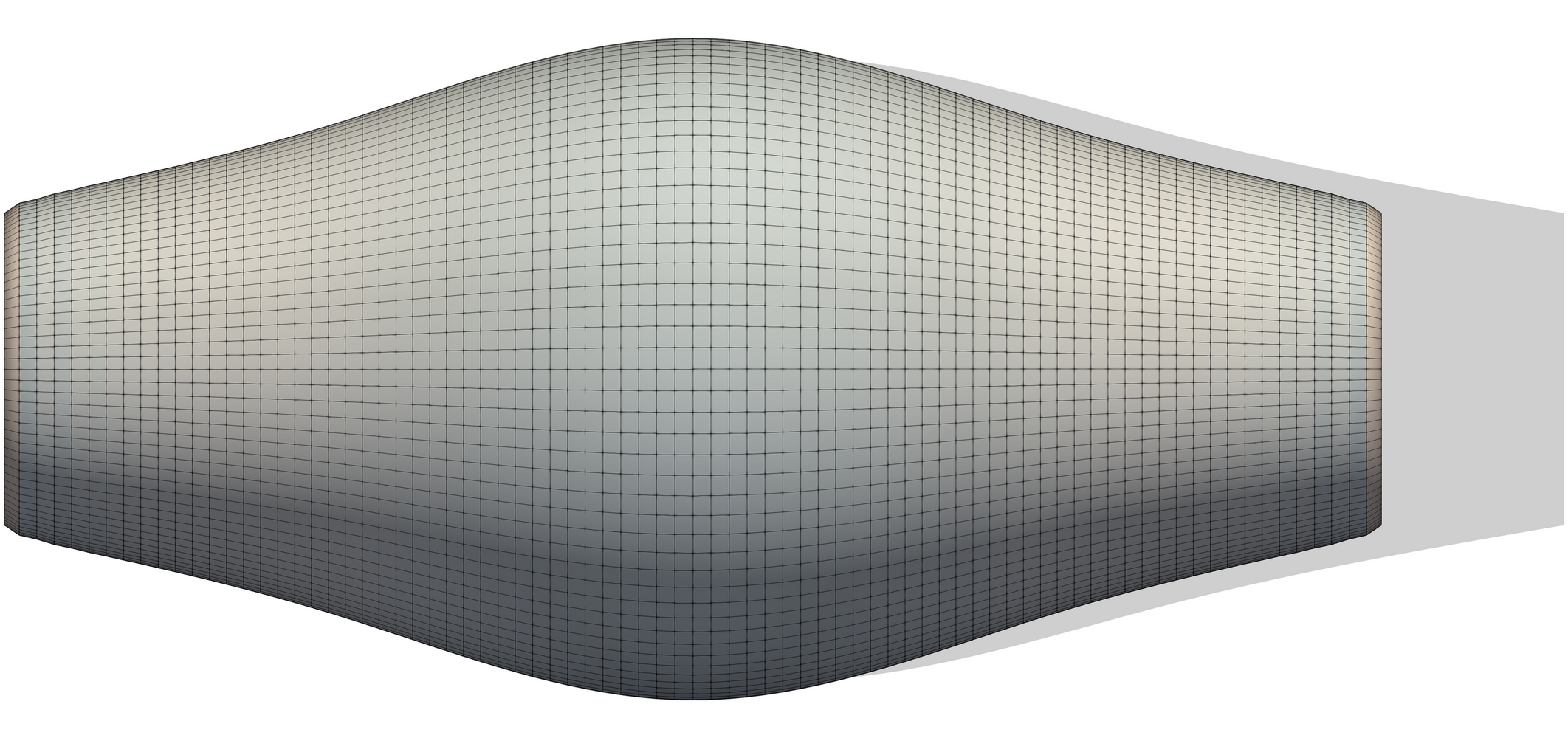}}
    \includegraphics[clip,trim=0 {.5\ht0} 0 0, width=\linewidth,left]{figures/fig_13_solid_cauchy_fib_dir_giantesio_n4_fc_element_0005.png}
    \endgroup 
    \\ 
    \begingroup\sbox0{\includegraphics{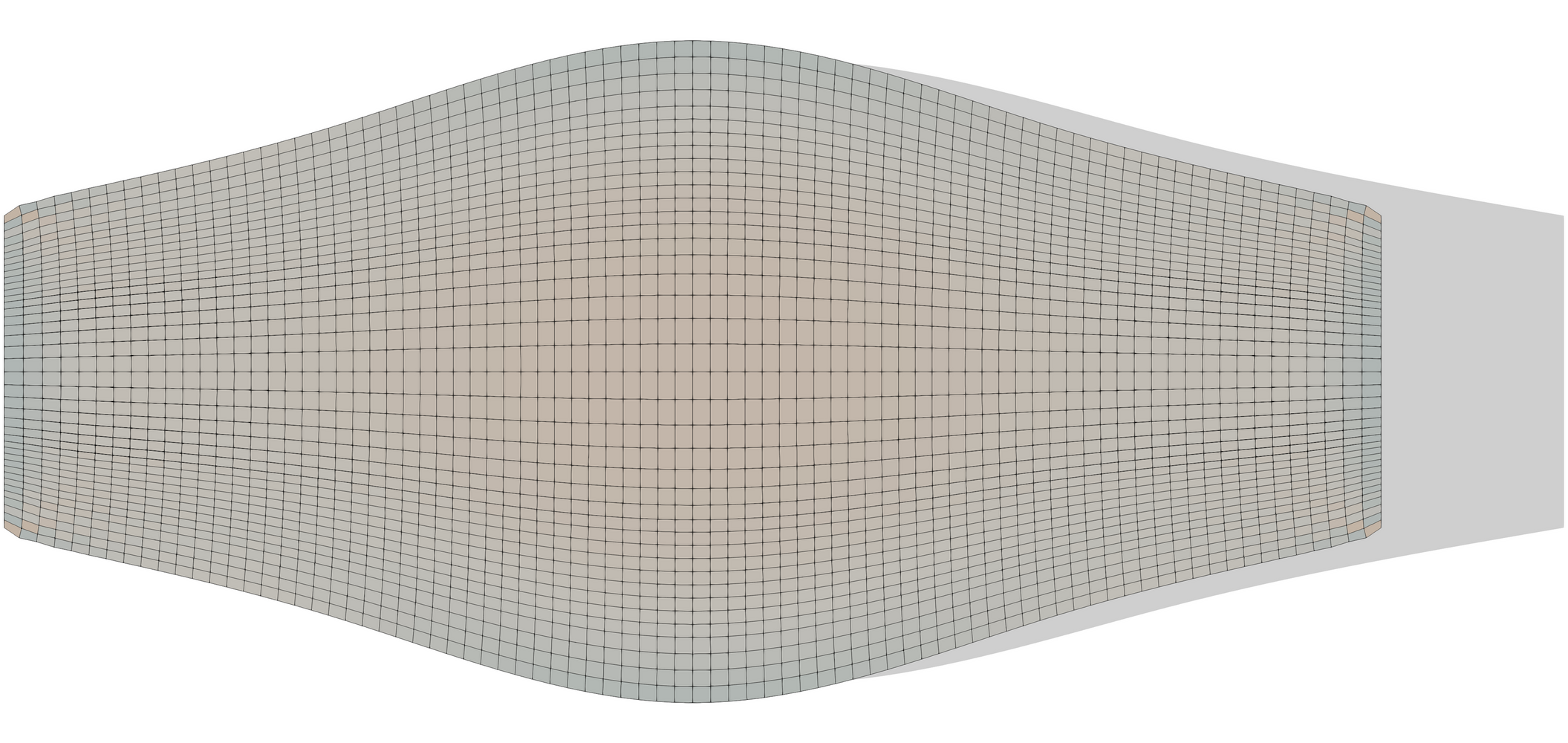}}
    \includegraphics[clip,trim=0 0 0 {.5\ht0}, width=\linewidth,left]{figures/fig_13_slice_zx_cauchy_fib_dir_giantesio_n4_fc_element_0005.png}
    \endgroup} \hspace*{6pt}
    \parbox{\LW}{ \begingroup\sbox0{\includegraphics{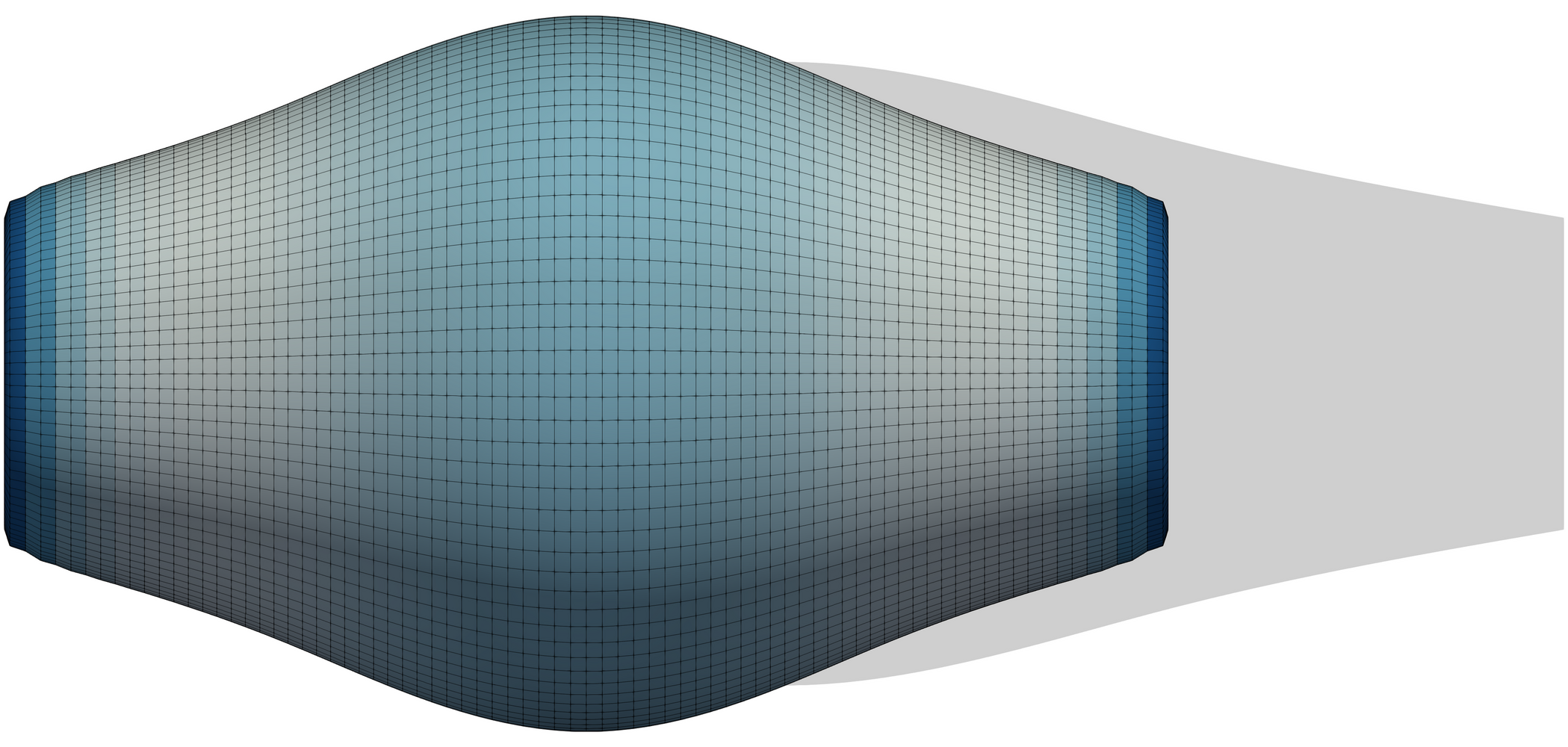}}
    \includegraphics[clip,trim=0 {.5\ht0} 0 0, width=\linewidth,left]{figures/fig_13_solid_cauchy_fib_dir_giantesio_n4_fc_element_0020.png}
    \endgroup
    \\ 
    \begingroup\sbox0{\includegraphics{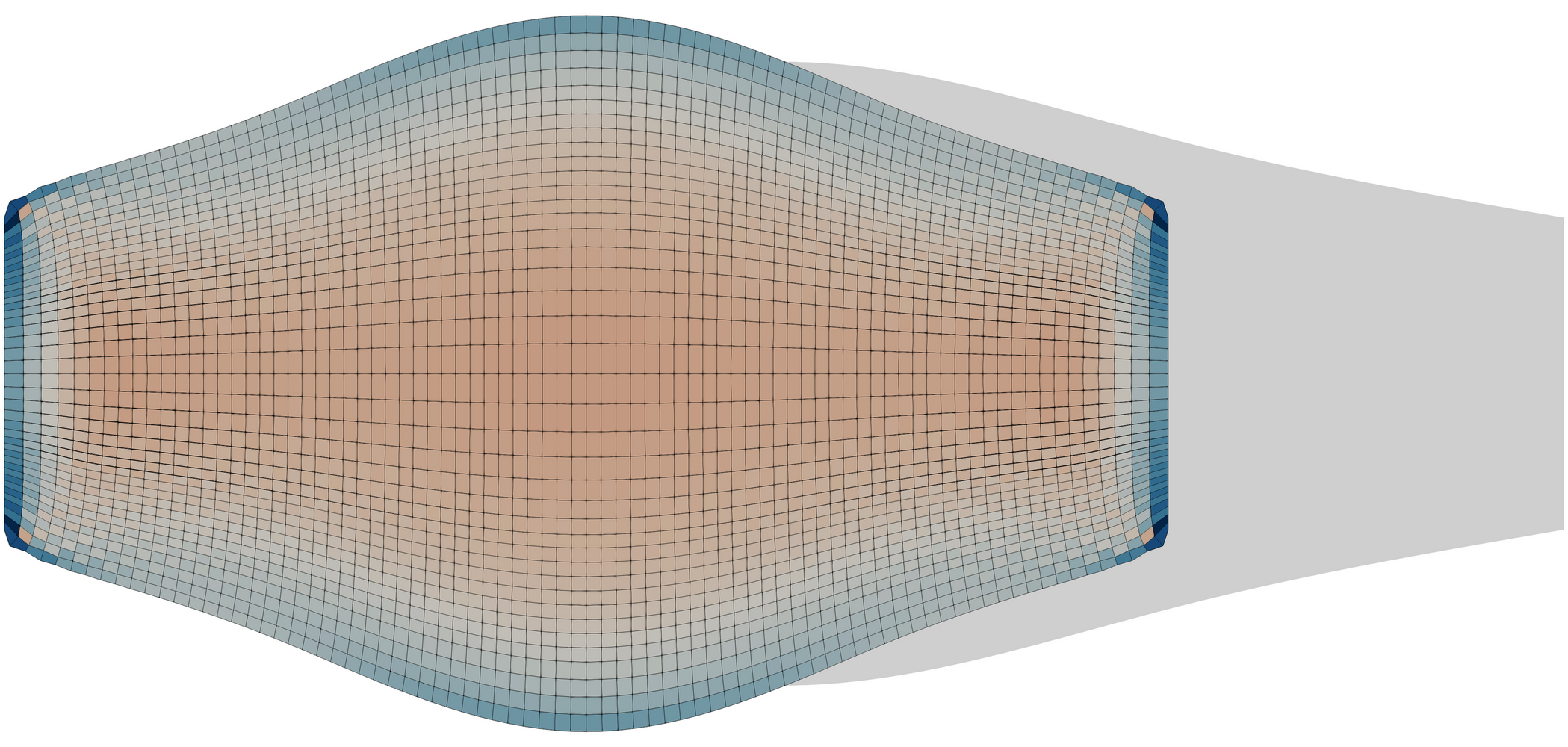}}
    \includegraphics[clip,trim=0 0 0 {.5\ht0}, width=\linewidth,left]{figures/fig_13_slice_zx_cauchy_fib_dir_giantesio_n4_fc_element_0020.png}
    \endgroup} \hspace*{6pt}
    \parbox{\LW}{ \begingroup\sbox0{\includegraphics{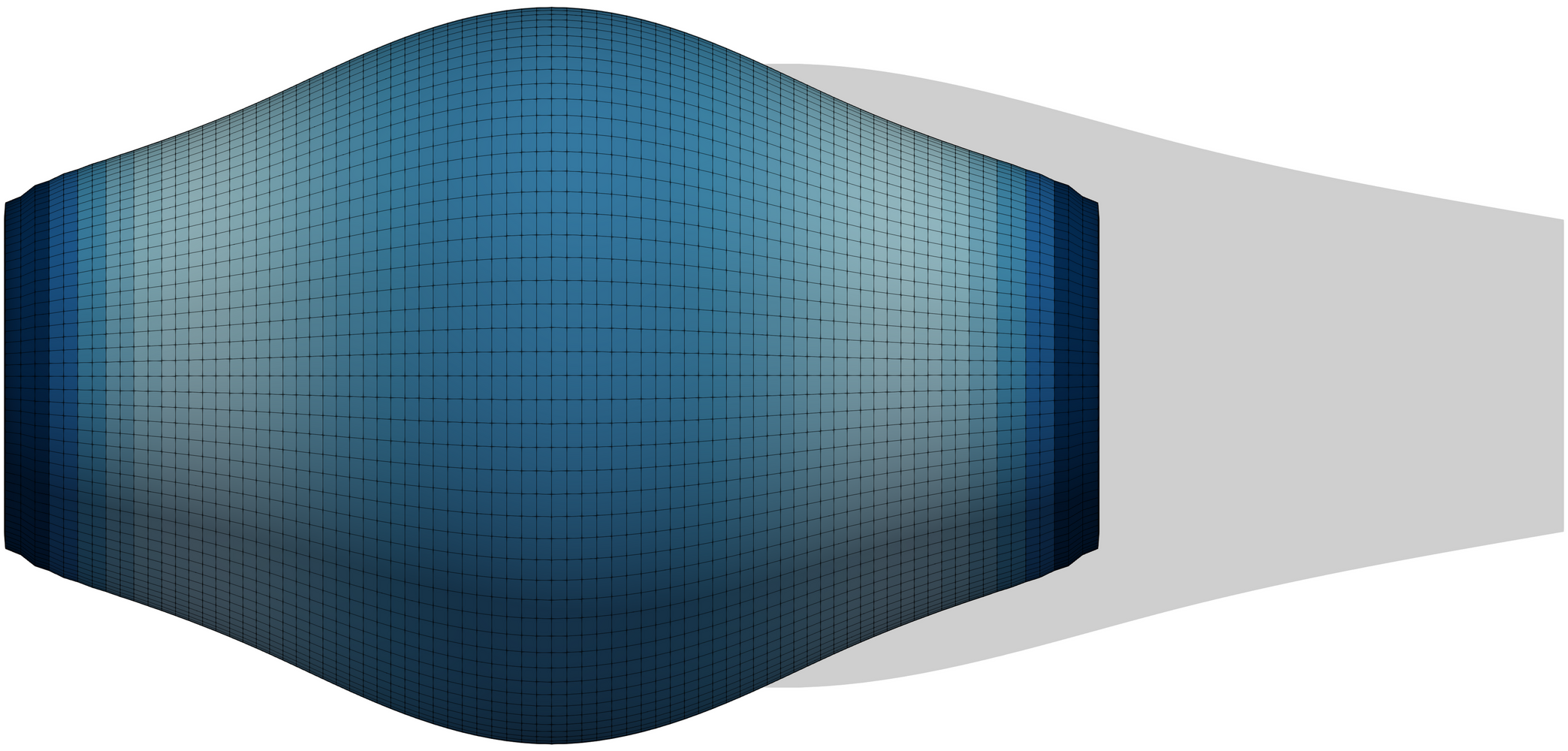}}
    \includegraphics[clip,trim=0 {.5\ht0} 0 0, width=\linewidth,left]{figures/fig_13_solid_cauchy_fib_dir_giantesio_n4_fc_element_0150.png}
    \endgroup
    \\ 
    \begingroup\sbox0{\includegraphics{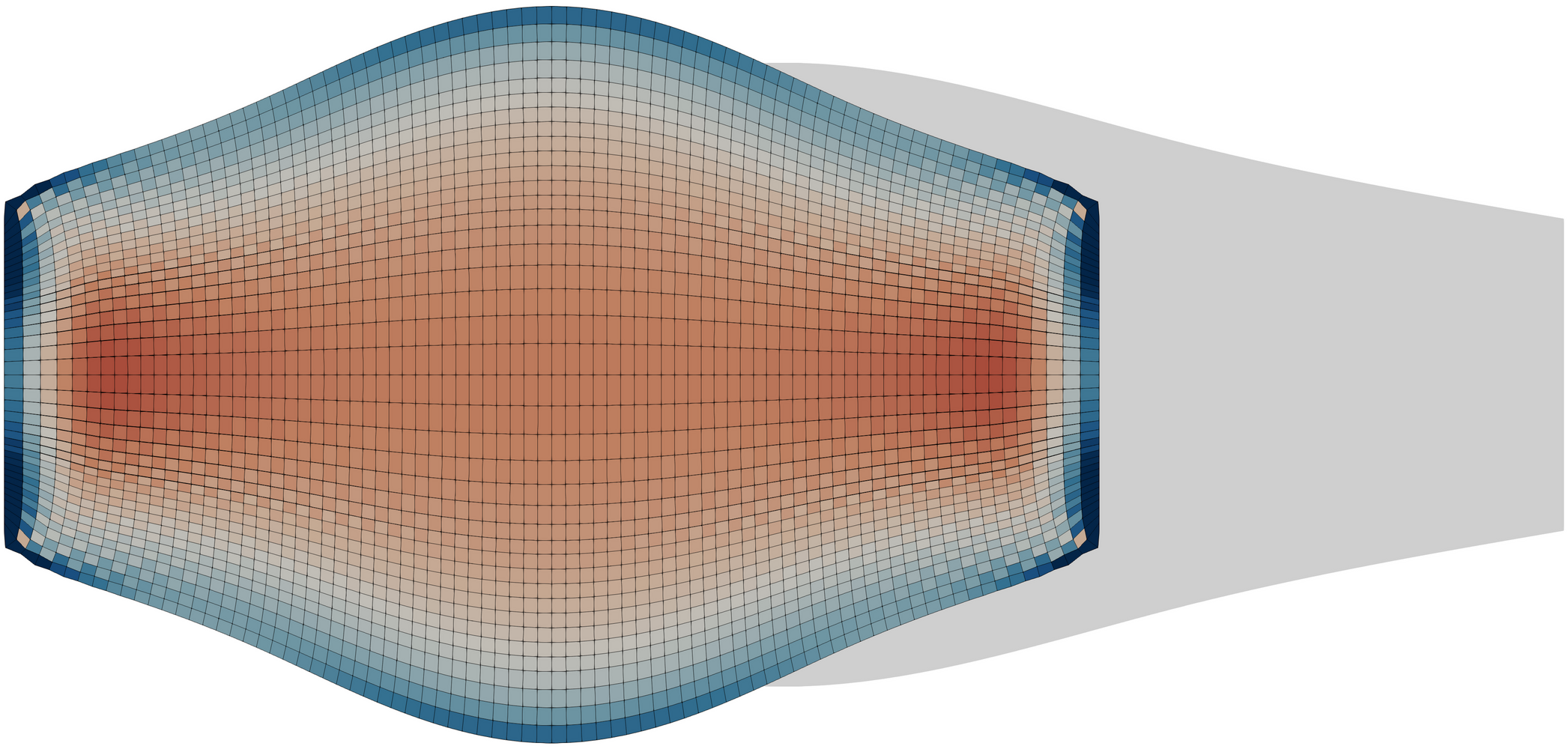}}
    \includegraphics[clip,trim=0 0 0 {.5\ht0}, width=\linewidth,left]{figures/fig_13_slice_zx_cauchy_fib_dir_giantesio_n4_fc_element_0150.png}
    \endgroup} \\
    \vspace*{6pt}
    \parbox{\LWCap}{\subcaption{\raggedright \WKM}} \hspace*{-10pt}
    \parbox{\LW}{ \begingroup\sbox0{\includegraphics{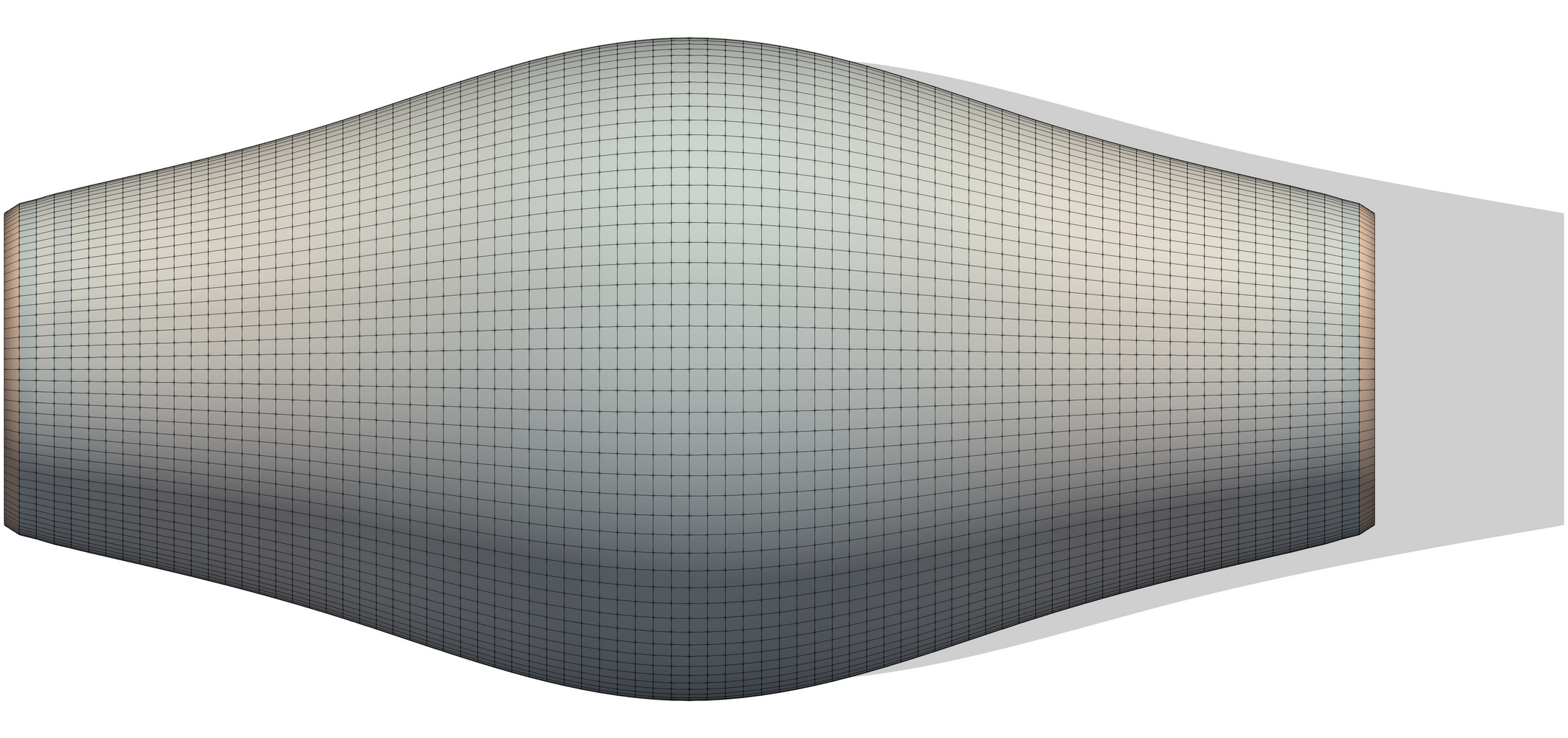}}
    \includegraphics[clip,trim=0 {.5\ht0} 0 0, width=\linewidth,left]{figures/fig_13_solid_cauchy_fib_dir_weickenmeier_n4_fc_element_0005.png}
    \endgroup 
    \\
    \begingroup\sbox0{\includegraphics{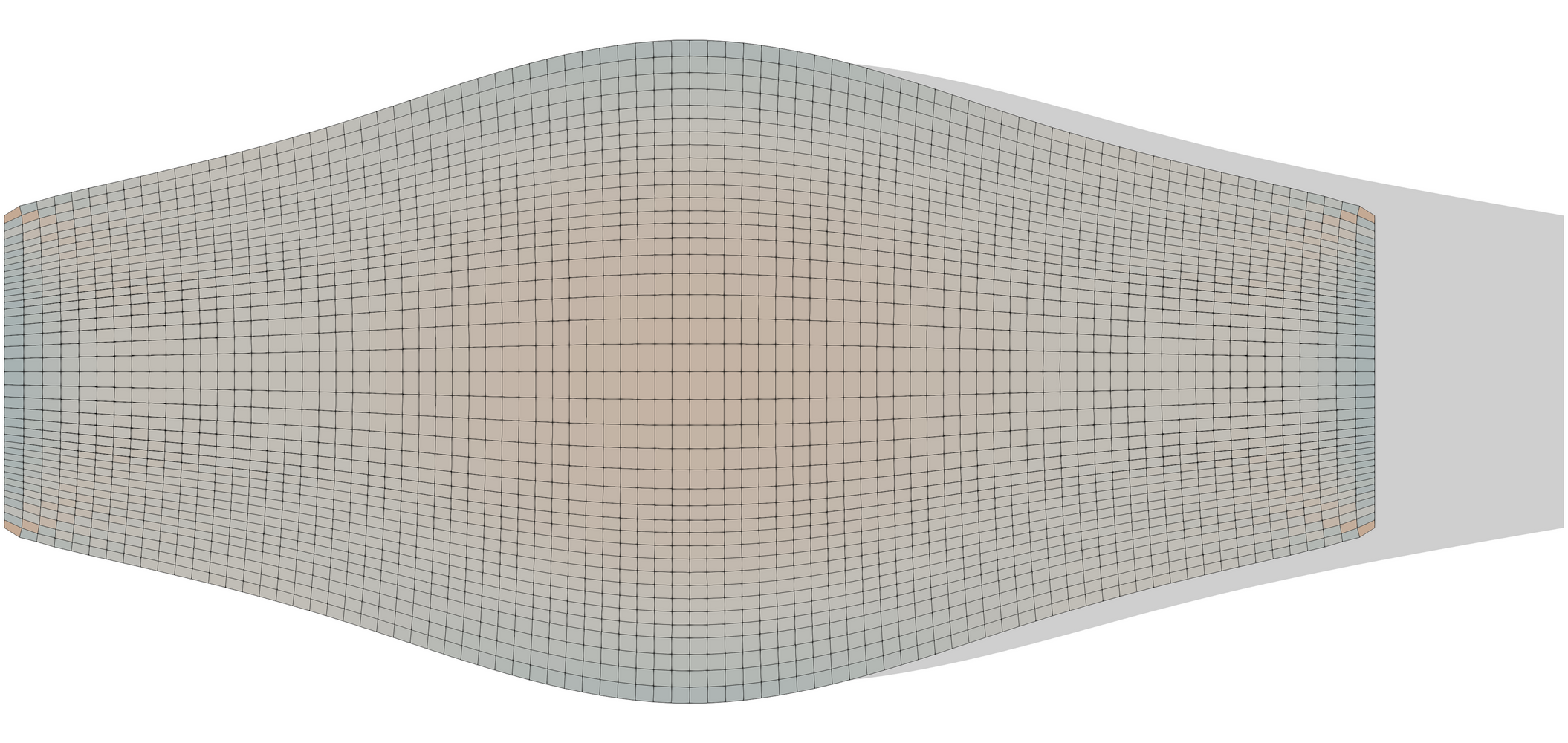}}
    \includegraphics[clip,trim=0 0 0 {.5\ht0}, width=\linewidth,left]{figures/fig_13_slice_zx_cauchy_fib_dir_weickenmeier_n4_fc_element_0005.png}
    \endgroup} \hspace*{6pt}
    \parbox{\LW}{ \begingroup\sbox0{\includegraphics{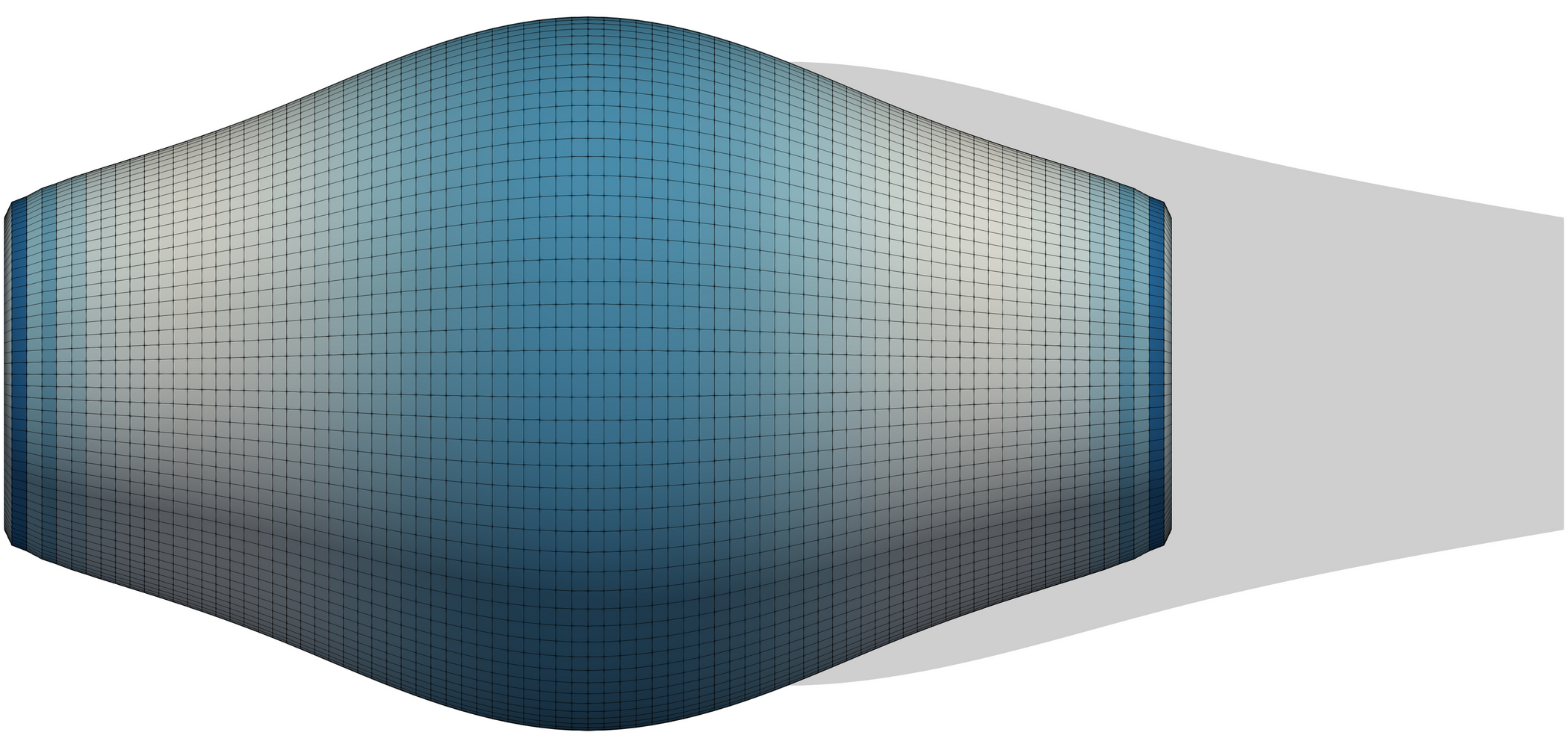}}
    \includegraphics[clip,trim=0 {.5\ht0} 0 0, width=\linewidth,left]{figures/fig_13_solid_cauchy_fib_dir_weickenmeier_n4_fc_element_0020.png}
    \endgroup
    \\
    \begingroup\sbox0{\includegraphics{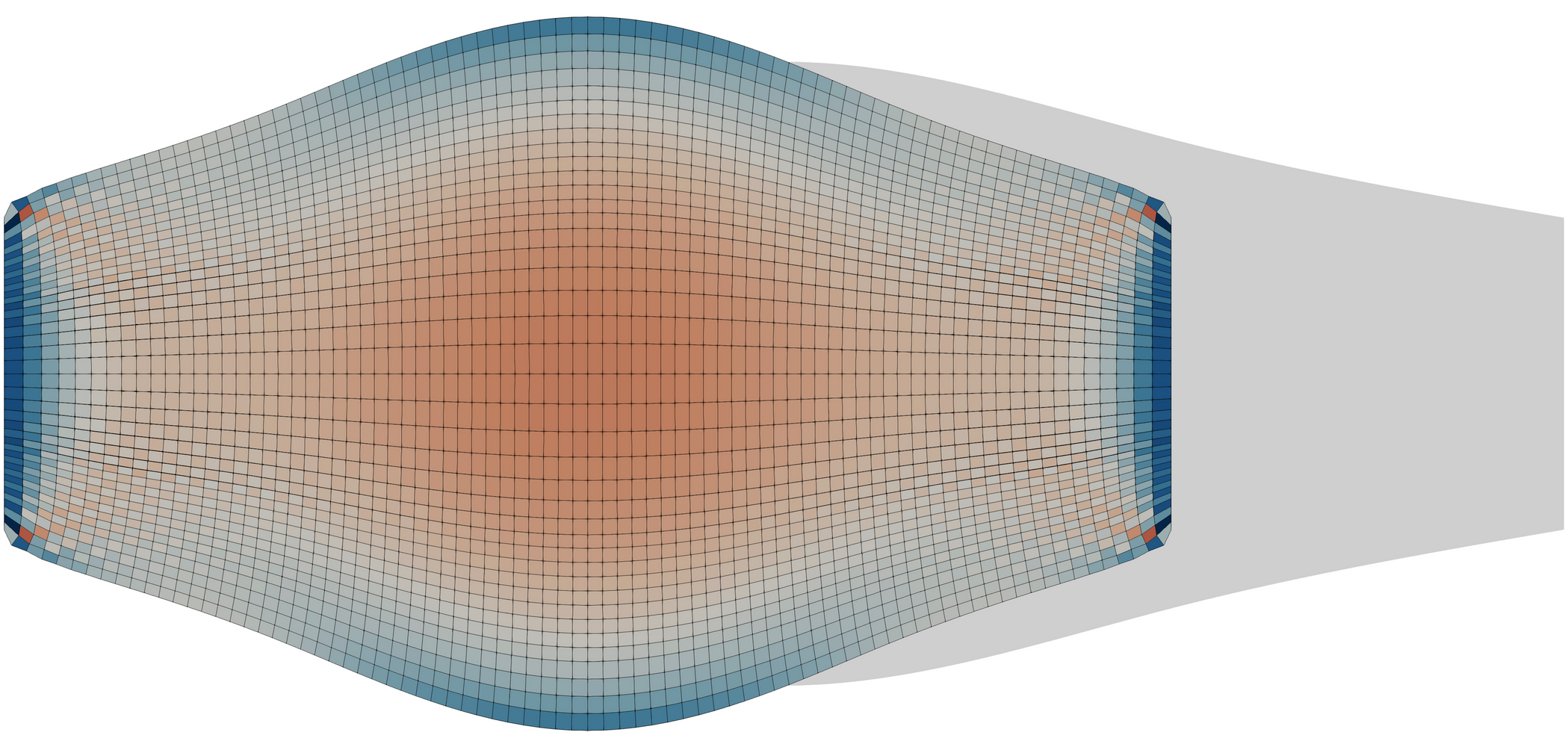}}
    \includegraphics[clip,trim=0 0 0 {.5\ht0}, width=\linewidth,left]{figures/fig_13_slice_zx_cauchy_fib_dir_weickenmeier_n4_fc_element_0020.png}
    \endgroup} \hspace*{6pt}
    \parbox{\LW}{ \begingroup\sbox0{\includegraphics{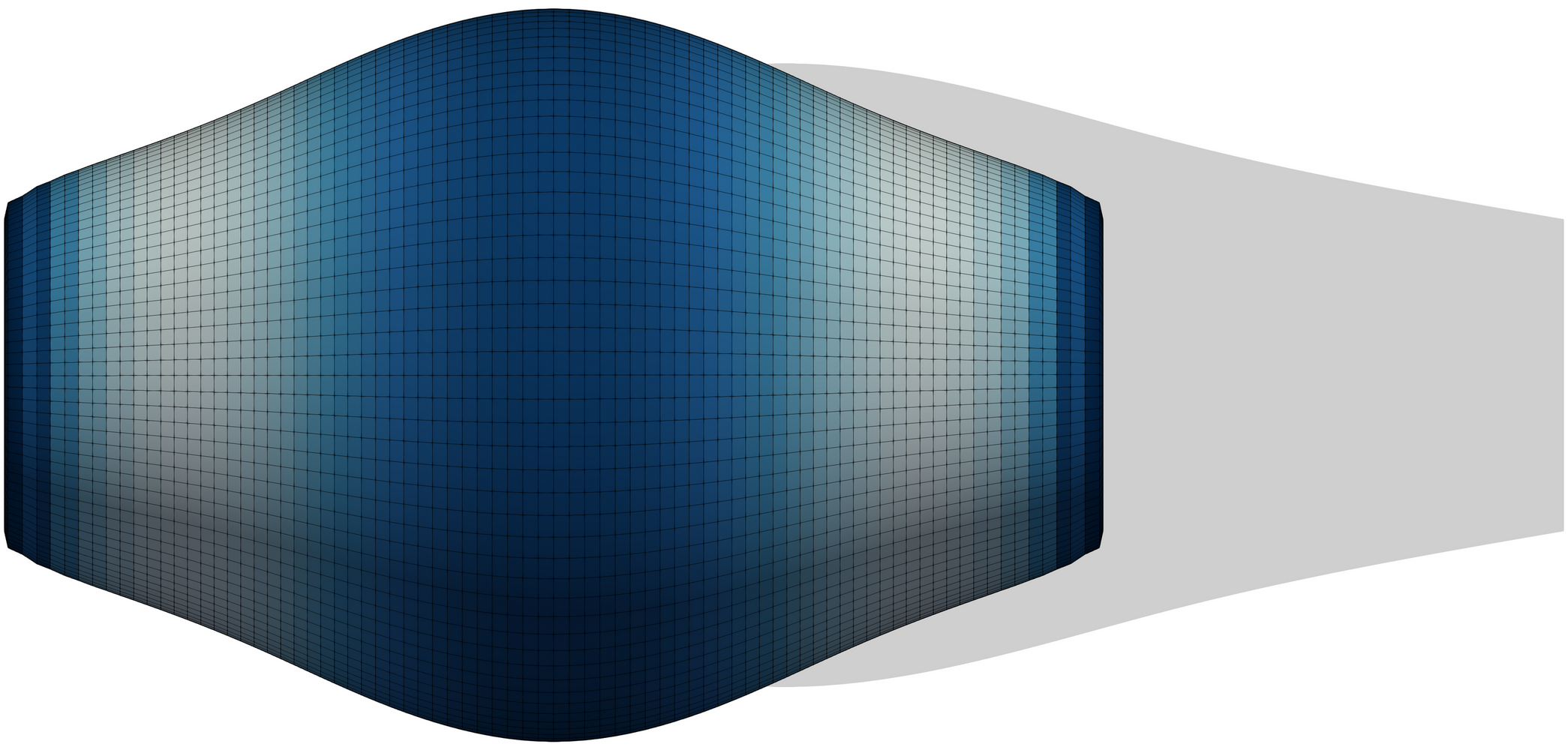}}
    \includegraphics[clip,trim=0 {.5\ht0} 0 0, width=\linewidth,left]{figures/fig_13_solid_cauchy_fib_dir_weickenmeier_n4_fc_element_0150.png}
    \endgroup
    \\
    \begingroup\sbox0{\includegraphics{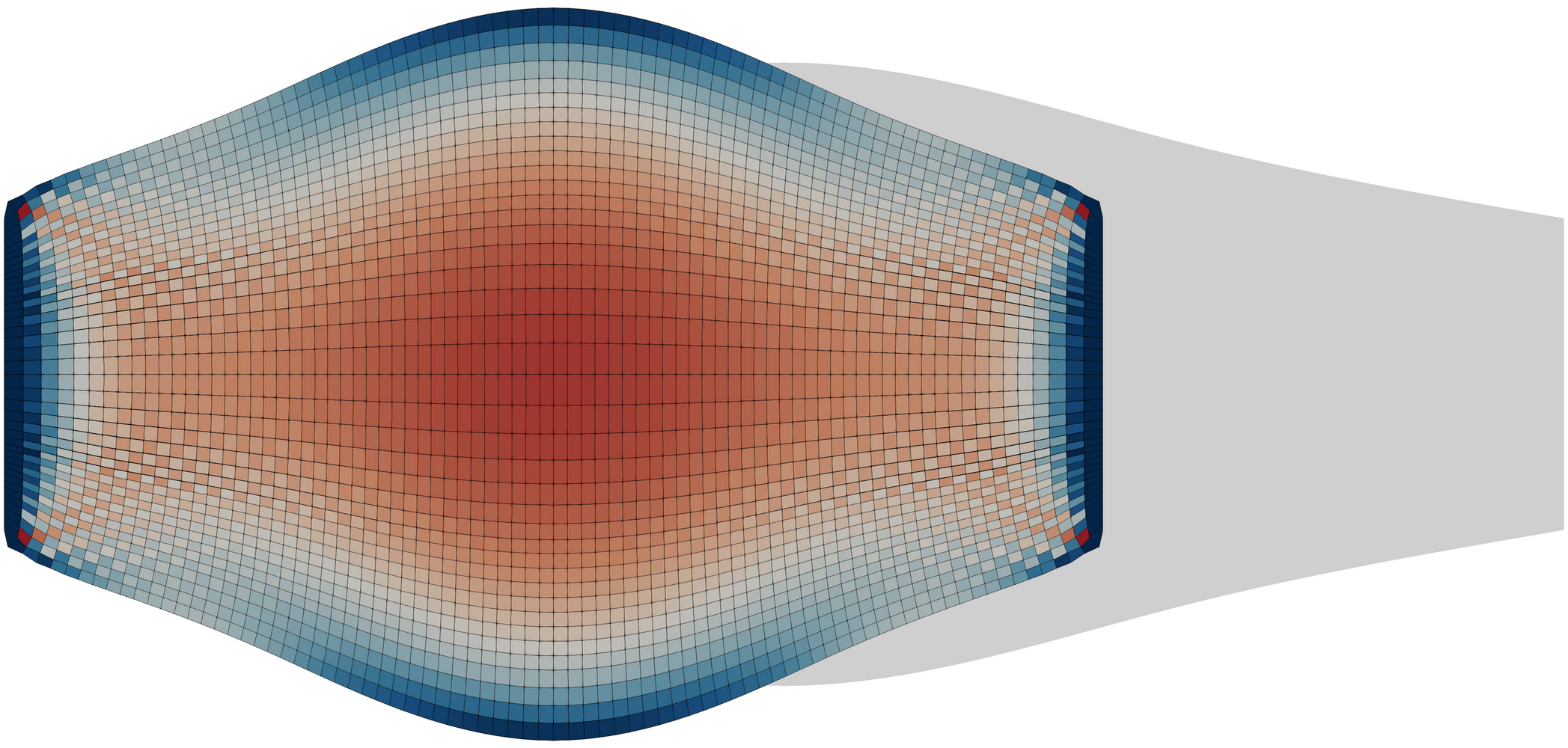}}
    \includegraphics[clip,trim=0 0 0 {.5\ht0}, width=\linewidth,left]{figures/fig_13_slice_zx_cauchy_fib_dir_weickenmeier_n4_fc_element_0150.png}
    \endgroup}\\
    \vspace*{6pt}
    \parbox{\LWCap}{\subcaption{\raggedright \COMBI}} \hspace*{-10pt}
    \parbox{\LW}{ \begingroup\sbox0{\includegraphics{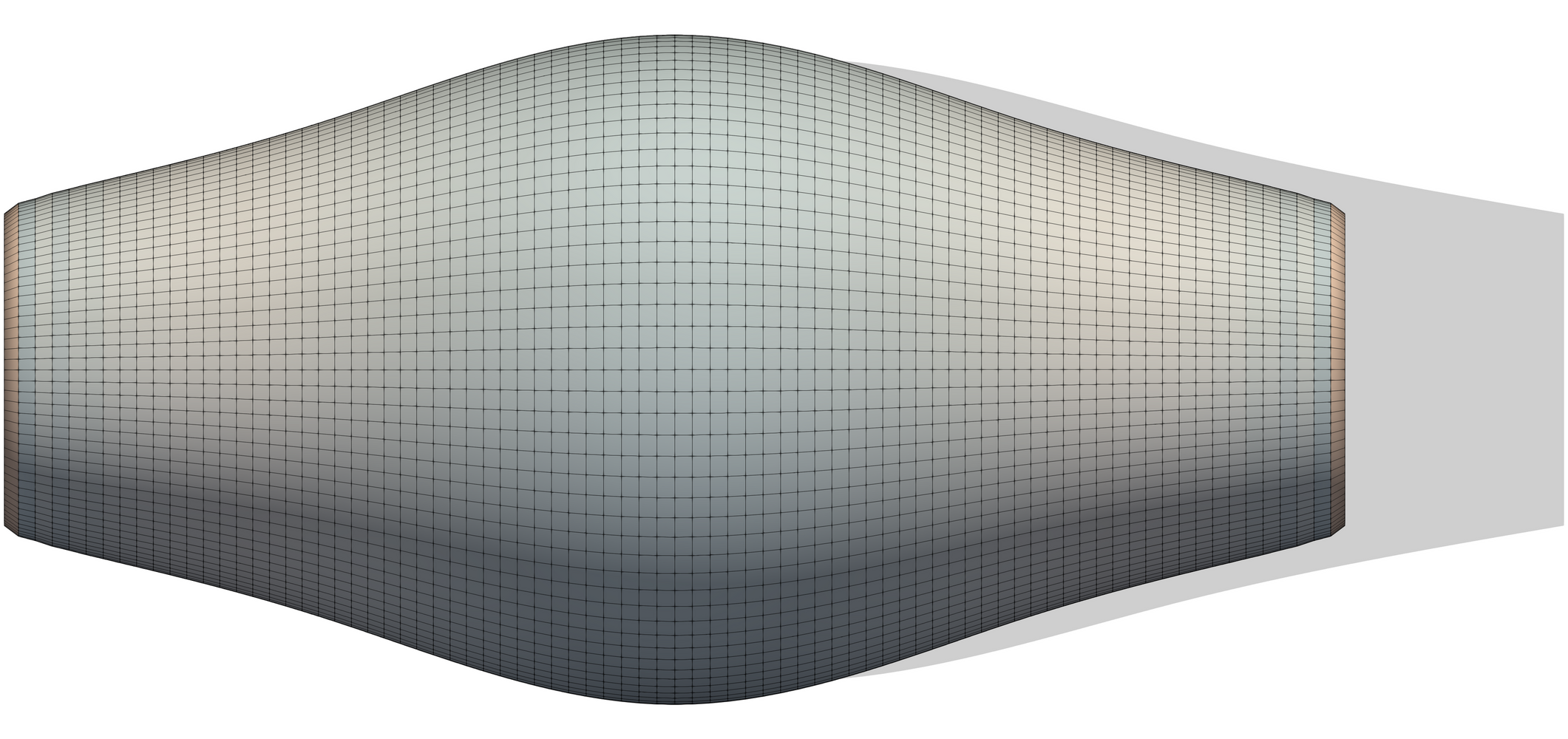}}
    \includegraphics[clip,trim=0 {.5\ht0} 0 0, width=\linewidth,left]{figures/fig_13_solid_cauchy_fib_dir_combi_n4_fc_element_0005.png}
    \endgroup 
    \\
    \begingroup\sbox0{\includegraphics{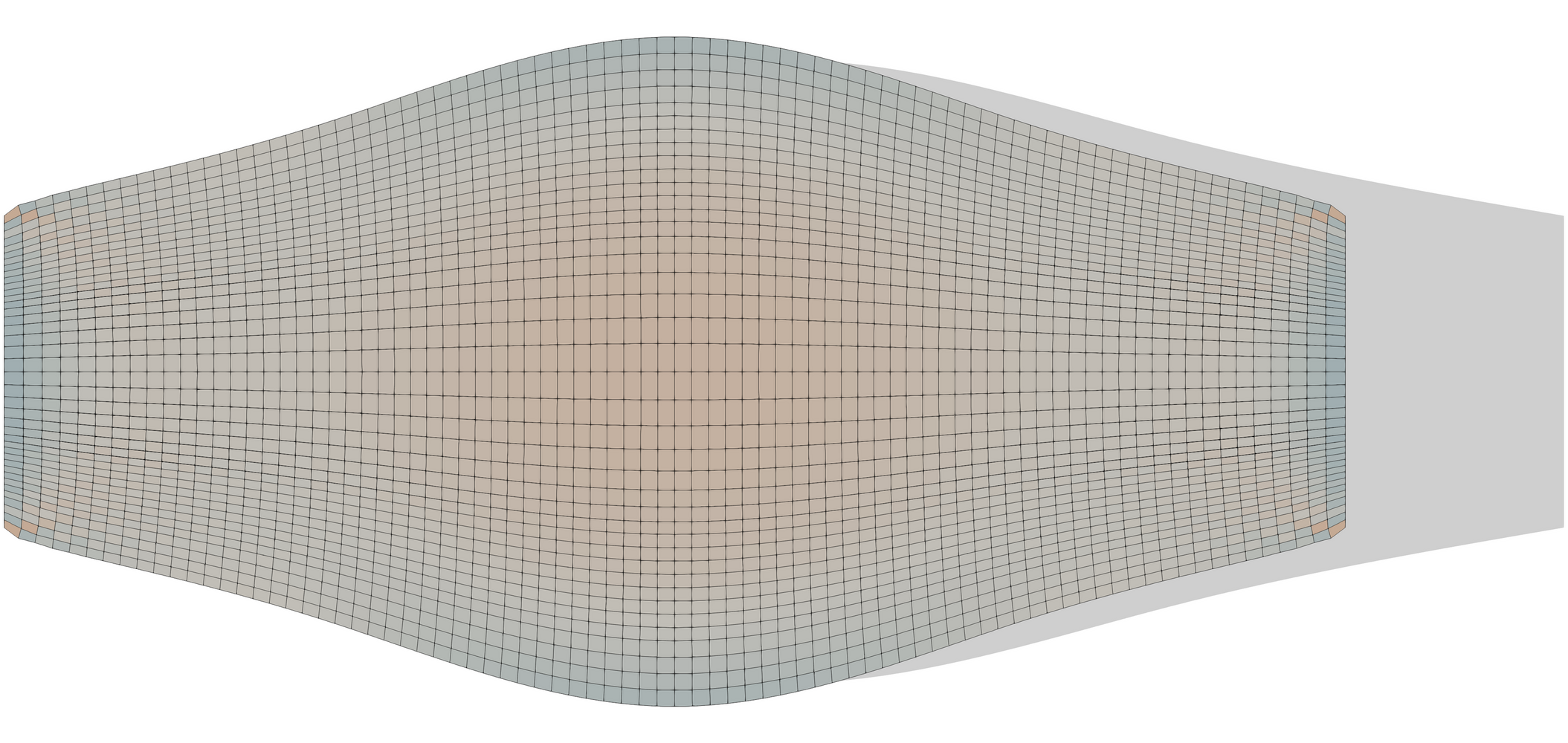}}
    \includegraphics[clip,trim=0 0 0 {.5\ht0}, width=\linewidth,left]{figures/fig_13_slice_zx_cauchy_fib_dir_combi_n4_fc_element_0005.png}
    \endgroup} \hspace*{6pt}
    \parbox{\LW}{ \begingroup\sbox0{\includegraphics{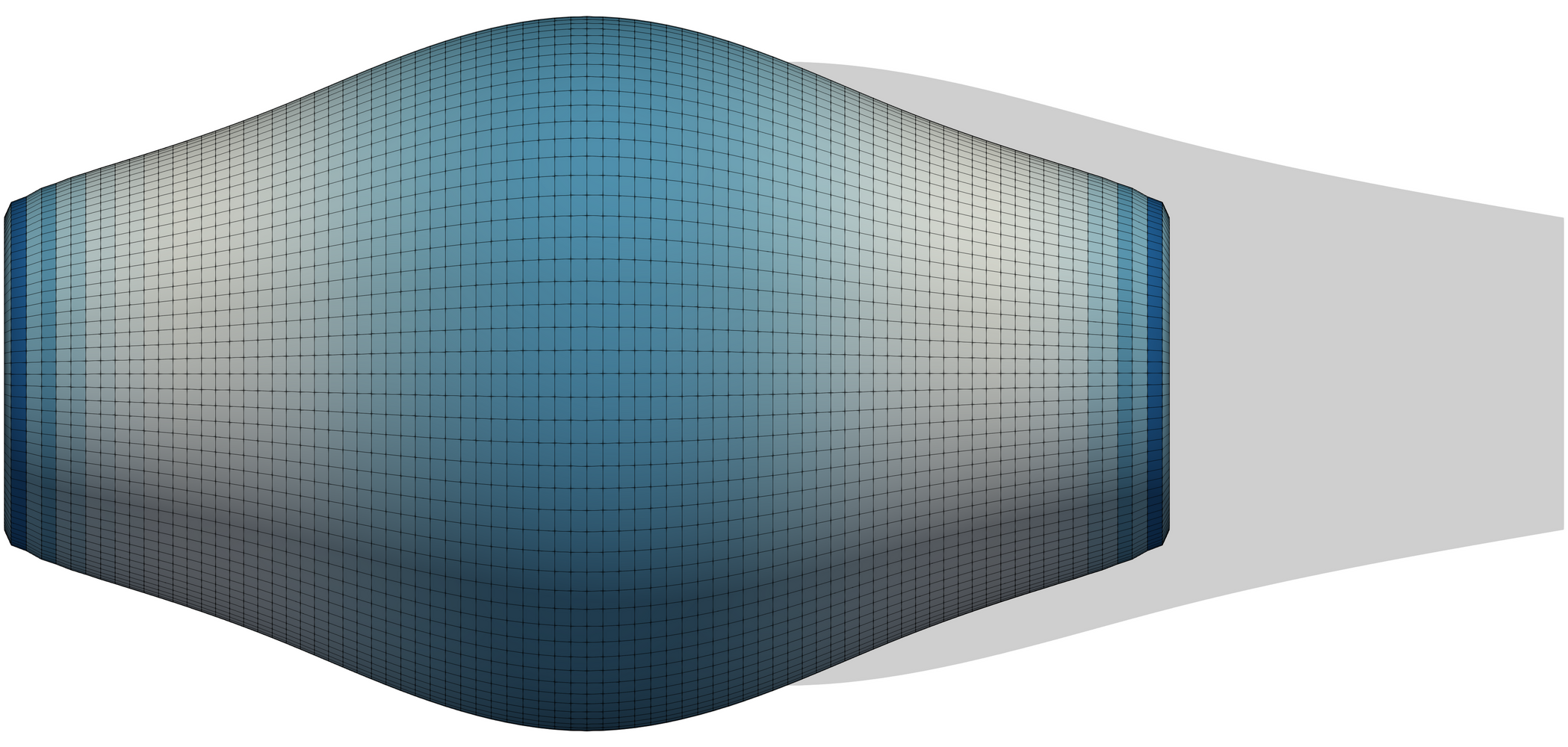}}
    \includegraphics[clip,trim=0 {.5\ht0} 0 0, width=\linewidth,left]{figures/fig_13_solid_cauchy_fib_dir_combi_n4_fc_element_0020.png}
    \endgroup
    \\
    \begingroup\sbox0{\includegraphics{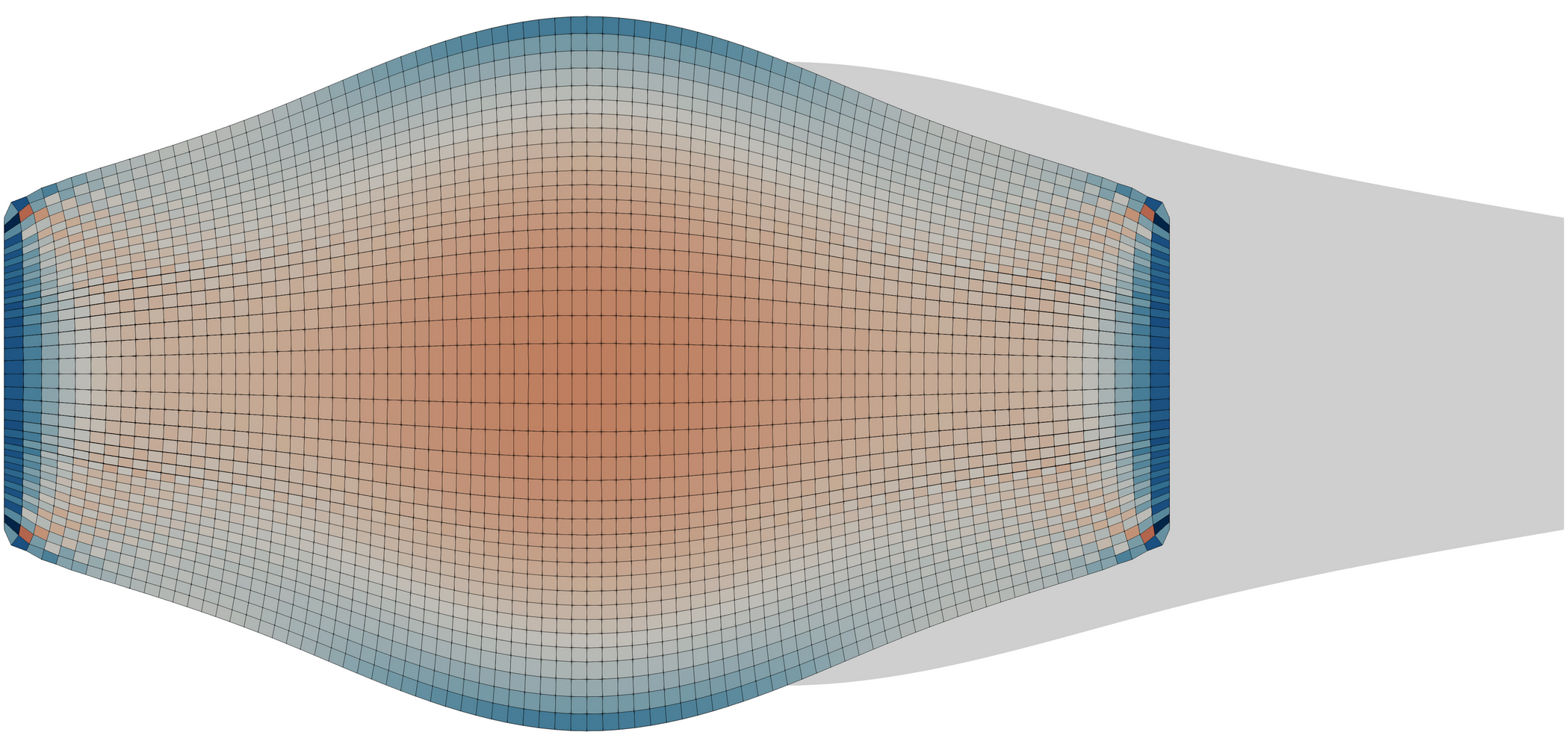}}
    \includegraphics[clip,trim=0 0 0 {.5\ht0}, width=\linewidth,left]{figures/fig_13_slice_zx_cauchy_fib_dir_combi_n4_fc_element_0020.png}
    \endgroup} \hspace*{6pt}
    \parbox{\LW}{ \begingroup\sbox0{\includegraphics{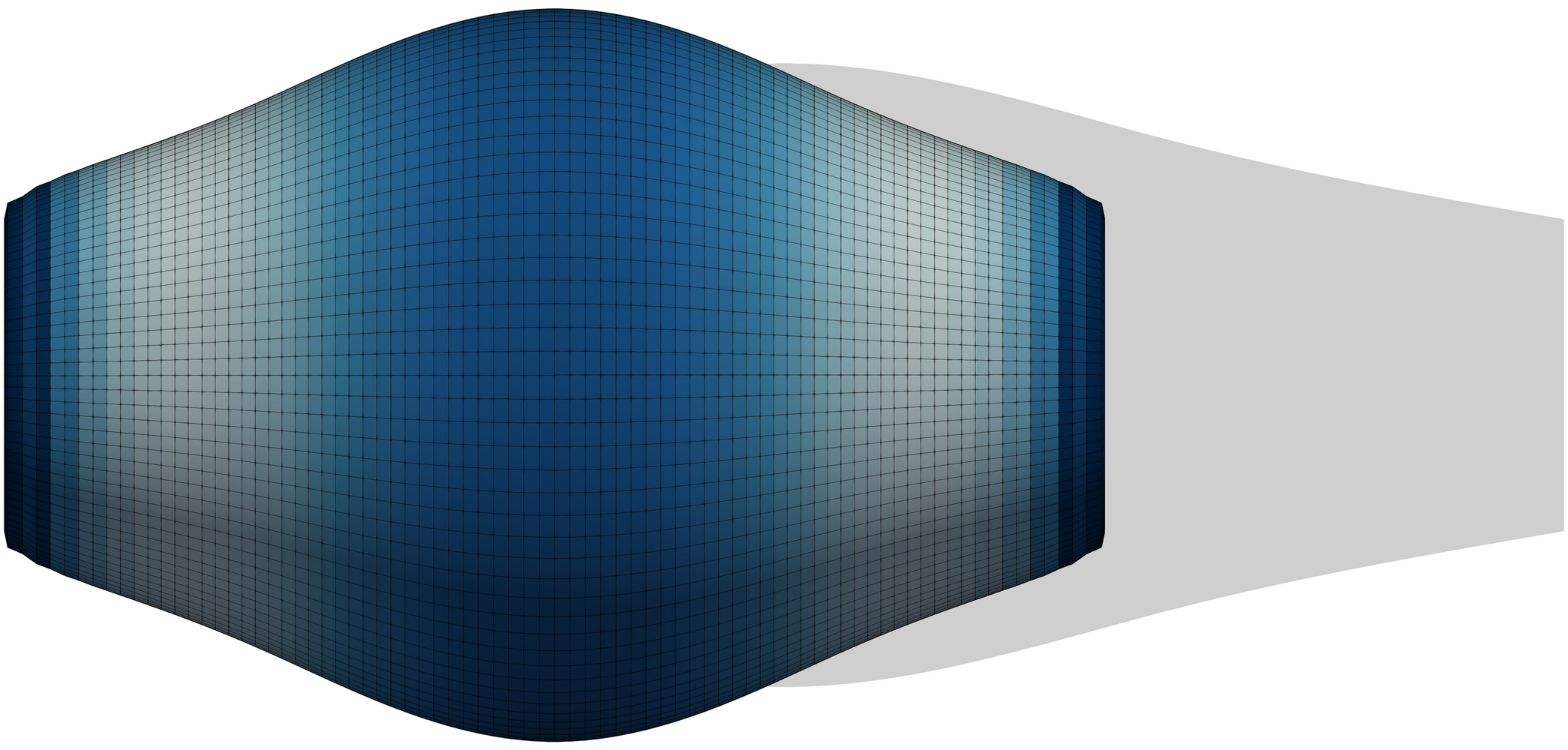}}
    \includegraphics[clip,trim=0 {.5\ht0} 0 0, width=\linewidth,left]{figures/fig_13_solid_cauchy_fib_dir_combi_n4_fc_element_0150.png}
    \endgroup
    \\
    \begingroup\sbox0{\includegraphics{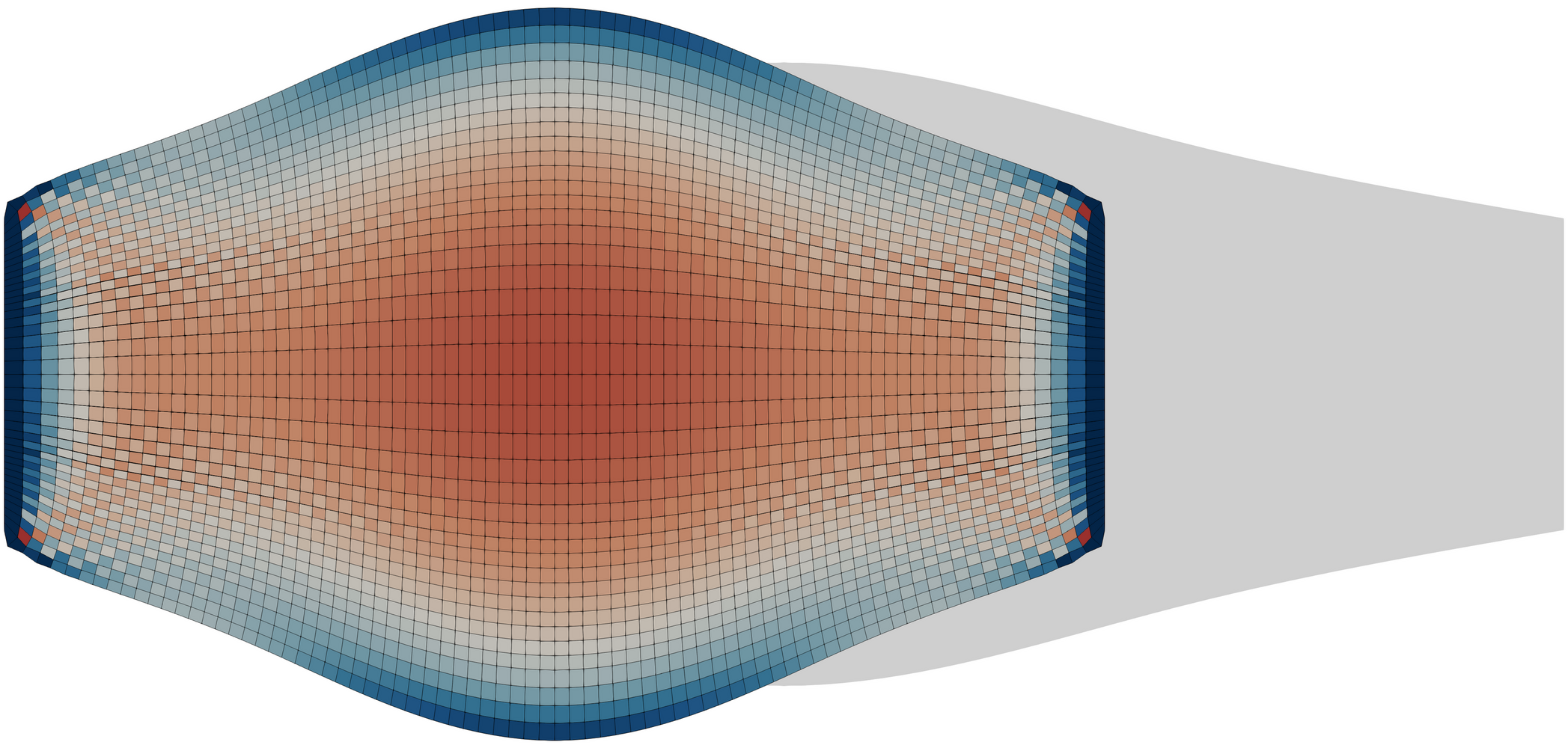}}
    \includegraphics[clip,trim=0 0 0 {.5\ht0}, width=\linewidth,left]{figures/fig_13_slice_zx_cauchy_fib_dir_combi_n4_fc_element_0150.png}
    \endgroup}
\end{minipage}
\hfill
\begin{minipage}[c]{1.29cm}
\raggedleft
\includegraphics[height=3cm]{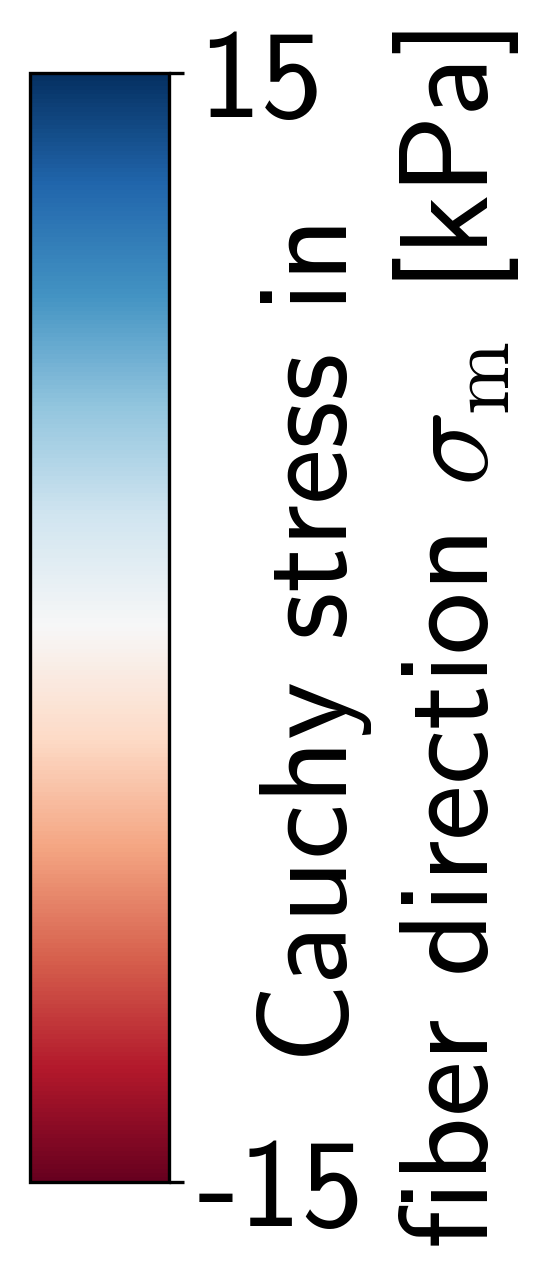}
\end{minipage}
    \caption{Cauchy stress in fiber direction $\sigma_\mathrm{m}$ for a free contraction of the fusiform muscle ($n=4$) at selected times. Results are visualized on the surface (top) and in the axial cross-section (bottom) in comparison to the initial configuration (grey).}
	\label{fig:fusi_free_cauchy_stress_vis}
\end{figure}

Second, we analyze local deformations and stress distributions. We evaluate the fiber stretch $\lambda$, the Cauchy stress in fiber direction $\sigma_\mathrm{m}$, and, as a measure for the combined stress, the von Mises stress $\sigma_\mathrm{v}$. 
For the isometric contraction, results are visualized in the final deformed configuration ($\lambda$ in Fig. \ref{fig:fusi_iso_stretch_vis}, $\sigma_\mathrm{m}$ in Fig. \ref{fig:fusi_iso_cauchy_stress_vis}, and $\sigma_\mathrm{v}$ in Fig. \ref{fig:fusi_iso_mises_stress_vis} in the Appendix). Results of the free contraction are shown at three selected points in time ($\lambda$ in Fig. \ref{fig:fusi_free_stretch_vis}, $\sigma_\mathrm{m}$ in Fig. \ref{fig:fusi_free_cauchy_stress_vis} and $\sigma_\mathrm{v}$ in Fig. \ref{fig:fusi_free_mises_stress_vis} in the Appendix). Since the conclusion drawn from the investigation of $\sigma_\mathrm{v}$ and $\sigma_\mathrm{m}$ align, we focus on a thorough examination of $\sigma_\mathrm{m}$.

\begin{figure}[htb]
    \centering 
    \begin{minipage}[b]{0.8\textwidth}
        \begin{subfigure}[t]{0.18\textwidth}
            \centering \includegraphics[height=3cm]{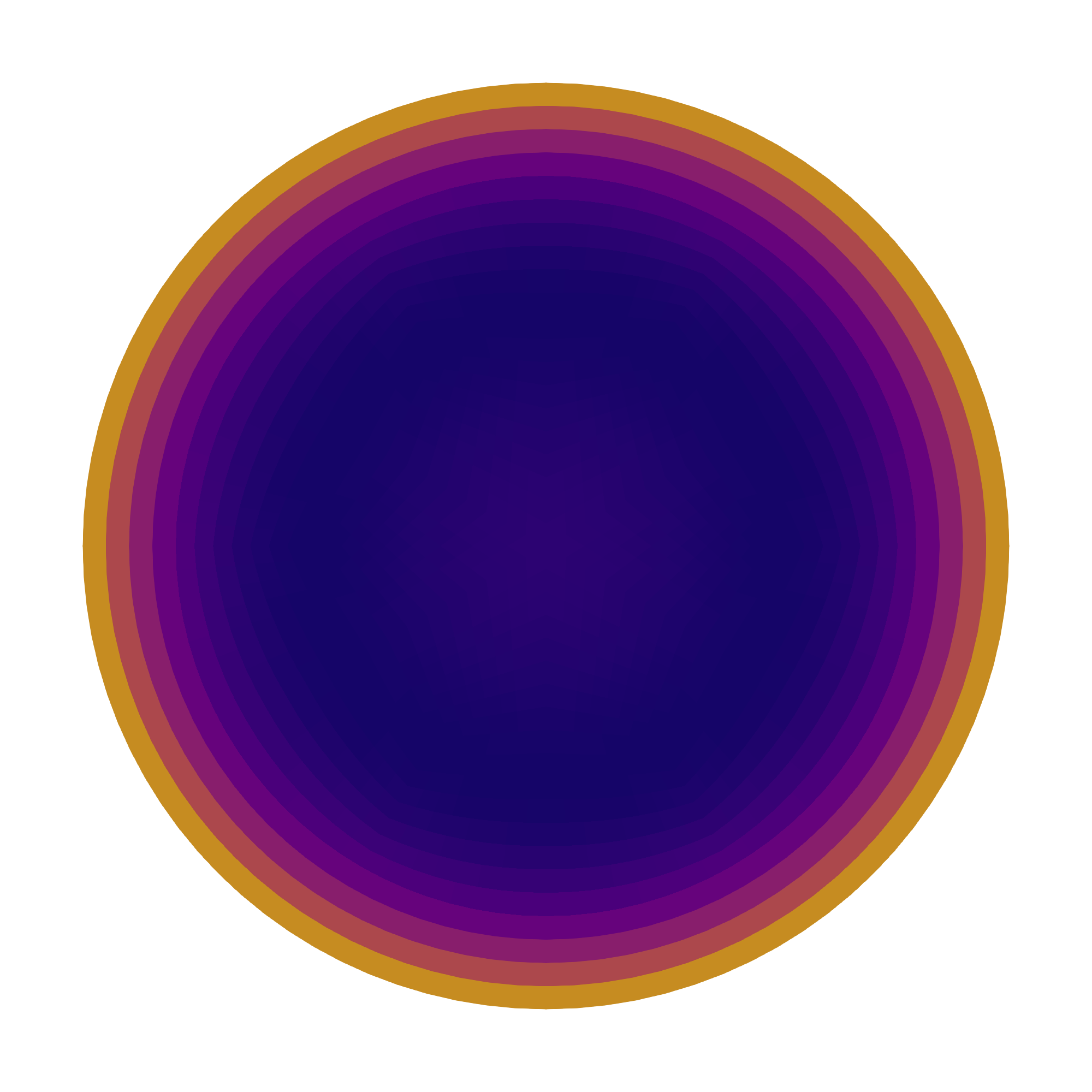}
        \end{subfigure} \hfill
        \begin{subfigure}[t]{0.18\textwidth}
        	\centering \includegraphics[height=3cm]{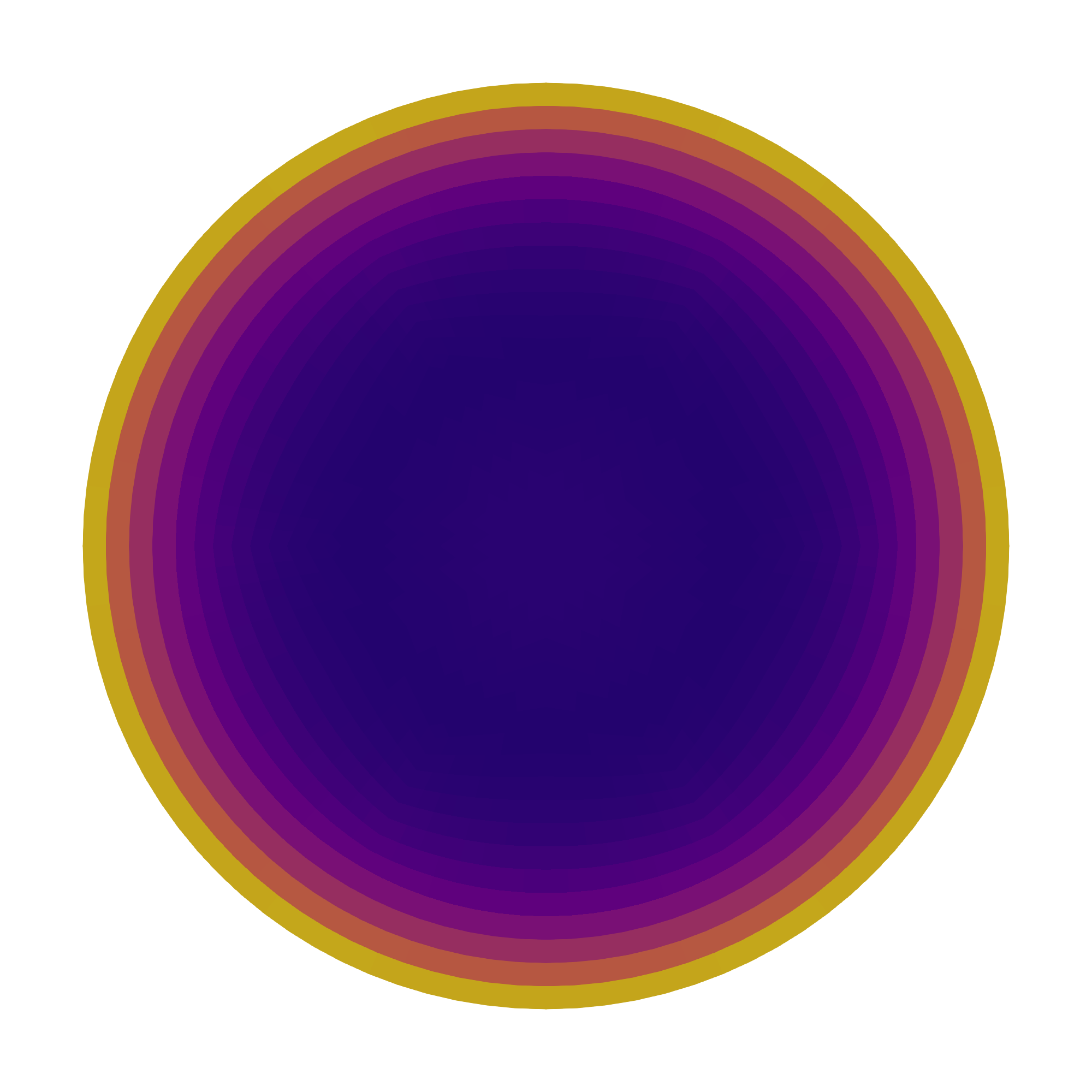}
        \end{subfigure} \hfill
        \begin{subfigure}[t]{0.18\textwidth}
        	\centering \includegraphics[height=3cm]{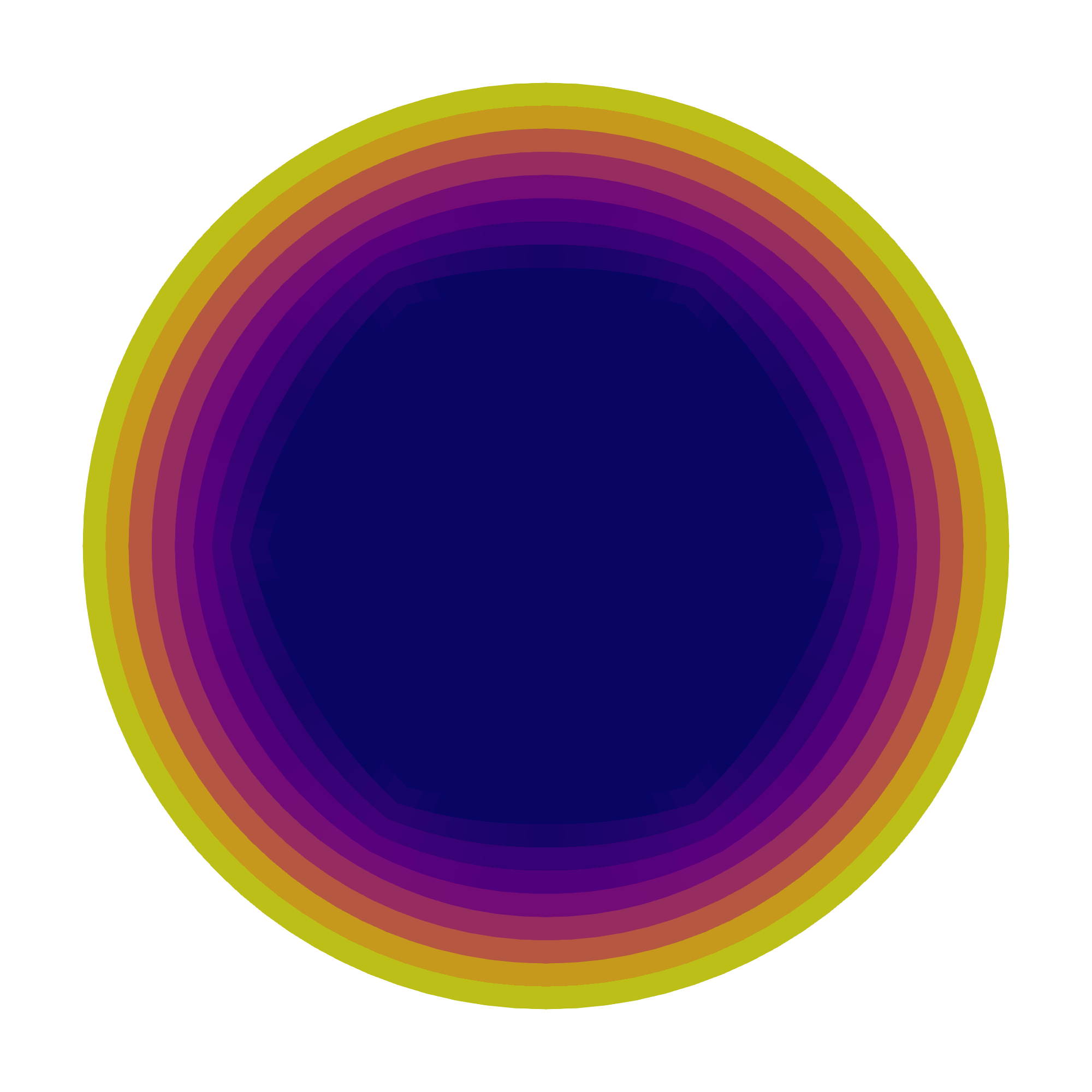}
        \end{subfigure} \hfill
        \begin{subfigure}[t]{0.18\textwidth}
            \centering \includegraphics[height=3cm]{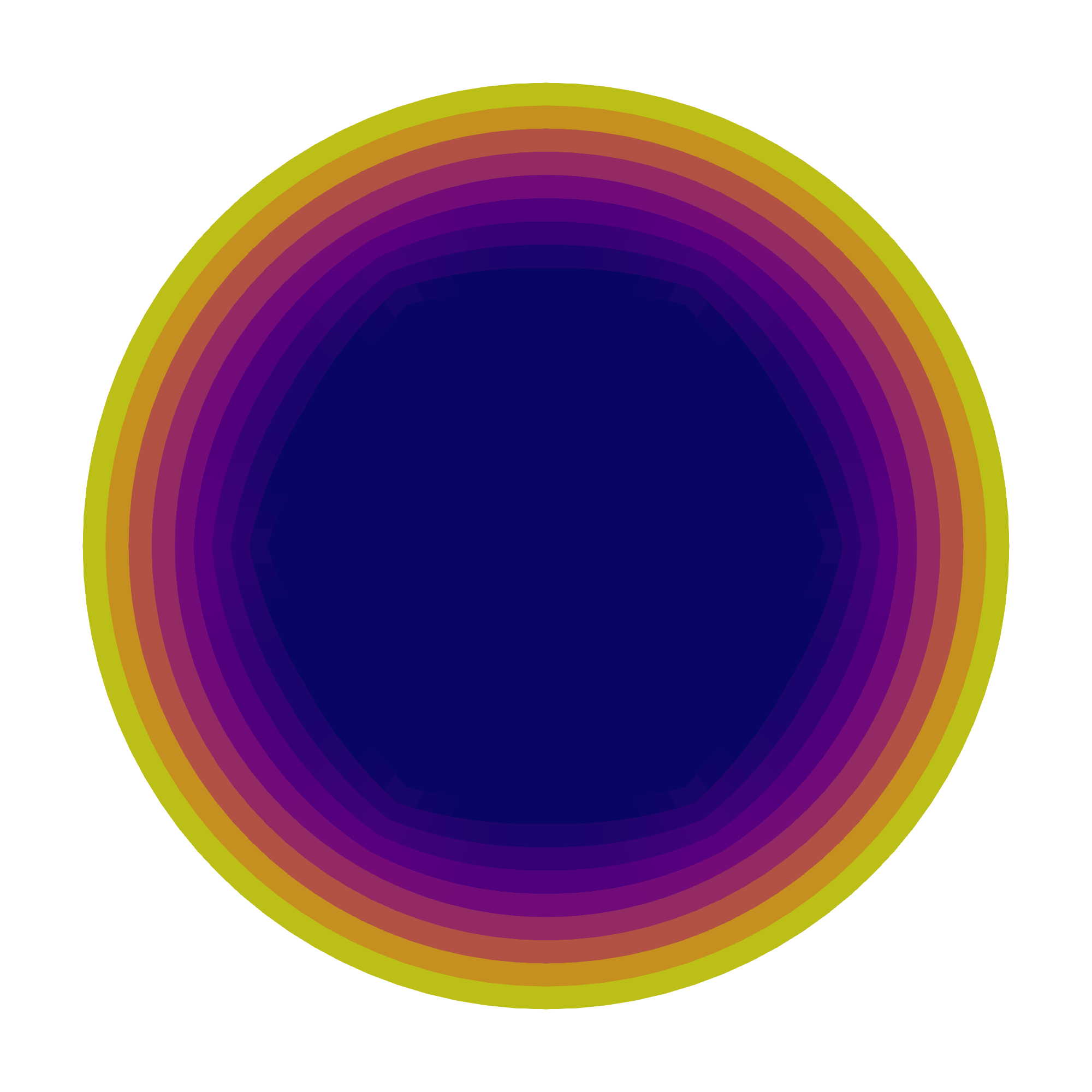}
        \end{subfigure}
    \\
        \begin{subfigure}[t]{0.18\textwidth}
        \centering
        \begingroup\sbox0{\includegraphics{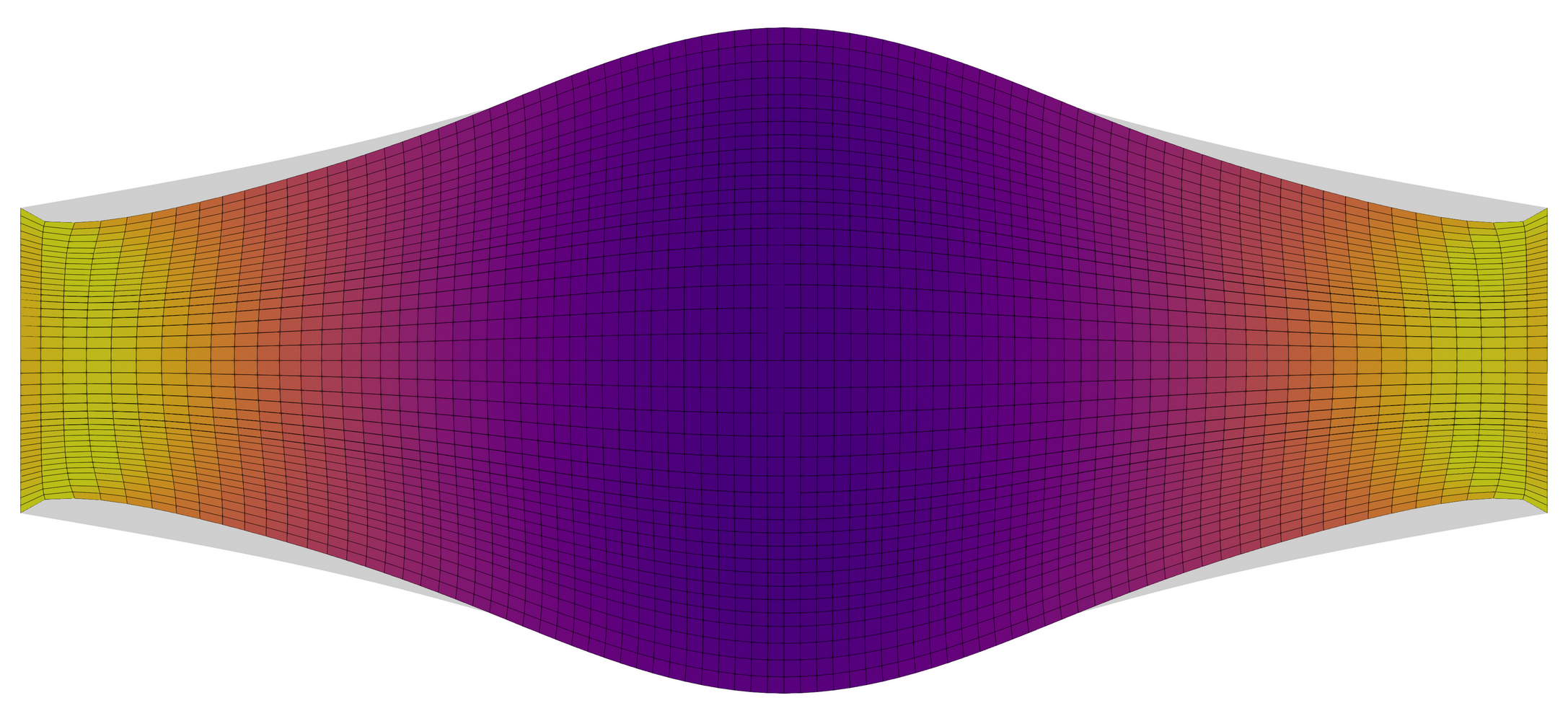}}
        \includegraphics[clip,trim={.5\wd0} 0 0 0, height=3cm]{figures/fig_11_slice_zx_cauchy_fib_dir_blemker_n4_ic_element.png}
        \endgroup 
        \caption{\BLE} \label{fig:fusi_iso_cauchy_stress_vis_ble}
        \end{subfigure} \hfill
        \begin{subfigure}[t]{0.18\textwidth}
        \centering \begingroup\sbox0{\includegraphics{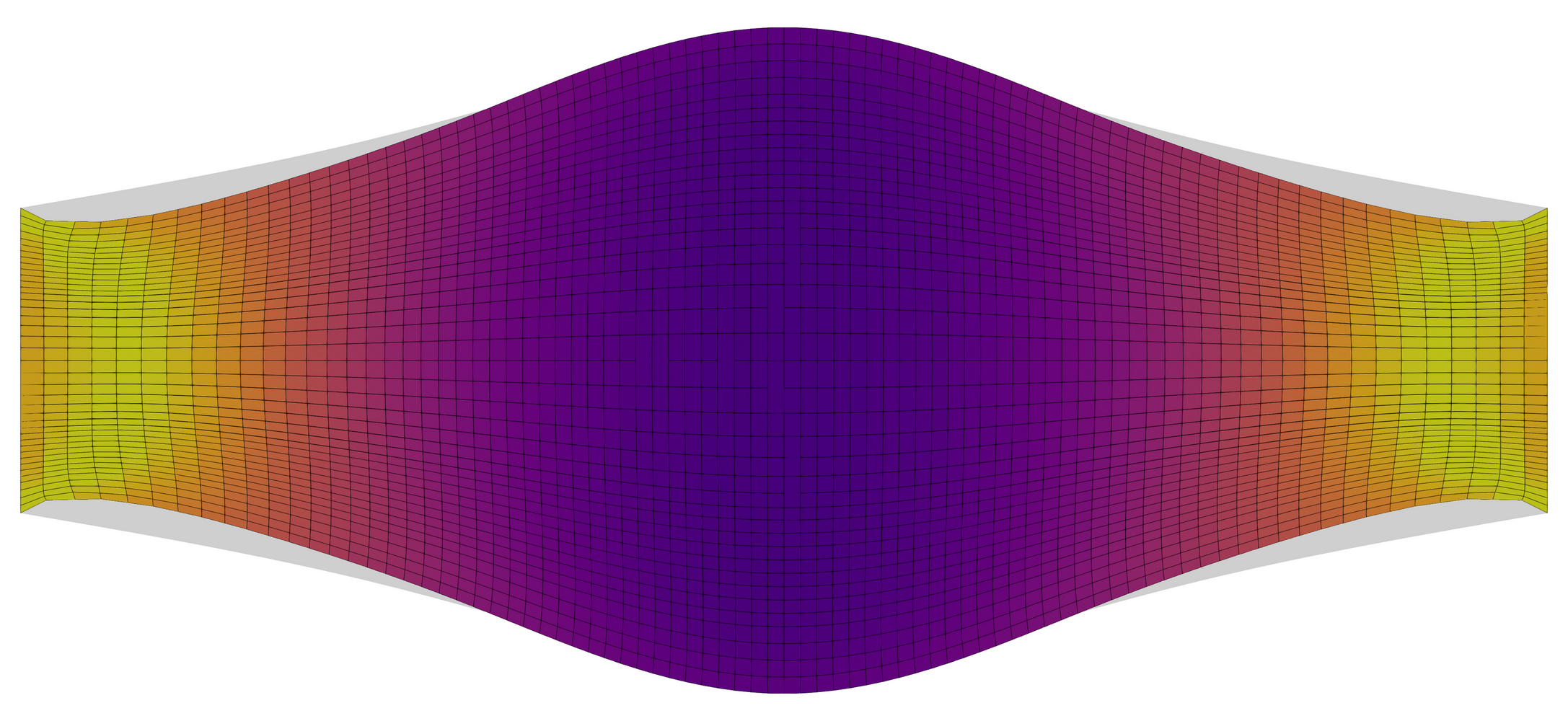}}
        \includegraphics[clip,trim={.5\wd0} 0 0 0, height=3cm]{figures/fig_11_slice_zx_cauchy_fib_dir_giantesio_n4_ic_element.png}
        \endgroup \caption{\GIANT} \label{fig:fusi_iso_cauchy_stress_vis_giant}
        \end{subfigure}  \hfill
        \begin{subfigure}[t]{0.18\textwidth}
        \centering \begingroup\sbox0{\includegraphics{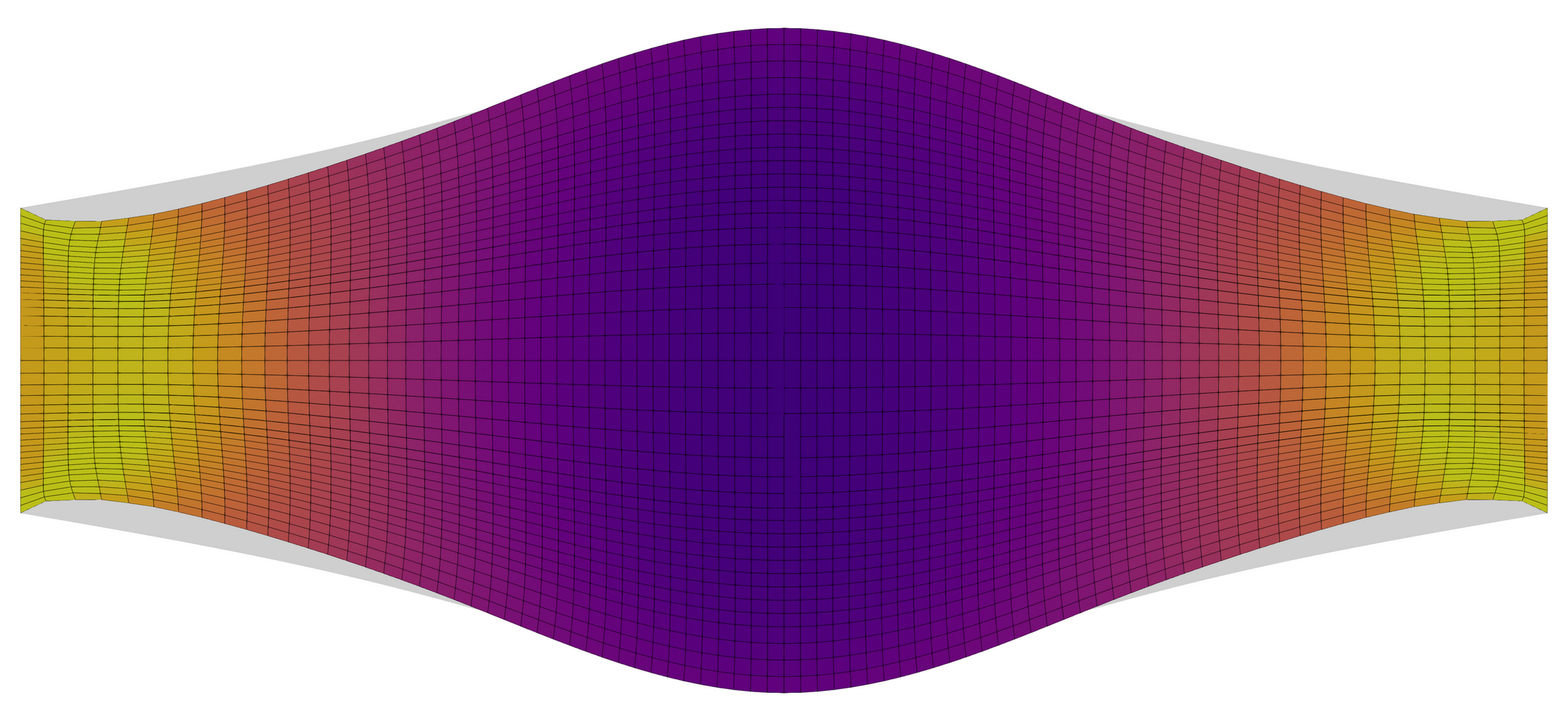}}
        \includegraphics[clip,trim={.5\wd0} 0 0 0, height=3cm]{figures/fig_11_slice_zx_cauchy_fib_dir_weickenmeier_n4_ic_element.png}
        \endgroup \caption{\WKM} \label{fig:fusi_iso_cauchy_stress_vis_wkm}
        \end{subfigure} \hfill
        \begin{subfigure}[t]{0.18\textwidth}
        \centering \begingroup\sbox0{\includegraphics{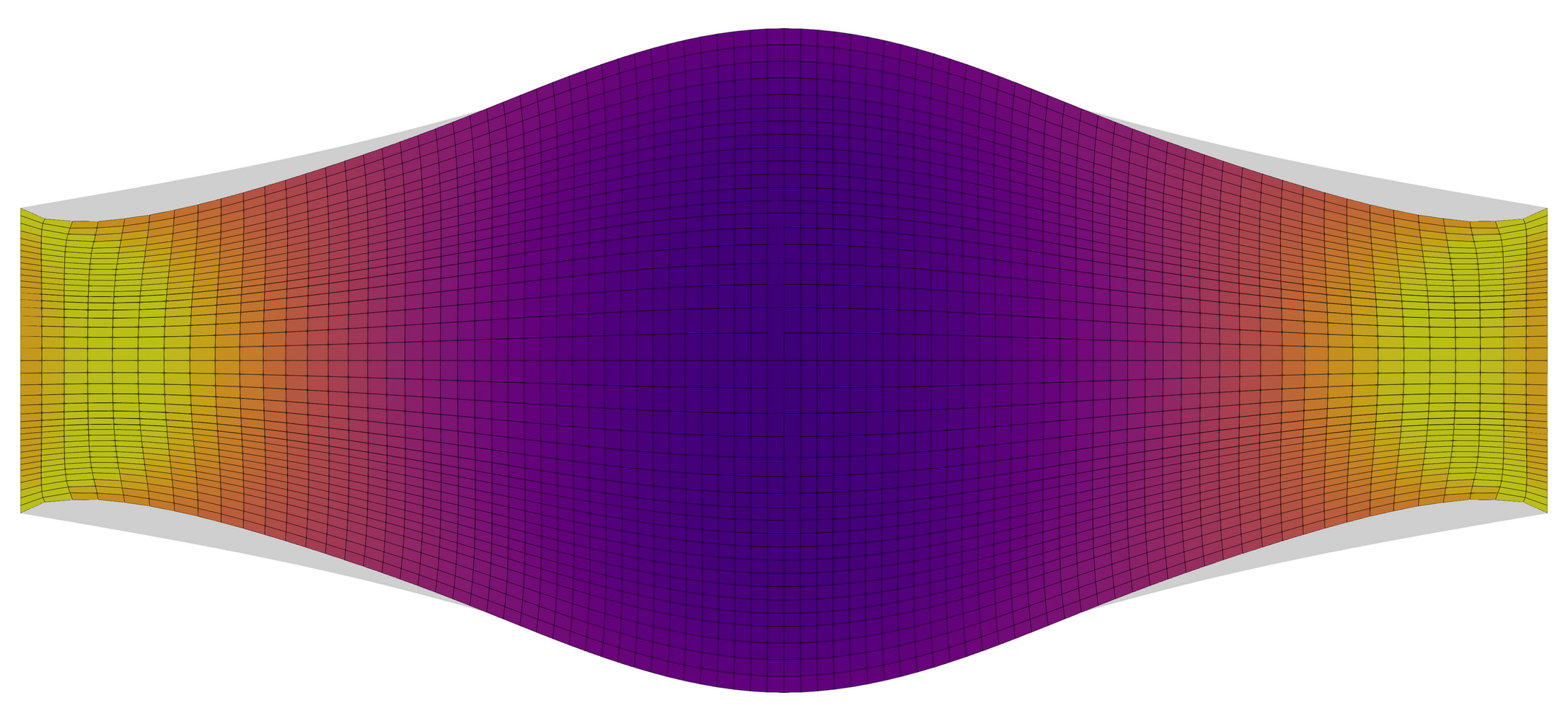}}
        \includegraphics[clip,trim={.5\wd0} 0 0 0, height=3cm]{figures/fig_11_slice_zx_cauchy_fib_dir_combi_n4_ic_element.png}
        \endgroup \caption{\COMBI} \label{fig:fusi_iso_cauchy_stress_vis_combi}
        \end{subfigure}
    \end{minipage}
    \hfill
    \begin{minipage}[b]{1.29cm}
        \raggedleft
    	\includegraphics[height=6cm]{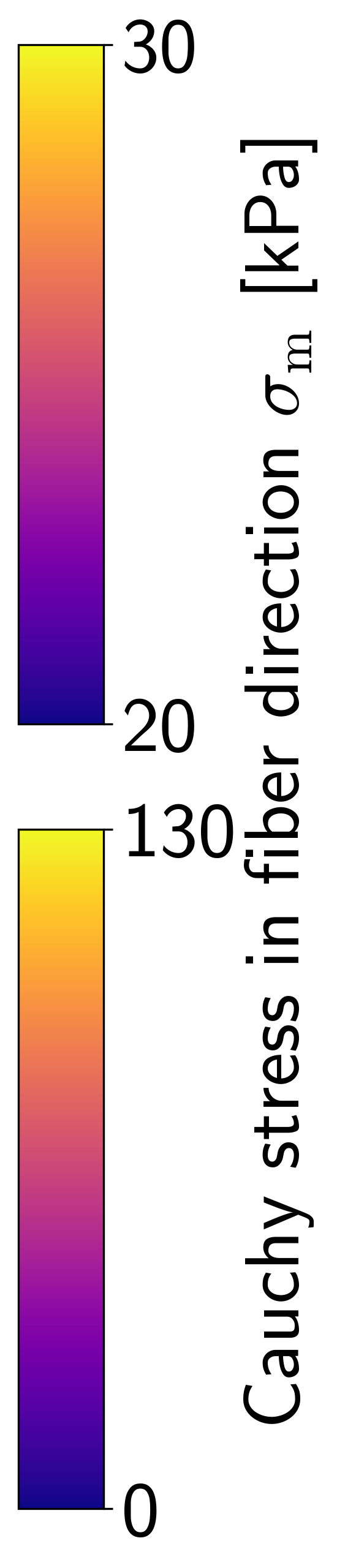}
    \end{minipage}
    \caption{Cauchy stress in fiber direction $\sigma_\mathrm{m}$ in the central cross-section (top) and in the axial cross-section (bottom) of the fusiform muscle ($n=4$) for an isometric contraction in the tetanized state at $t=\SI{0.15}{\s}$. The initial configuration is displayed in grey. Only half the symmetric muscle is visualized.}
	\label{fig:fusi_iso_cauchy_stress_vis}
\end{figure}

\begin{figure}[htb]
	\centering
	\includegraphics[height=4.5cm]{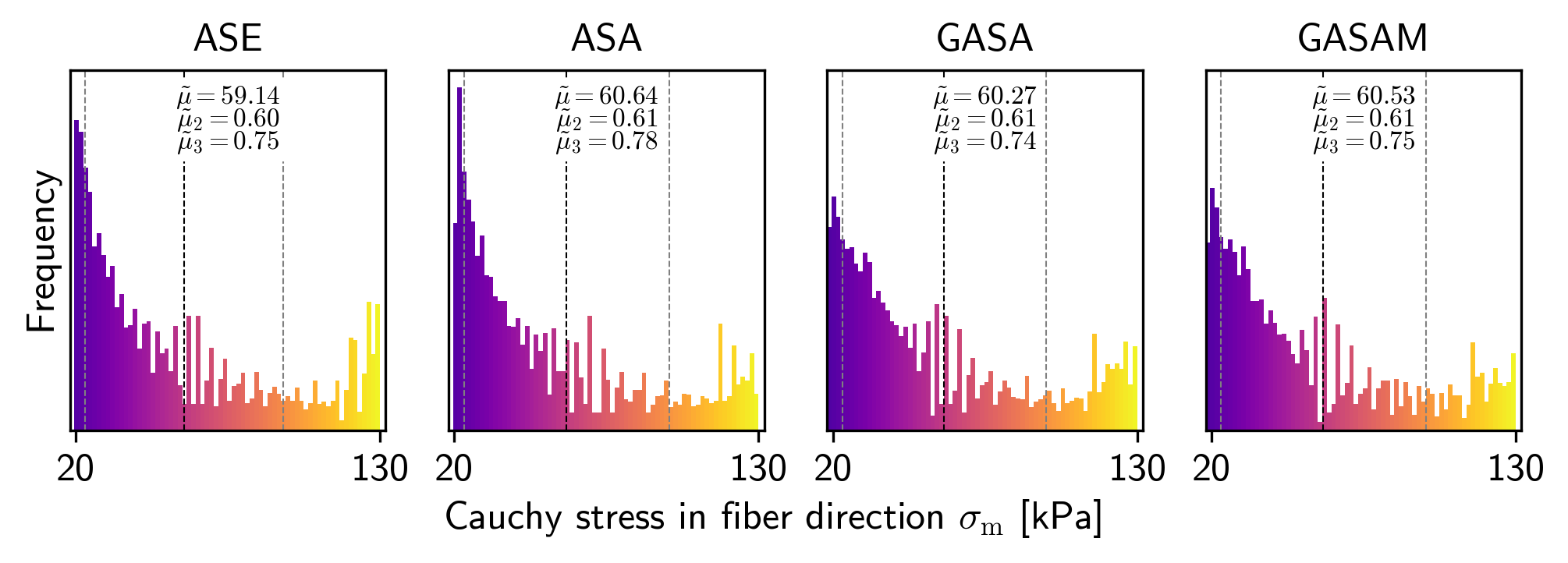}
	\caption{Distribution of Cauchy stresses in fiber direction $\sigma_\mathrm{m}$ over the entire domain of the fusiform muscle ($n=4$) for an isometric contraction in the tetanized state at $t=\SI{0.15}{\s}$. The plotted frequency represents the number of elements (bar height) with stress values within a defined range (bar width). While the average $\Tilde{\mu}^\BleName < \Tilde{\mu}^\GiantName \approx \Tilde{\mu}^\WkmName \approx \Tilde{\mu}^\CombiName$, the variances  $\Tilde{\mu}_2$ are similar. The skewness $\Tilde{\mu}_3$ is most pronounced for the \GIANT- and least pronounced for the \WKM-material model.}
	\label{fig:fusi_iso_cauchy_stress_histogram}
\end{figure}

\noindent\subparagraph{Isometric contraction}
The isometric muscle force $F_{33}$ increases over time up to the tetanic force maximum (see Fig. \ref{fig:fusiform_plot_iso_free}). Between the investigated material models, the computed force maxima are close to equal. For the \GIANT- and \WKM-model, the separate force peaks caused by the superposition of the individual twitches in $\ft$ are clearly visible. Due to the use of the smooth function $\factive$, the force increases continuously for the \BLE- and \COMBI-models. 
A detailed look at the stress distribution in the radial direction (in the center cross-section), reveals that $\sigma_\mathrm{m}$ is more evenly distributed for the \BLE- and \GIANT-models, while the \WKM- and \COMBI-models reveal a larger radial gradient (see Fig. \ref{fig:fusi_iso_cauchy_stress_vis}). A closer inspection of $\sigma_\mathrm{m}$ over the entire fusiform continuum in Fig. \ref{fig:fusi_iso_cauchy_stress_histogram} in Section \ref{sec:appendix_fusi_vis_distribution} the Appendix confirms those small deviations.

In the following, we provide a brief explanation as to why the muscle forces $F_{33}$ coincide while the distributions of the stress $\sigma_\mathrm{m}$ vary. 
Due to the $\mathbf{e}_{1}$-$\mathbf{e}_{2}$-symmetry, the central cross-section at $l=\SI{5}{cm}$ does not deform in $\mathbf{e}_{3}$-direction and fibers remain aligned in $\mathbf{e}_{3}$-direction. 
Consequently, $\sigma_{33}$ equals $ \sigma_\mathrm{m}$, and $F_{33}$ equals the force acting in fiber direction (i.e., the total muscle force). 
According to the force-stretch dependency, $\sigma_\mathrm{m}$ depends on the fiber stretch $\lambda$. $\lambda$, in turn, is not solely determined by the material model's active and passive stiffness in fiber direction but is rather a result of the complex three-dimensional deformation state. 
Because the material models exhibit different stiffnesses to shear and deformations transverse to the fiber direction, the distribution of $\sigma_\mathrm{m}$ differs, even though stiffnesses in compression and tension along the fiber direction coincide (see Fig. \ref{fig:fitting_active} in the relevant stretch ratio range $\epsilon=\SIrange{0.7}{1.3}{}$).
Since the incompressibility assumption limits the transverse expansion, and the isometric constraint restricts the overall deformation in the $\mathbf{e}_{3}$-direction, in this case, differences in the deformation and stress distribution are not very pronounced. 
Integrated over the cross-section, the remaining differences in $\sigma_\mathrm{m}$ balance out such that the calculated forces are close to equal. For other load cases, where shear or deformation transverse to the fiber direction is more pronounced, the generated muscle force may vary considerably more among the material models.

The local distribution of fiber stretches and stresses in the axial direction does not differ noticeably between the four constitutive laws (see Fig. \ref{fig:fusi_iso_stretch_vis} and Fig. \ref{fig:fusi_iso_cauchy_stress_vis}, respectively). 
In all cases, the muscle center is compressed ($\lambda<1$), while the origin and insertion regions, where the deformation is constrained by the Dirichlet conditions, are stretched ($\lambda>1$). 
The observed stress $\sigma_\mathrm{m}$ is positive in the entire muscle continuum, with values increasing towards the lateral ends. It may seem counterintuitive that although we observe compressive and tensile deformation states, $\sigma_\mathrm{m}$  is always positive. Yet this is easily explained: In contrast to a purely passive material, compression and tension are not specifically related to stresses smaller and larger than zero. Instead, the active contribution shifts the root of the stress-strain curve towards stretches $\lambda<1$ (see Fig. \ref{fig:fitting_active}). Accordingly, positive stress may occur even for compressive stretches. 

\noindent\subparagraph{Free contraction}
During the free contraction, as expected, the deviation from the reference stretch ratio $\epsilon=1$ increases with increasing activation, i.e., the muscle shortens (see Fig. \ref{fig:fusiform_plot_iso_free}). For the \GIANT-, \WKM-, and \COMBI-models, the total shortening is close to equal. In comparison, we observe a higher shortening for the \BLE-model. 

To explain this observation, we consider two factors: different minimal active fiber stretches and different passive resistances in compression. First, the models use different force-stretch dependencies associated with different minimal fiber stretches $\lambdaMin$. 
For $\lambda < \lambdaMin$, the generated active contribution is zero. While for the \WKM-, \GIANT- and \COMBI-model $\lambdaMin = 0.5680$, this value is $0.4\lambdaOpt^\mathrm{ASE} = 0.4906$ for the \BLE-model. Hence, the \BLE-model generates active stresses even for lower fiber stretches such that a larger muscle contraction is to be expected. Second, the \BLE-model exhibits a lower passive resistance against compression in fiber direction (see Fig. \ref{fig:fitting_passive_UTCAF}). 
Both these effects accumulate in the active stress response (see Fig. \ref{fig:fitting_active}). Consider the free contraction of a simple unit cube. In the absence of body forces and external loads, the system is in static equilibrium when it is in its stress-free state. Considering the stress-stretch response for UTCAF in Fig. \ref{fig:fitting_active}, the \BLE-model reaches a stress-free configuration when $\lambda\approx 0.68$ while this is the case for $\lambda\approx 0.71$ for the \WKM-, \GIANT- and \COMBI-models. Of course, stress states are more complex for the three-dimensional fusiform muscle geometry, but this simple analogy explains the observed differences in total shortening well.

In three dimensions we observe the expected compression along the fiber direction ($\lambda<1$ in the entire continuum) and the related transverse expansion. Qualitatively, the distribution of $\lambda$ and $\sigma_\mathrm{m}$ is similar for all material models (see Fig. \ref{fig:fusi_free_stretch_vis} and Fig. \ref{fig:fusi_free_cauchy_stress_vis}). As for the isometric contraction, we observe slight variations that can be attributed to different stiffnesses in shear and compression transverse to the fiber direction. Quantitatively, $\lambda$ and $\sigma_\mathrm{m}$ are lower for the \BLE-model, for the reasons already explained.

In conclusion, all material models are suitable for simulating realistic muscle contractions. However, we observe slight variations in the local deformations and stress distributions. Without additional information about the active stiffness, e.g., transversal to the fiber direction or due to shear load, it is impossible to identify the material model with the most realistic characteristics. Nonetheless, we assume that models based on the generalized active strain approach (\WKM and \COMBI) represent the stress distribution and deformation most accurately, as this activation concept correctly models the serial and parallel coupling between active and passive mechanical properties in reality. Depending on the application scenario, the local deviations may be negligible, e.g., if solely the movement of an adjacent bone -- determined by the global muscle contraction -- is of interest. In a biomechanical analysis of the complex shoulder system, this is likely not the case. The local material characteristics are certainly important for the computation of contact between the joint components, and complex geometries may amplify the variations.

\begin{figure}[tbh]
\centering
\includegraphics[height=5cm]{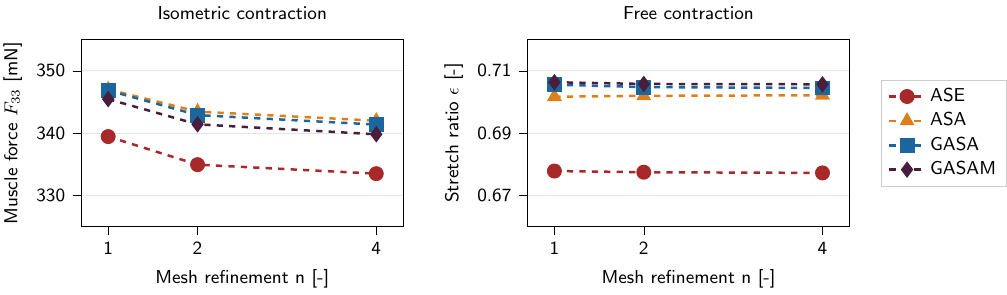}
\captionof{figure}{Influence of the mesh refinement $n$: Simulated muscle force $F_{33}$ and stretch ratio $\epsilon$ at $t=\SI{0.15}{\s}$ for an isometric and a free contraction of the fusiform muscle.}
\label{fig:resolution_influence}
\end{figure}

\paragraph{Influence of the mesh resolution}
As mesh resolution can significantly impact the prediction, we repeated the simulations for a series of mesh refinements $n=1,2,4$. Fig. \ref{fig:resolution_influence} investigates the influence of $n$ on the isometric muscle force $F_{33}$ and the stretch ratio $\epsilon$ for the free contraction. 
The quantities are evaluated for the final configuration at $t=\SI{0.15}{\s}$. 

$F_{33}$ slightly decreases with increasing $n$. However, the influence remains below a maximal deviation of $\SI{1.79}{\percent}$ between the results for $n=1$ and $n=4$ using the \BLE-material model. 
Deviations of $\epsilon$ are close to zero. The \GIANT-model exhibits the most significant though minor deviation ($\SI{0.09}{\percent}$ between the results for $n=1$ and $n=4$).
 
\paragraph{Influence of the incompressibility}
Solving the linear system of equations arising from the FEM discretization requires an iterative scheme for large-size problems. The convergence rate of such iterative solvers strongly depends on the conditioning of the coefficient matrix $\mathbf{A}$. A strictly enforced incompressibility (achieved through a high incompressibility parameter value) can worsen the system's conditioning and thus be difficult to prescribe. In contrast, an insufficiently high incompressibility parameter inadequately approximates the material's incompressible characteristics, leading to a change in volume. 

To analyze the influence of the incompressibility parameter ($\kappa$ for the \GIANT-, \WKM-, and \COMBI-model, and $\kappaBle$ for the \BLE-model) and determine an appropriate value for the simulation of a larger scale problem, we repeat the simulations for the fusiform muscle. Since mesh refinement has no notable effect on the volume change, we restrict the investigation to simulations with mesh refinement $n=1$. 
The incompressibility parameter is varied by the scaling factor $s$ such that $\kappa = s \cdot \kappa_\mathrm{ref}$ with $\kappa_\mathrm{ref}=\SI{1000}{}$, and  $\kappaBle = s \cdot \kappaBle_\mathrm{ref}$ with $\kappaBle_\mathrm{ref} = \SI{10}{\mega \Pa}$. 

We evaluate the percentage volume change $\Delta V = \frac{V-V_{0}}{V_{0}}$ in the final configuration at $t=\SI{0.15}{\s}$. As a measure for the conditioning of the linear system matrix $\mathbf{A}$, we approximate its condition number $k(\mathbf{A}) = \left| \frac{\mathrm{ev}_{\max}(\mathbf{A})}{\mathrm{ev}_{\min}(\mathbf{A})} \right|$ by the ratio of maximal to minimal eigenvalues $\mathrm{ev}$. The closer $k$ is to 1, the better the conditioning and, consequently, the better the convergence rate of the iterative solver.

\begin{figure}[h!]
\centering
\includegraphics[height=8.7cm]{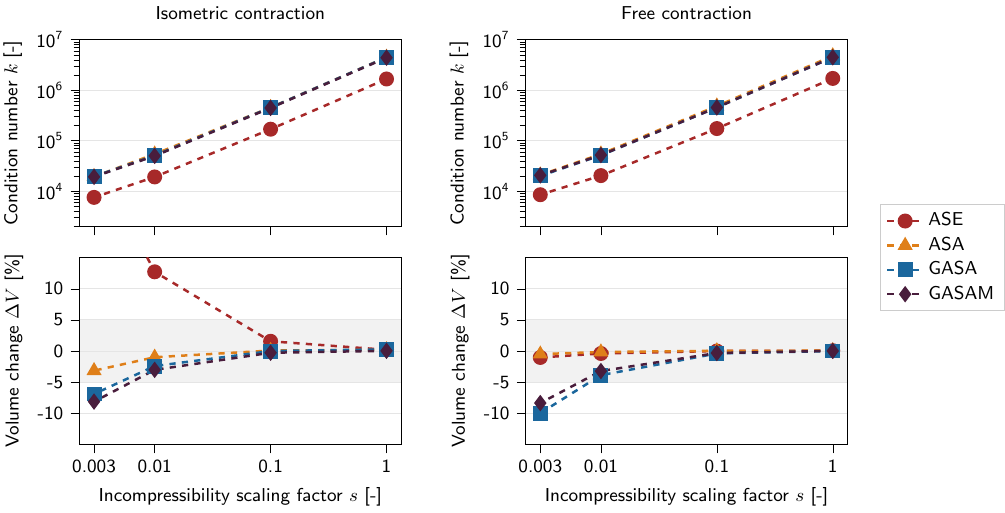}
\captionof{figure}{Influence of the incompressibility parameters $\kappa$ and $\kappaBle$: Simulation results for an isometric and a free contraction of the fusiform muscle ($n=1$). The reference values $\kappa_\mathrm{ref}=\SI{1000}{}$ and $\kappaBle_\mathrm{ref} = \SI{10000}{\kPa}$ were scaled by the factor $s$. The volume change is evaluated in the final configuration at $t=\SI{0.15}{\s}$, and the condition number is plotted for the last Newton iteration of the second time step.}
\label{fig:incomp_influence}
\end{figure}

Fig. \ref{fig:incomp_influence} shows the computed quantities for the free and isometric contraction simulation. As expected, a lower incompressibility penalty leads to higher absolute volume changes, but lower condition numbers. 
Since we consider an absolute volume change of $|\Delta V| = \SI{5}{\percent}$ acceptable, we regard $\kappaBle = 0.1 \cdot \SI{10}{\mega \Pa} = \SI{1}{\mega \Pa}$, $\kappa^{\GiantName} = 0.003 \cdot \SI{1000}{}= \SI{3}{}$, and $\kappa^{\WkmName} = \kappa^{\CombiName} = 0.01 \cdot \SI{1000}{} = \SI{10}{}$ as sufficient for large-scale simulations.

\subsection{Spatiotemporally varying activation and contact interactions in a muscle-bone model} \label{sec:musclebone}
As an intermediate step towards applying the modified and improved \COMBI-material model in a full continuum-mechanical model of the human shoulder, we first consider a simplified model involving two components: the humerus bone and the deltoid muscle. The deltoid serves as the prime mover during arm abduction, thereby lifting the humerus away from the body. Inspired by this scenario, we simulate the deltoid's contraction while accounting for the contact interaction between the two components. 

\paragraph{Geometry and mesh}
Our model is based on the humerus and deltoid geometries (data version 4.3) provided by the BodyParts3D database \cite{mitsuhashi_bodyparts3d_2009}. Further smoothing operations and geometry adaptations are performed using Materialise 3-matic \cite{materialise_3-matic_version_nodate}. Both parts are meshed separately using Gmsh (version 2.12.0) \cite{gmsh}. We convert the obtained linear tetrahedral elements to quadratic tetrahedrons with a custom Python script, resulting in a total of 195189 nodes and 127019 elements. To compute the muscle fiber directions $\mathbf{m}$, we follow the same approach as described for the fusiform muscle example.

\paragraph{Simulation scenario}
To simulate the ball-and-socket-type glenohumeral joint, we fix the center node of the humeral head in space. We fix the deltoid's origin surface nodes and apply tied constraints to connect the deltoid's insertion surface nodes to the humerus. Considering potential contact between the deltoid's and humerus' outer surfaces, we apply a penalty regularization strategy for constraint enforcement.

The humerus bone is modeled using a linear St. Venant-Kirchhoff relation with Young's modulus \(E_\mathrm{b} = \SI{0.1}{\giga\pascal}\), consistent with values reported in the literature (see \cite{carey_situ_2000, smith_tensile_2008}). For the deltoid muscle, we use the \COMBI-model and the parameters specified in Table \ref{tab:fitting_all} (except $\kappa=10$).

\begin{figure}[htb]
\begin{minipage}[c]{0.6\textwidth}
    \begin{subfigure}[b]{0.8\textwidth}
        \begin{subfigure}[t]{0.18\textwidth}
        \centering
        \includegraphics[height=2.25cm]{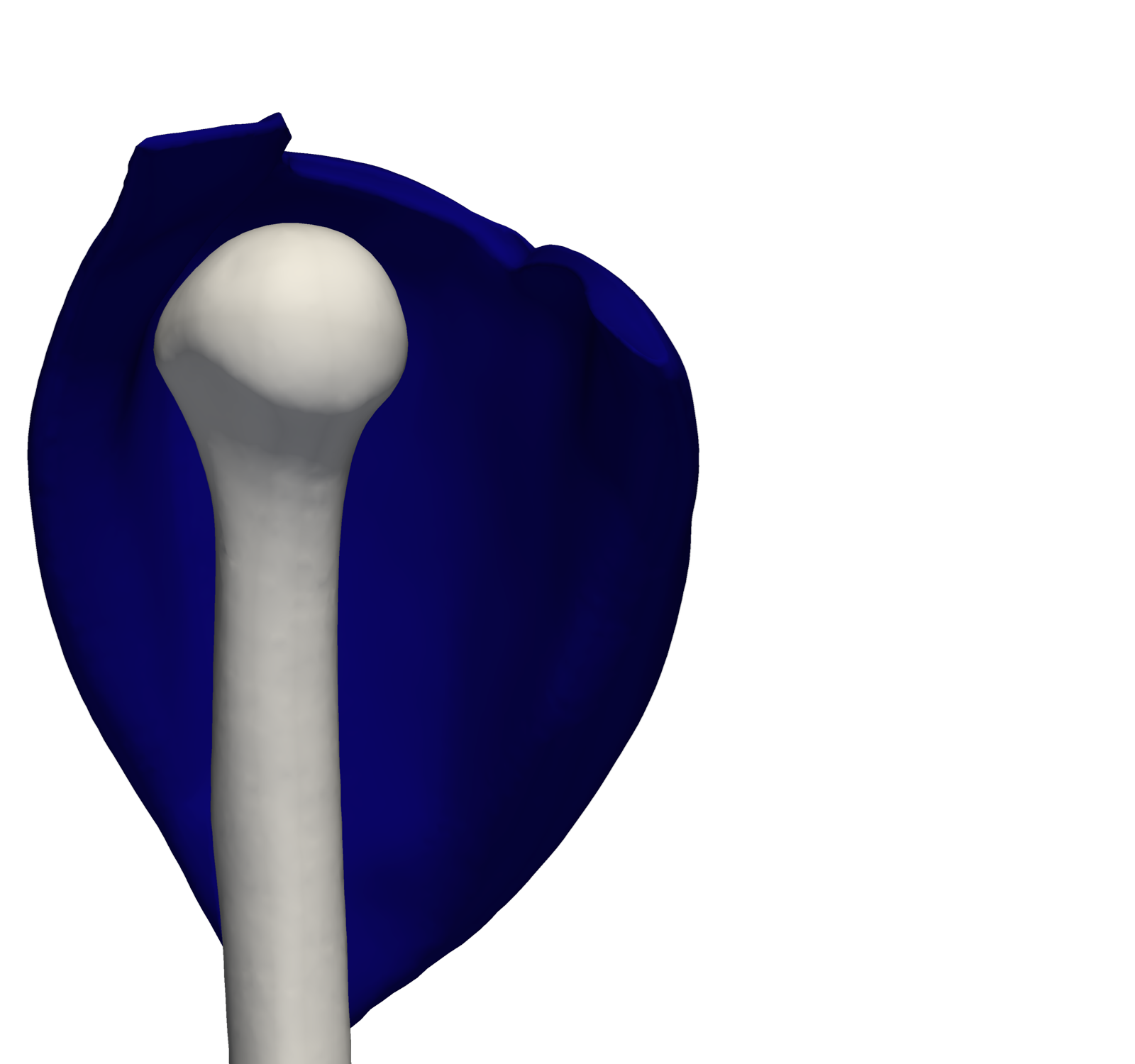}
        \subcaption*{$t = \SI{0.0}{\s}$}
        \end{subfigure} \hfill
        \begin{subfigure}[t]{0.18\textwidth}
        \centering
        \includegraphics[height=2.25cm]{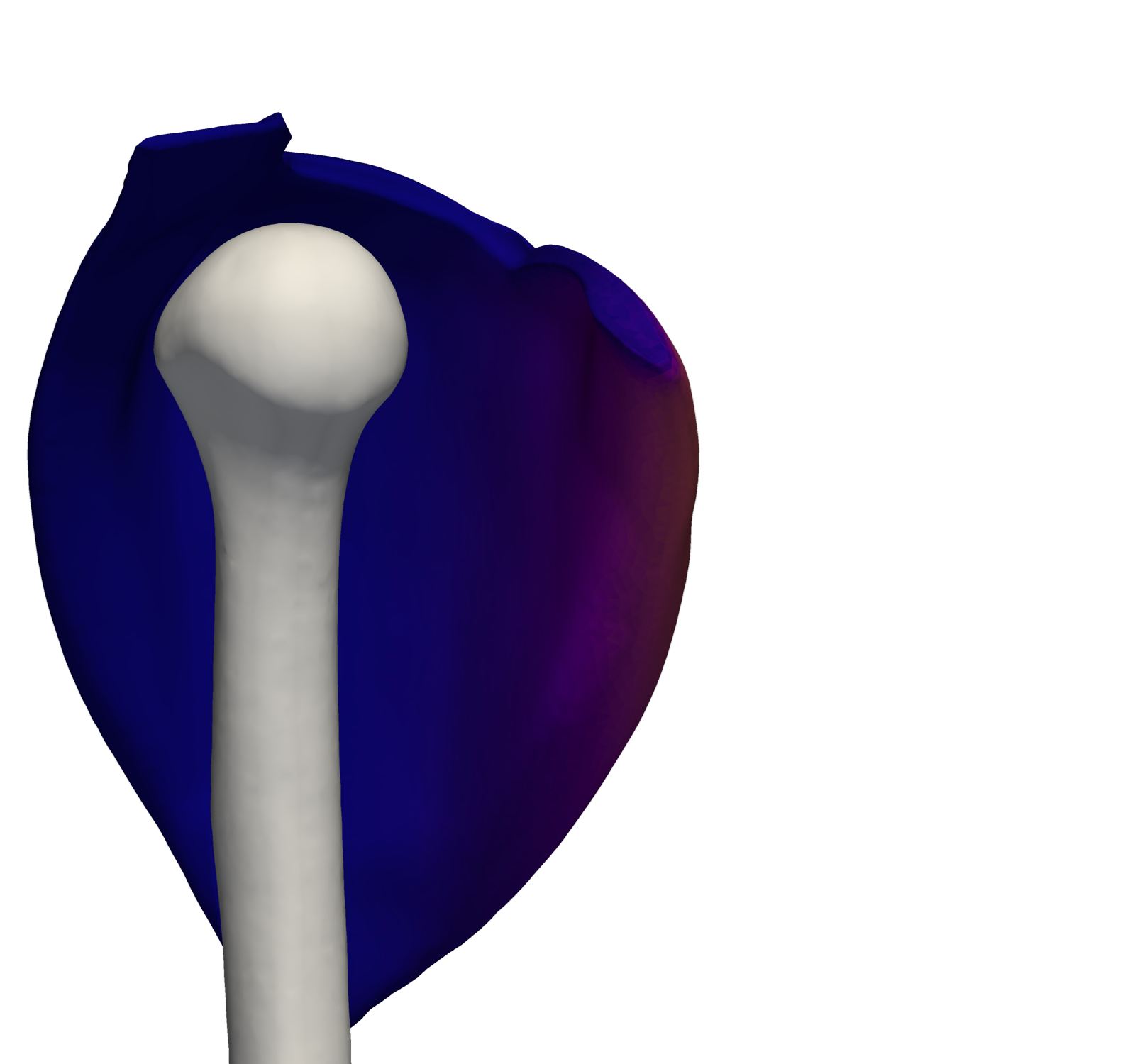}
        \subcaption*{$t = \SI{0.3}{\s}$}
        \end{subfigure} \hfill
        \begin{subfigure}[t]{0.18\textwidth}
        \centering
        \includegraphics[height=2.25cm]{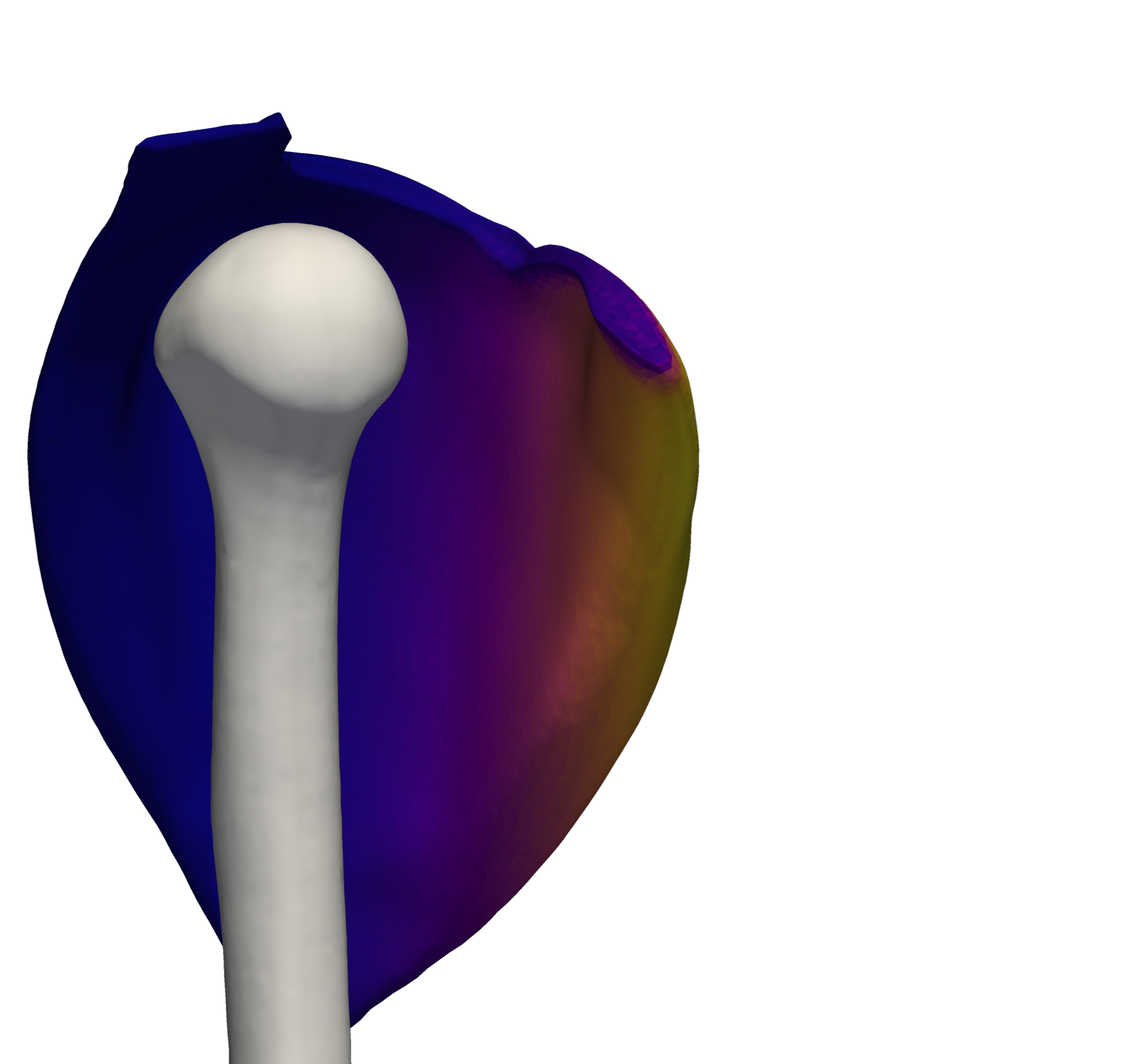}
        \subcaption*{$t = \SI{0.4}{\s}$}
        \end{subfigure} \hfill
        \begin{subfigure}[t]{0.18\textwidth}
        \centering
        \includegraphics[height=2.25cm]{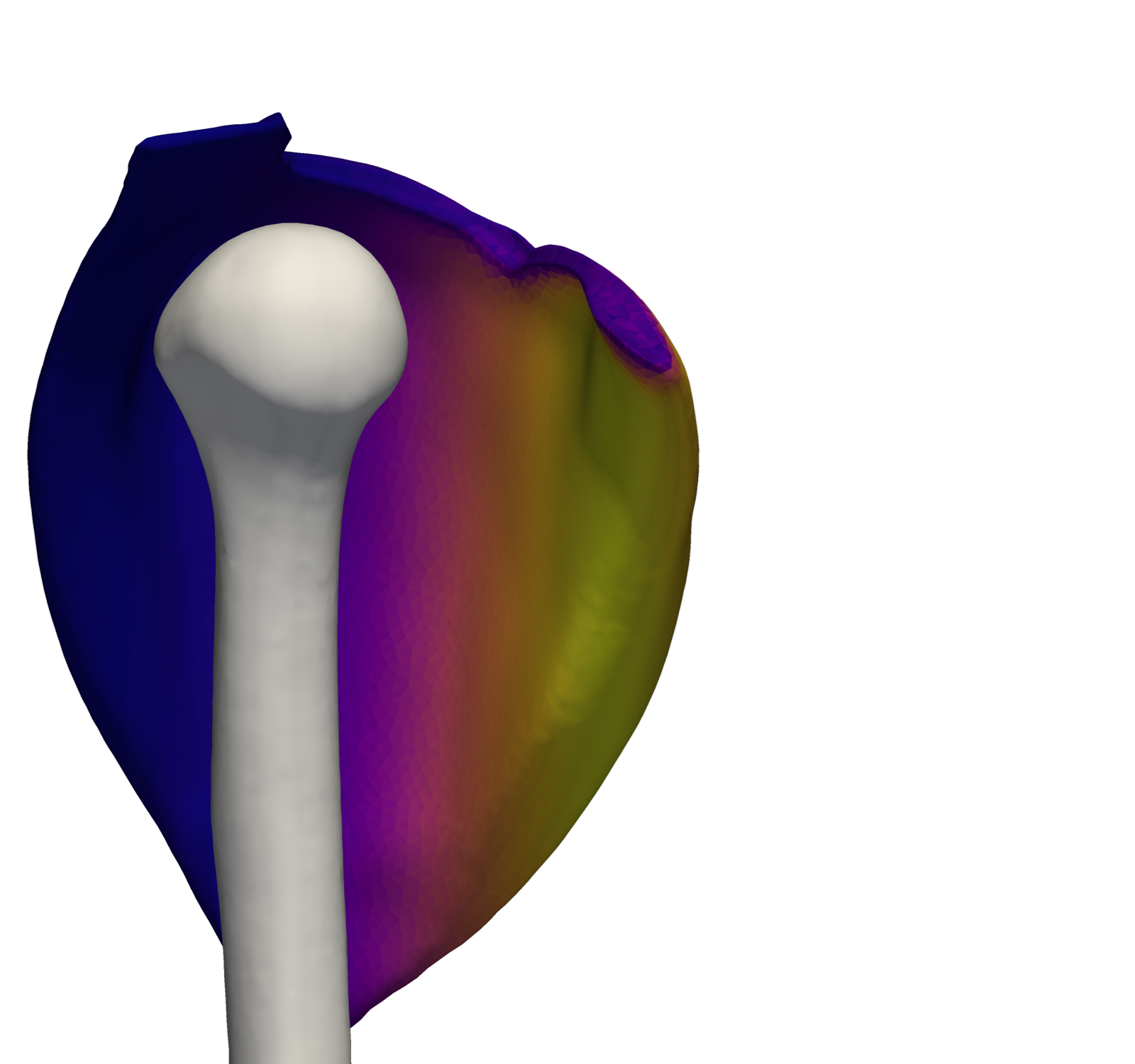}
        \subcaption*{$t = \SI{0.5}{\s}$}
        \end{subfigure}
    \subcaption{Prescribed activation scaling factor}
    \label{fig:results_muscle_shoulder_activation_input}
    \end{subfigure}
        \hfill
        \begin{subfigure}[b]{1.215cm}
        \raggedleft
            \includegraphics[height=3cm]{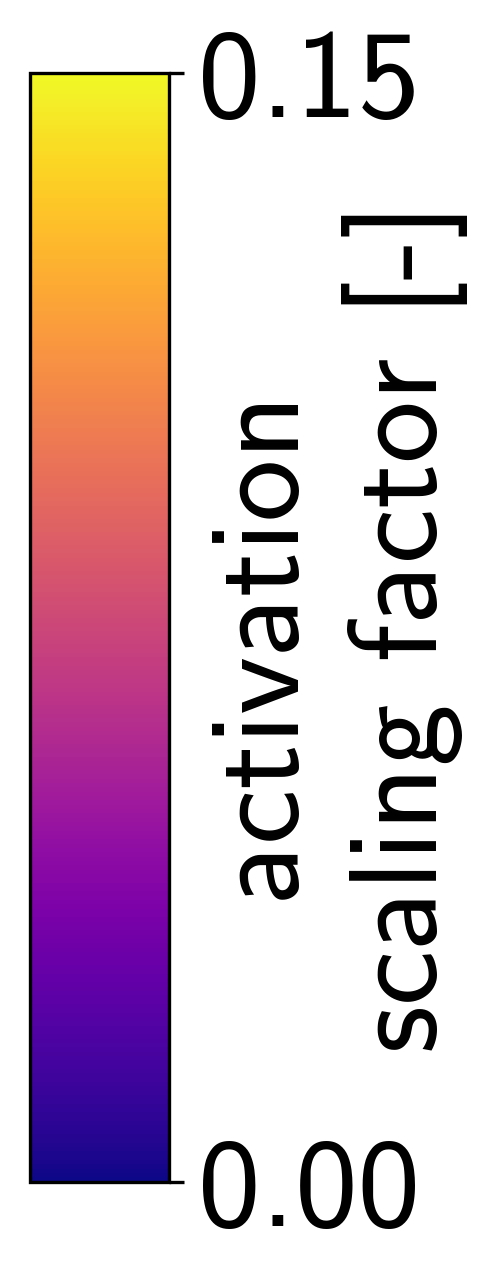}
            \subcaption*{}
        \end{subfigure}\\
        
    \begin{subfigure}[t]{0.8\textwidth}
        \begin{subfigure}[t]{0.18\textwidth}
        \centering
        \includegraphics[height=4.5cm]{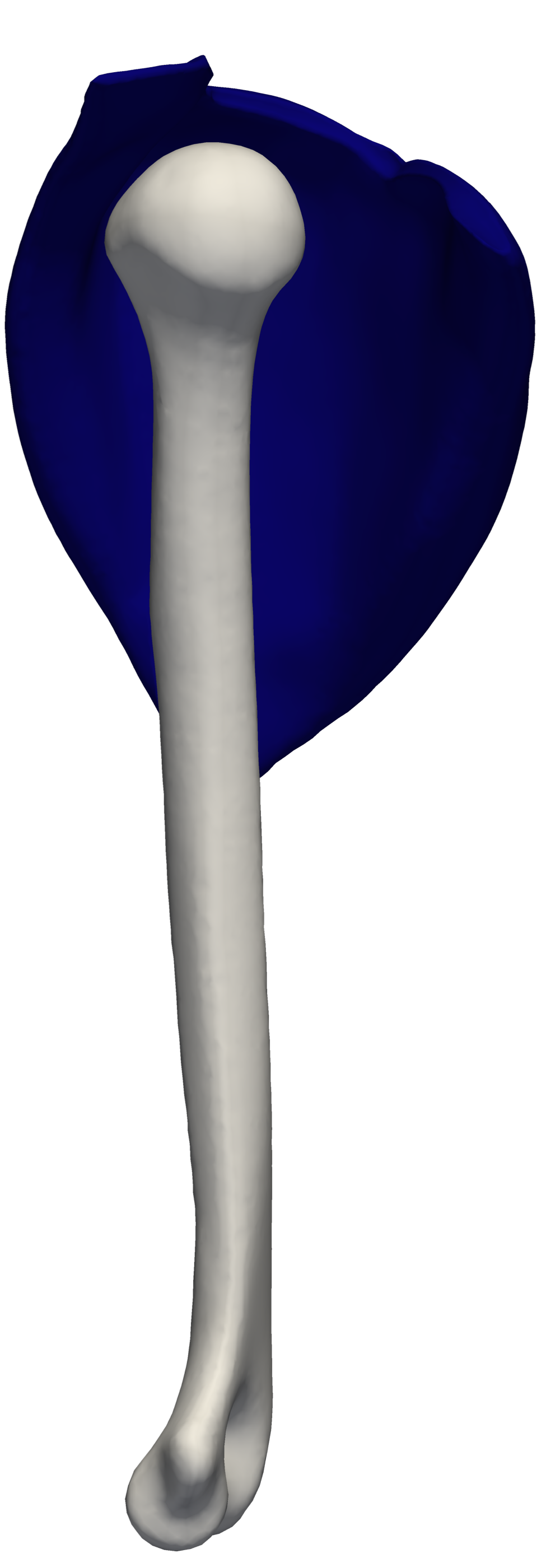}
        \subcaption*{$t = \SI{0.0}{\s}$}
        \end{subfigure} \hfill
        \begin{subfigure}[t]{0.18\textwidth}
        \centering
        \includegraphics[height=4.5cm]{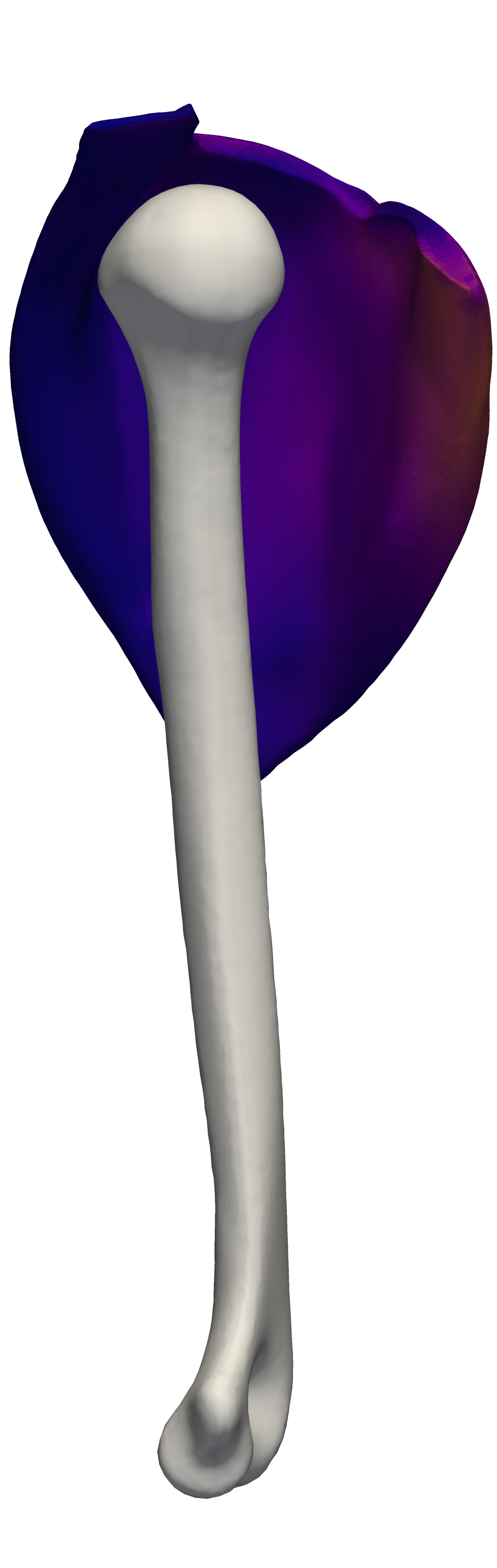}
        \subcaption*{$t = \SI{0.3}{\s}$}
        \end{subfigure} \hfill
        \begin{subfigure}[t]{0.18\textwidth}
        \centering
        \includegraphics[height=4.5cm]{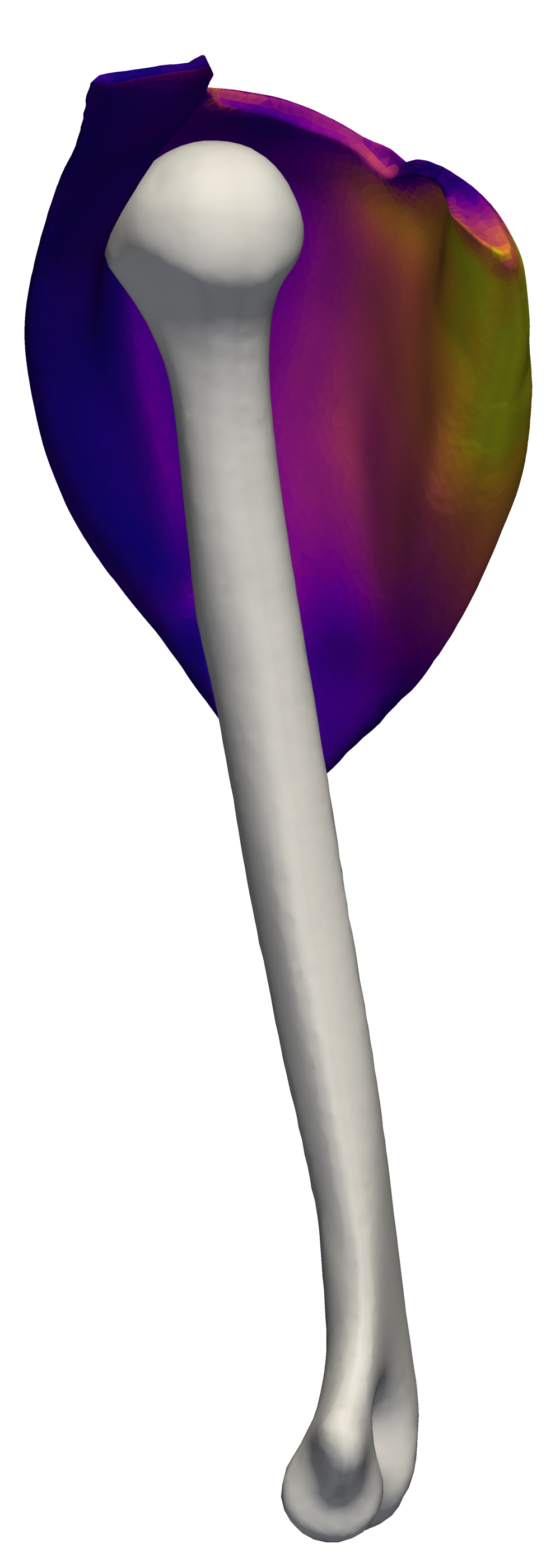}
        \subcaption*{$t = \SI{0.4}{\s}$}
        \end{subfigure} \hfill
        \begin{subfigure}[t]{0.18\textwidth}
        \centering
        \includegraphics[height=4.5cm]{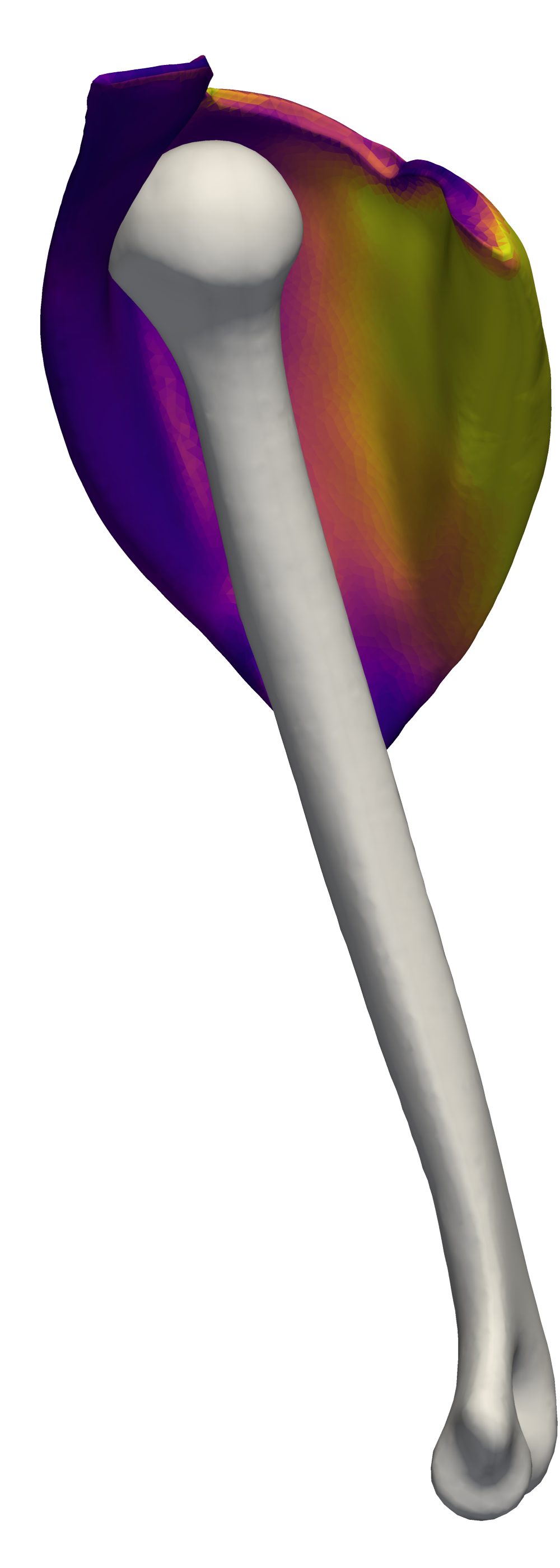}
        \subcaption*{$t = \SI{0.5}{\s}$}
        \end{subfigure}
    \subcaption{Simulated deformation states with von Mises strain $\epsilon_\mathrm{v}$}
    \label{fig:results_muscle_shoulder_strain}
    \end{subfigure}
        \hfill
        \begin{subfigure}[b]{1.215cm}
        \raggedleft
            \includegraphics[height=3cm]{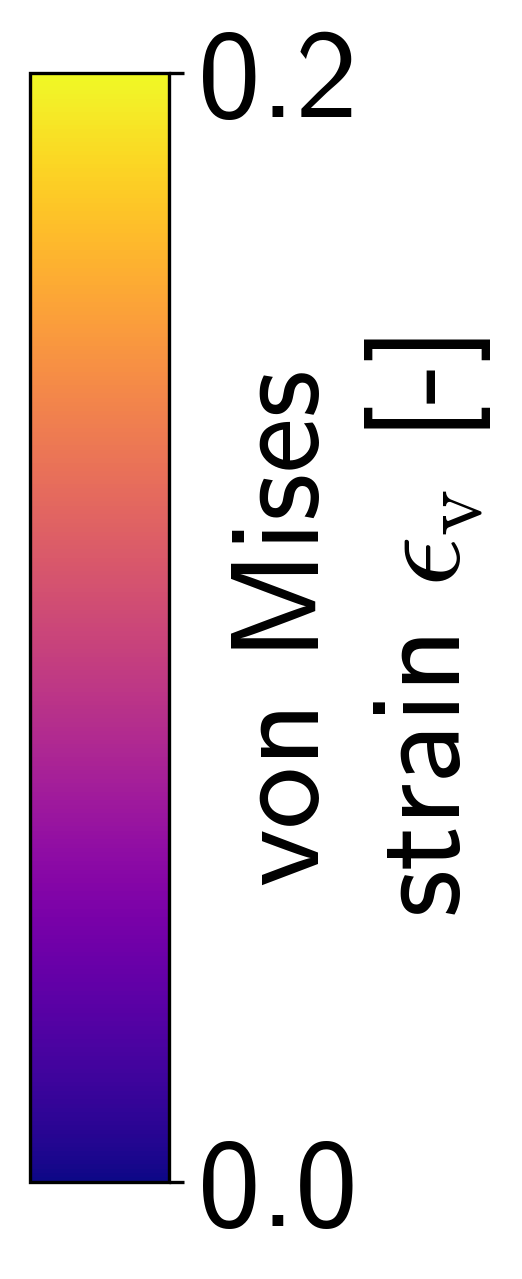}
            \subcaption*{}
        \end{subfigure}
    \end{minipage}\hfill
    \begin{subfigure}[c]{0.35\textwidth}
        \includegraphics[height=5cm]{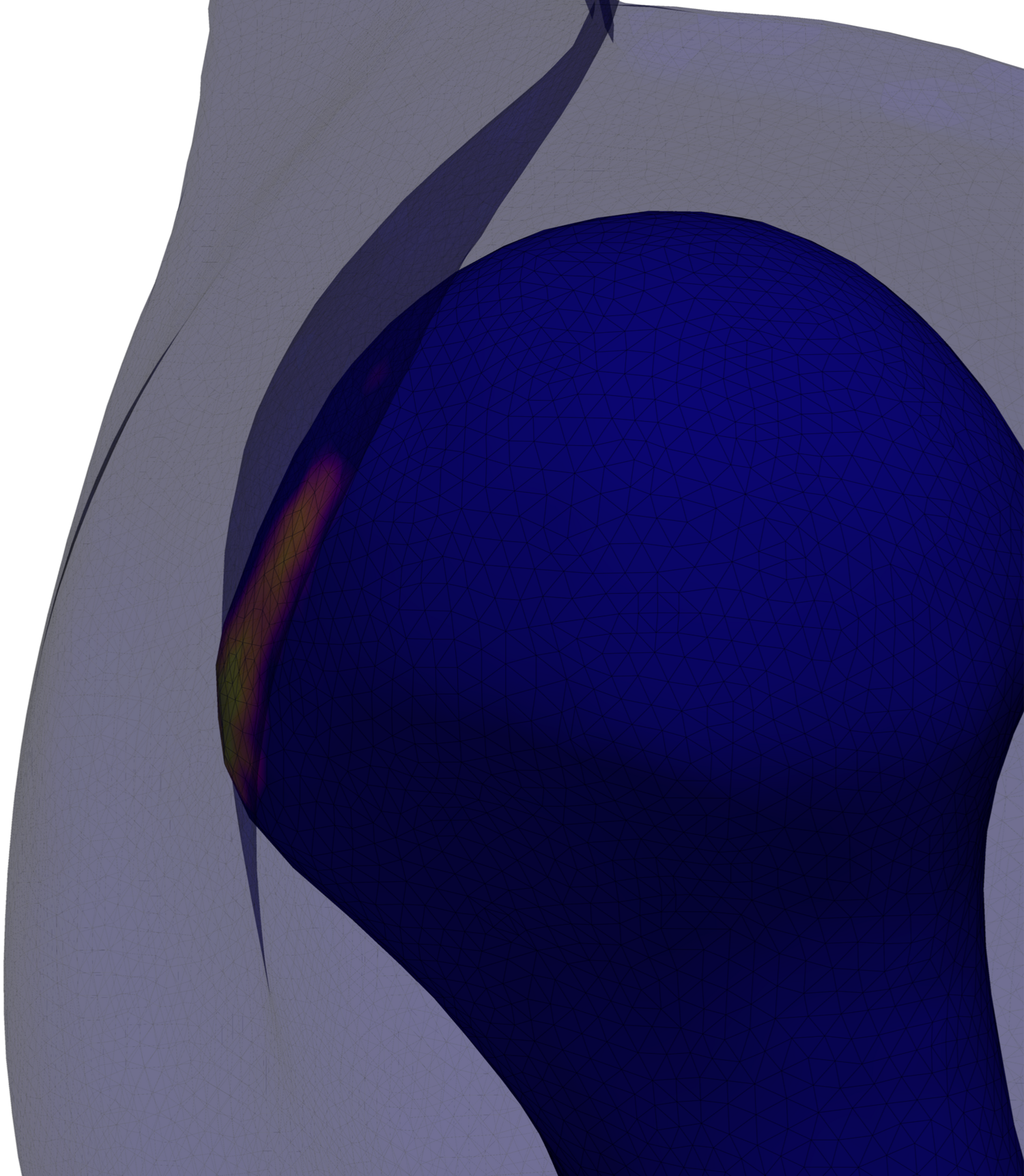}
        \hfill
        \includegraphics[height=3cm]{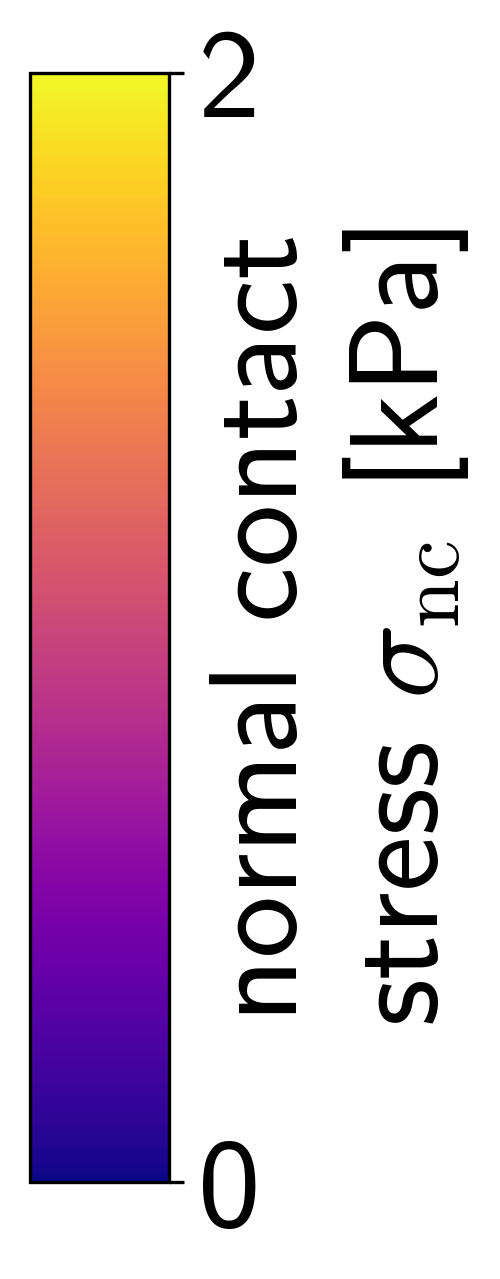}
        \subcaption{Detailed view of the normal contact stress $\sigma_\mathrm{nc}$ between the humerus and the clavicular part of the deltoid at $t = \SI{0.5}{\s}$. For visualization purposes, the deltoid muscle is displayed slightly transparent.}
        \label{fig:results_muscle_shoulder_contact}
    \end{subfigure}
    \caption{Simulation results at selected points in time for a spatiotemporally varying activation in a two-component model of the humerus bone and deltoid muscle. \textbf{(a)}: The activation scaling factor is prescribed element-wise. Starting from the spinal part of the deltoid (right), the activation increases over time and moves towards the acromial part (center). \textbf{(b)}: The simulated contraction of the deltoid causes an abduction of the humerus in the spinal direction \textbf{(c)}: As the humerus is abducted in the spinal direction, the clavicular part of the deltoid is drawn towards the humeral head, leading to contact between the muscle and bone.}
    \label{fig:results_muscle_shoulder}
\end{figure}

During a physiological muscle contraction, activation is generally not uniform throughout the muscle, but varies in different locations (spatially) and over time (temporally). To model such complex activation patterns, we introduce a scaling factor that modifies $\Popt$. Prescribing this scaling factor for each element and time step allows us to model temporally and spatially varying activation. 
Fig. \ref{fig:results_muscle_shoulder_activation_input} shows the scaling factors applied in this scenario at four distinct points in time. While activation increases over time, the region of maximal activation progresses from the deltoid's spinal part towards its acromial part.

As with the fusiform muscle example, we perform a quasi-static simulation and neglect inertia effects.

\paragraph{Simulation results}
As a measure for the combined strain, we evaluate the von Mises strains $\epsilon_\mathrm{v}$. The results are shown in Fig. \ref{fig:results_muscle_shoulder_strain} in the deformed configurations corresponding to the activation patterns presented in Fig. \ref{fig:results_muscle_shoulder_activation_input}. 
Over time, increasing activation in the spinal part of the deltoid causes the muscle to contract, resulting in the humerus being lifted in the spinal direction. A closer inspection reveals that rising activation in the deltoid's acromial region also induces a slight rotation of the humerus towards the acromial part. As expected, areas with higher activation experience greater strains.

Through the abduction of the humerus in the spinal direction, the clavicular part of the deltoid is pulled towards the humeral head. At $t =\SI{0.325}{\s}$, the muscle and the bone first make contact. Fig. \ref{fig:results_muscle_shoulder_contact}  illustrates the resulting normal contact stresses at $t=\SI{0.5}{\s}$ when the contact area is at its maximum.

To summarize, this example thus demonstrates a first application of the modified material model in a simple musculoskeletal system, incorporating complex activation patterns and accounting for contact interactions. 

\FloatBarrier
\subsection{Dynamic stabilization of the shoulder through rotator cuff contraction} \label{sec:shoulder}
As initially outlined, our primary goal is to identify a constitutive law suitable for modeling muscle tissue in a continuum mechanical shoulder model. To demonstrate the applicability of the adapted \COMBI-model in such a scenario, we present a third numerical example. For this purpose, we use a self-created FE model comprising the skeletal structure and the essential muscles surrounding the glenohumeral joint. 

As motivation for our simulation, we consider the concavity compression mechanism of the glenohumeral joint. Concavity compression is a dynamic stabilizing mechanism in which the active rotator cuff muscles tightly compress the humeral head against the glenoid fossa, thereby increasing resistance against translating forces. In the following, we simulate such a contraction of the rotator cuff and the resulting contact between the glenoid fossa and the humeral head.

\paragraph{Geometry and mesh} 
The Visible Human Project \cite{VM} provides an image data set of cross-sectional cryosections of a human male and female cadaver. We select the male data set for the manual segmentation with Materialise Mimics/3-matic \cite{materialise_3-matic_version_nodate} as individual components appear more clearly distinguishable. 

Our segmented model includes the shoulder joint's bones (\textit{humerus}, \textit{clavicula}, and \textit{scapula}), the cartilaginous \textit{glenoid labrum}, the three-parted deltoid muscle (\textit{deltoideus spinalis, acromialis, and clavicularis}), and the rotator cuff muscles, i.e., the  \textit{teres minor}, \textit{infraspinatus}, \textit{supraspinatus}, and \textit{subscapularis}. In this example, we omit the clavicula and treat the three-parted deltoid as one continuum. 

\begin{figure}[htb]
	\centering
    \includegraphics[width=0.7\textwidth]{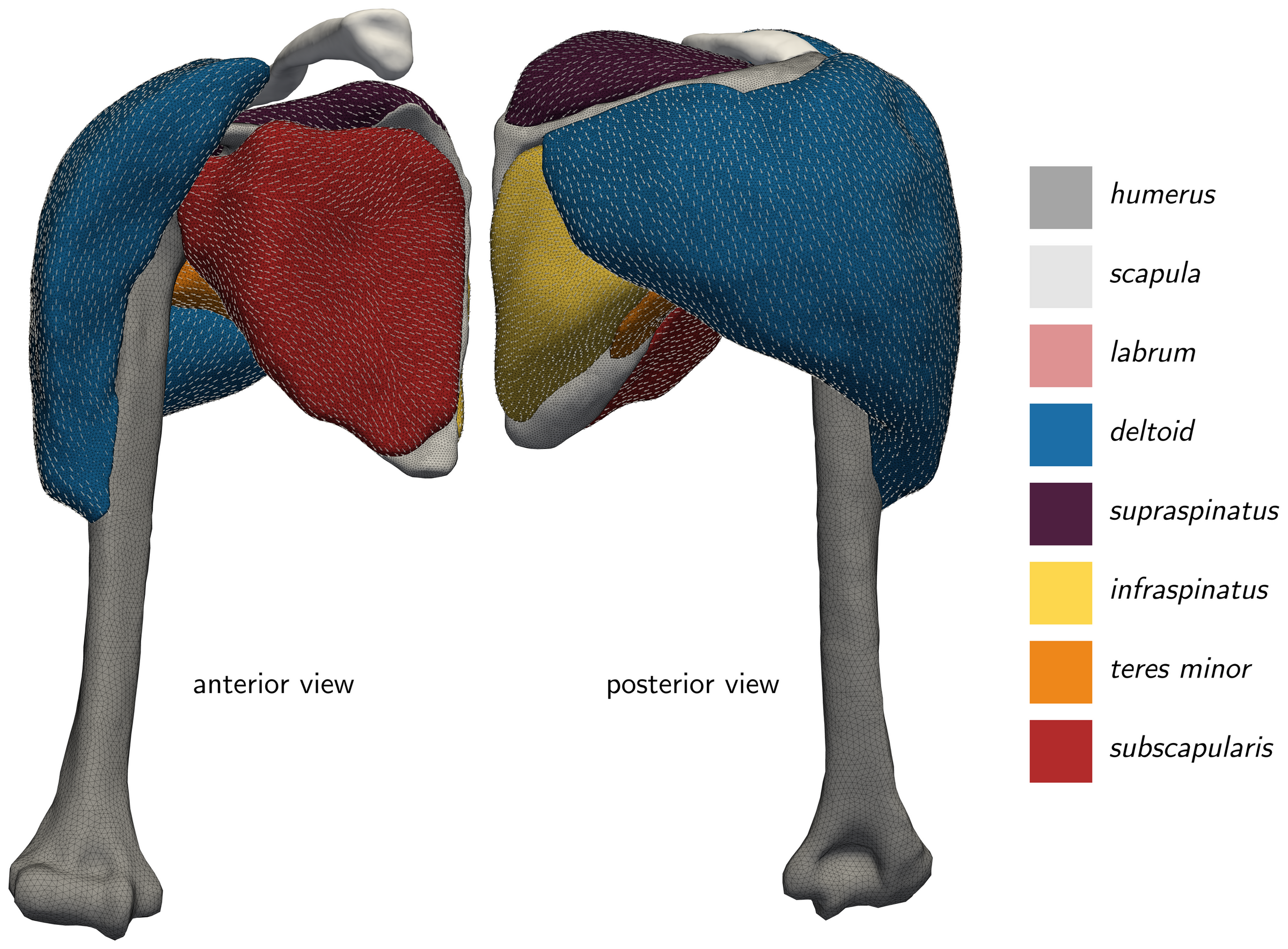}
    \caption{Shoulder model with fiber directions indicated by white arrows. The geometry is meshed with quadratic tetrahedral elements and comprises 659901 nodes and 448858 elements in total. The clavicula is not simulated in this example and hence displayed unmeshed.}
    \label{fig:shoulder_model}
\end{figure}

All anatomical parts are meshed separately with Gmsh (version 2.12.0) \cite{gmsh}. Scapula and labrum are meshed as one entity and thus coupled via shared nodes. We convert the created linear tetrahedral elements to quadratic tetrahedrons using a custom Python script. In total our model comprises 659901 nodes and 448858 elements.
As for the fusiform muscle example, we compute the normalized muscle fiber directions $\mathbf{m}$ as the solution of the Laplacian problem. Fig. \ref{fig:shoulder_model} depicts the model and the computed fiber directions.

\paragraph{Constitutive desciptions}
Muscles are modeled with the \COMBI-material and the parameters identified in Section \ref{sec:fitting}. By scaling $\Popt$, we prescribe a \SI{5}{\percent} activation of the rotator cuff. The deltoid remains passive.
For the stiff bones, we use a Young's modulus $E_\mathrm{b}$ within the ranges reported in the literature \cite{carey_situ_2000, smith_tensile_2008}. To model the much softer labrum, we apply a lower Young's modulus $E_\mathrm{c}$ (see \cite{hart_nh_mechanical_2017}).
The mass densities $\rho$ are chosen to align with literature values (see \cite{mendez_density_1960} for muscle, and \cite{gatti_development_2010} for the labrum). The bone mass density is computed based on literature data available for the humerus as described in detail in Section \ref{sec:appendix_bone_density} in the Appendix.
Table \ref{tab:materials_used_shoulder} provides a summary of the defined constitutive descriptions and parameters. 

\begin{table}[b]
	\centering
    \caption{Material models and parameters defined for the concavity compression simulation.}
    \begin{tabular}{lll} \toprule
        Part & Material & Parameters \\ \midrule
        Bones & St. Venant-Kirchhoff & $E_\mathrm{b}=\SI{0.1}{\giga \pascal}$, $\rho_\mathrm{b}=\SI{0.76}{\gram \per \cm^3}$ \\
        Labrum & St. Venant-Kirchhoff & $E_\mathrm{c}=\SI{5}{\mega \pascal}$, $\rho_\mathrm{c}=\SI[per-mode=fraction]{1.2}{\gram \per \cm^3}$\\
        Rotator cuff & active \COMBI-model & Table \ref{tab:fitting_all}, $\Popt = \SI{3.234}{\kilo \pascal}$, $\kappa = 10$, $\rho_\mathrm{m}=\SI{1.06}{\gram \per \cm^3}$\\
        Deltoid & passive \COMBI-model & Table \ref{tab:fitting_all}, $\Popt = \SI{0}{\kilo \pascal}$, $\kappa = 10$, $\rho_\mathrm{m}=\SI{1.06}{\gram \per \cm^3}$ \\
        \bottomrule
    \end{tabular}
    \label{tab:materials_used_shoulder}
\end{table}

\paragraph{Boundary, contact and meshtying conditions}
To fix the structure in space, zero Dirichlet boundary conditions are prescribed to the inner nodes of the scapula volume and the muscle's origin surfaces (where they connect to scapula and clavicula).

The FE meshes of the individual muscles are tied to the FE meshes of the humerus at the respective insertion surfaces via tied constraints. In contrast to coupling via shared nodes, this approach allows connecting dissimilar meshes, such that the mesh size of each anatomical part can be chosen individually and is not constricted to the mesh size of adjoining parts. 
For constraint enforcement in the weak sense, we apply the mortar method with dual Lagrange multipliers discretized by segment-based quadratic shape functions \cite{popp_dual_2012}. By choosing \textit{dual} Lagrangean multipliers, the additionally introduced degrees of freedom can be eliminated by a condensation approach -- the size of the global linear systems of equations thus remains the same. To solve the system of equations with inequality constraints, we employ a semi-smooth Newton method utilizing a primal-dual active set strategy \cite{popp_finite_2009}. 

To prevent penetration and account for three-dimensional interactions between individual components, we prescribe frictionless contact (Karush-Kuhn-Tucker conditions) for muscle-bone, muscle-muscle, and bone-bone surface pairs. For simplification reasons, contact between the individual rotator cuff muscles is neglected. Contact constraints are enforced using a penalty regularization approach. 

A comprehensive description of the surfaces defined for the application of the boundary conditions is provided in Fig. \ref{fig:surface_volume_shoulder} and Table \ref{tab:surface_volume_nds_ids_shoulder} in the Appendix. Surfaces fixed by Dirichlet conditions are summarized in Table \ref{tab:dirichlet_bcs_shoulder}, and meshtying and contact surface pairs in Table \ref{tab:meshtying_bcs_shoulder} and \ref{tab:contact_bcs_shoulder}, respectively.

\paragraph{Solution strategy}
We apply the Generalized-alpha time integration method in combination with a standard Newton-Raphson scheme to solve the nonlinear structural dynamics problem. The resulting linear system of equations is solved iteratively using the Generalized Minimal RESidual method (GMRES) in combination with an algebraic multigrid preconditioner, implemented in the software packages Trilinos Belos \cite{belos} and Trilinos MueLu \cite{muelu, mueluURL}, respectively. 

We simulate 160 time steps with a step size of $\SI{2.5e-4}{\s}$ on 64 Intel Xeon E5-2630 v3 processors (12 cores, 2.5 GHz, 64 GB RAM) of our Linux cluster, resulting in a total computation time of $\SI{45}{\hour}$. 

\begin{figure}[htb]
    \begin{subfigure}[t]{0.42\textwidth}
        \begin{subfigure}[c]{0.32\textwidth}
        \centering
        \includegraphics[height=4.5cm]{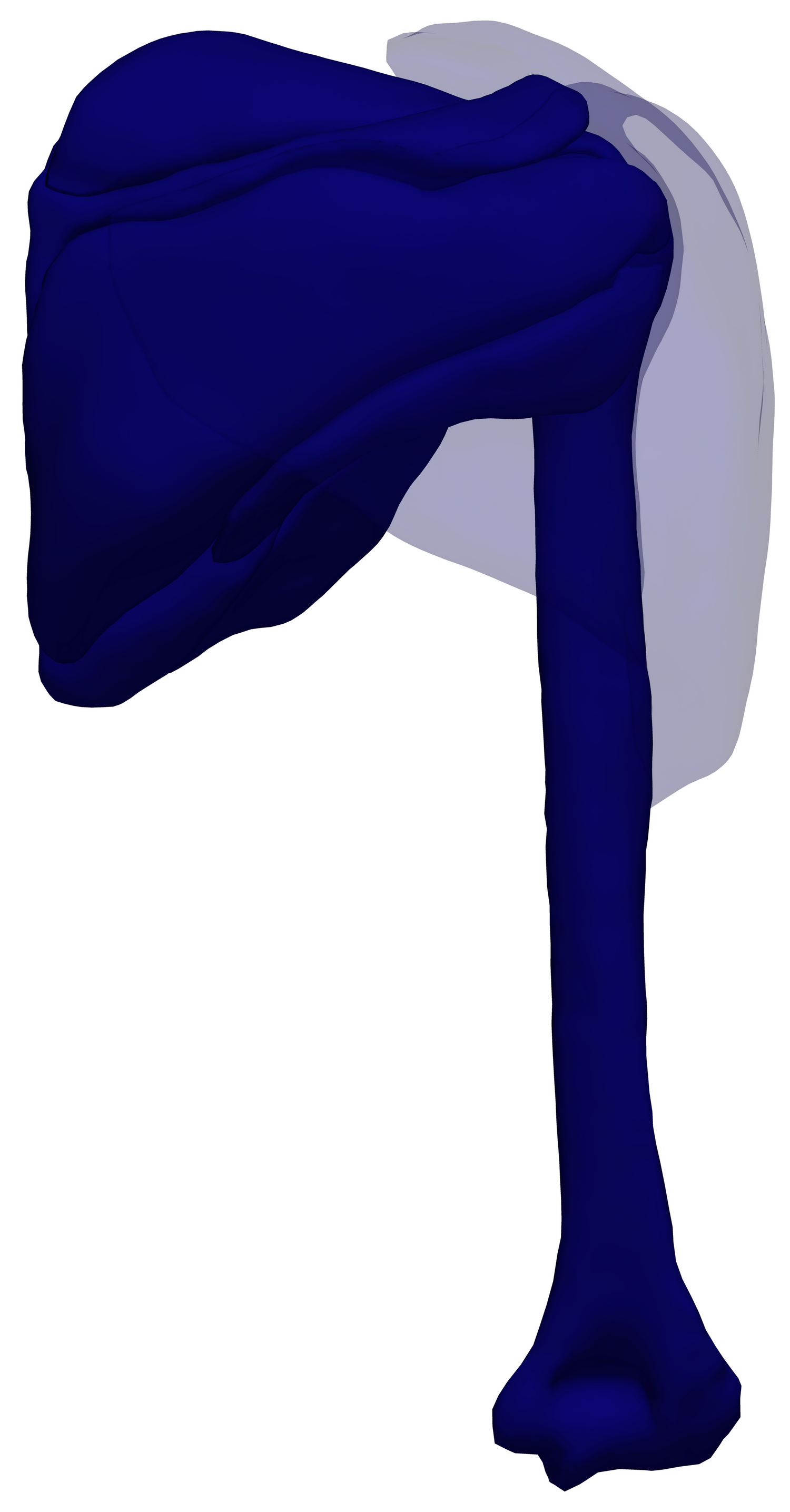}
        \subcaption*{$t = \SI{0.0}{\s}$}
        \label{fig:displ_t_start}
        \end{subfigure} \hfill
        \begin{subfigure}[c]{0.32\textwidth}
        \centering
        \includegraphics[height=4.5cm]{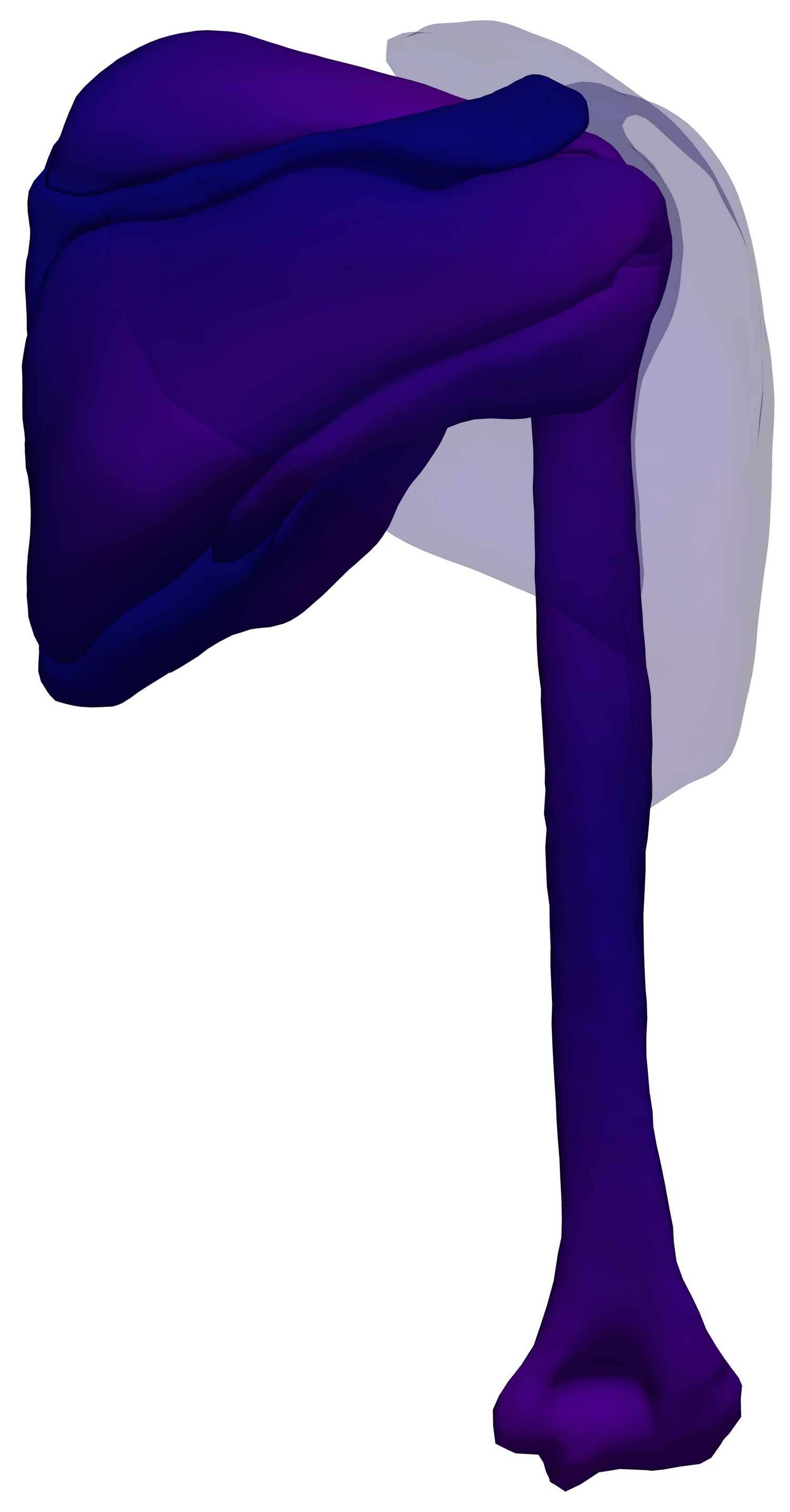}
        \subcaption*{$t = \SI{0.02}{\s}$}
        \label{fig:displ_t_middle}
        \end{subfigure} \hfill
        \begin{subfigure}[c]{0.32\textwidth}
        \centering
        \includegraphics[height=4.5cm]{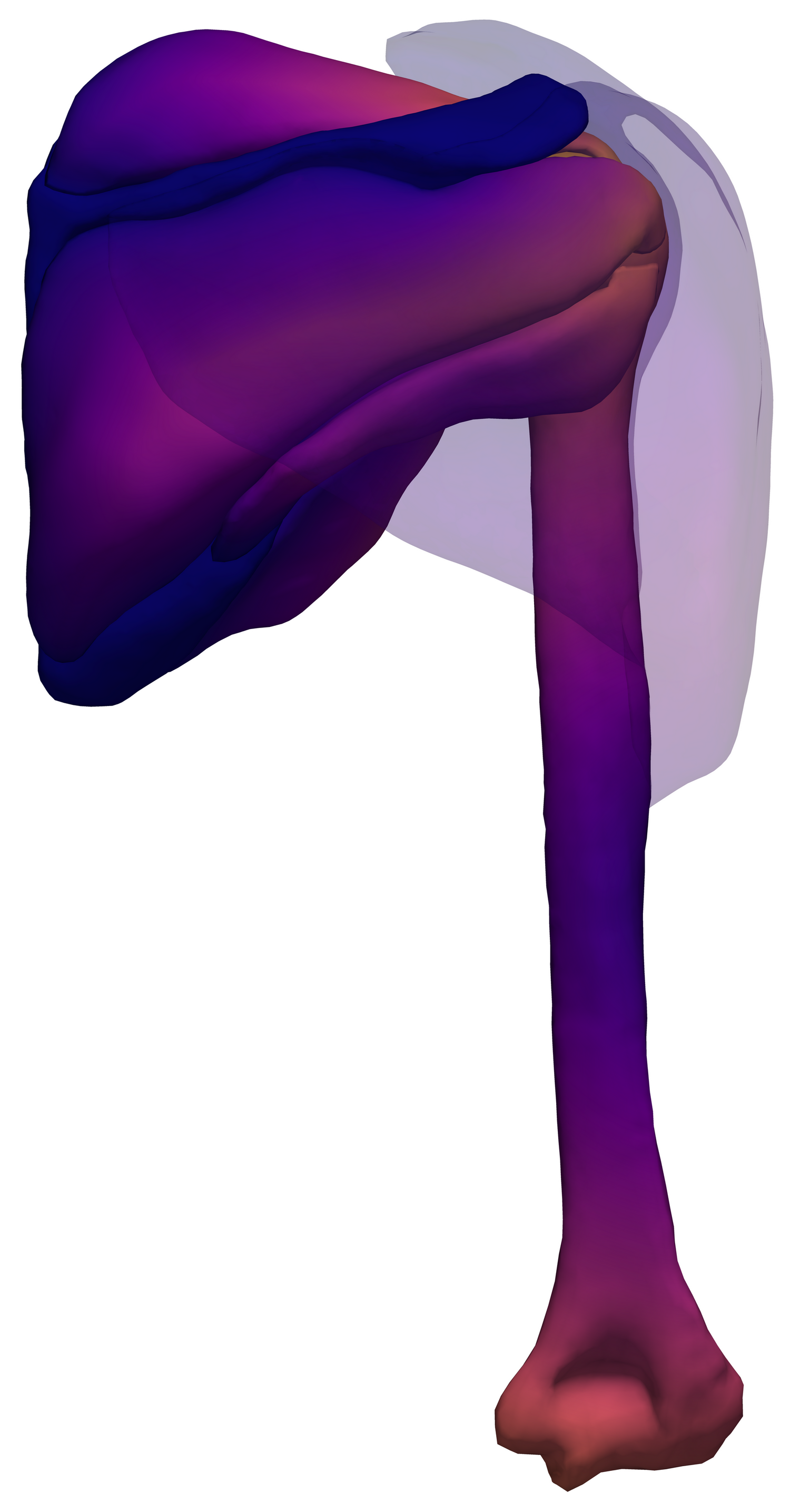}
        \subcaption*{$t = \SI{0.04}{\s}$}
        \label{fig:displ_t_end}
        \end{subfigure}
    \subcaption{Posterior view}
    \end{subfigure}
    \hfill
    \begin{subfigure}[t]{0.44\textwidth}
        \begin{subfigure}[c]{0.32\textwidth}
        \centering
        \includegraphics[height=4.5cm]{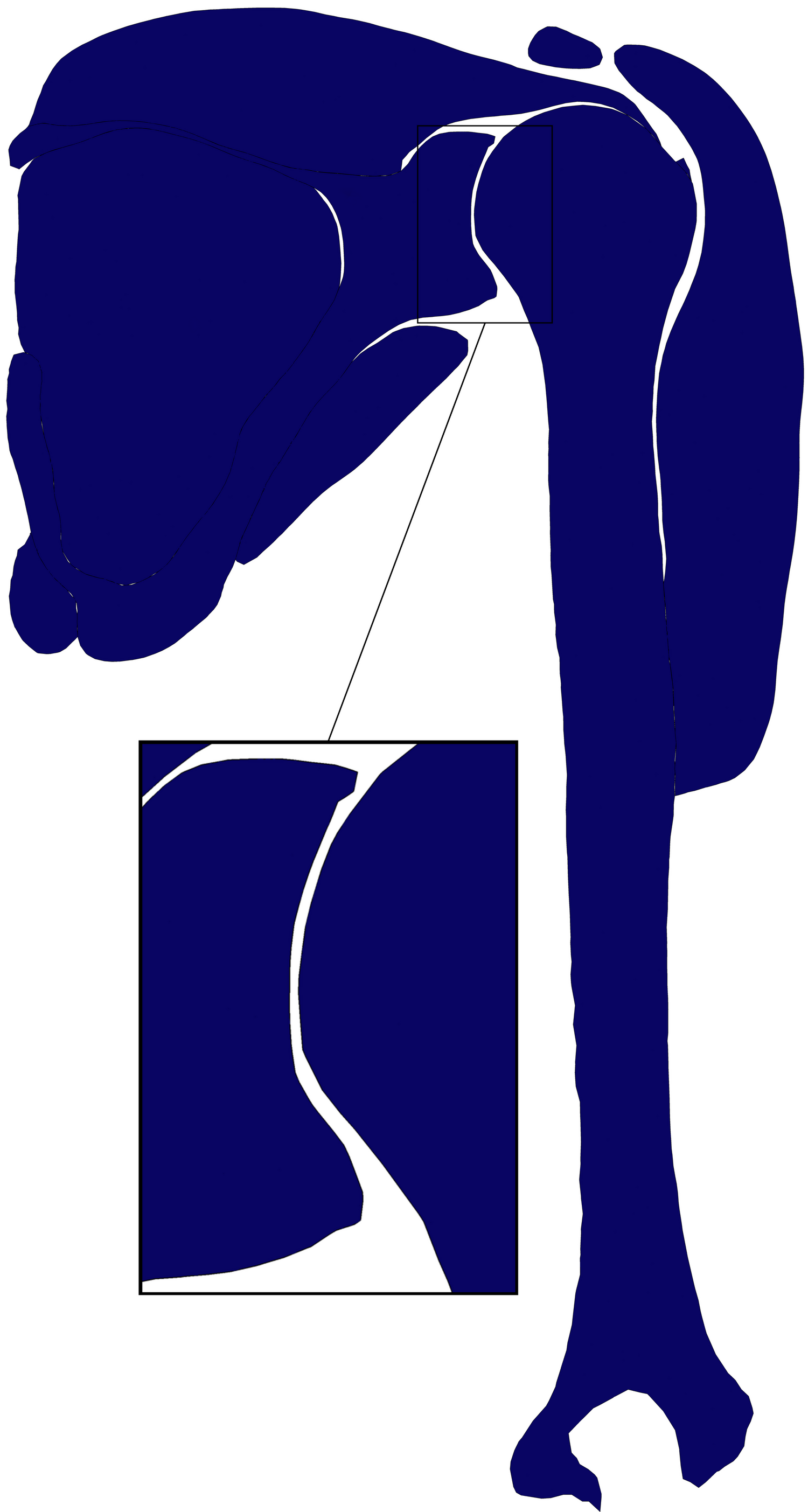}
        \subcaption*{$t = \SI{0.0}{\s}$}
        \label{fig:slice_displ_t_start}
        \end{subfigure} \hfill
        \begin{subfigure}[c]{0.32\textwidth}
        \centering
        \includegraphics[height=4.5cm]{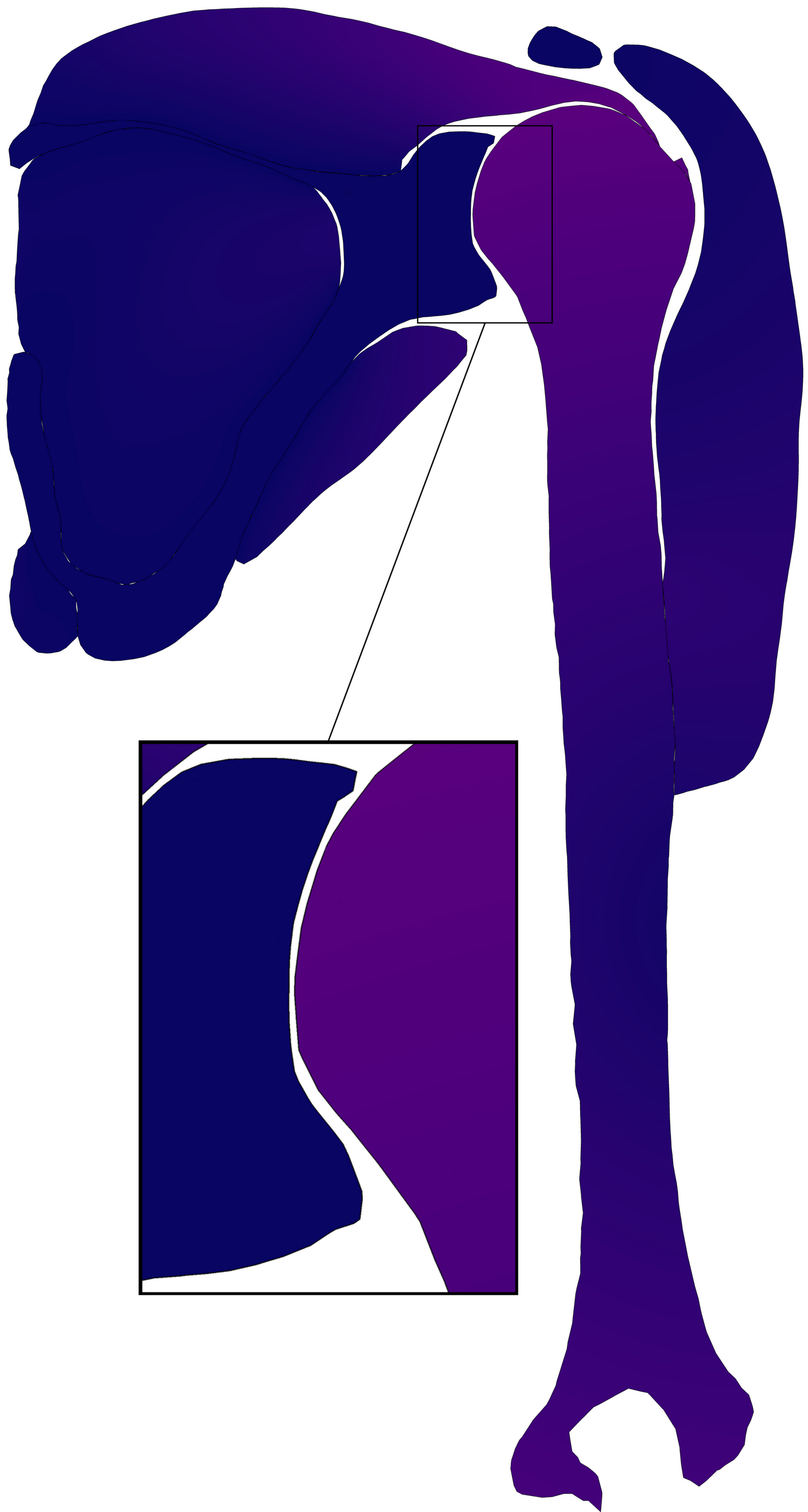}
        \subcaption*{$t = \SI{0.02}{\s}$}
        \label{fig:slice_displ_t_middle}
        \end{subfigure} \hfill
        \begin{subfigure}[c]{0.32\textwidth}
        \centering
        \includegraphics[height=4.5cm]{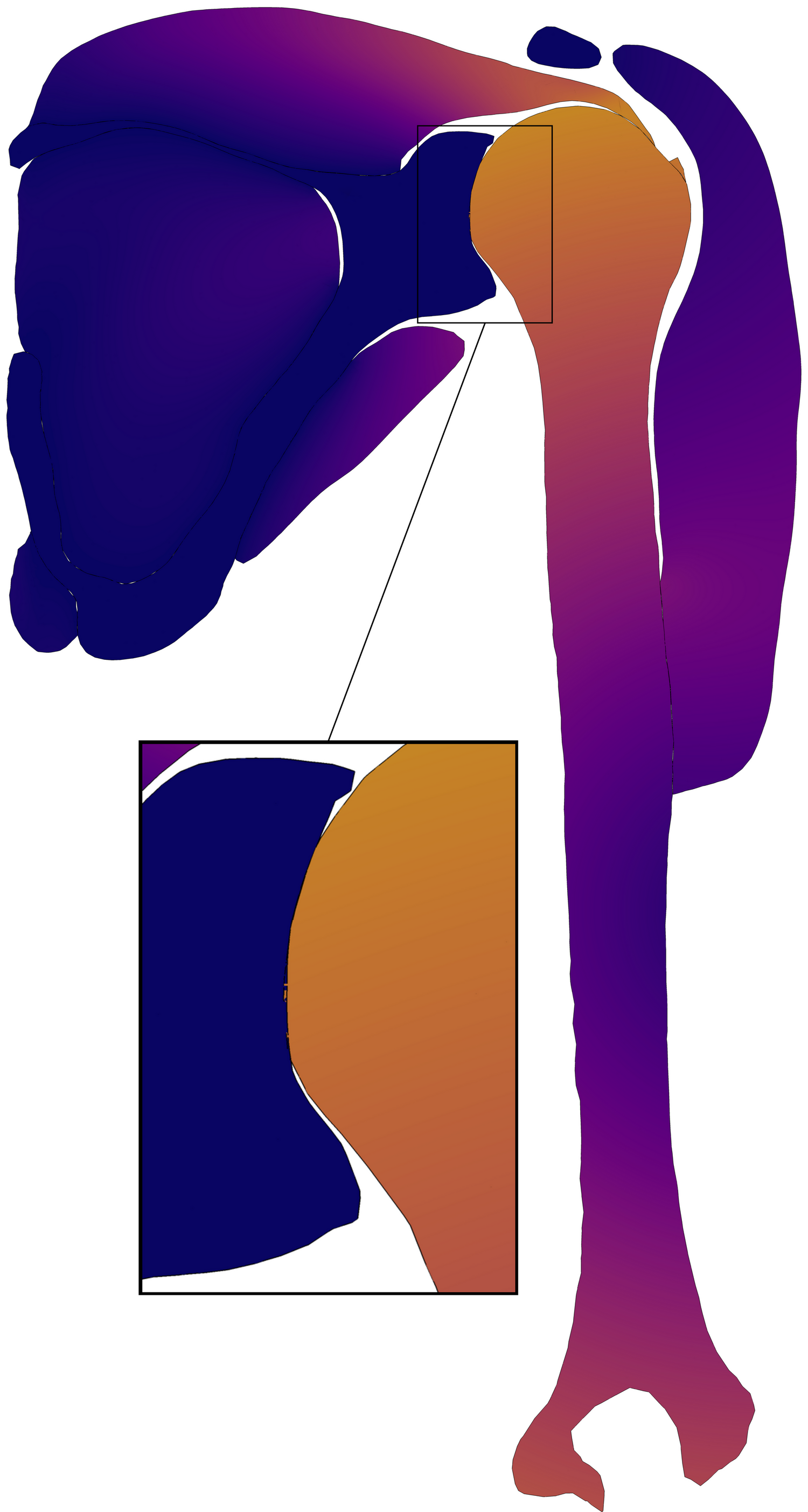}
        \subcaption*{$t = \SI{0.04}{\s}$}
        \label{fig:slice_displ_t_end}
        \end{subfigure}
    \subcaption{Cross-sectional view through humeral head center with a detailed view on the glenohumeral joint space}
    \end{subfigure}
    \hfill
    \begin{subfigure}[c]{1.29cm}
    \raggedleft
        \includegraphics[height=3cm]{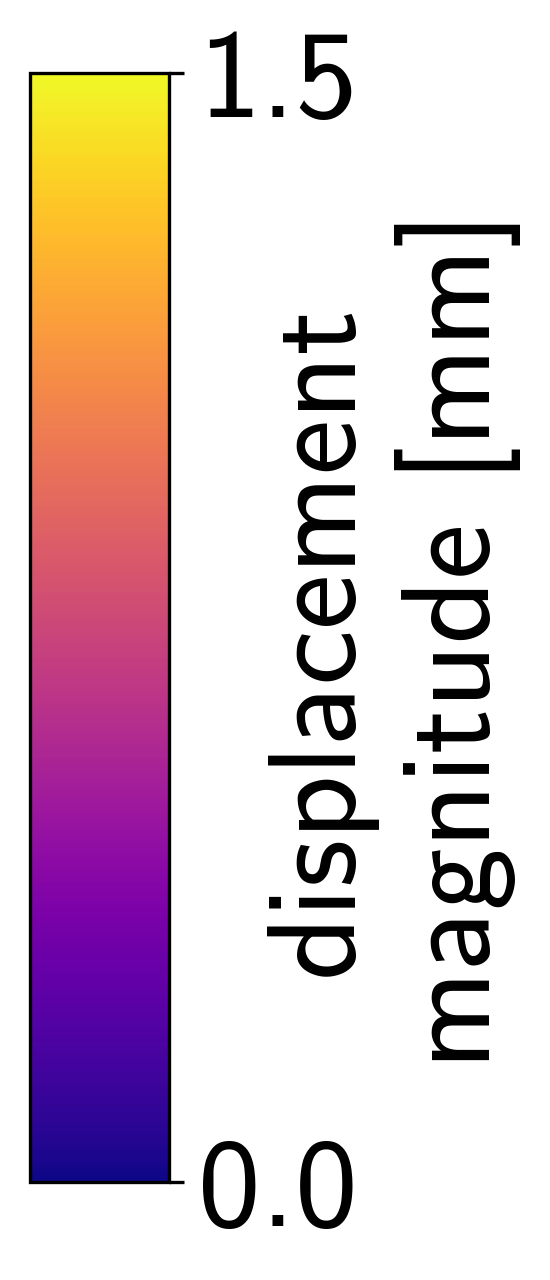}
        \subcaption*{}
    \end{subfigure}
    \caption{Displacement magnitude at selected points in time for the simulation of rotator cuff activation in a model of the human shoulder. For visualization purposes, the deltoid muscle is displayed transparently. The rotator cuff muscle contraction pulls the humeral head medially towards the glenoid fossa such that the joint space closes.}
    \label{fig:displacements_shoulder}
\end{figure}

\paragraph{Simulation results}
Fig. \ref{fig:displacements_shoulder} shows the simulated displacements at selected points in time. 
As expected, the rotator cuff's activation causes the muscles to contract. As a result, the humeral head is pulled towards the glenoid fossa, and the joint space closes. 

\begin{figure}[htb]
\begin{minipage}[c]{0.5\textwidth}
    \begin{subfigure}[t]{\textwidth}
        \begin{subfigure}[c]{0.32\textwidth}
        \centering
        \includegraphics[height=2.5cm]{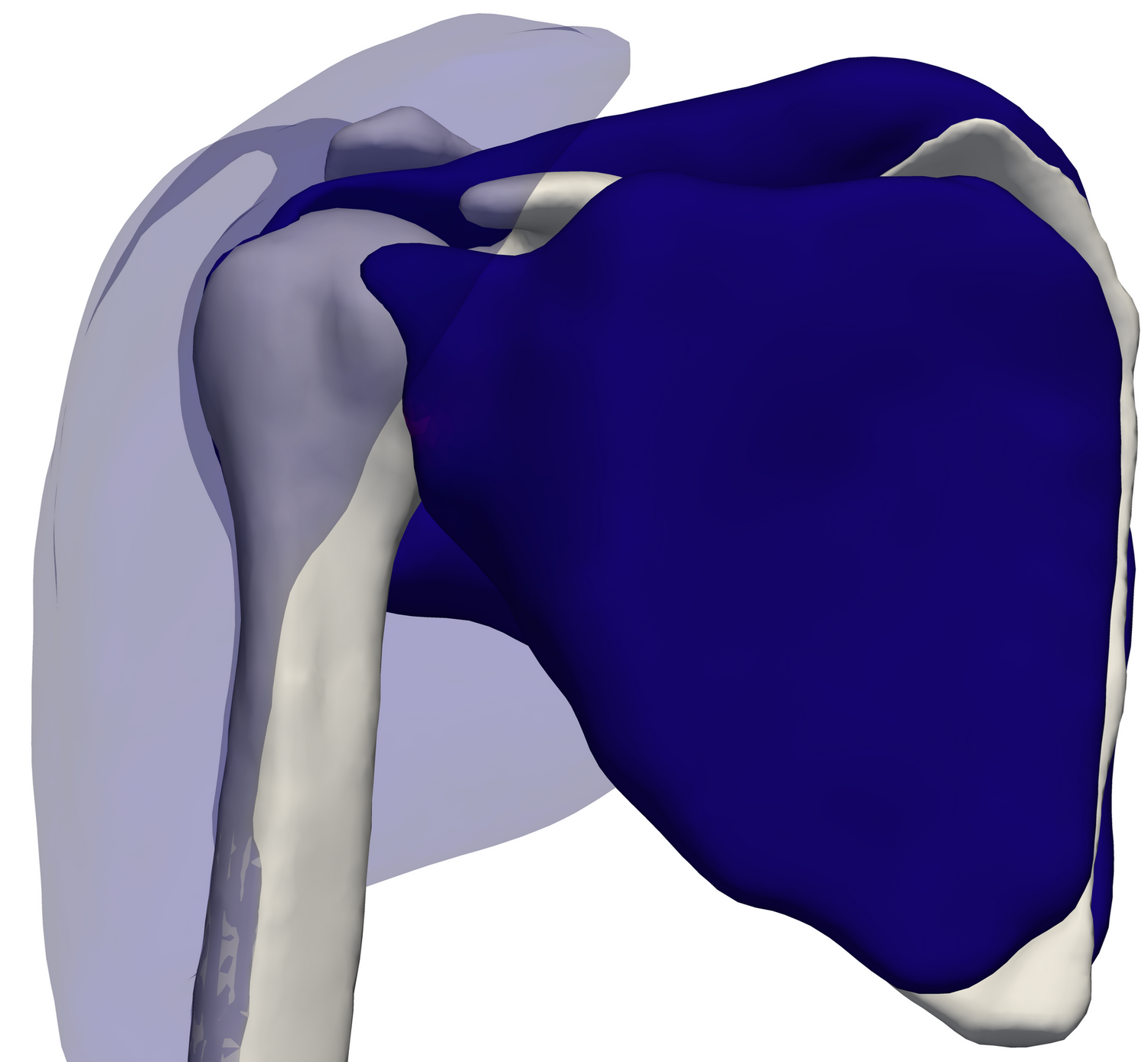}
        \subcaption*{$t = \SI{0.0}{\s}$}
        \label{fig:stress_t_start_ant}
        \end{subfigure} \hfill
        \begin{subfigure}[c]{0.32\textwidth}
        \centering
        \includegraphics[height=2.5cm]{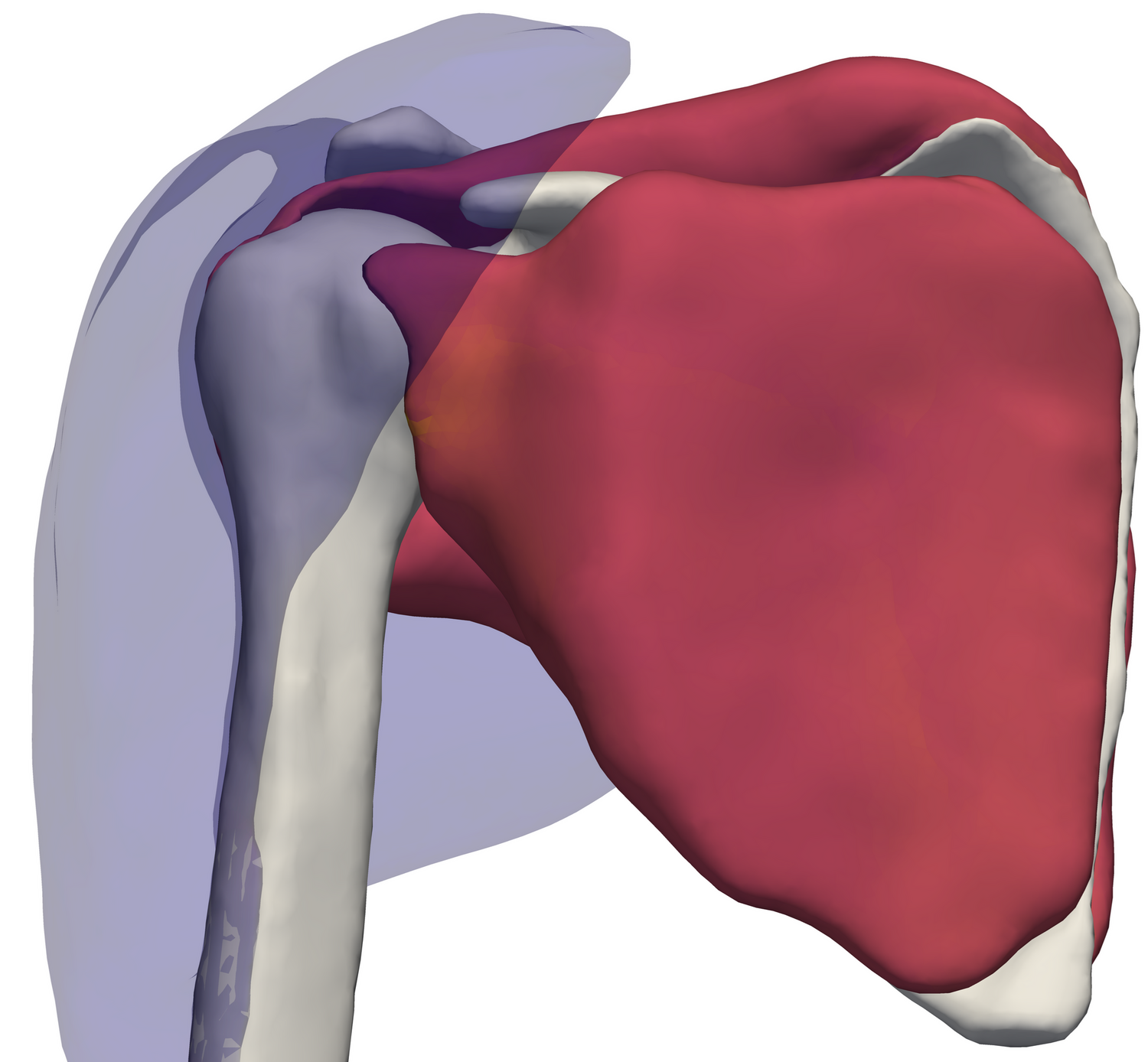}
        \subcaption*{$t = \SI{0.02}{\s}$}
        \label{fig:stress_t_middle_ant}
        \end{subfigure} \hfill
        \begin{subfigure}[c]{0.32\textwidth}
        \centering
        \includegraphics[height=2.5cm]{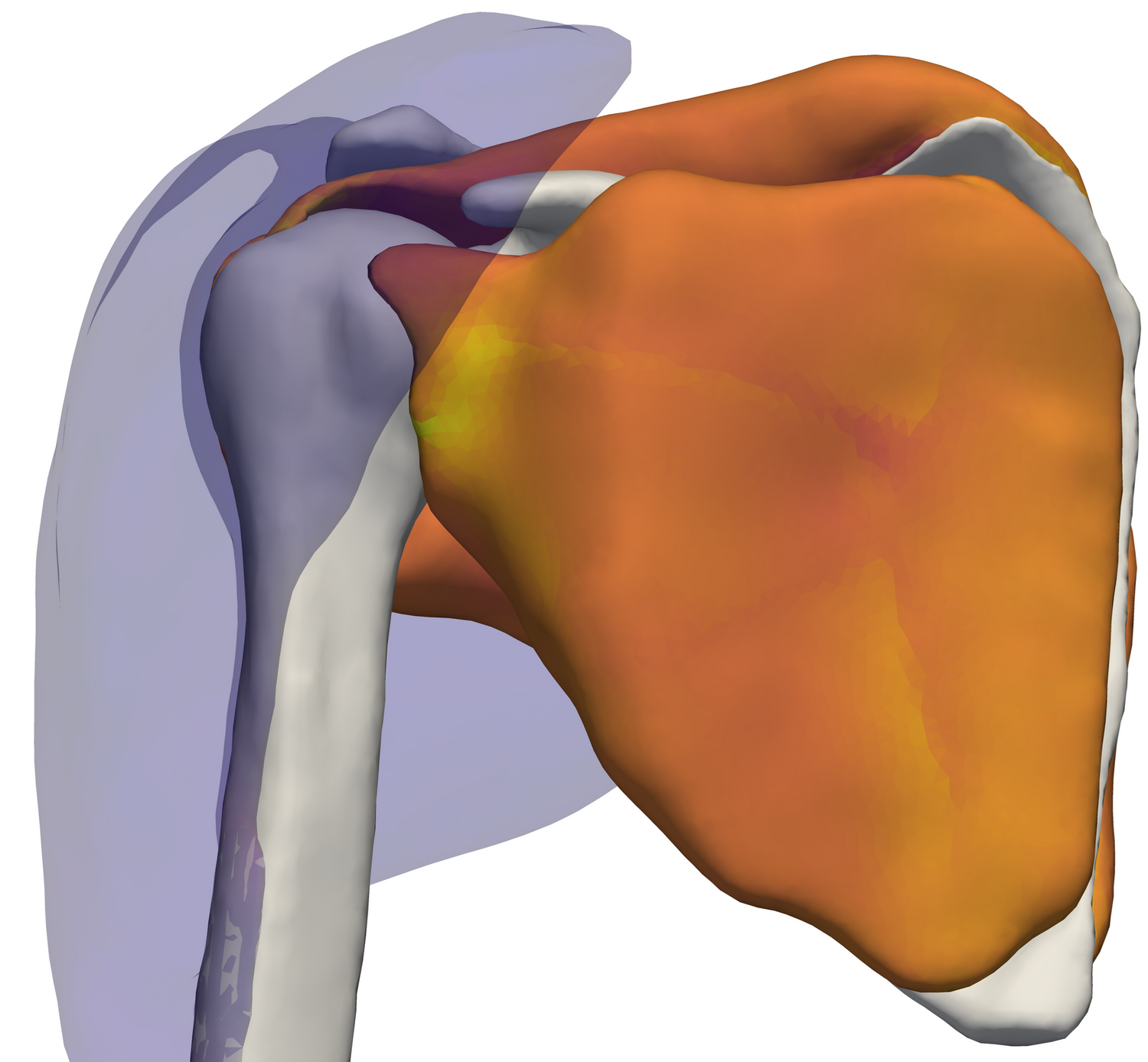}
        \subcaption*{$t = \SI{0.04}{\s}$}
        \label{fig:stress_t_end_ant}
        \end{subfigure}
    \subcaption{Anterior view}
    \label{fig:anterior_view_stress}
    \end{subfigure}\\
    
    \vspace{1em}
    \begin{subfigure}[b]{\textwidth}
        \begin{subfigure}[c]{0.32\textwidth}
        \centering
        \includegraphics[height=2.5cm]{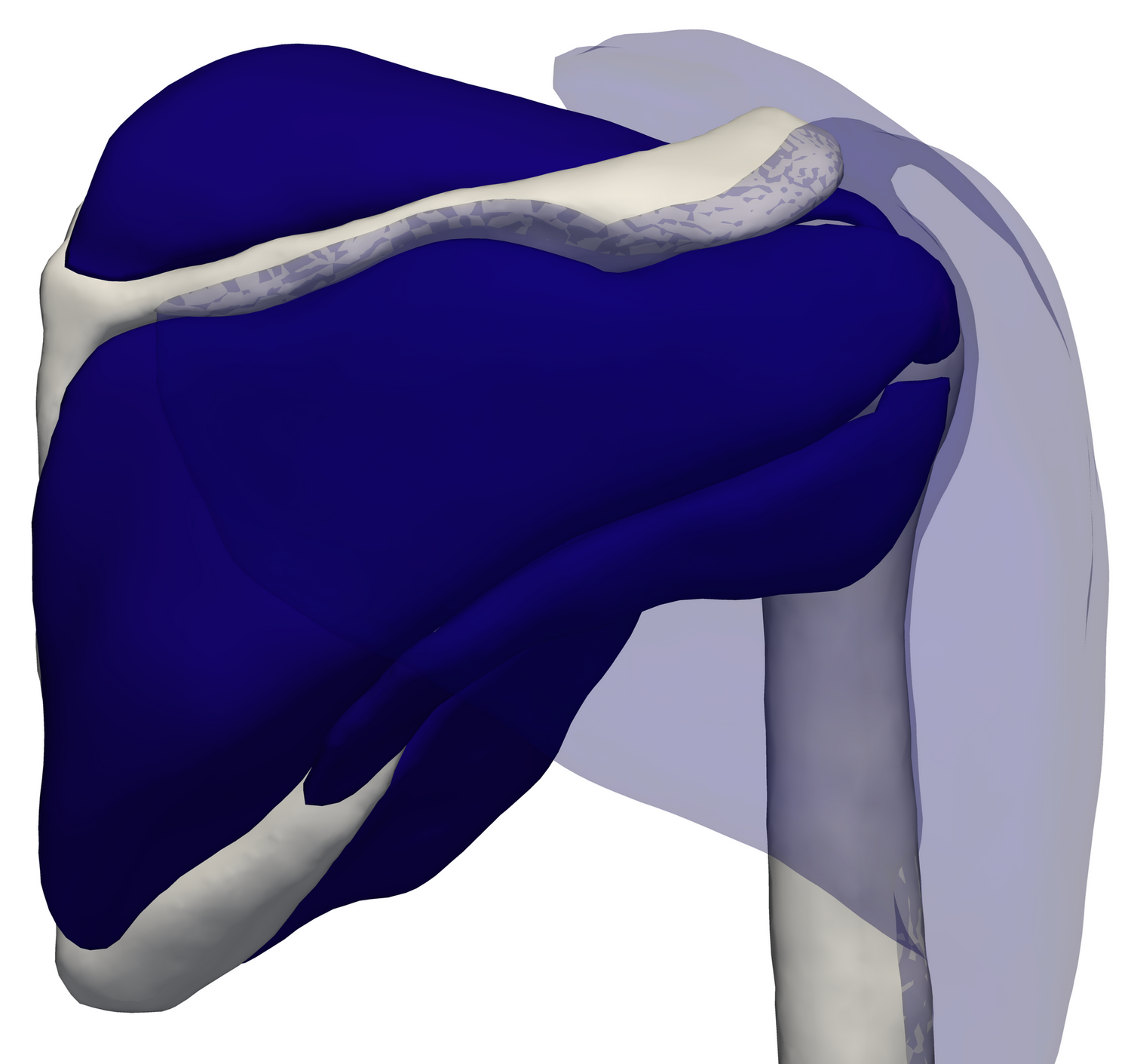}
        \subcaption*{$t = \SI{0.0}{\s}$}
        \label{fig:stress_t_start_post}
        \end{subfigure} \hfill
        \begin{subfigure}[c]{0.32\textwidth}
        \centering
        \includegraphics[height=2.5cm]{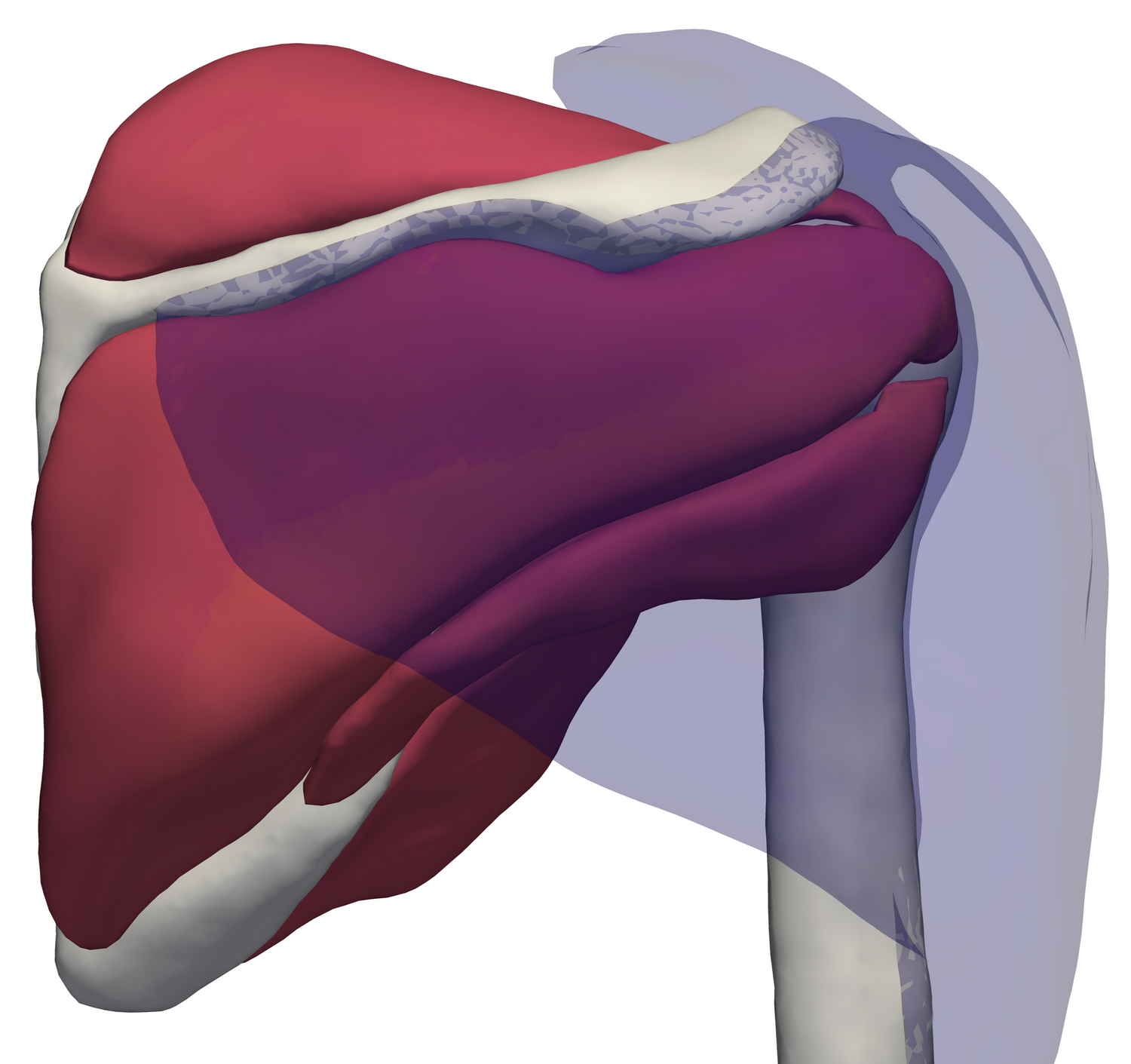}
        \subcaption*{$t = \SI{0.02}{\s}$}
        \label{fig:stress_t_middle_post}
        \end{subfigure} \hfill
        \begin{subfigure}[c]{0.32\textwidth}
        \centering
        \includegraphics[height=2.5cm]{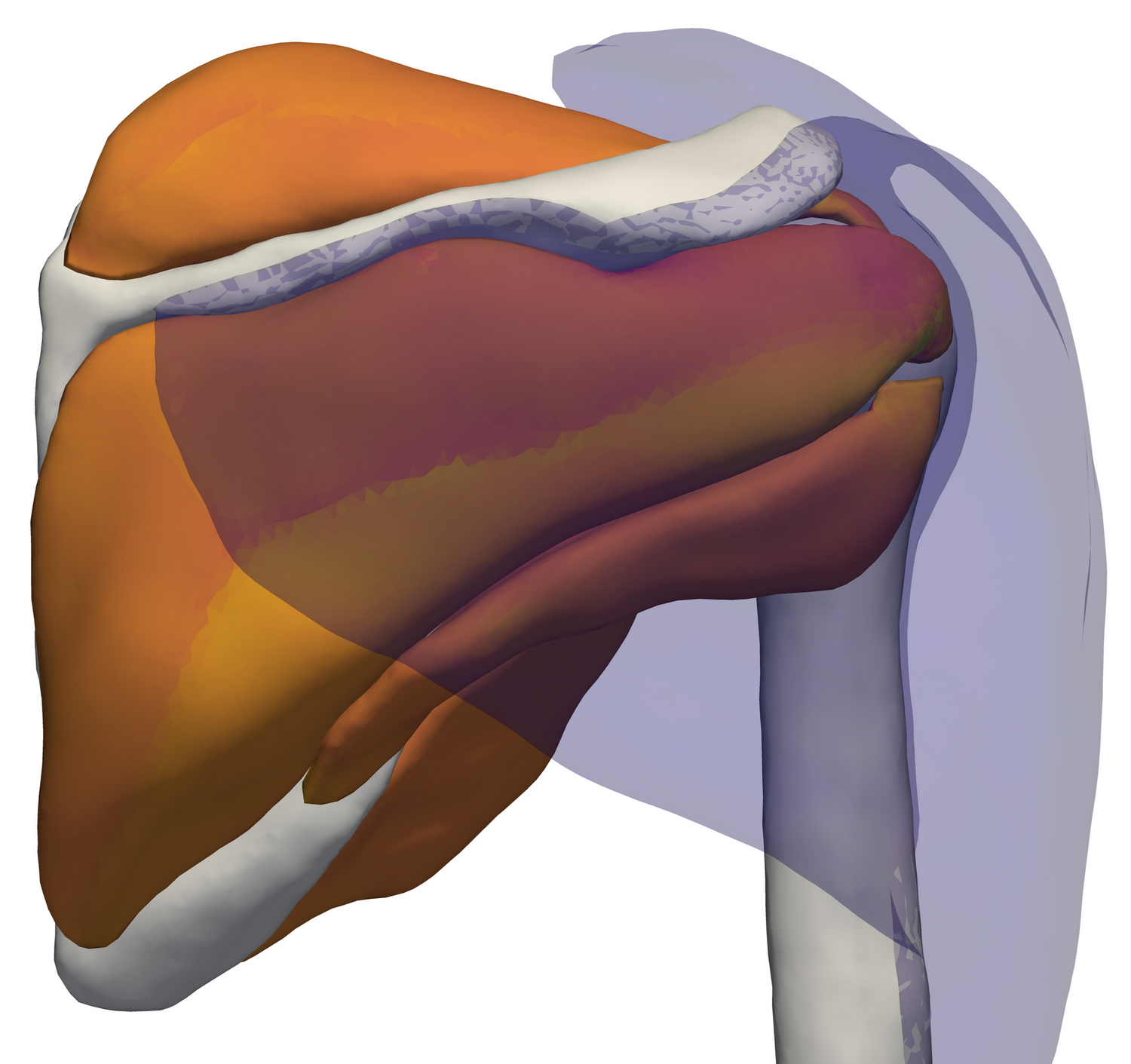}
        \subcaption*{$t = \SI{0.04}{\s}$}
        \label{fig:stress_t_end_post}
        \end{subfigure}
    \subcaption{Posterior view}
    \label{fig:post_view_stress}
    \end{subfigure}
    \end{minipage}
    \hfill
    \begin{subfigure}[c]{0.36\textwidth}
        \begin{subfigure}[c]{1.0\textwidth}
        \centering
        \includegraphics[width=0.95\textwidth]{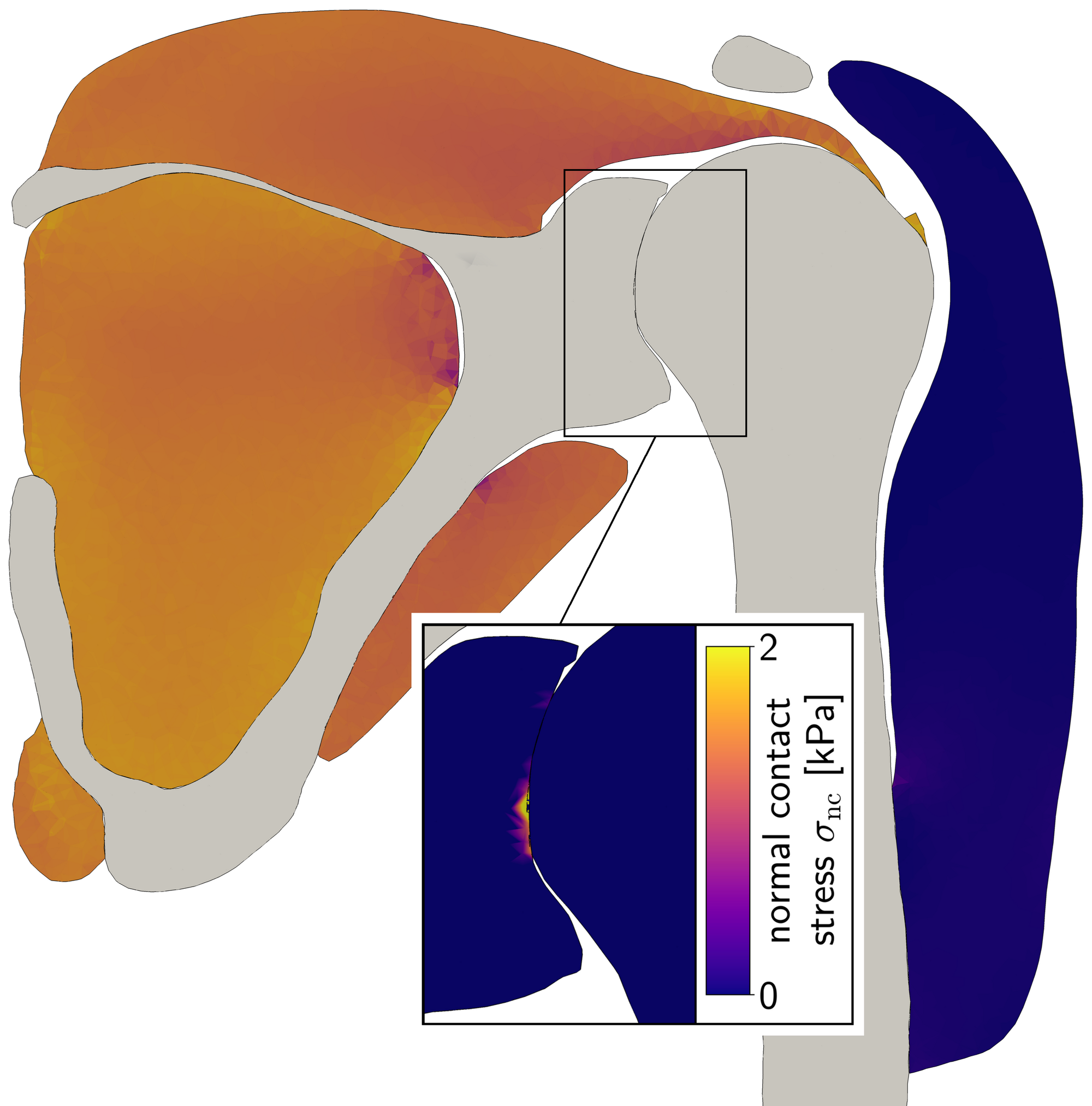}
        \subcaption*{$t = \SI{0.04}{\s}$}
        \label{fig:stress_slice_and_contact_t_end}
        \end{subfigure}
    \subcaption{Cross-sectional view through humeral head center with a detailed view of the normal contact stresses $\sigma_\mathrm{nc}$ in the joint space}
    \end{subfigure}
    \hfill
    \begin{subfigure}[c]{1.215cm}
    \raggedleft
        \includegraphics[height=3cm]{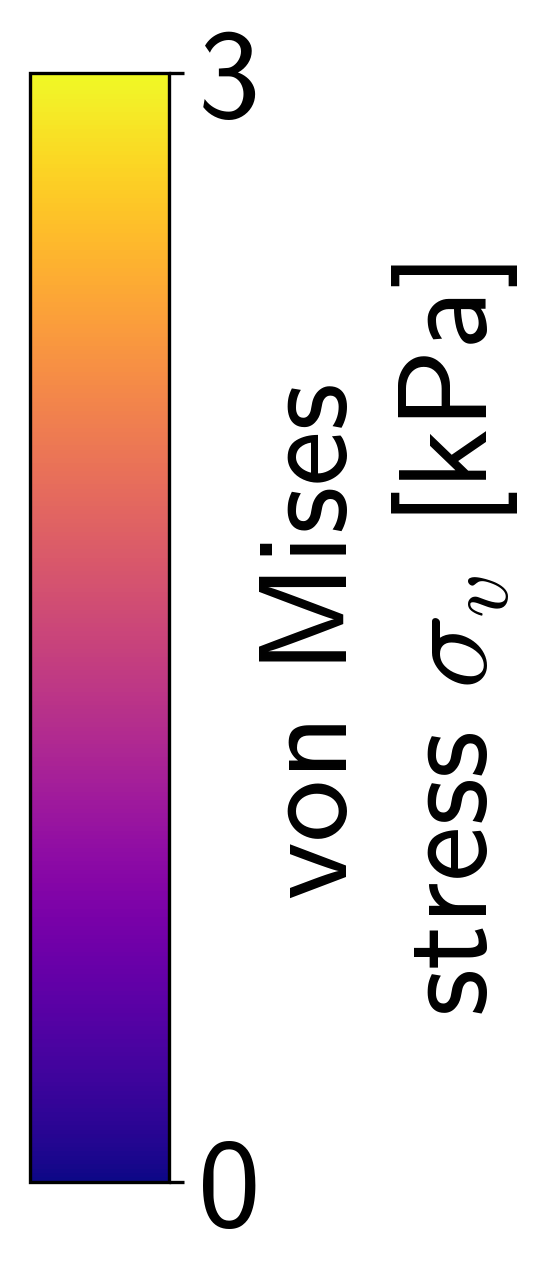}
        \subcaption*{}
    \end{subfigure}
    \caption{Von Mises stresses $\sigma_\mathrm{v}$ at selected points in time for the simulation of rotator cuff activation in a model of the human shoulder. For visualization purposes, the deltoid muscle is displayed transparently. \textbf{(a)} and \textbf{(b)}: Increasing rotator cuff activation causes increasing stress in the muscle continuum. \textbf{(c)}: Once the glenohumeral joint space closes, contact stresses develop between the glenoid fossa and the humeral head.}
    \label{fig:vm_stress_shoulder}
\end{figure}

As a measure for the three-dimensional stress distribution, we evaluate the von Mises stress $\sigma_\mathrm{v}$ in Fig. \ref{fig:vm_stress_shoulder}. Over time, stresses in the activated rotator cuff increase. Since the rotator cuff deforms only slightly, we conclude that the resulting stresses are primarily caused by muscular activation. Contrarily, stresses in the passive deltoid muscle exclusively develop due to its deformation and thus are close to zero.

\begin{figure}[htb]
\centering
\begin{minipage}[b]{.62\textwidth}
  \raggedright
    \begin{subfigure}[b]{0.19\linewidth}
        \centering
        \includegraphics[height=2.5cm]{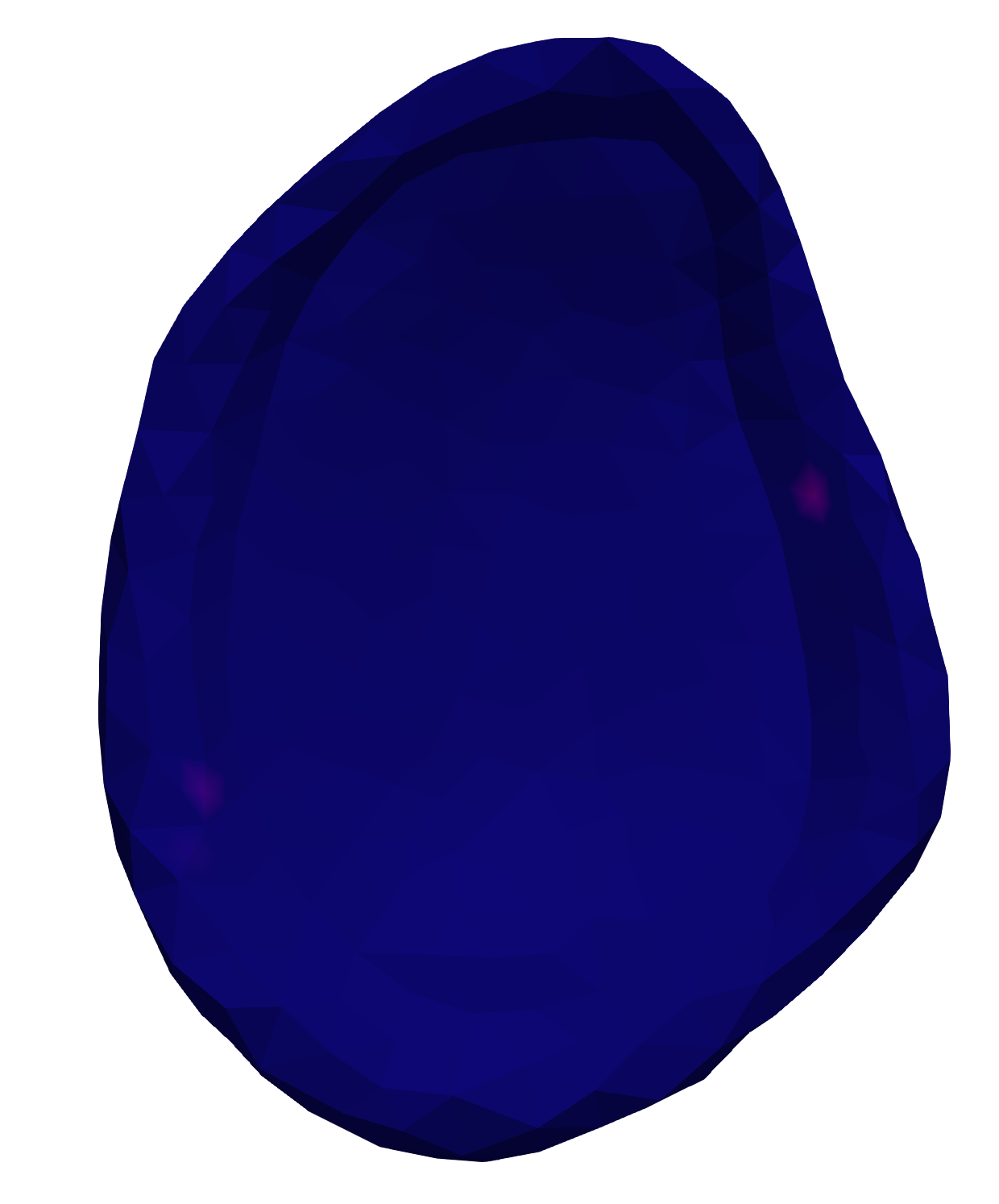}
        \subcaption*{$t = \SI{0.25}{\s}$}
        \label{fig:stress_t_start_post}
    \end{subfigure} \hfill
    \begin{subfigure}[b]{0.19\linewidth}
        \centering
        \includegraphics[height=2.5cm]{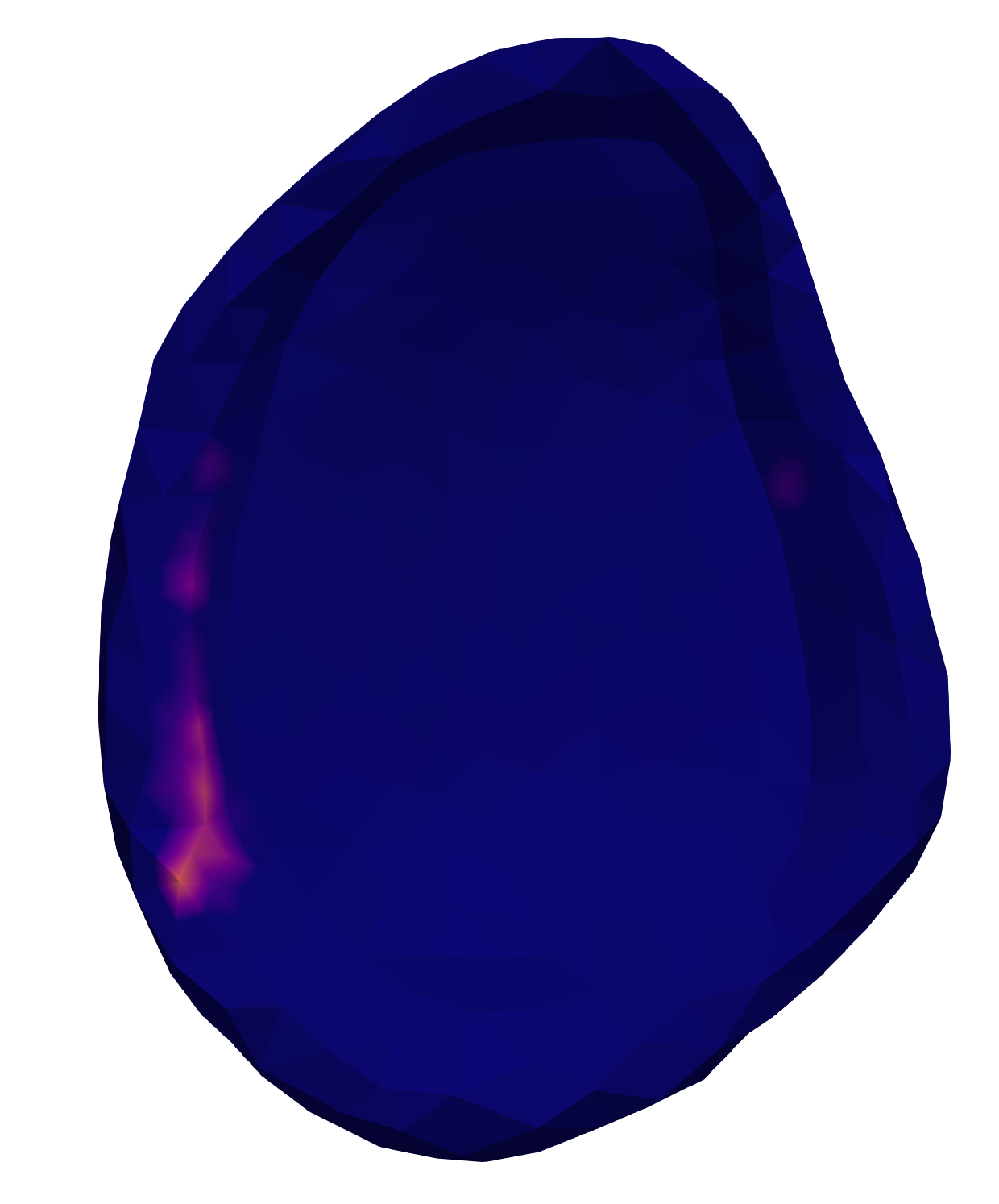}
        \subcaption*{$t = \SI{0.3}{\s}$}
        \label{fig:stress_t_start_post}
    \end{subfigure} \hfill
    \begin{subfigure}[b]{0.19\linewidth}
        \centering
        \includegraphics[height=2.5cm]{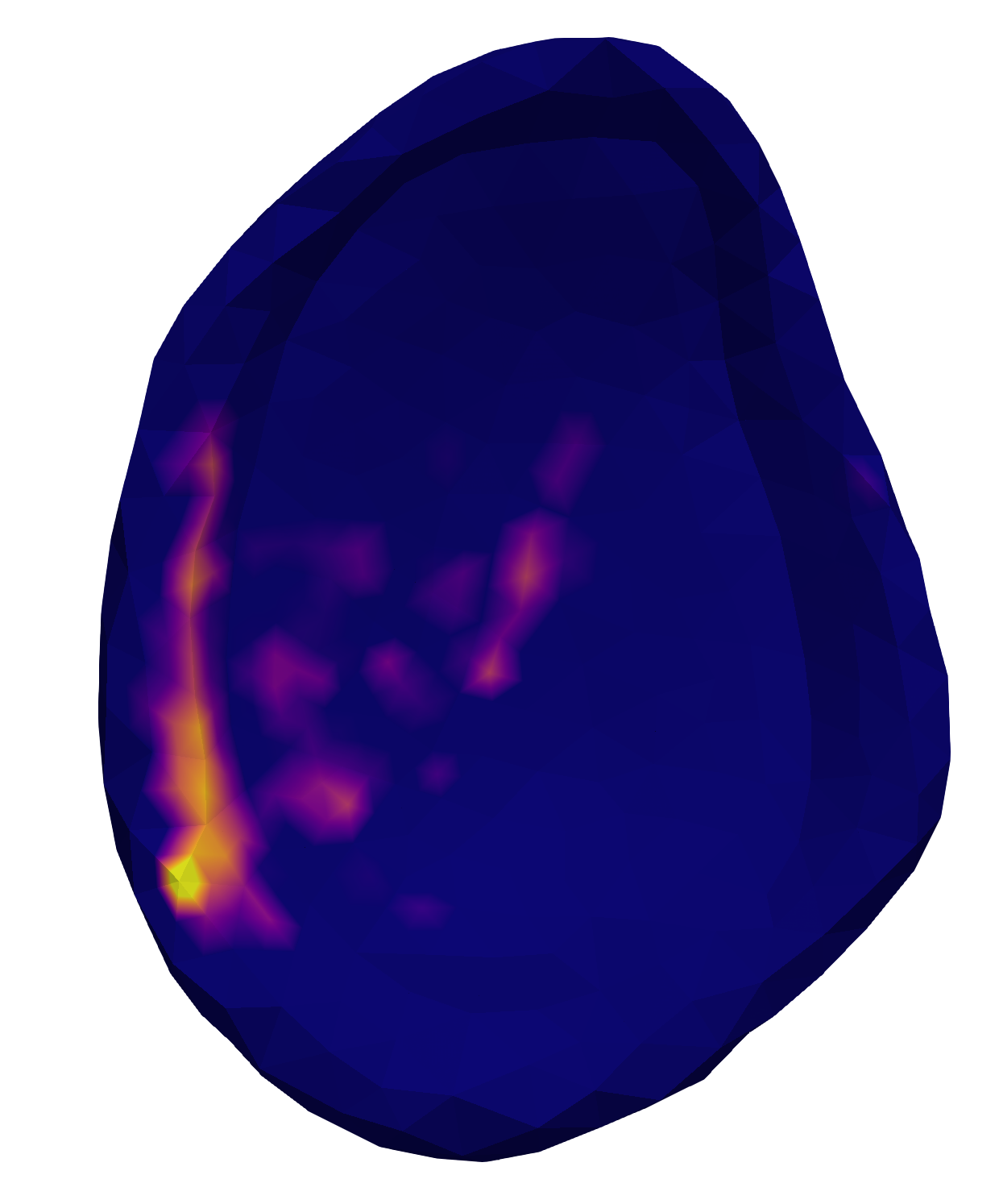}
        \subcaption*{$t = \SI{0.35}{\s}$}
        \label{fig:stress_t_middle_post}
    \end{subfigure} \hfill
    \begin{subfigure}[b]{0.19\linewidth}
        \centering
        \includegraphics[height=2.5cm]{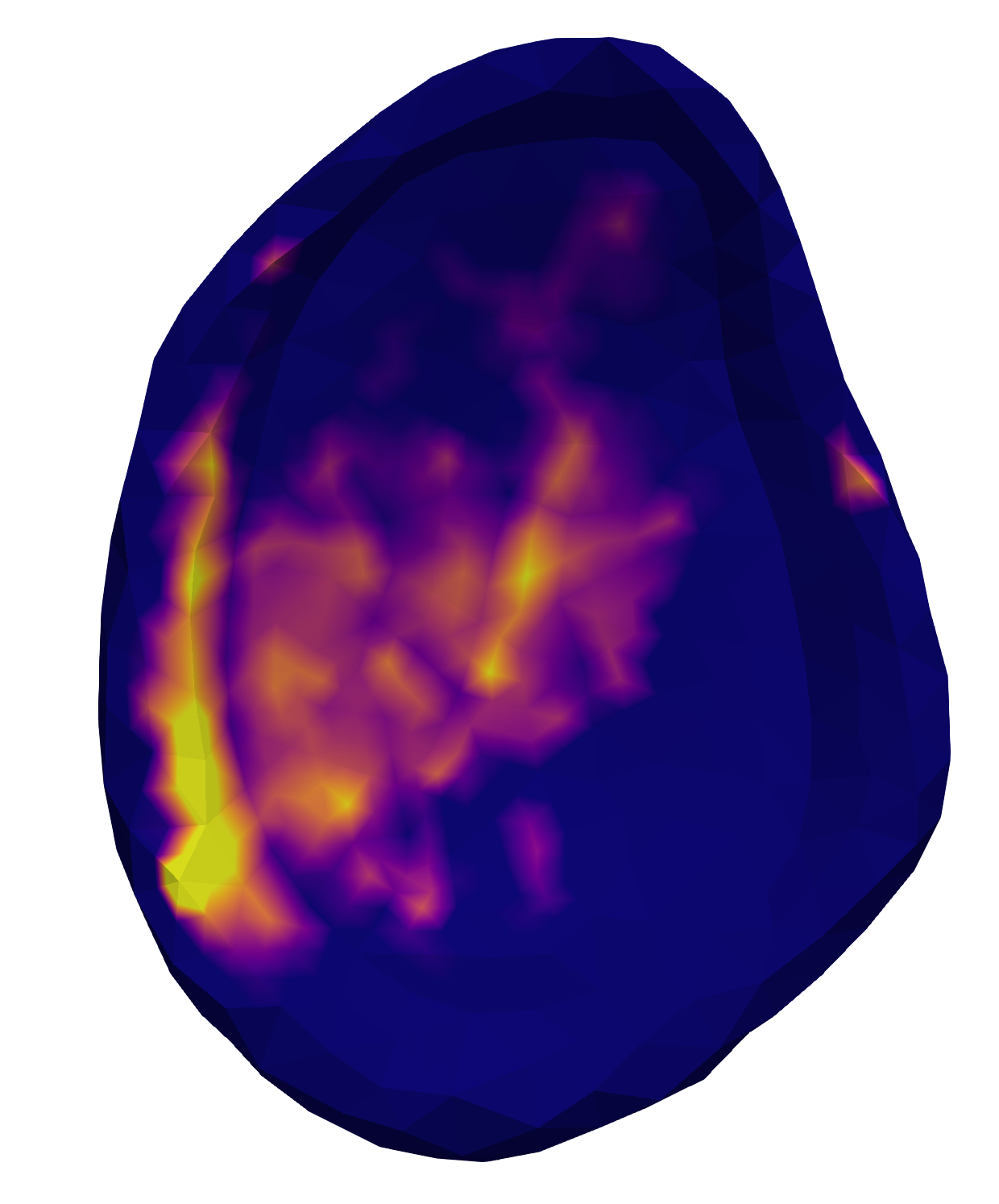}
        \subcaption*{$t = \SI{0.04}{\s}$}
    \label{fig:stress_t_end_post}
    \end{subfigure}\hfill
    \begin{subfigure}[b]{0.18\linewidth}
    \raggedleft
    \includegraphics[height=3cm]{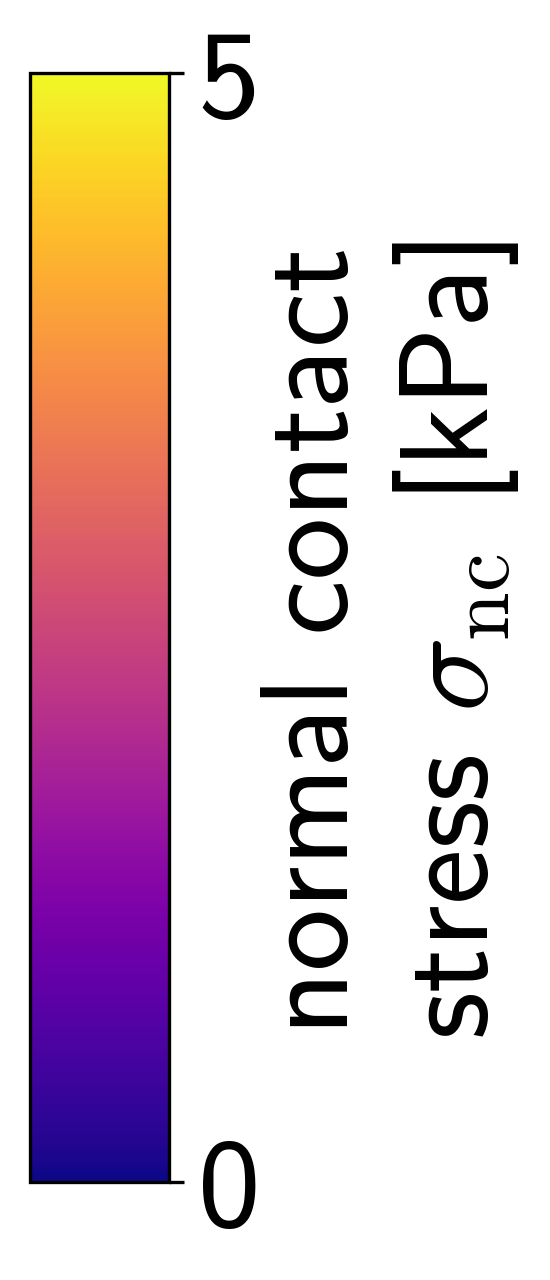}
    \end{subfigure}
    \caption{Normal contact stresses $\sigma_\mathrm{nc}$ visualized on the glenoid fossa surface. After initial contact with the humeral head is made at \mbox{$t=\SI{0.24}{s}$}, the contact area and normal contact stress increase due to the continuous contraction and associated pulling force of the rotator cuff.}
    \label{fig:contact_stresses_shoulder}
\end{minipage}\hfill
\begin{minipage}[b]{.35\textwidth}
  \centering
    \includegraphics[height=3.902cm]{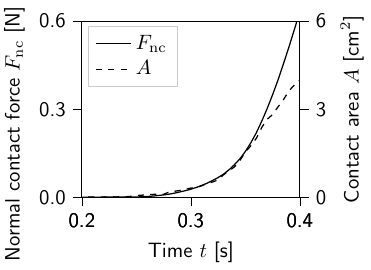}
    \caption{Evolution of the normal contact force $F_\mathrm{nc}$ and the associated contact area $A$ between glenoid fossa and humeral head over time.}
    \label{fig:contact_force_shoulder}
\end{minipage}
\end{figure}

Initial contact between the humeral head and the glenoid fossa is made at $t=\SI{0.24}{\s}$. With ongoing time and a steadily growing pulling force of the rotator cuff, the contact area $A$ and the normal contact stresses $\sigma_\mathrm{nc}$ increase, as depicted in Fig. \ref{fig:contact_stresses_shoulder}. Consequently, the normal contact force $F_\mathrm{nc}$, evaluated as the integral of $\sigma_\mathrm{nc}$ over $A$, rises as well, see Fig. \ref{fig:contact_force_shoulder}.

Due to a lack of suitable validation data, we here refrain from an in-depth quantitative analysis. However, our presented results are qualitatively plausible and showcase the applicability of the modified \COMBI-material model within a large-scale continuum-mechanical shoulder model.

\section{Conclusion}
The objective of this work was to identify a constitutive model that accurately represents both active and passive muscle characteristics within continuum-mechanical models for complex musculoskeletal systems, particularly for the human shoulder.
Therefore, we conducted a comprehensive review of active skeletal muscle constitutive laws and identified three commonly used activation concepts: active strain, active stress, and generalized active strain. Corresponding to those concepts, we selected three material models from the reviewed literature for a thorough comparison. We identified differences considering both the active and the passive material characteristics, including the applied force-stretch- and time-activation-dependencies, the computational efficiency of the activation level computation, the mathematical properties of the underlying activation concepts, and the assumed coupling of passive and active mechanics. 
Based on these findings, we proposed a fourth material model, combining the most promising features from \cite{weickenmeier_physically_2014} and \cite{giantesio_strain-dependent_2017}. This approach couples active and passive mechanics in a generalized active strain approach, thus representing actual muscle tissue in a physiologically realistic manner. Stresses are consistently derived from the strain energy function, the explicit computation of the activation level enhances computational efficiency and the applied force-stretch-dependency aligns with empirical data throughout the entire stretch regime. 

As a basis for a numerical comparison, we fitted the stress responses to experimental data obtained under one active and six passive load conditions. 
Depending on the load case, one or the other material model approximated the experimental data better, but overall, the approximations were equally satisfying. Our analysis further underscored the importance of considering multiple load cases to uniquely determine the material parameters. 
We applied the material models to simulate fusiform muscle activation in an isometric and a free concentric contraction case. Our results show that the different activation concepts affect shearing and deformation transversal to the fiber direction, even though the material characteristics along the fiber direction may coincide. Although not proven by empirical data, we argued that the more realistic modeling of the coupling between active and passive elements through the generalized active strain approach -- as applied in our modified material model -- yields the most realistic results. 
We utilized the proposed modified material model to simulate the abduction of the humerus bone by the deltoid muscle within a simplified two-component muscle-bone model.
Providing first insights into the concavity compression mechanism of the glenohumeral joint, we finally demonstrated the application of our modified material model in an example simulation of rotator cuff activation within a continuum mechanical model of the human shoulder.

Despite the comprehensive investigation, our work comes with limitations. As we focused on purely hyperelastic approaches, we did not consider viscous phenomena and history-dependent activation properties, such as force enhancement and depression. Depending on the particular problem, those effects can be of significant influence and may need to be incorporated into the material model (e.g., done in \cite{seydewitz_three-dimensional_2019}). 
When comparing the model responses with experimental data, we considered several load cases for the passive muscle but only data from uniaxial compression/tension tests for the active muscle. Our preference for the generalized active strain approach is based on theoretical considerations regarding the modeling aspect but has not yet been confirmed by empirical data. Experimental measurements of transverse and shear stress-stretch responses for active muscle tissue could further support our argument. The verification of the passive material response could be extended by experimental data from other load cases, e.g., planar biaxial loading \cite{wheatley_investigating_2020}.
The presented continuum shoulder model already incorporates various physiologically relevant anatomical components, contact interactions and material properties. However, there is potential for even further improvement in achieving a more accurate and realistic representation of the shoulder complex. Possible enhancements include the incorporation of tendons and ligaments, image data-based fiber architectures, and more sophisticated boundary conditions (e.g., to account for scapulothoracic gliding). Not being the focus of this work, we will address these aspects in future research.
We applied the proposed active muscle material model in a simulation of the shoulder model with relatively modest (contractile) deformation and movement. Further investigations are imperative to ascertain if the material model (with the identified parameters) is suited to simulate broader ranges of motion in a physiologically realistic manner. A comparison of the resulting three-dimensional deformations and stresses against dynamically acquired image data (e.g., dynamic MRI, shear wave elastography measurements) can help to uncover potential drawbacks. Previously neglected but potentially relevant effects can then be identified and incorporated into the constitutive model. 
\section*{Acknowledgements}
We thank our colleagues Dr.-Ing. Maximilian Grill and  Amadeus Gebauer for technical discussions and support, as well as Lisa Engstler and Moritz Brüch for their contributions during their student projects.

\section*{Declarations}
\paragraph{Data availability} All data generated or analyzed during this study are included in this published article.\vspace{-\baselineskip}
\paragraph{Conflict of interest} The authors declare that they have no competing interests.\vspace{-\baselineskip}
\paragraph{Ethical approval} Not applicable.\vspace{-\baselineskip}

\renewcommand*{\bibfont}{\normalfont\small}
\printbibliography[heading=bibintoc, title={References}]

\setcounter{section}{0}
\renewcommand{\thesection}{}
\renewcommand{\thesubsection}{\Roman{subsection}}
\section*{Appendix}
\subsection{Stress tensor deviations and supplementary equations}
\subsubsection*{Active stress approach (\BLE)}
The second Piola-Kirchhoff stress tensor is assembled similarly to the strain-energy function in Eq. \eqref{eq:Psi_Ble} into an isochoric and a volumetric contribution as
\begin{equation}
	\SecPK = \SecPKiso + \SecPKvol.
\end{equation}
Using the fictitious second Piola-Kirchhoff stress \mbox{$\modS = 2\frac{\partial \PsiIso(\modC)}{\partial \modC}$} and the projection tensor $\PP = \mathbb{I} - \frac{1}{3} \invC \otimes \C$, the contributions are computed to
\begin{equation}
	\SecPKiso = J^{-2/3} \PP : \modS \qquad \text{and} \qquad \SecPKvol = \kappaBle \ln{(J)} \invC.
\end{equation}
Following the active stress approach, $\modS$ involves an active and a passive contribution according to
\begin{equation}
	\modS = \modS_{\mathrm{a}} + \modS_{\mathrm{p}} = \modgammafour^{\mathrm{a}} \M + \left( \modgammaone \I +
	\modgammafour^{\mathrm{p}} \M + 
	\modgammafive \left( \M \modC^T + \modC^T \M \right) \right)
\end{equation}
with the pre-factors
\begin{equation*}
	\modgammaone = 2 G_{2} A_2{\modIfour},
	\qquad 
	\modgammafour^{\mathrm{a}} = \frac{\sigmafibact}{\modIfour},
	\qquad 
	\modgammafour^{\mathrm{p}} = -4 G_{1} \frac{\modIfive}{{\modIfour}^3}
	+ 2 G_{2} A_2 \left(\modIone -A_1 \right) + \frac{\sigmafibpas}{\modIfour},
	\qquad 
	\modgammafive =  \frac{2 G_{1}}{{\modIfour}^2}
	-2 G_{2} A_2,
	\label{eq:modgammas}
\end{equation*}
and the helper quantities
\begin{equation}
	A_1 = \frac{\modIone \modIfour - \modIfive}{2 \modIfour},
	\qquad \qquad
	A_2 = \frac{1}{\sqrt{\modIfour}} \frac{\acosh ({A_1 \sqrt{\modIfour}})}{\sqrt{A_1^2 \modIfour -1}},
	\qquad \qquad
	A_3 = \frac{A_1 A_2 \modIfour - 1}{A_1^2 \modIfour-1}.
	\label{eq:helper_As}
\end{equation}

\subsubsection*{Generalized active strain approach (\WKM)}
The passive first generalized invariant for uniaxial tension and its derivative w.r.t. the fiber stretch read
\begin{equation}
\Iptilde = \lambda^2 \left(\tfrac{1 + 2\omegazero}{3} \left(\lambda^{-3} -1\right)\right) 
\quad \text{and} \quad
\Iptilde' = \DIptildeDlambdaM = 2\lambda \left(1-\tfrac{\omegazero}{3} \left(\lambda^{-3} + \tfrac{2}{3} \right) \right). \label{eq:Iptilde}
\end{equation}
The second Piola-Kirchhoff stress is derived from Eq. \eqref{eq:Psi_WKM} as
\begin{equation}
  \SecPK = \tfrac{\gamma}{2} \left[
  \eexp{\alpha(\Itilde-1)} \left( \Ltilde + \omegaa \M \right) - \eexp{\beta(\Jtilde-1)} \det (\C) \invC \Ltilde \invC +
  \left( \Jtilde \eexp{\beta(\Jtilde-1)} - \det(\C)^{-\kappa} \right) \invC \right] .
  \label{eq:S_WKM}
\end{equation}

\subsubsection*{Active strain approach (\GIANT)}
The generalized elastic invariants for uniaxial tension are 
\begin{equation}
\Ietilde(\lambda, \omegaa)=\tfrac{2\omegazero(1-\omegaa)}{3\lambda}+\left(1-\tfrac{2\omegazero}{3}\right) \tfrac{\lambda^2}{(1-\omegaa)^2}
\quad \text{and} \quad
\Jetilde(\lambda, \omegaa)= \tfrac{2\omegazero \lambda}{3(1-\omegaa)}+\left(1-\tfrac{2\omegazero}{3}\right) \tfrac{(1-\omegaa)^2}{\lambda^2} ,
\end{equation}
with their passive counterparts $\Iptilde=\Ietilde(\lambda, 0)$ and $\Jptilde=\Jetilde(\lambda, 0)$.
\newline
The elastic second Piola-Kirchhoff stress $\SecPKe$ results from the elastic deformation $\defgrade$ and is thus computed based on Eq. \eqref{eq:Psi_Giant} as 
\begin{equation}
\SecPKe = 2 \frac{\partial \Psie}{\partial \Ce} = \tfrac{\gamma}{2} \left[
	\eexp{\alpha(\Ietilde-1)} \Ltilde - \eexp{\beta(\Jetilde-1)} \det (\Ce) \invCe \Ltilde \invCe +
	\Jetilde \eexp{\beta(\Jetilde-1)} \invCe \right].
\end{equation}

The second Piola-Kirchhoff stress $\SecPK$ comprises the derivatives of the elastic and the volumetric part of the strain-energy function with respect to $\C$. Since the included activation level $\omegaa$ depends on the fiber stretch and is a function of $\C$, the term $\SecPK_2$ complements the conventional contribution $\SecPK_1$. With the volumetric stress $\SecPKvol$ the stress tensor reads
\begin{equation}
\SecPK =  \SecPK_1 + \SecPK_2 + \SecPKvol = 2 \frac{\partial \Psie}{\partial \C} + 2 \frac{\partial \Psie}{\partial \omegaa} \frac{\partial \omegaa}{\partial \C} + 2 \frac{\partial \PsiVol}{\partial \C}.
\label{eq:S_Giant}
\end{equation}
The respective contributions therein are derived as
\begin{equation}
\SecPK_1 = \invdefgrada \SecPKe \invdefgrada,
\label{eq:S_Giant_conventional}
\end{equation}
\begin{equation}
\SecPK_2 = - 2 \left( \SecPK_1 : \left(\C \invdefgrada 
\DdefgradaDomegaa \right)\right) \DomegaaDC = - \frac{1}{\lambda} \left(\SecPK_1 : \left(\C \invdefgrada \DdefgradaDomegaa \right)\right) \M,
\label{eq:S_Giant_omegaa}
\end{equation}
\begin{equation}
\SecPKvol = -\tfrac{\gamma}{2} \det(\C)^{-\kappa} \invC.
\label{eq:S_Giant_vol}
\end{equation}
Eq. \eqref{eq:S_Giant_omegaa} comprises the derivatives
\begin{equation}
\DdefgradaDomegaa = - \M - \tfrac{1}{2} (1-\omegaa)^{-\frac{3}{2}} (\I - \M) \qquad \text{and} \qquad
\DomegaaDC = \DomegaaDlambdaM \frac{\partial\lambda}{\partial\C} = \frac{1}{2\lambda} 
\DomegaaDlambdaM \M.
\label{eq:DomegaaDC}
\end{equation}
Considering the implicit definition of $\omegaa$ in Eq. \eqref{eq:waimplicit}, $\DomegaaDlambdaM$ in Eq. \eqref{eq:DomegaaDC} can be approximated by a central finite differences scheme.

\subsection{Parameter identification}
\subsubsection*{Boundary conditions and analytical stress responses for investigated load scenarios}

\newcolumntype{Y}{>{\centering\arraybackslash}X}
\begin{table}[htb]
\centering
\caption{Load cases and associated Dirichlet boundary conditions applied in the simulation of the unit cubes during the material parameter fitting. The constraints are applied as indicated by the sketched support joints. Blue arrows indicate the direction of the prescribed displacement $\hat{u} = d (\lambdaBC-1)$ and $\hat{u} = d \nuBC$ (only for SAF) enforcing the deformation $\defgrad$ in Table \ref{tab:defgrad_and_loading_sketches}. Red arrows illustrate the muscle fiber direction.}
\begin{tabularx}{\textwidth}{p{2cm}YYY}
\toprule
     & UTCAF & UTCTF & SAF \\ \midrule
    \begin{minipage}{.1\textwidth}
      \centering \includegraphics[width=\linewidth]{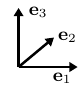}
    \end{minipage}
    &\begin{minipage}{.2\textwidth}\centering
      \includegraphics[width=\linewidth, height=\linewidth]{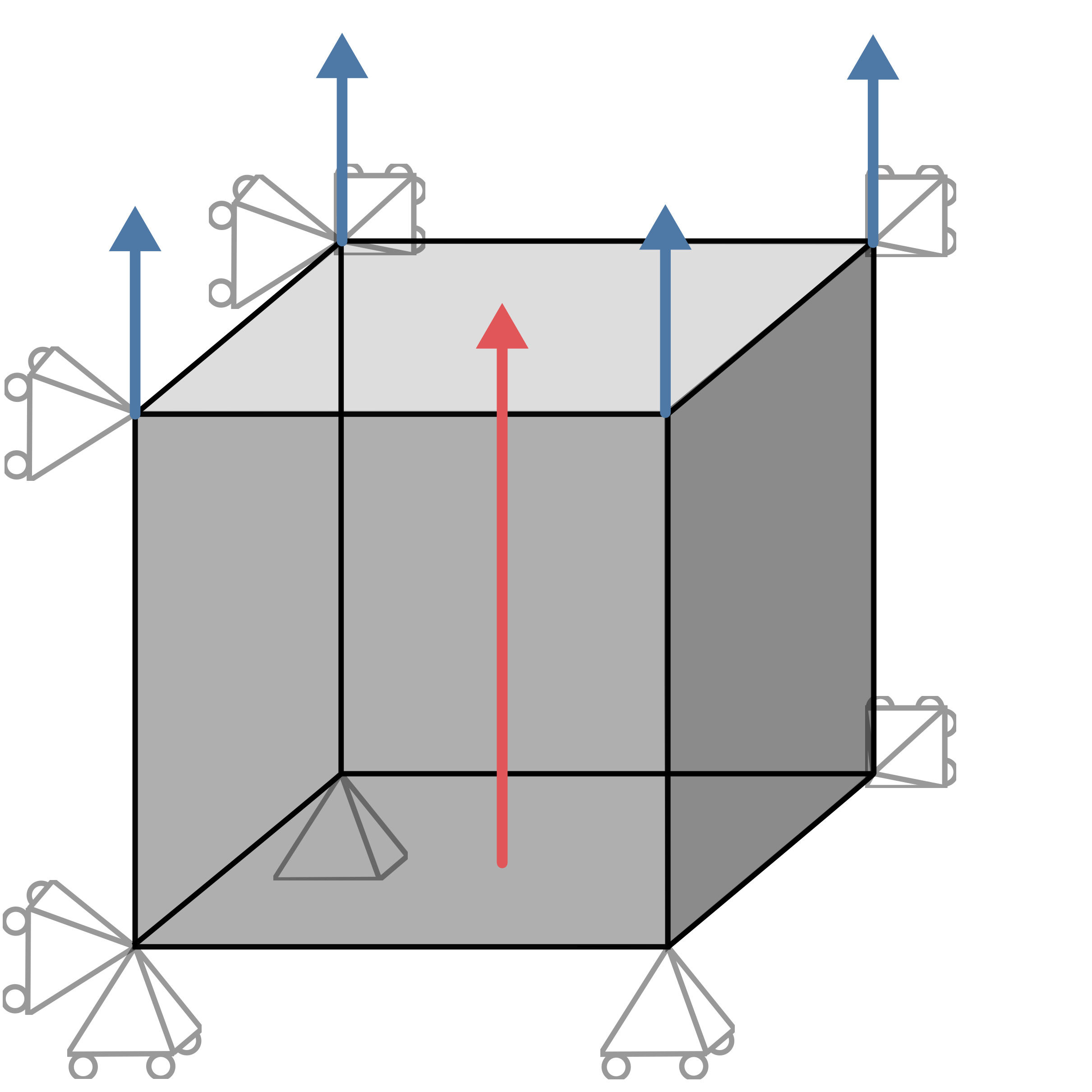}
    \end{minipage} 
    &\begin{minipage}{.2\textwidth}\centering
      \includegraphics[width=\linewidth, height=\linewidth]{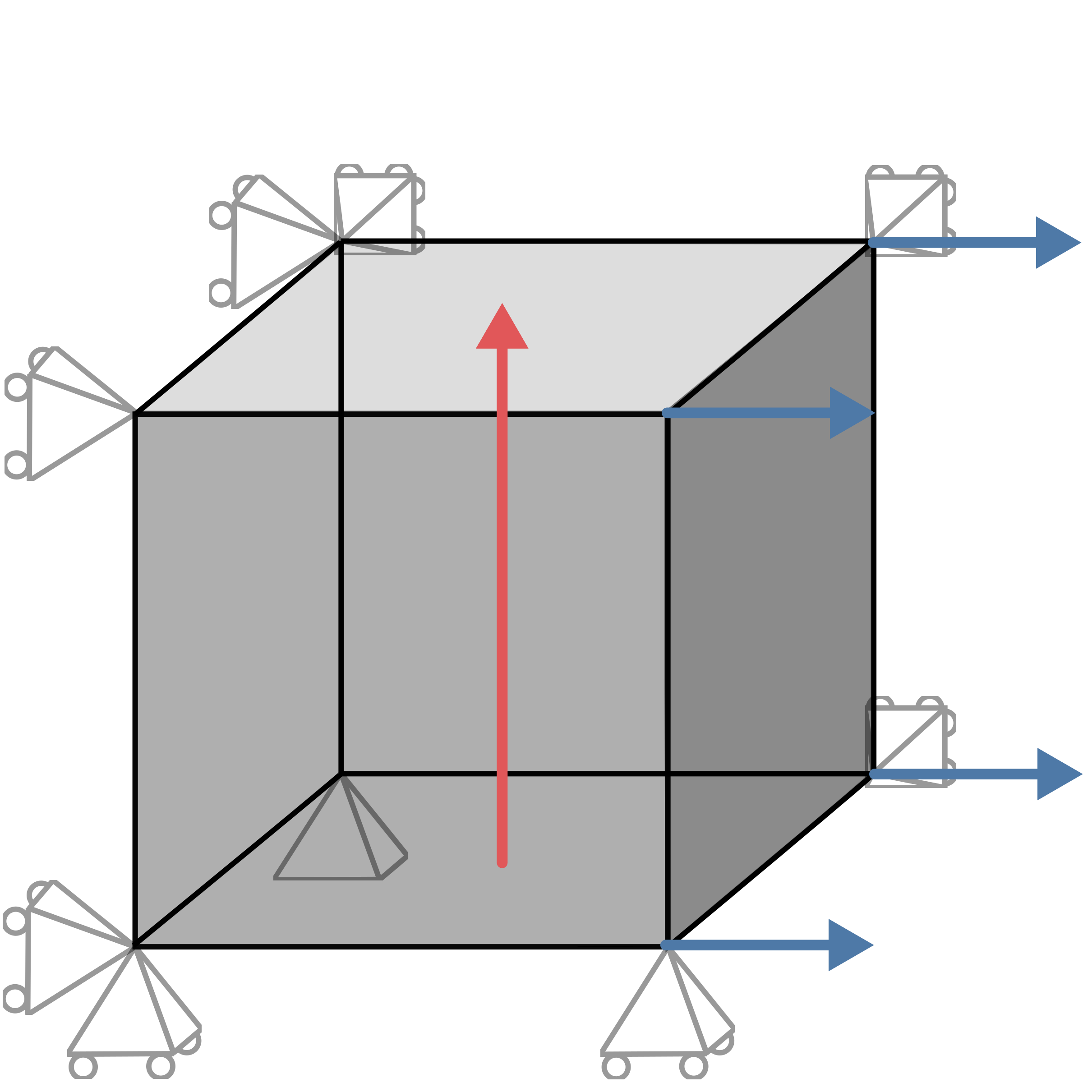}
    \end{minipage}
    &\begin{minipage}{.2\textwidth}\centering
      \includegraphics[width=\linewidth, height=\linewidth]{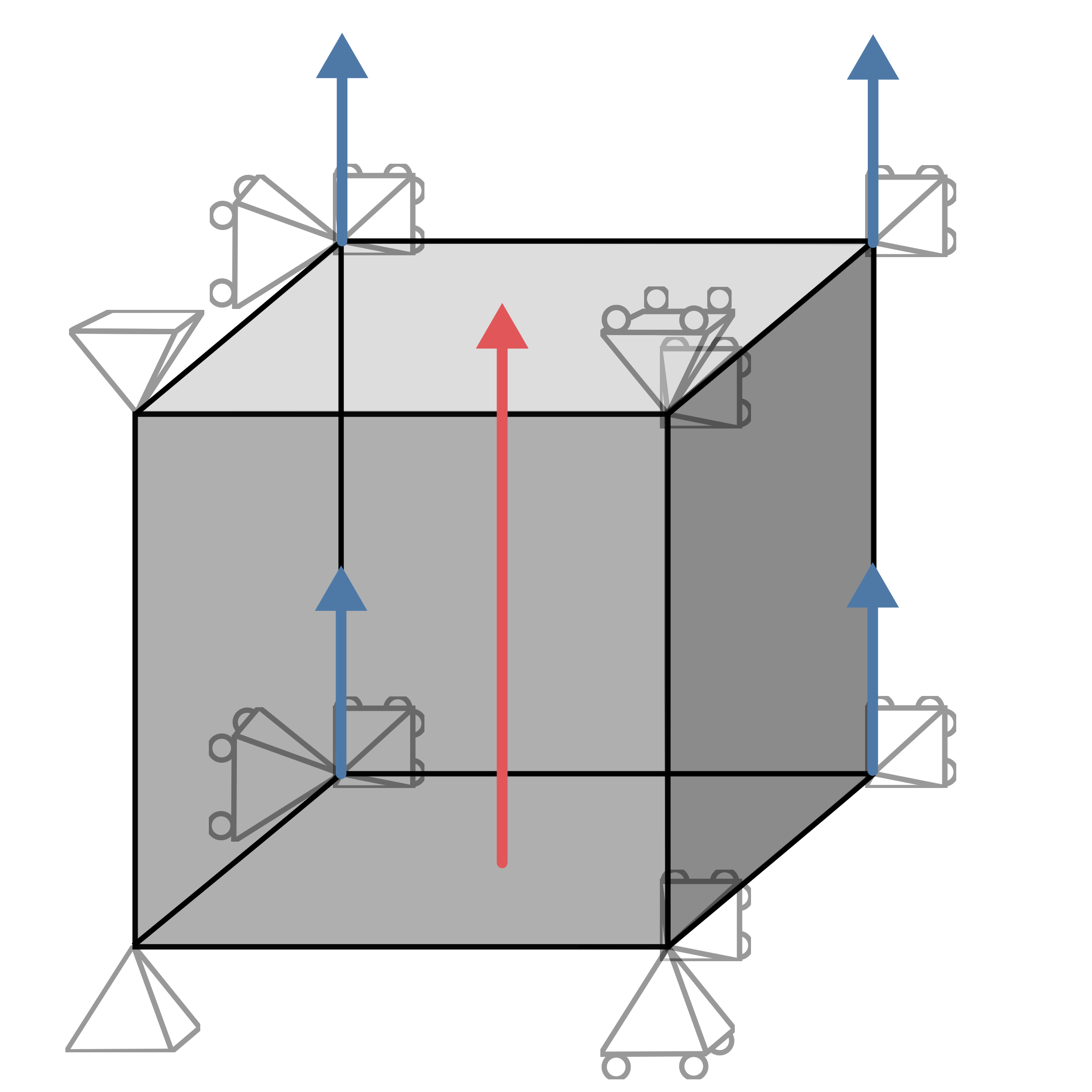}
    \end{minipage}\\ \addlinespace[10pt] \midrule
    & PSAF & PSTF & PSTIF \\  \midrule
    \begin{minipage}{.1\textwidth}
      \centering \includegraphics[width=\linewidth]{figures/tab_9_coordinate_system}
    \end{minipage}
    &\begin{minipage}{.2\textwidth}\centering
      \includegraphics[width=\linewidth, height=\linewidth]{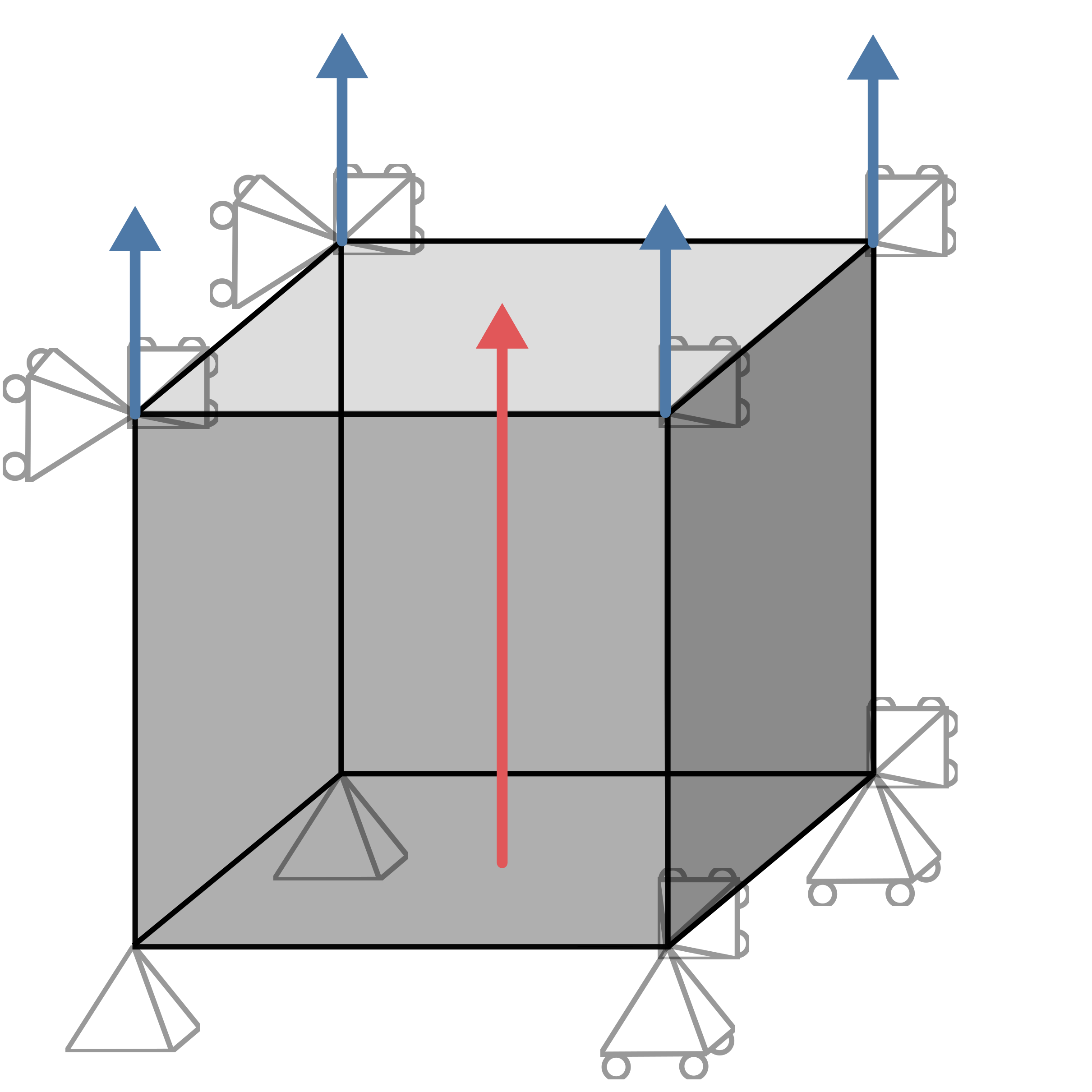}
    \end{minipage}
    &\begin{minipage}{.2\textwidth}\centering
      \includegraphics[width=\linewidth, height=\linewidth]{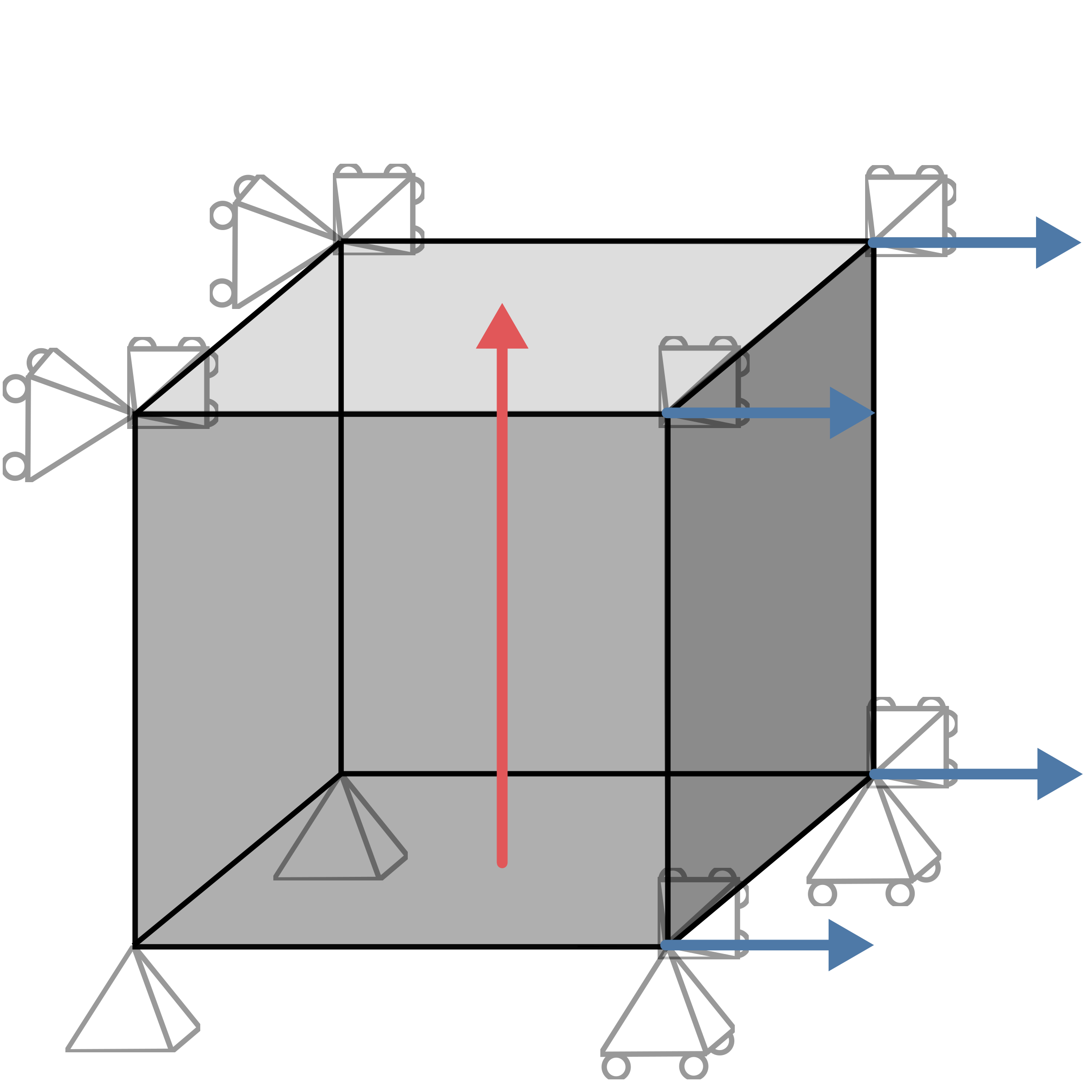}
    \end{minipage}
    &\begin{minipage}{.2\textwidth}\centering
      \includegraphics[width=\linewidth, height=\linewidth]{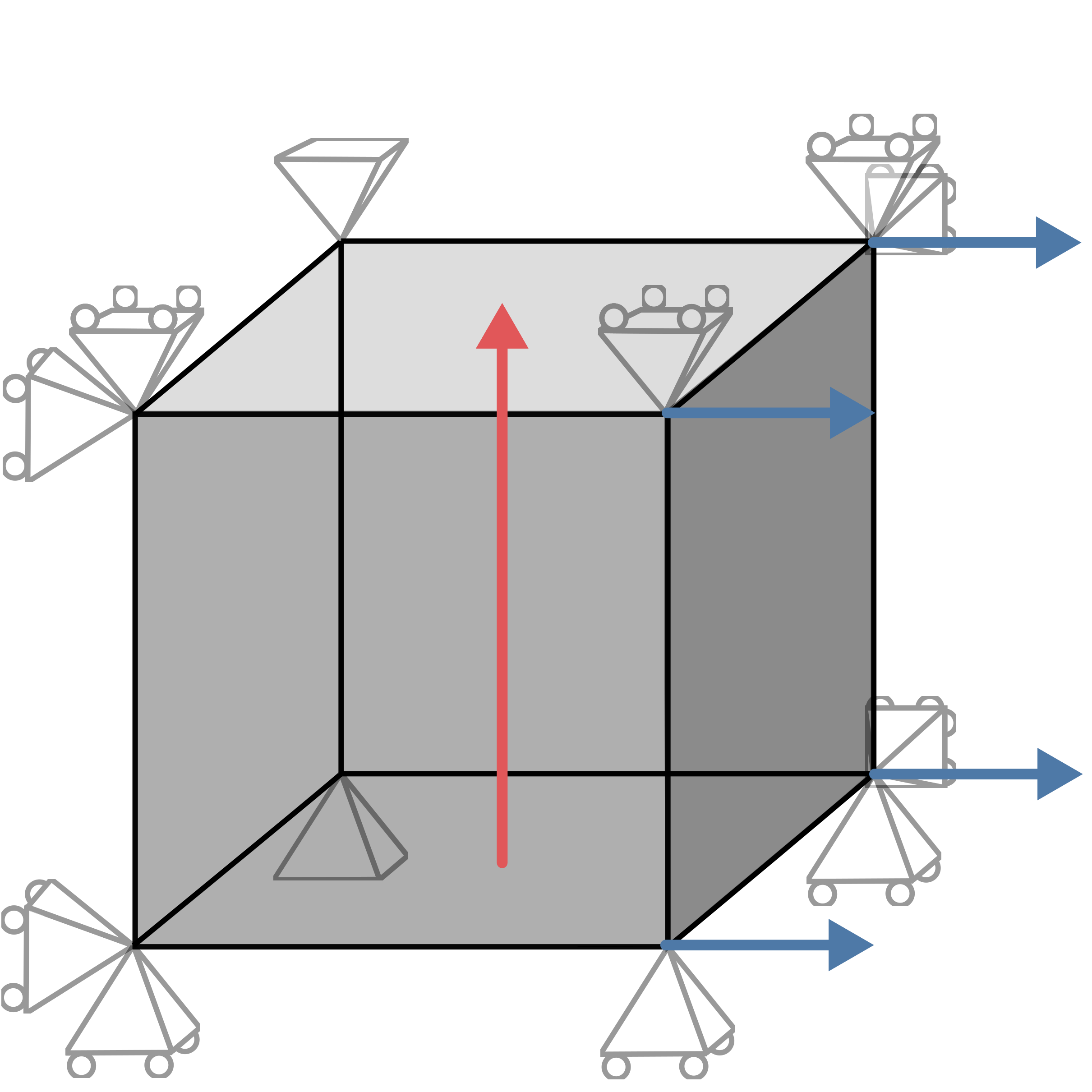}
    \end{minipage} \\
\bottomrule
\end{tabularx}
\label{tab:constraints_cube}
\end{table}

\begin{table}[htb]
\centering
\caption{Analytical expressions of the first Piola-Kirchhoff stress for the load cases defined in Table \ref{tab:overview_loading_abbreviatons}. The \BLE-terms include the Neo-Hookean contribution. The passive muscle response is obtained by setting the respective activation parameters to zero. In that case, the stress responses of the \GIANT-, \WKM-, and \COMBI-model coincide, and the elastic invariants $\Ietilde$ and $\Jetilde$ (\GIANT) are equal to the passive invariants $\Iptilde$ and $\Jptilde$ (\WKM, \COMBI).}
\begin{tabularx}{\textwidth}{p{1.2cm}p{0.2cm}X}
\toprule
\multicolumn{3}{l}{\textbf{Active muscle state}} \\\midrule
\multicolumn{3}{l}{\textbf{UTCAF}: Load in fiber direction in $\mathbf{e}_3$-direction} \\ \arrayrulecolor{white}\midrule
$P_{33}^{\BleName}$ & $=$ &  ${\lambdaBC^{\shortminus 1}} \sigmafibtot\left(\lambdaBC\right) + \mu \left(\lambdaBC  - {\lambdaBC^{\shortminus 2}}\right)$\\\noalign{\vspace{3pt}}
$P_{33}^{\GiantName}$ & $=$ & $- \tfrac{\gamma}{6} 
\left[\left( \omegazero \left(2 + {\lambdaBC^{\shortminus 3}}\eta^{3} \right)  - 3\right) \lambdaBC \eta^{\shortminus 3} \eexp{\alpha(\Ietilde-1)} - \left( \omegazero \left( 2+ \eta^{\shortminus 3}\lambdaBC^{3} \right) - 3 \right) \lambdaBC^{\shortminus 3} \eta \eexp{\beta(\Jetilde-1)}
\right] \left(\eta + \DomegaaDlambdaM \lambdaBC \right) $ \newline
with $\eta=1-\omegaa$, $\Ietilde = \lambdaBC^2 \eta \left(\tfrac{2}{3}\omegazero \left(\lambdaBC^{\shortminus 3} - \eta^{\shortminus 3} \right) + \eta\right)$, and $\Jetilde =   \lambdaBC^{\shortminus 2} \eta^{\shortminus 1} \left( \tfrac{2}{3} \omegazero  \left( \lambdaBC^{3}   - \eta^3 \right) + \eta^3 \right)$
\\\noalign{\vspace{3pt}}
$P_{33}^{\WkmName}$ & $=$ & $ - \tfrac{\gamma}{6} \left[ \left( \omegazero\left(2 + {\lambdaBC^{\shortminus 3}}\right) - 3\left(1+ \omegaa\right) \right) \lambdaBC \eexp{\alpha(\Itilde-1)}
 - \left(\omegazero \left(1 + {2\lambdaBC^{\shortminus 3}}\right) - {3\lambdaBC^{\shortminus 3}}\right)  \eexp{\beta(\Jtilde-1)} \right]$ \newline with $\Itilde = \lambdaBC^2 \left(\tfrac{2}{3} \omegazero \left(\lambdaBC^{\shortminus 3} -1\right) + 1 + \omegaa\right)$, and $\Jtilde = \lambdaBC^{\shortminus 2} \left(\tfrac{2}{3} \omegazero \left(\lambdaBC^3 - 1\right) + 1\right)$ \\\noalign{\vspace{3pt}}
$P_{33}^{\CombiName}$ & $=$ & $P_{33}^{\WkmName} + \tfrac{\gamma}{4} \lambdaBC^2  \eexp{\alpha(\Itilde-1)} \DomegaaDlambdaM$ \tabularnewline \noalign{\vspace{6pt}}
\arrayrulecolor{black}\midrule
\multicolumn{3}{l}{\textbf{Passive muscle state}} \\
\arrayrulecolor{black}\midrule
\multicolumn{3}{l}{\textbf{UTCTF}: Load transversal to the fiber direction in $\mathbf{e}_1$-direction} \\ \arrayrulecolor{white}\midrule
$P_{11}^{\BleName}$ & $=$ &  ${3\lambdaBC^{\shortminus 1}} \Gtwo
  \acosh \left(\tfrac{1}{2}{{\lambdaBC}^{\frac1{2}}} \left({\lambdaBC^{\shortminus 1}} + \lambdaBC^2 \right)\right) \sgn(\lambdaBC^3-1)
 - \tfrac{1}{2} \lambdaBC^{\shortminus 1} \sigmafibtot\left( {\lambdaBC}^{\frac1{2}} \right) + \mu \lambdaBC \left(1 -{\lambdaBC^{\shortminus 3}}\right)$\\\noalign{\vspace{0pt}}
$P_{11}^{\footnotesize\substack{\GiantName\\\WkmName\\\CombiName}}$ & $=$ & $ 
        \tfrac{\gamma}{6} \omegazero \left(1 -{\lambdaBC^{\shortminus 3}}\right)  \left( \lambdaBC \eexp{\alpha(\Itilde-1)} +\eexp{\beta(\Jtilde-1)} \right)$ \newline with $\Iptilde = {\lambdaBC^{\shortminus 1}} \left(\tfrac{1}{3} \omegazero \left(\lambdaBC^{3} -1\right) + 1\right)$, and $\Jptilde = \lambdaBC \left(\tfrac{1}{3} \omegazero \left({\lambdaBC^{\shortminus 3}} - 1\right) + 1\right)$
\tabularnewline \noalign{\vspace{4pt}}
\arrayrulecolor{gray}\midrule
\multicolumn{3}{l}{\textbf{SAF}: Load in fiber direction in $\mathbf{e}_3$-direction, fixed $\mathbf{e}_2$-direction} \\ \arrayrulecolor{white}\midrule
$P_{33}^{\BleName}$ & $=$ &  $ 2 \nuBC \Gone + \nuBC \mu$\\\noalign{\vspace{0pt}}
$P_{33}^{\footnotesize\substack{\GiantName\\\WkmName\\\CombiName}}$ & $=$ & $ 
        \tfrac{\gamma}{6}
         \nuBC \left[ \omegazero  \eexp{\alpha(\Itilde-1)} - \left(2  \omegazero - 3\right) \left(\nuBC^2 + 1\right)  \eexp{\beta(\Jtilde-1)}
        \right]$ \newline with $\Iptilde = \tfrac{1}{3} \omegazero \nuBC^2 + 1$, and $\Jptilde = \left(1-\tfrac{2}{3} \omegazero\right) \nuBC^2 + 1$ 
\tabularnewline \noalign{\vspace{4pt}}
\arrayrulecolor{gray}\midrule
\multicolumn{3}{l}{\textbf{PSAF}: Load in fiber direction in $\mathbf{e}_3$-direction, fixed $\mathbf{e}_2$-direction} \\ \arrayrulecolor{white}\midrule
$P_{33}^{\BleName}$ & $=$ &  $2 \Gtwo \left(1 - {\lambdaBC^{\shortminus 1}} \right) + {\lambdaBC^{\shortminus 1}} \sigmafibtot\left(\lambdaBC\right) +  \mu \lambdaBC\left(1-{\lambdaBC^{\shortminus 4}}\right)$ \\\noalign{\vspace{0pt}}
$P_{33}^{\footnotesize\substack{\GiantName\\\WkmName\\\CombiName}}$ & $=$ & $ 
        - \tfrac{\gamma}{6} \lambdaBC \left[ \left(\omegazero \left(2 + {\lambdaBC^{\shortminus 4}} \right)  - 3\right)  \eexp{\alpha(\Itilde-1)} - \left( \omegazero \left(1 + {2\lambdaBC^{\shortminus 4}}\right) - {3\lambdaBC^{\shortminus 4}} \right) \eexp{\beta(\Jtilde-1)} \right]$ \newline with $\Iptilde = \lambdaBC^{2} \left(\tfrac{1}{3} \omegazero \left(\lambdaBC^{\shortminus 4} + \lambdaBC^{\shortminus 2} - 2\right) + 1\right)$, and $\Jptilde = \lambdaBC^{\shortminus 2} \left(\tfrac{1}{3} \omegazero \left(\lambdaBC^{4} + \lambdaBC^{2} - 2\right) + 1\right)$ 
\tabularnewline \noalign{\vspace{4pt}}
\arrayrulecolor{gray}\midrule
\multicolumn{3}{l}{\textbf{PSTF}: Load transversal to the fiber direction in $\mathbf{e}_1$-direction, fixed $\mathbf{e}_2$-direction} \\ \arrayrulecolor{white}\midrule
$P_{11}^{\BleName}$ & $=$ &  $
	2 \Gtwo \acosh \left(\tfrac{1}{2} \left({\lambdaBC^{\shortminus 1}} + \lambdaBC \right)\right) {\lambdaBC^{\shortminus 1}} \sgn\left(\lambdaBC^2-1\right)
	-{\lambdaBC^{\shortminus 1}} \sigmafibtot\left({\lambdaBC^{\shortminus 1}}\right) + 
\mu \lambdaBC\left(1-{\lambdaBC^{\shortminus 4}}\right)$\\\noalign{\vspace{0pt}}
$P_{11}^{\footnotesize\substack{\GiantName\\\WkmName\\\CombiName}}$ & $=$ & $ 
        \tfrac{\gamma}{6} \lambdaBC \left[ \left(\omegazero \left(1 + {2\lambdaBC^{\shortminus 4}}\right) - {3\lambdaBC^{\shortminus 4}}\right)  \eexp{\alpha(\Itilde-1)} - \left(\omegazero \left(2 + {\lambdaBC^{\shortminus 4}}\right)-3\right)  \eexp{\beta(\Jtilde-1)} \right]$ \newline with $\Iptilde = \lambdaBC^{\shortminus 2} \left(\tfrac{1}{3} \omegazero \left(\lambdaBC^{4} + \lambdaBC^{2} - 2\right) + 1\right)$, and $\Jptilde = \lambdaBC^{2} \left(\tfrac{1}{3} \omegazero \left(\lambdaBC^{\shortminus 4} + \lambdaBC^{\shortminus 2} - 2\right) + 1\right)$ 
\tabularnewline \noalign{\vspace{4pt}}
\arrayrulecolor{gray}\midrule
\multicolumn{3}{l}{\textbf{PSTIF}: Load transversal to the fiber direction in $\mathbf{e}_1$-direction, fixed $\mathbf{e}_3$-direction} \\ \arrayrulecolor{white}\midrule
$P_{11}^{\BleName}$ & $=$ &  ${4\lambdaBC^{\shortminus 1}} \Gtwo \acosh \left(\tfrac{1}{2} \left({\lambdaBC^{\shortminus 2}} + {\lambdaBC^2} \right)\right) \sgn\left(\lambdaBC^4-1\right) + \mu \lambdaBC\left(1-{\lambdaBC^{\shortminus 4}}\right)$ \\\noalign{\vspace{0pt}}
$P_{11}^{\footnotesize\substack{\GiantName\\\WkmName\\\CombiName}}$ & $=$ & $\tfrac{\gamma}{6} \lambdaBC \omegazero \left(1 - {\lambdaBC^{\shortminus 4}}\right) \left( \eexp{\alpha(\Itilde-1)} + \eexp{\beta(\Jtilde-1)} \right)$ with $\Iptilde$ and $\Jptilde$ as for the PSTF case
\tabularnewline \noalign{\vspace{4pt}}
\arrayrulecolor{black}\bottomrule
\end{tabularx}
\label{tab:analytical_responses}
\end{table}

\FloatBarrier

\subsubsection*{Error measures}
With the \(L_\infty\) norm providing insight into the maximum absolute deviation, we define the relative \(L_\infty\) norm error 
\begin{equation}
    \varepsilon_\infty = \frac{L_\infty(x - x^*)}{L_\infty(x^*)} \quad \text{with} \quad L_\infty(\tilde{x}) = \|\tilde{x}\|_\infty = \max_i |\tilde{x}_i|. 
    \label{eq:varepsiloninf}
\end{equation}
We further measure the total absolute difference between \(x\) and \(x^*\) using the \(L_1\) norm and define the relative \(L_1\) norm error
\begin{equation} 
\varepsilon_1 = \frac{L_1(x - x^*)}{L_1(x^*)} \quad \text{with} \quad L_1(\tilde{x}) = \|\tilde{x}\|_1 = \sum_i |\tilde{x}_i|. 
\label{eq:varepsilon1}
\end{equation} 

\subsection{Fusiform muscle contraction} 
\subsubsection*{Stress distribution over the muscle continuum} \label{sec:appendix_fusi_vis_distribution}

\subsubsection*{Visualizations of von Mises stress} \label{sec:appendix_fusi_vis}
%
%
\begin{figure}[h]
    \centering 
    \begin{minipage}[t]{0.8\textwidth}
     \raggedright
    \begin{subfigure}[t]{0.18\textwidth}
    \centering \begingroup\sbox0{\includegraphics{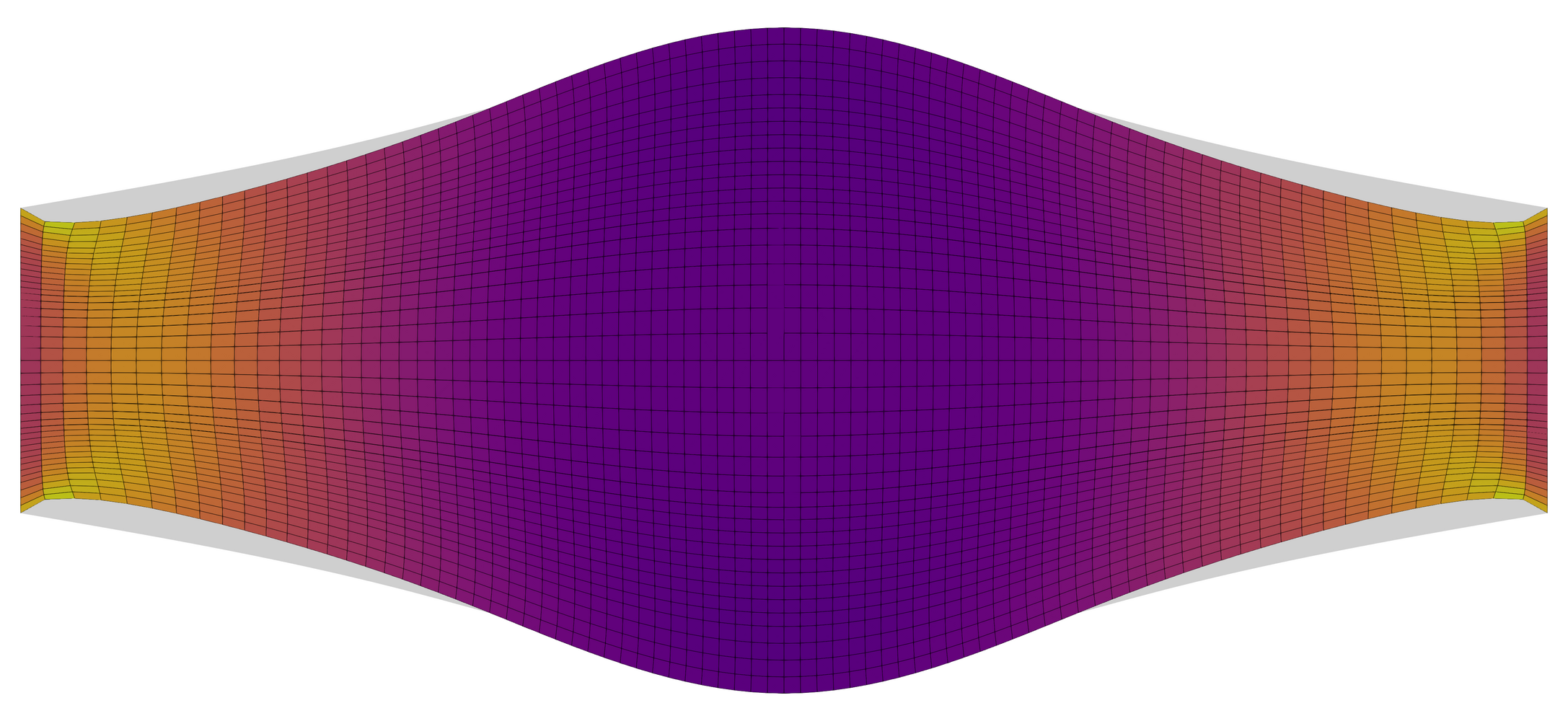}}
    \includegraphics[clip,trim={.5\wd0} 0 0 0, height=3cm]{figures/fig_23_slice_zx_von_mises_blemker_n4_ic_element.png}
    \endgroup \caption{\BLE} \label{fig:unknown}
    \end{subfigure}\hfill
    \begin{subfigure}[t]{0.18\textwidth}
    \centering \begingroup\sbox0{\includegraphics{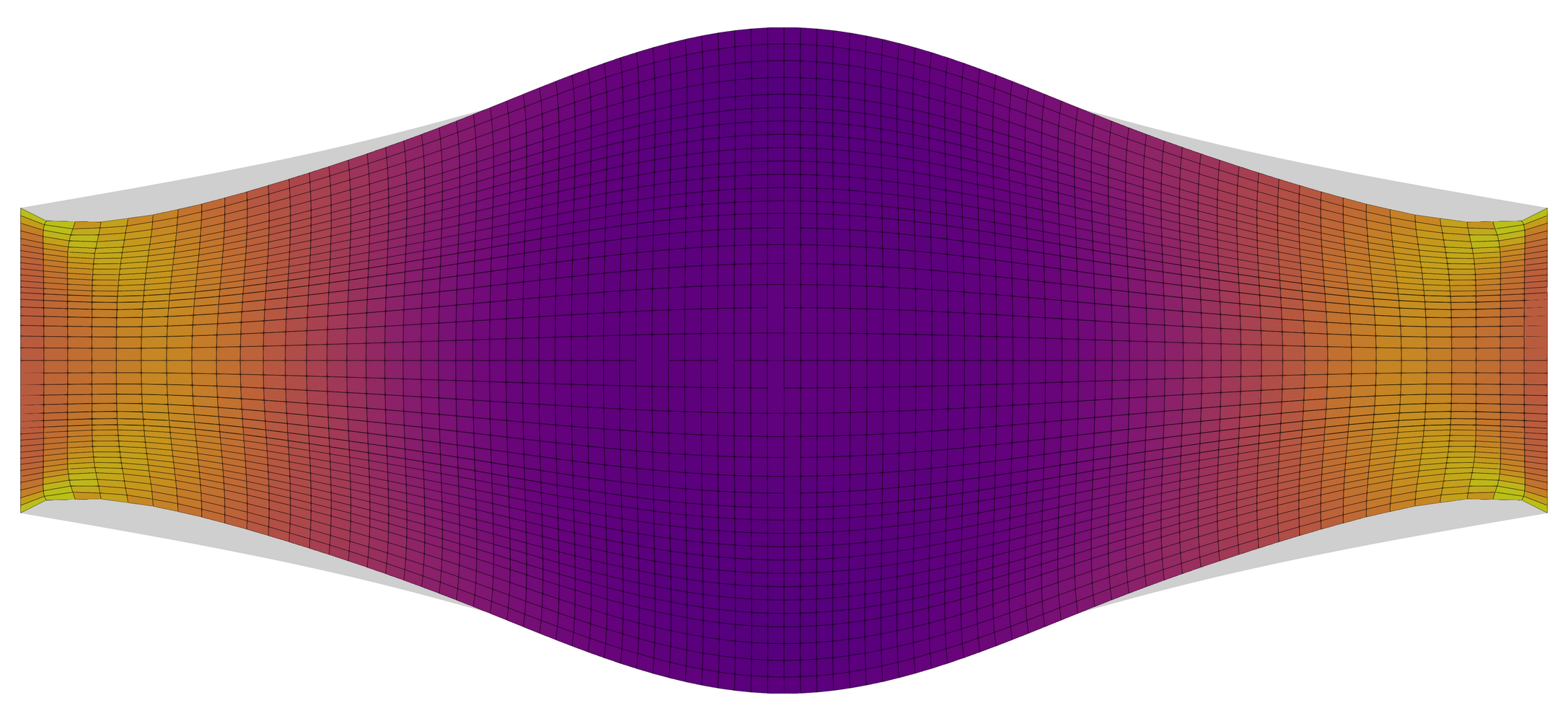}}
    \includegraphics[clip,trim={.5\wd0} 0 0 0, height=3cm]{figures/fig_23_slice_zx_von_mises_giantesio_n4_ic_element.png}
    \endgroup \caption{\GIANT} \label{fig:unknown}
    \end{subfigure}\hfill
    \begin{subfigure}[t]{0.18\textwidth}
    \centering \begingroup\sbox0{\includegraphics{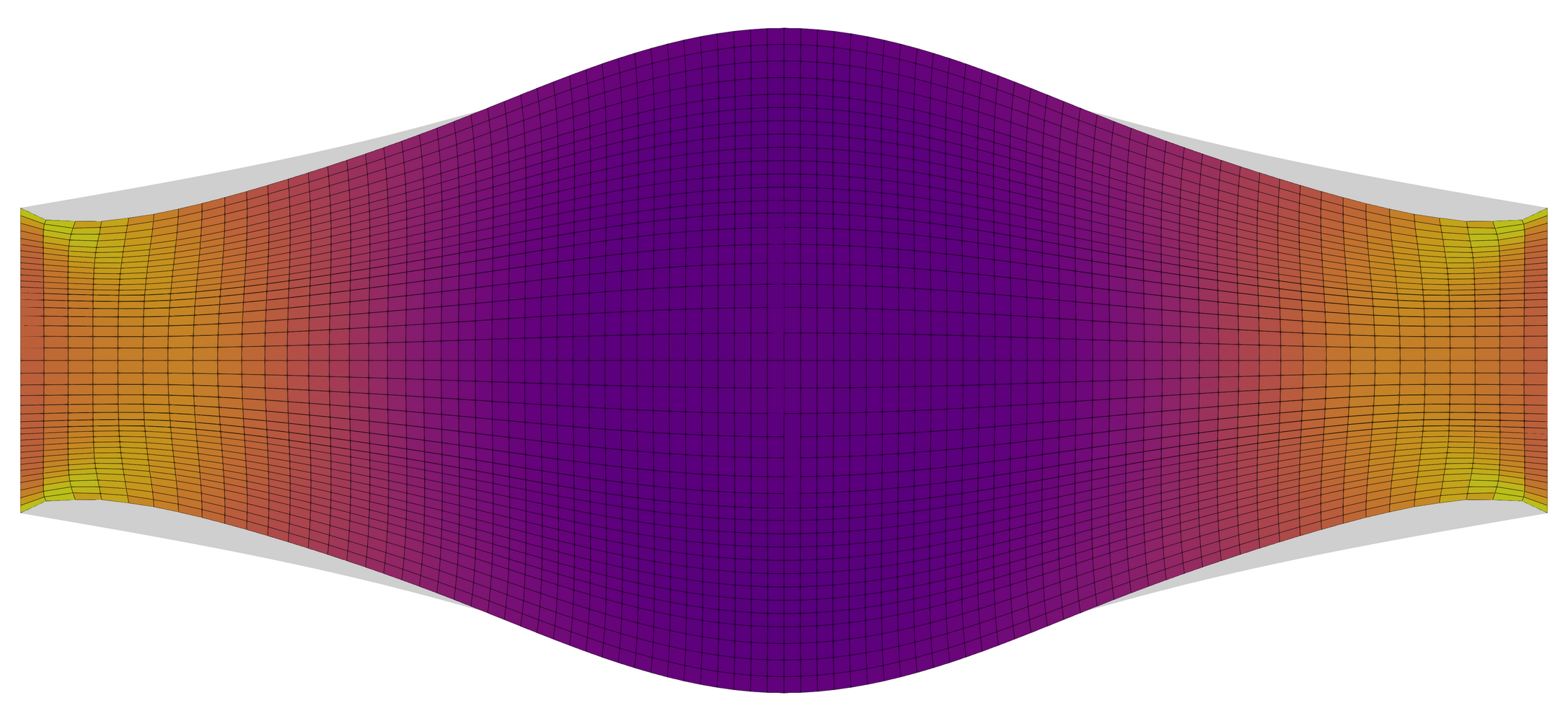}}
    \includegraphics[clip,trim={.5\wd0} 0 0 0, height=3cm]{figures/fig_23_slice_zx_von_mises_weickenmeier_n4_ic_element.png}
    \endgroup \caption{\WKM} \label{fig:unknown}
    \end{subfigure}\hfill
    \begin{subfigure}[t]{0.18\textwidth}
    \centering \begingroup\sbox0{\includegraphics{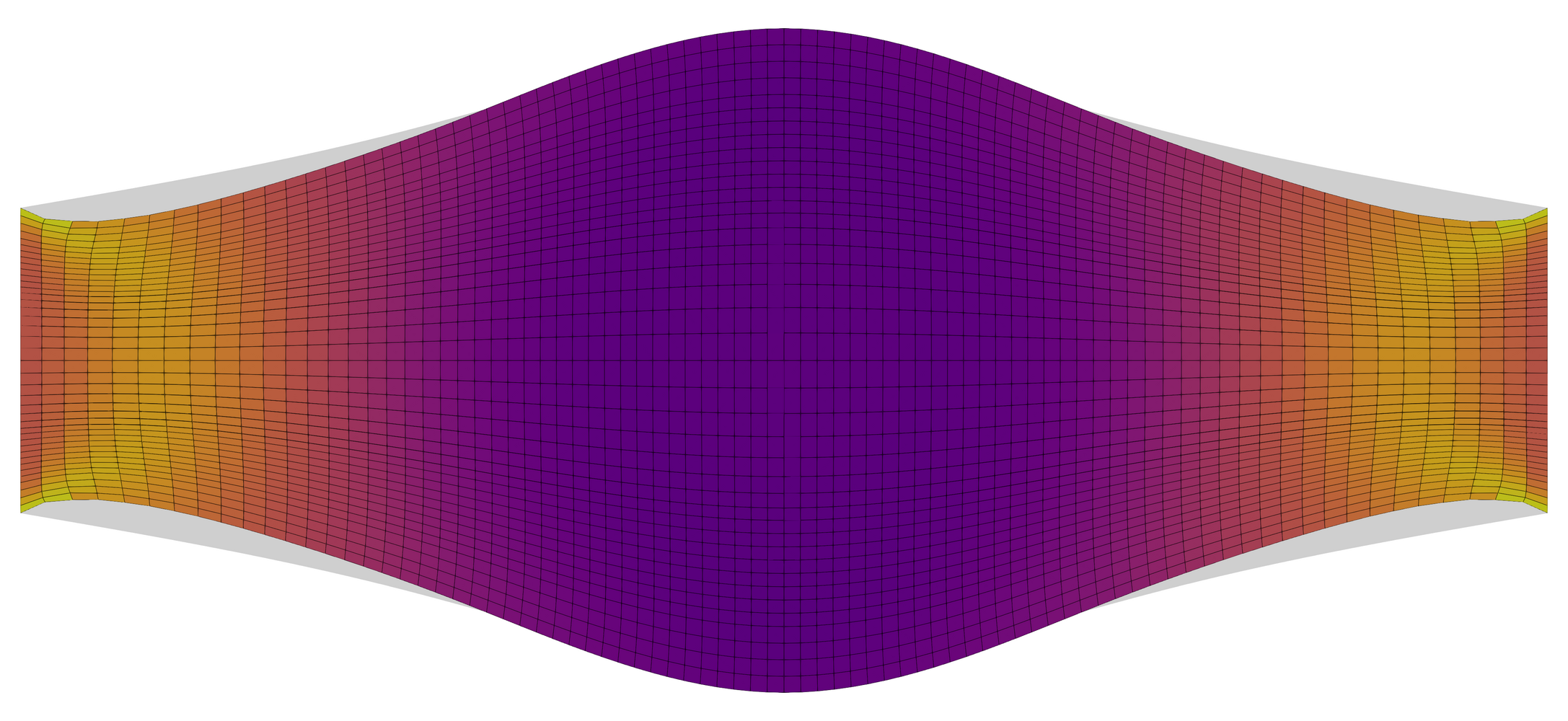}}
    \includegraphics[clip,trim={.5\wd0} 0 0 0, height=3cm]{figures/fig_23_slice_zx_von_mises_combi_n4_ic_element.png}
    \endgroup \caption{\COMBI} \label{fig:unknown}
    \end{subfigure}
    \end{minipage}\hfill
    \begin{minipage}[t]{1.29cm}
        \raggedleft 
    	\includegraphics[height=3cm]{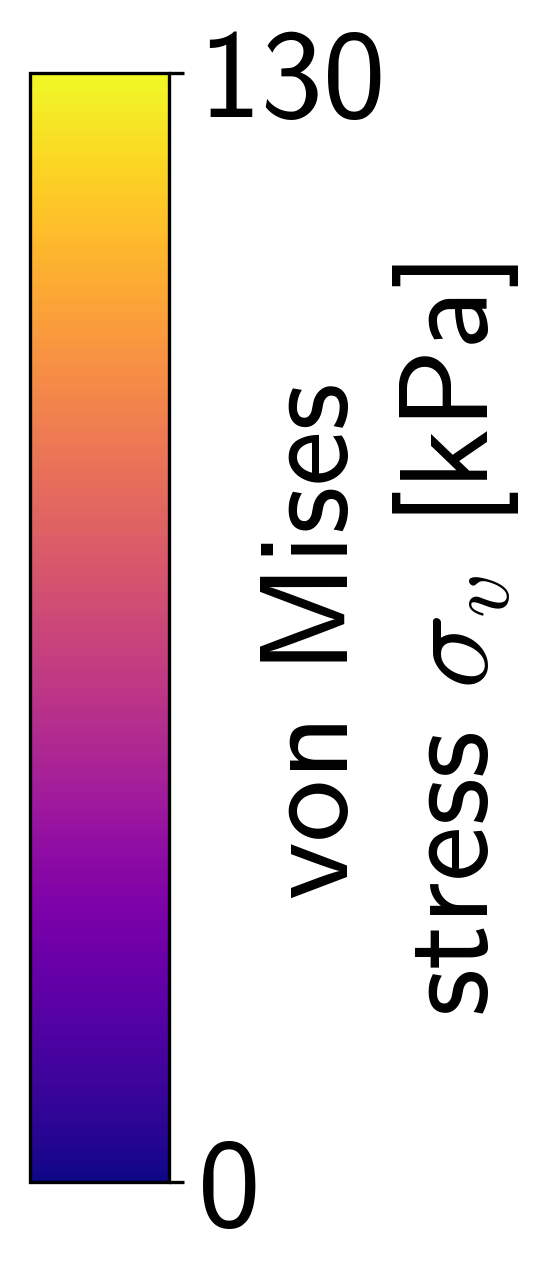}
    \end{minipage}
    \caption{Von Mises stress $\sigma_\mathrm{v}$ in the axial cross section of the fusiform muscle ($n=4$) for an isometric contraction in the tetanized state at $t=\SI{0.15}{\s}$. The initial configuration is displayed in grey. For symmetry reasons, only half the muscle is visualized.}
	\label{fig:fusi_iso_mises_stress_vis}
\end{figure}

\begin{figure}[h]
\centering
\begin{minipage}{0.86\textwidth}
\raggedright
    \parbox{\LWCap}{\centering \phantom{phantom text}} \hspace*{-10pt}
    \parbox{\LW}{\centering $t=\SI{0.005}{s}$ \\ 
    \vspace*{10pt}}
    \hspace*{6pt}
    \parbox{\LW}{\centering $t=\SI{0.02}{s}$ \\ 
    \vspace*{10pt}}
    \hspace*{6pt}
    \parbox{\LW}{\centering $t=\SI{0.15}{s}$ \\ 
    \vspace*{10pt}}
    \\ 
    \vspace*{6pt}
    \parbox{\LWCap}{\subcaption{\raggedright \BLE}} \hspace{-10pt}
    \parbox{\LW}{ \begingroup
    \sbox0{\includegraphics{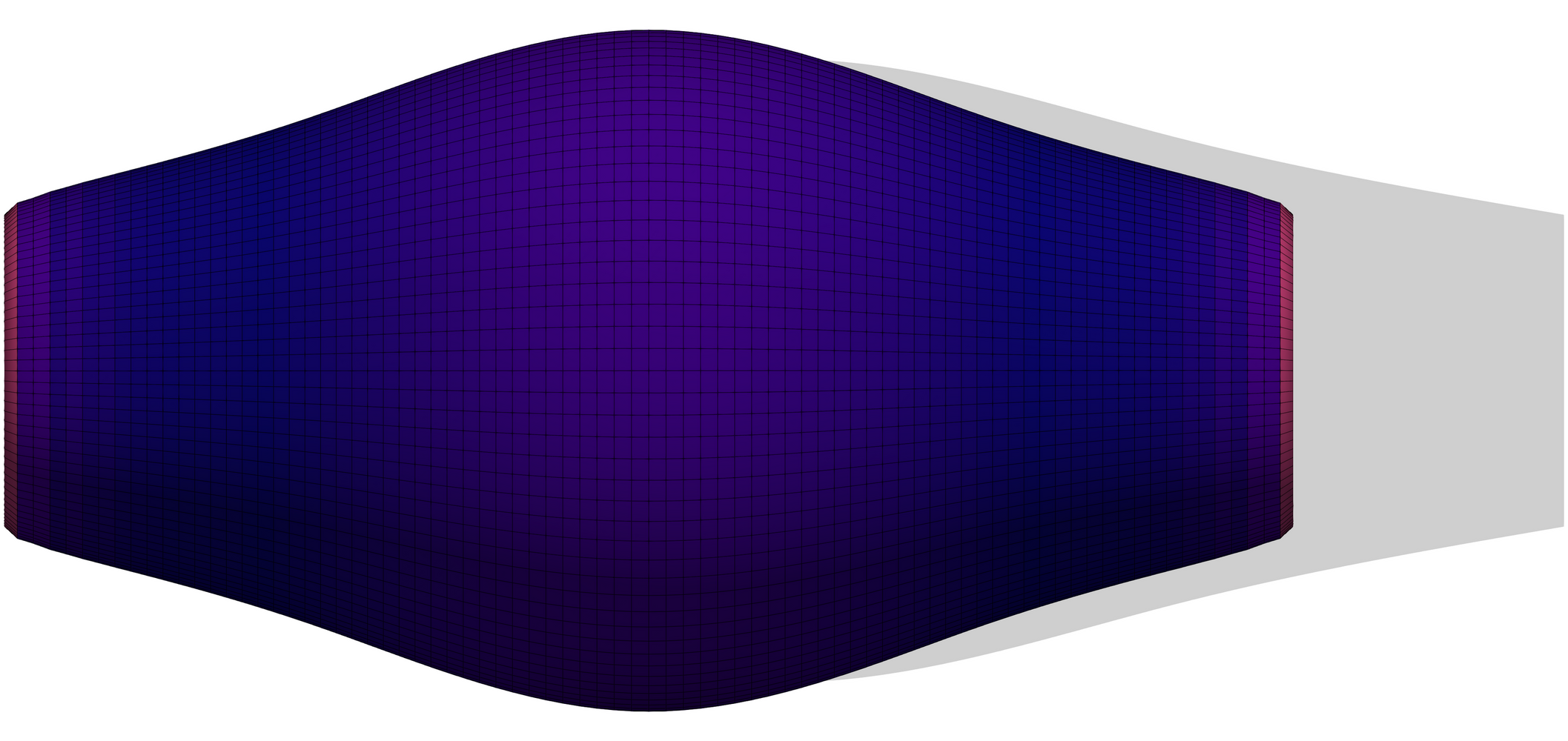}}
    \includegraphics[clip,trim=0 {.5\ht0} 0 0, width=\linewidth,left]{figures/fig_24_solid_von_mises_blemker_n4_fc_element_0005.png}
    \endgroup 
    \\
    \begingroup
    \sbox0{\includegraphics{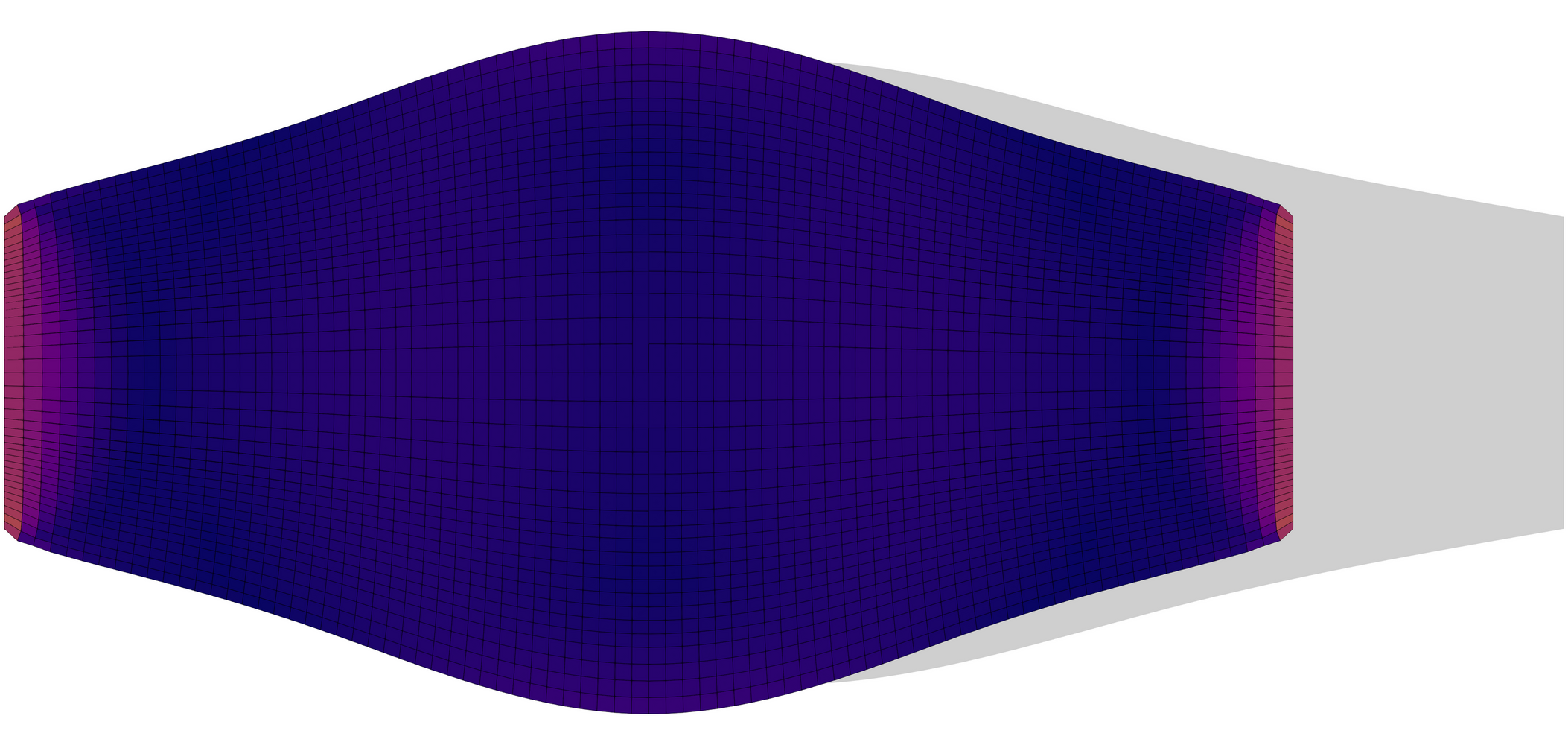}}
    \includegraphics[clip,trim=0 0 0 {.5\ht0}, width=\linewidth,left]{figures/fig_24_slice_zx_von_mises_blemker_n4_fc_element_0005.png}
    \endgroup} \hspace{6pt}
    \parbox{\LW}{ \begingroup
    \sbox0{\includegraphics{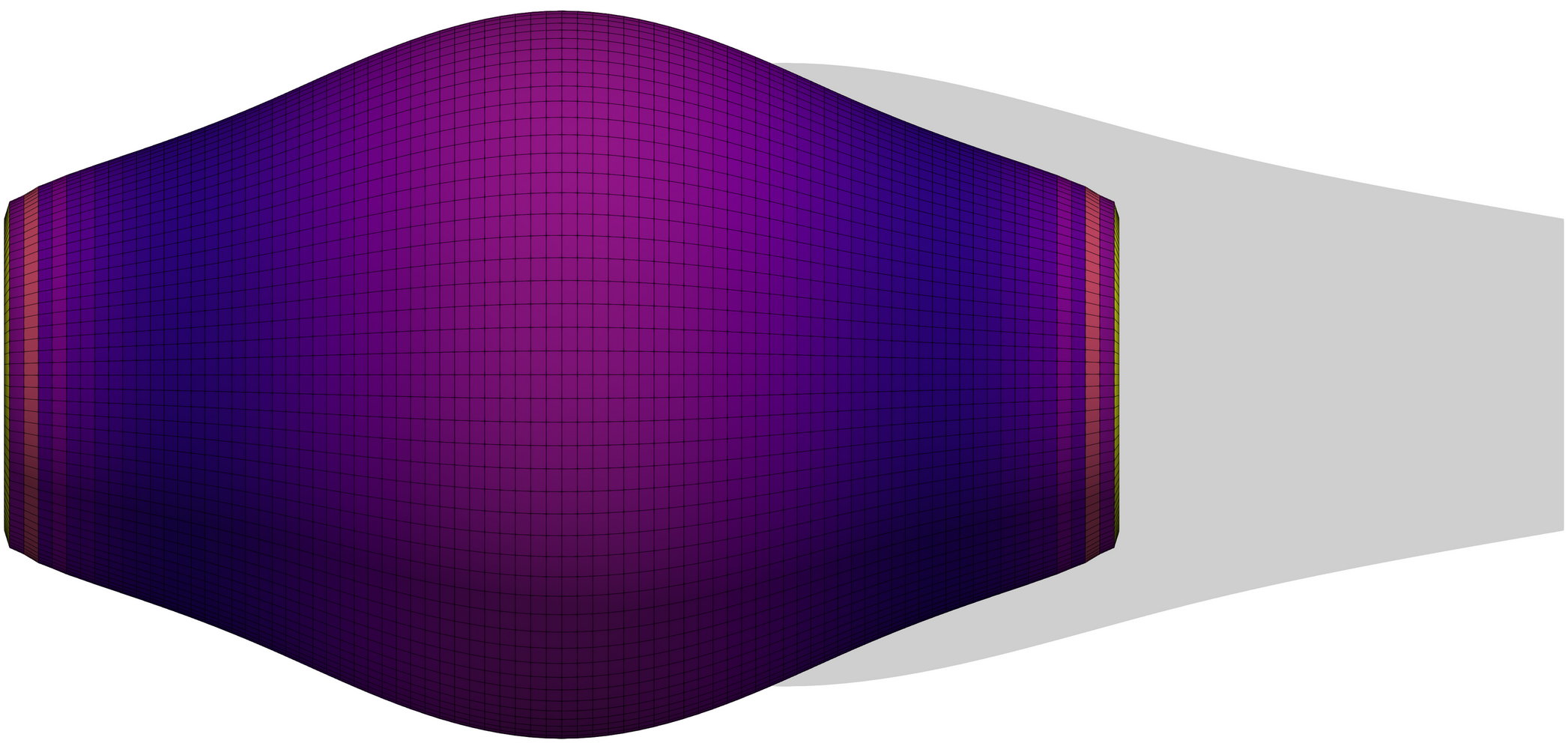}}
    \includegraphics[clip,trim=0 {.5\ht0} 0 0, width=\linewidth,left]{figures/fig_24_solid_von_mises_blemker_n4_fc_element_0020.png}
    \endgroup
    \\
    \begingroup
    \sbox0{\includegraphics{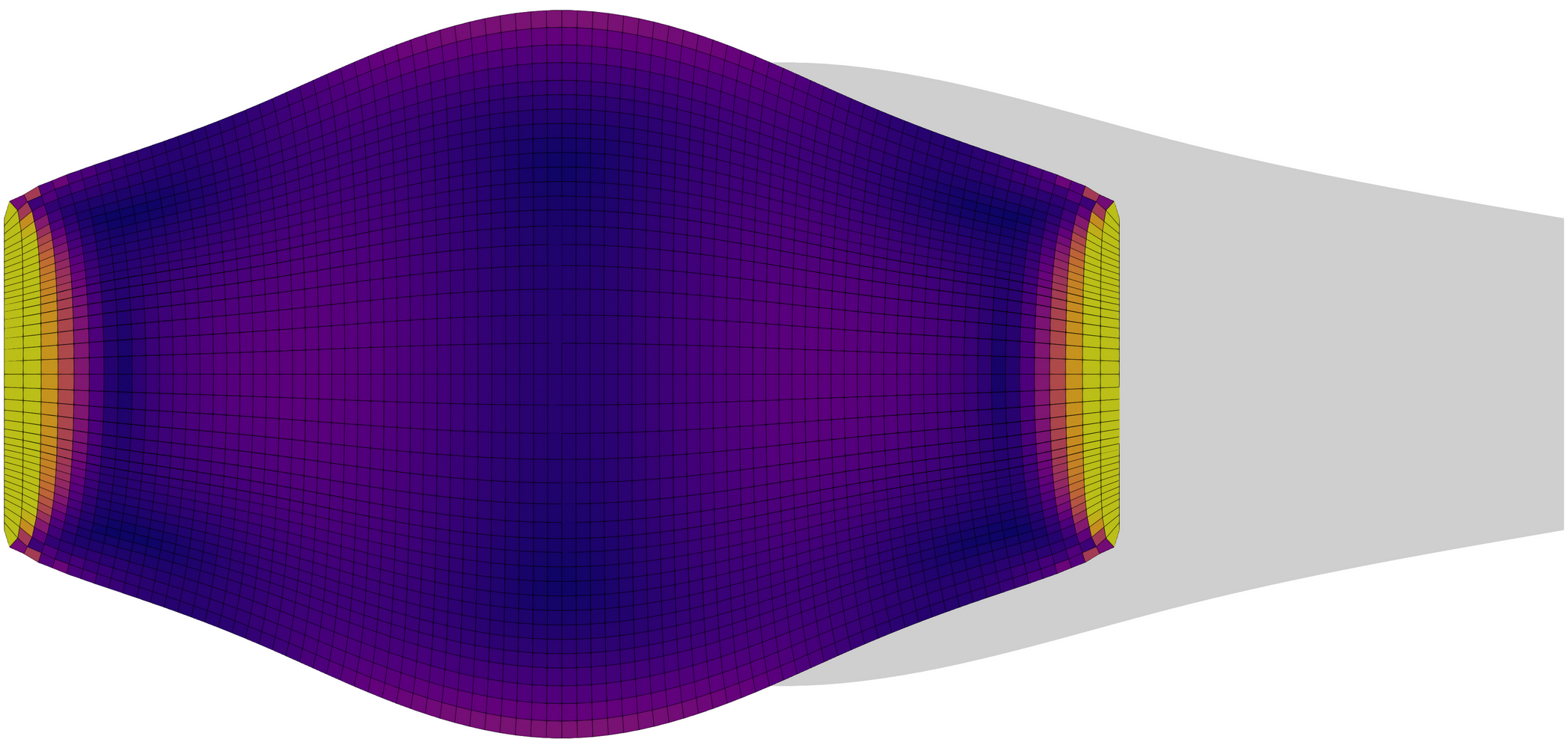}}
    \includegraphics[clip,trim=0 0 0 {.5\ht0}, width=\linewidth,left]{figures/fig_24_slice_zx_von_mises_blemker_n4_fc_element_0020.png}
    \endgroup} \hspace{6pt}
    \parbox{\LW}{ \begingroup
    \sbox0{\includegraphics{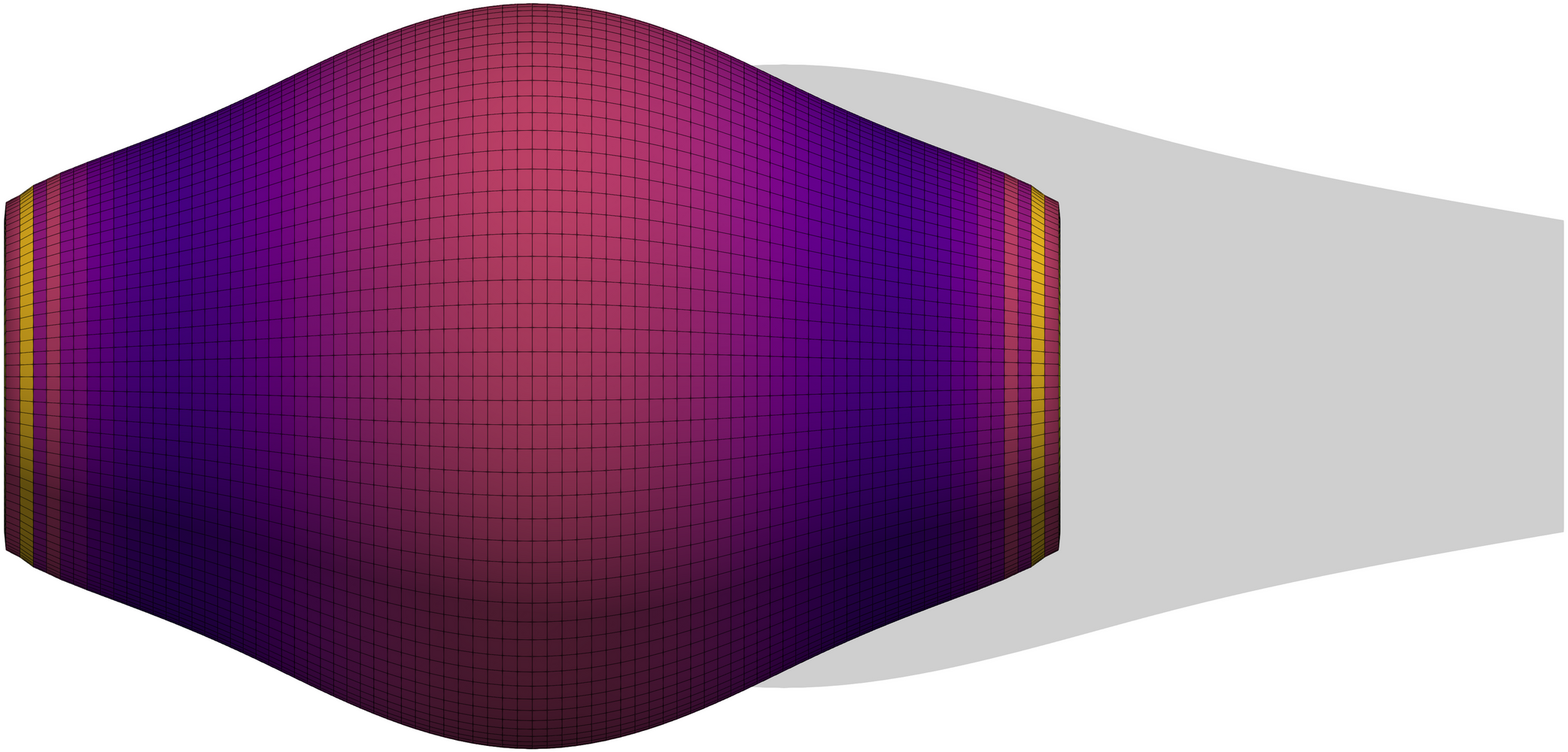}}
    \includegraphics[clip,trim=0 {.5\ht0} 0 0, width=\linewidth,left]{figures/fig_24_solid_von_mises_blemker_n4_fc_element_0150.png}
    \endgroup
    \\
    \begingroup
    \sbox0{\includegraphics{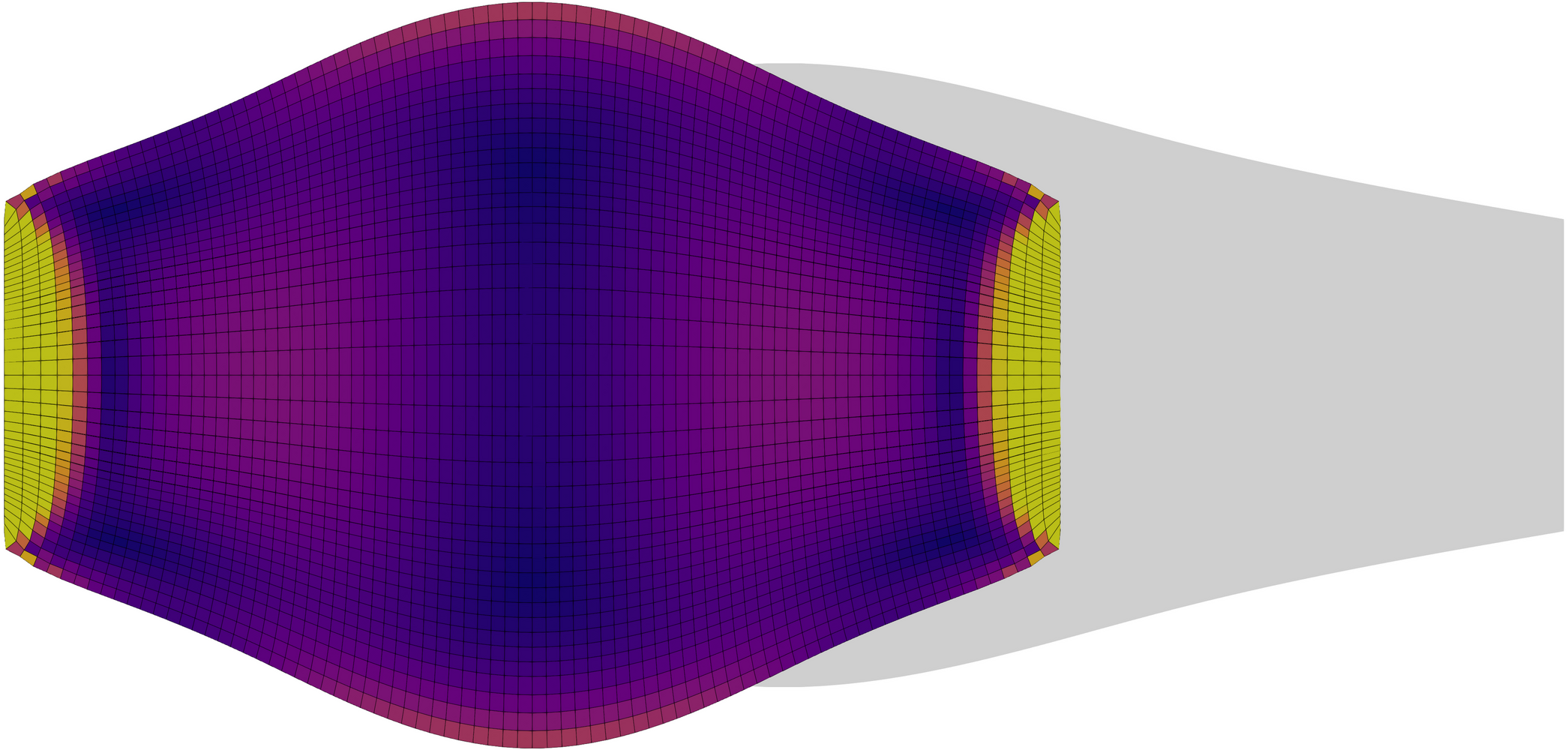}}
    \includegraphics[clip,trim=0 0 0 {.5\ht0}, width=\linewidth,left]{figures/fig_24_slice_zx_von_mises_blemker_n4_fc_element_0150.png}
    \endgroup} \\ 
    \vspace{6pt}
    \parbox{\LWCap}{\subcaption{\raggedright \GIANT}} \hspace{-10pt}
    \parbox{\LW}{ \begingroup
    \sbox0{\includegraphics{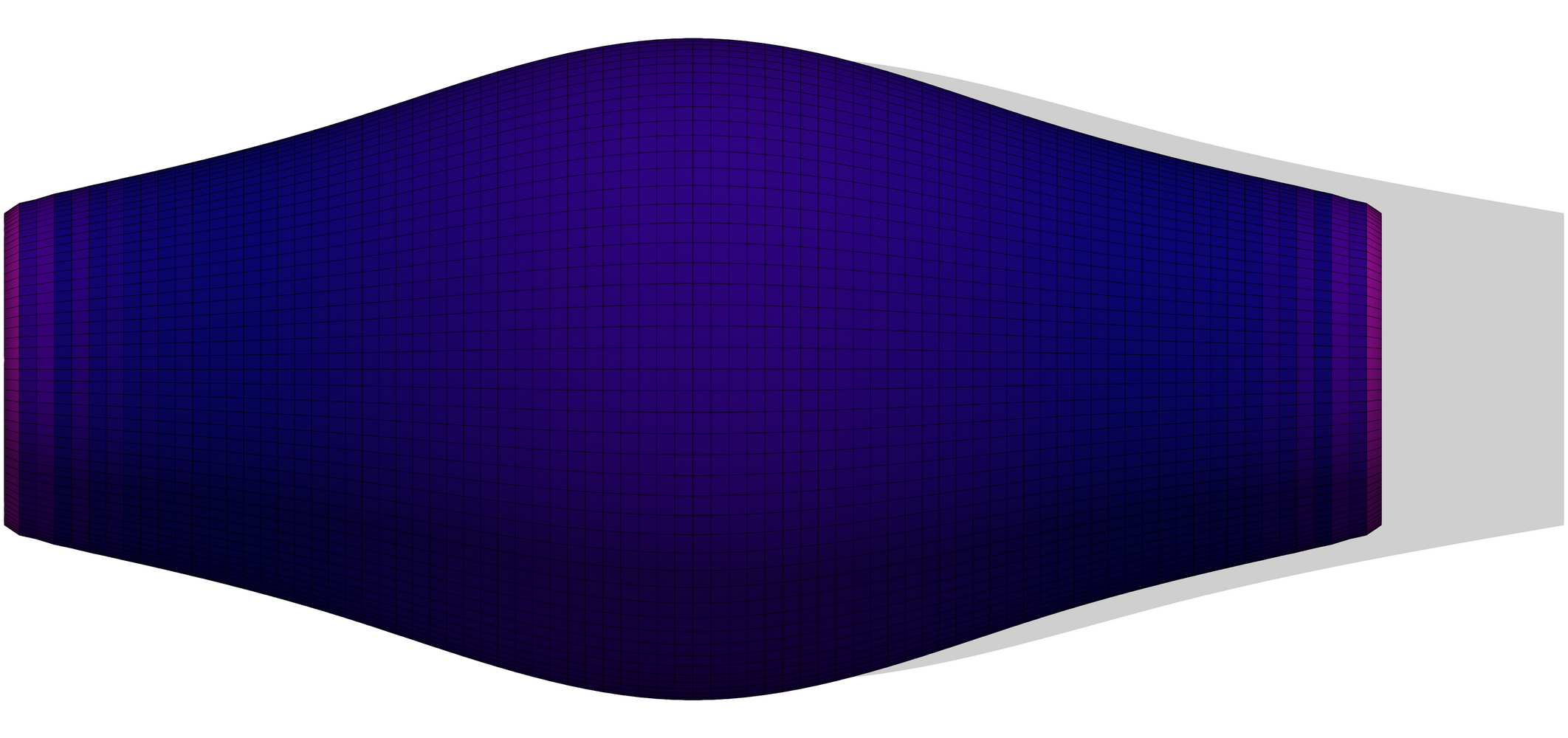}}
    \includegraphics[clip,trim=0 {.5\ht0} 0 0, width=\linewidth,left]{figures/fig_24_solid_von_mises_giantesio_n4_fc_element_0005.png}
    \endgroup 
    \\ 
    \begingroup
    \sbox0{\includegraphics{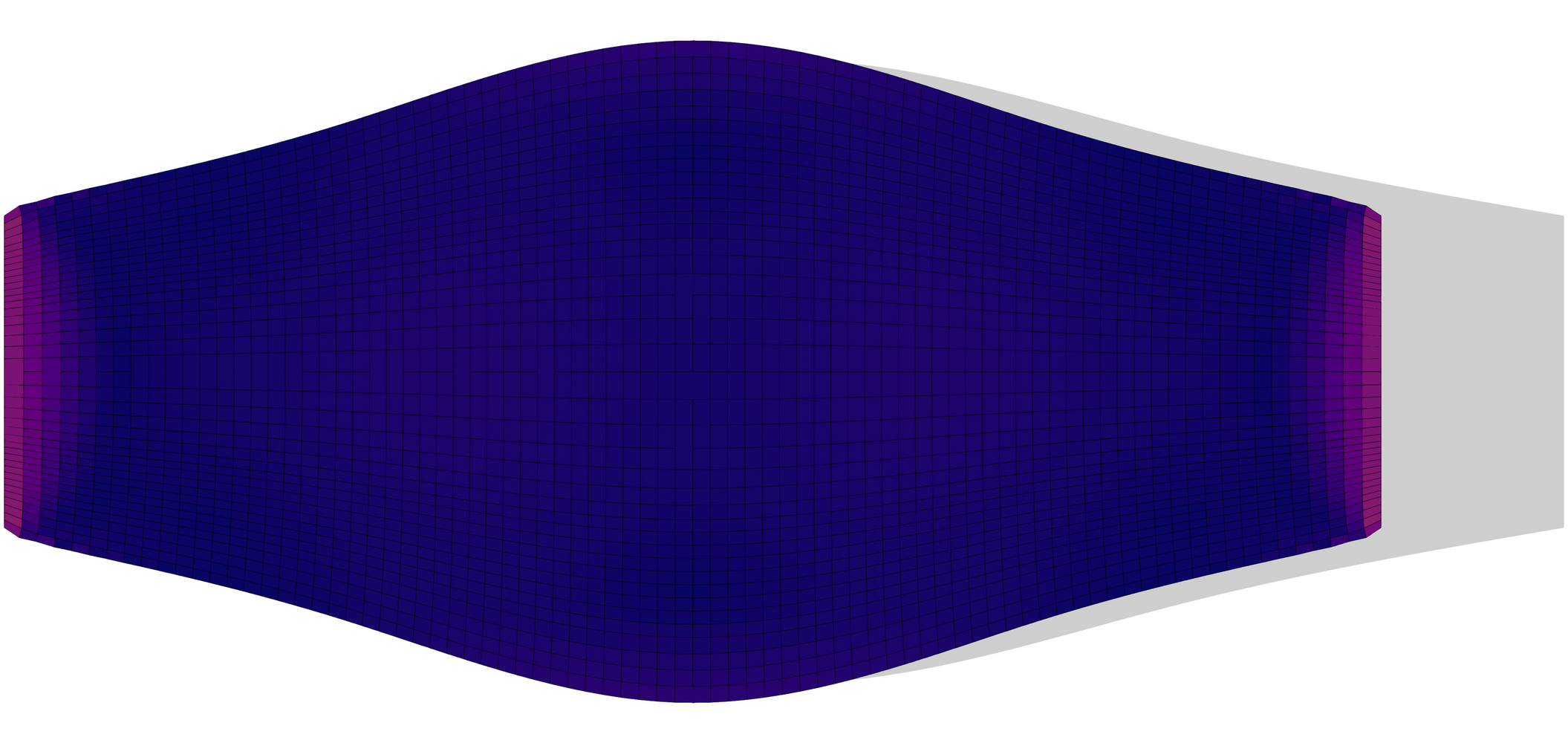}}
    \includegraphics[clip,trim=0 0 0 {.5\ht0}, width=\linewidth,left]{figures/fig_24_slice_zx_von_mises_giantesio_n4_fc_element_0005.png}
    \endgroup} \hspace{6pt}
    \parbox{\LW}{ \begingroup
    \sbox0{\includegraphics{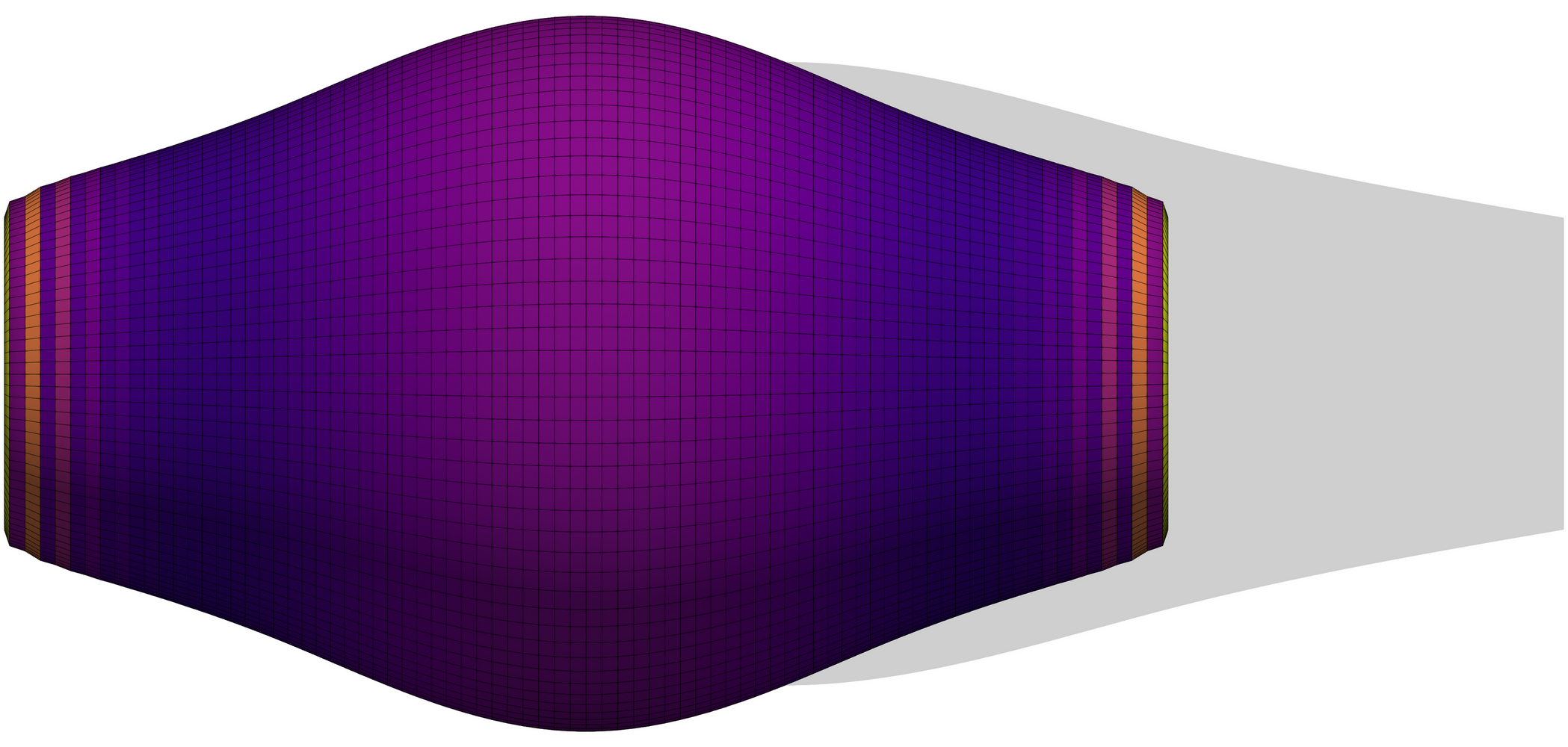}}
    \includegraphics[clip,trim=0 {.5\ht0} 0 0, width=\linewidth,left]{figures/fig_24_solid_von_mises_giantesio_n4_fc_element_0020.png}
    \endgroup
    \\ 
    \begingroup
    \sbox0{\includegraphics{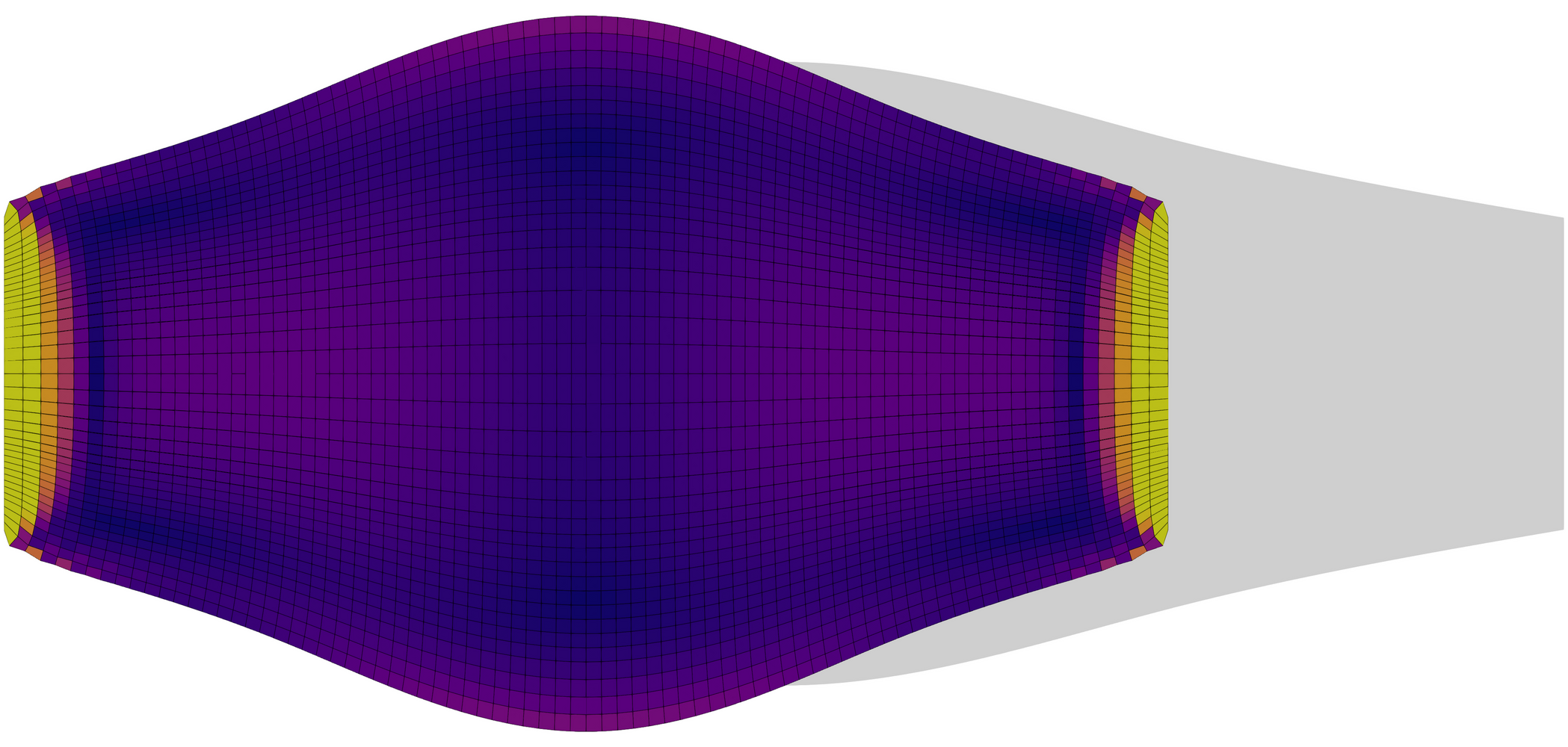}}
    \includegraphics[clip,trim=0 0 0 {.5\ht0}, width=\linewidth,left]{figures/fig_24_slice_zx_von_mises_giantesio_n4_fc_element_0020.png}
    \endgroup} \hspace{6pt}
    \parbox{\LW}{ \begingroup
    \sbox0{\includegraphics{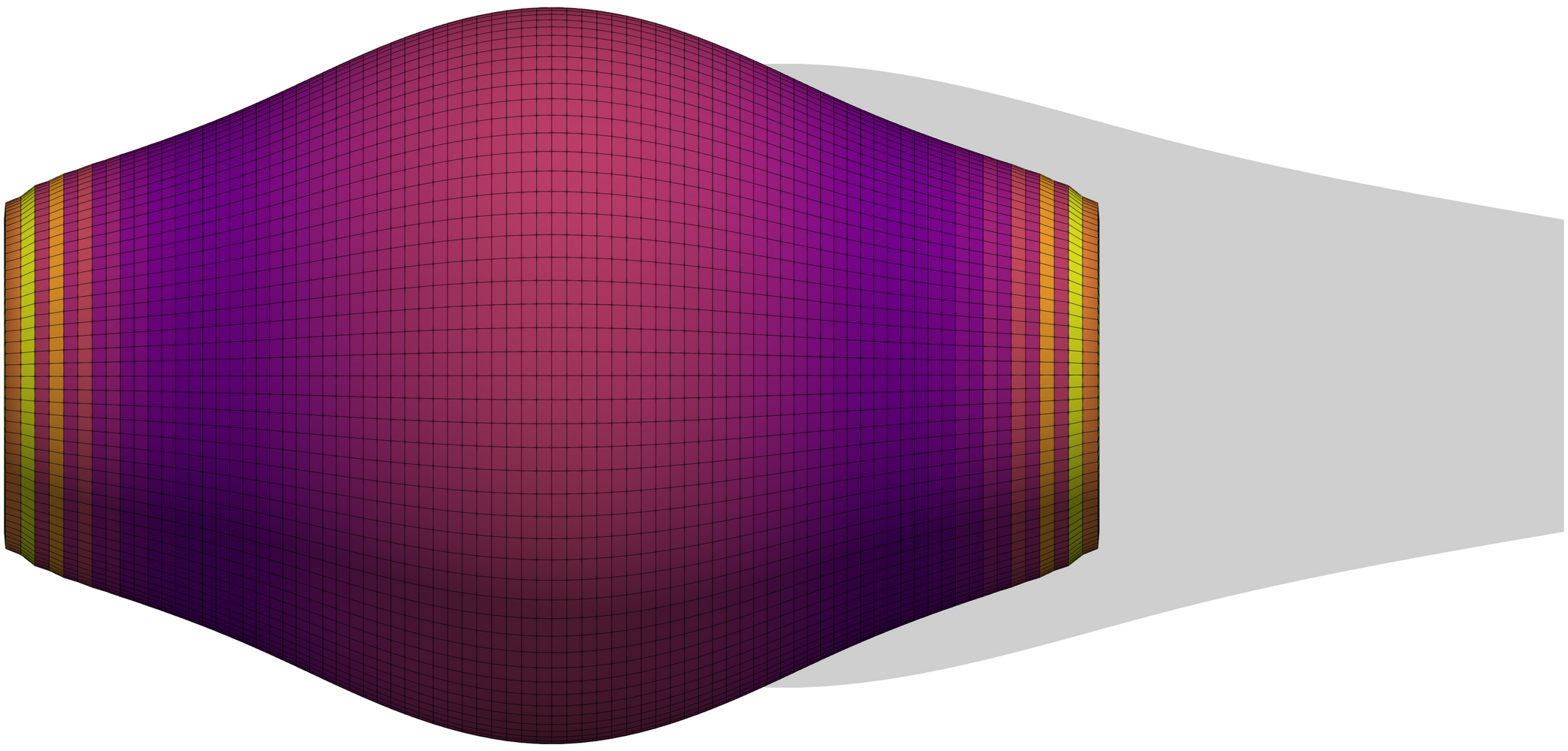}}
    \includegraphics[clip,trim=0 {.5\ht0} 0 0, width=\linewidth,left]{figures/fig_24_solid_von_mises_giantesio_n4_fc_element_0150.png}
    \endgroup
    \\ 
    \begingroup
    \sbox0{\includegraphics{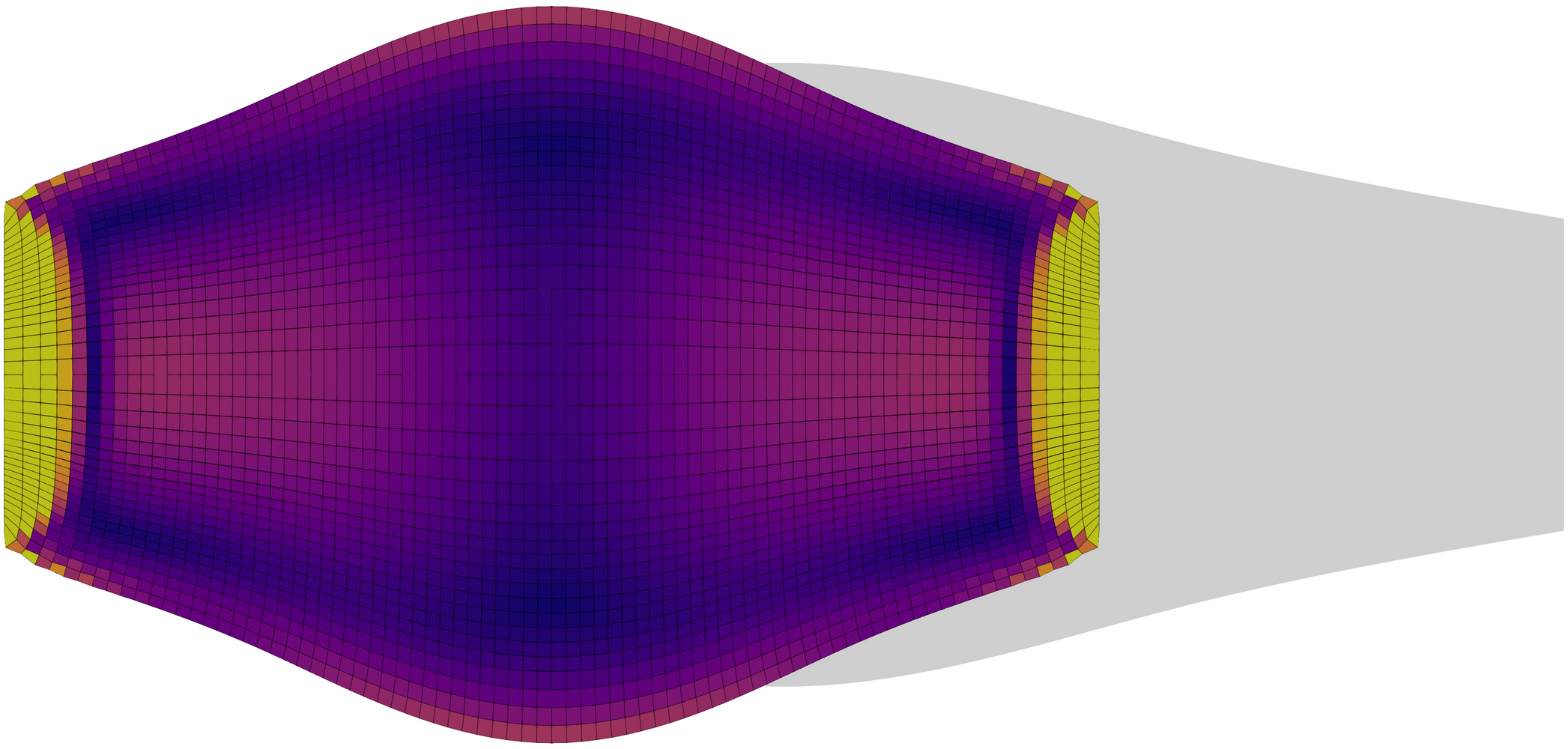}}
    \includegraphics[clip,trim=0 0 0 {.5\ht0}, width=\linewidth,left]{figures/fig_24_slice_zx_von_mises_giantesio_n4_fc_element_0150.png}
    \endgroup} \\
    \vspace{6pt}
    \parbox{\LWCap}{\subcaption{\raggedright \WKM}} \hspace{-10pt}
    \parbox{\LW}{ \begingroup
    \sbox0{\includegraphics{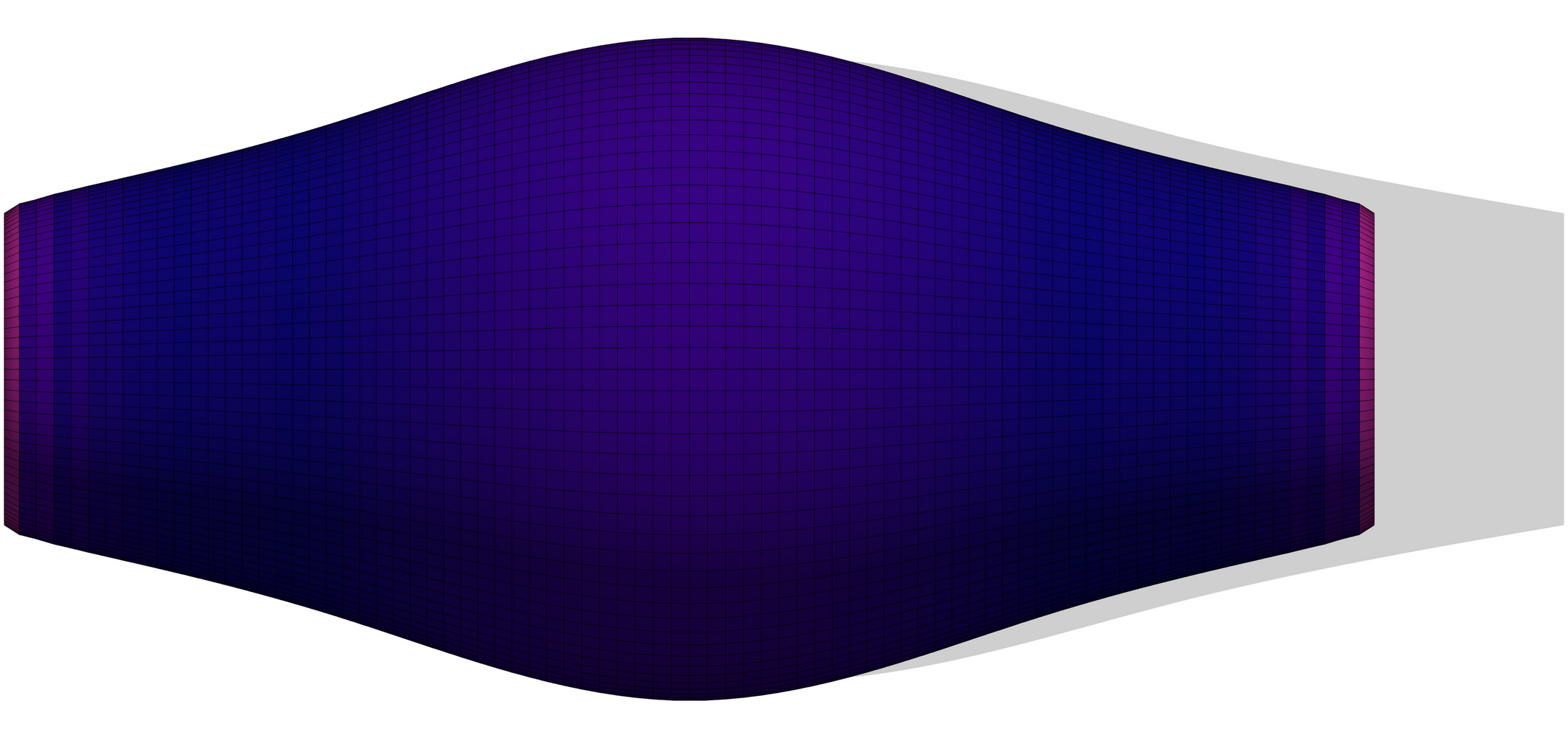}}
    \includegraphics[clip,trim=0 {.5\ht0} 0 0, width=\linewidth,left]{figures/fig_24_solid_von_mises_weickenmeier_n4_fc_element_0005.png}
    \endgroup 
    \\
    \begingroup
    \sbox0{\includegraphics{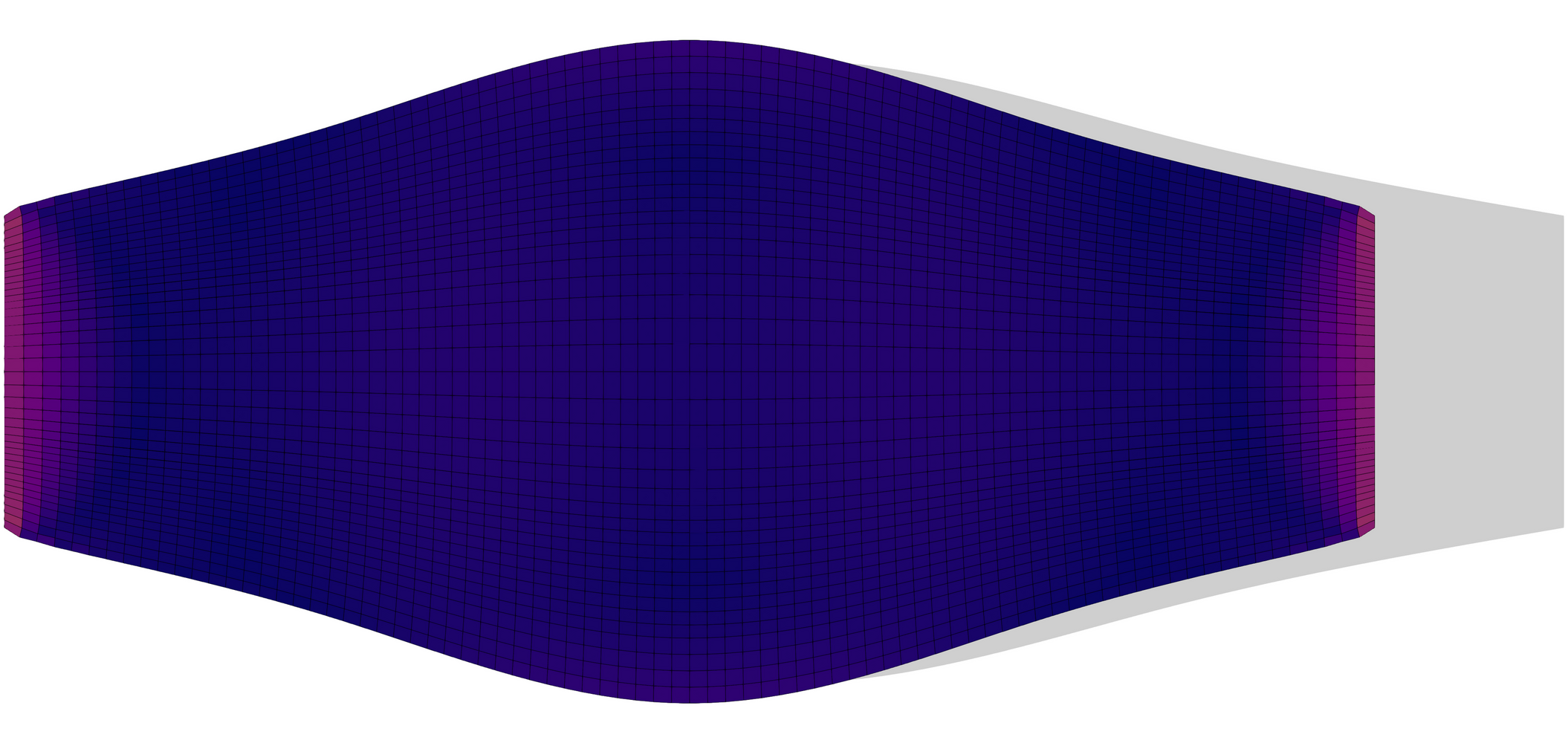}}
    \includegraphics[clip,trim=0 0 0 {.5\ht0}, width=\linewidth,left]{figures/fig_24_slice_zx_von_mises_weickenmeier_n4_fc_element_0005.png}
    \endgroup} \hspace{6pt}
    \parbox{\LW}{ \begingroup
    \sbox0{\includegraphics{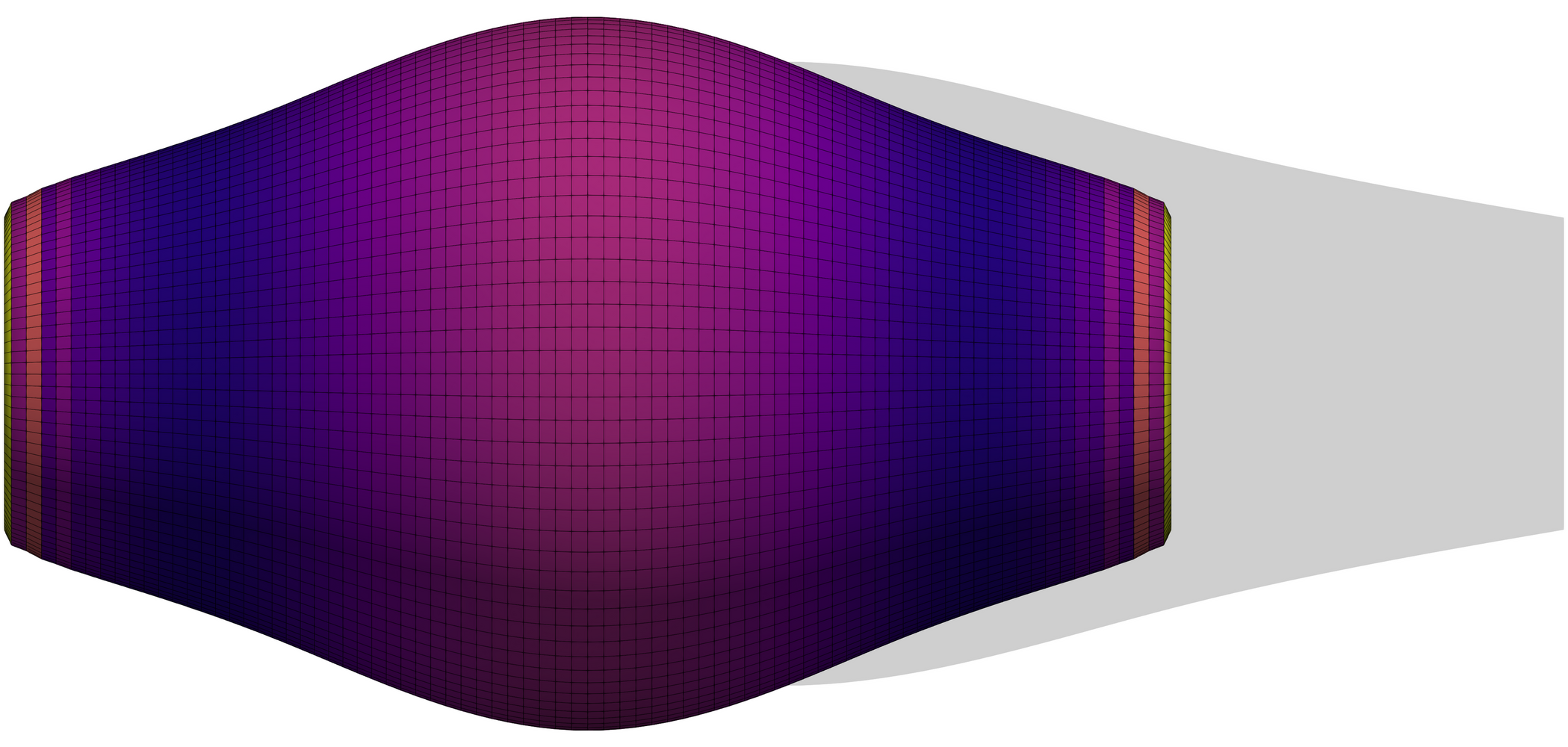}}
    \includegraphics[clip,trim=0 {.5\ht0} 0 0, width=\linewidth,left]{figures/fig_24_solid_von_mises_weickenmeier_n4_fc_element_0020.png}
    \endgroup
    \\
    \begingroup
    \sbox0{\includegraphics{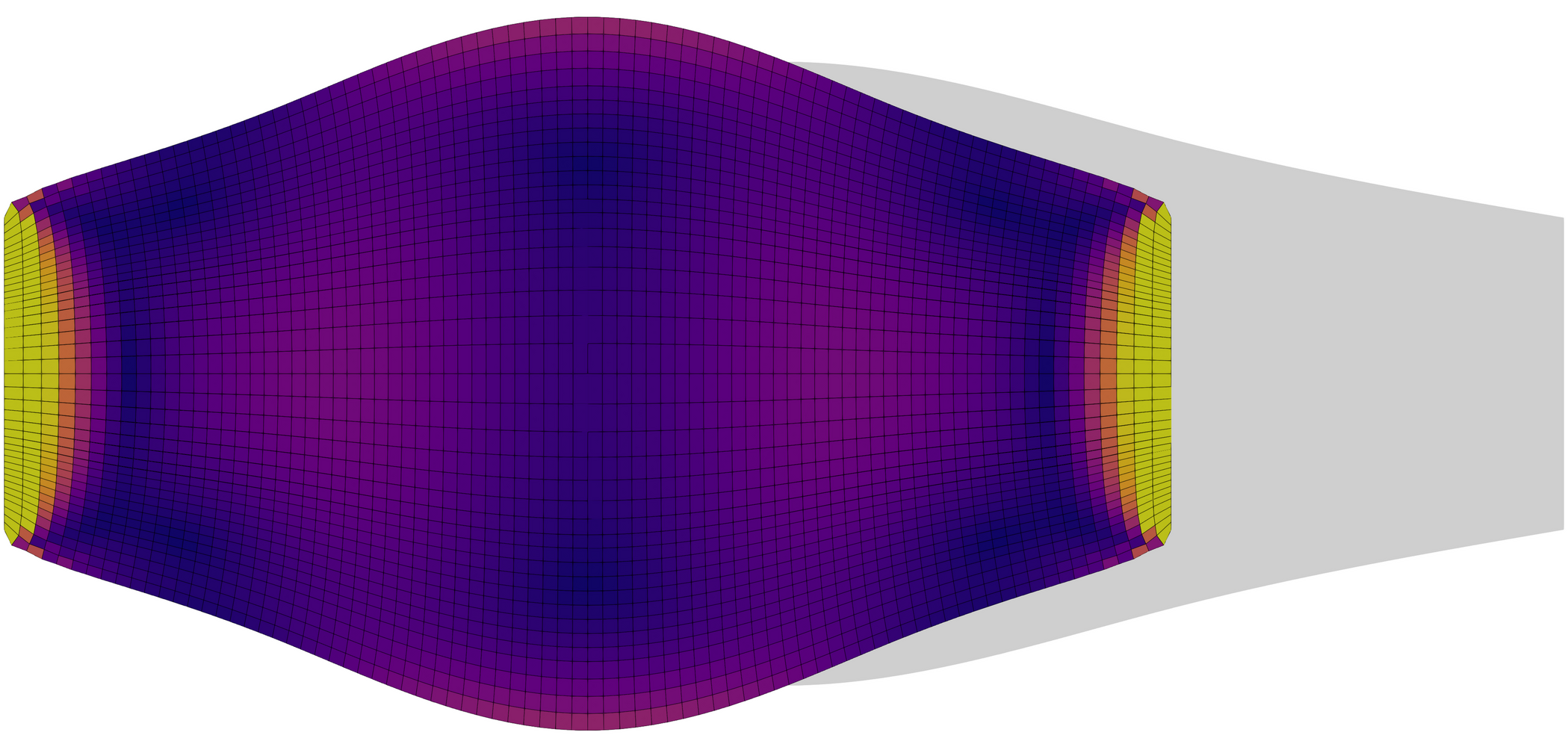}}
    \includegraphics[clip,trim=0 0 0 {.5\ht0}, width=\linewidth,left]{figures/fig_24_slice_zx_von_mises_weickenmeier_n4_fc_element_0020.png}
    \endgroup} \hspace{6pt}
    \parbox{\LW}{ \begingroup
    \sbox0{\includegraphics{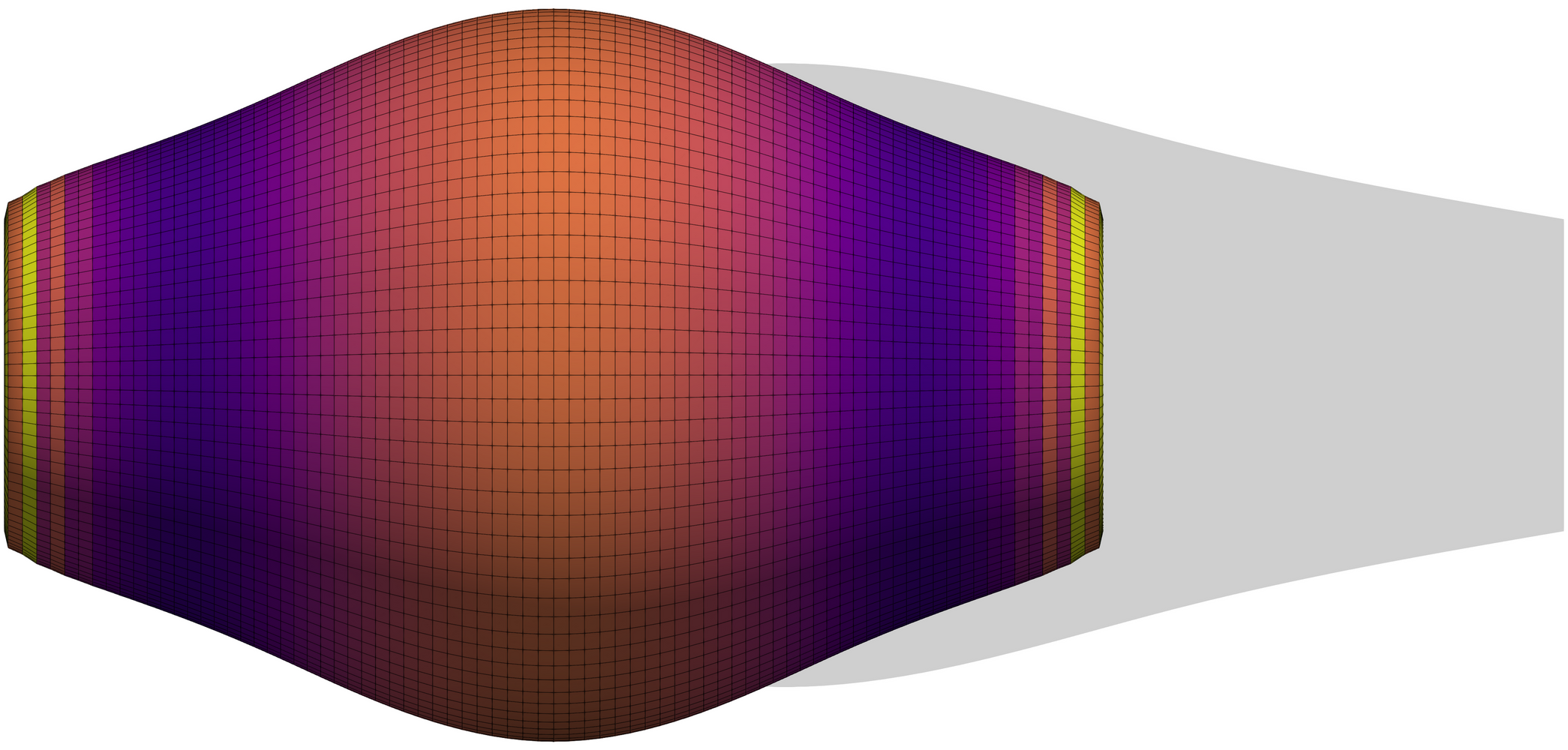}}
    \includegraphics[clip,trim=0 {.5\ht0} 0 0, width=\linewidth,left]{figures/fig_24_solid_von_mises_weickenmeier_n4_fc_element_0150.png}
    \endgroup
    \\
    \begingroup
    \sbox0{\includegraphics{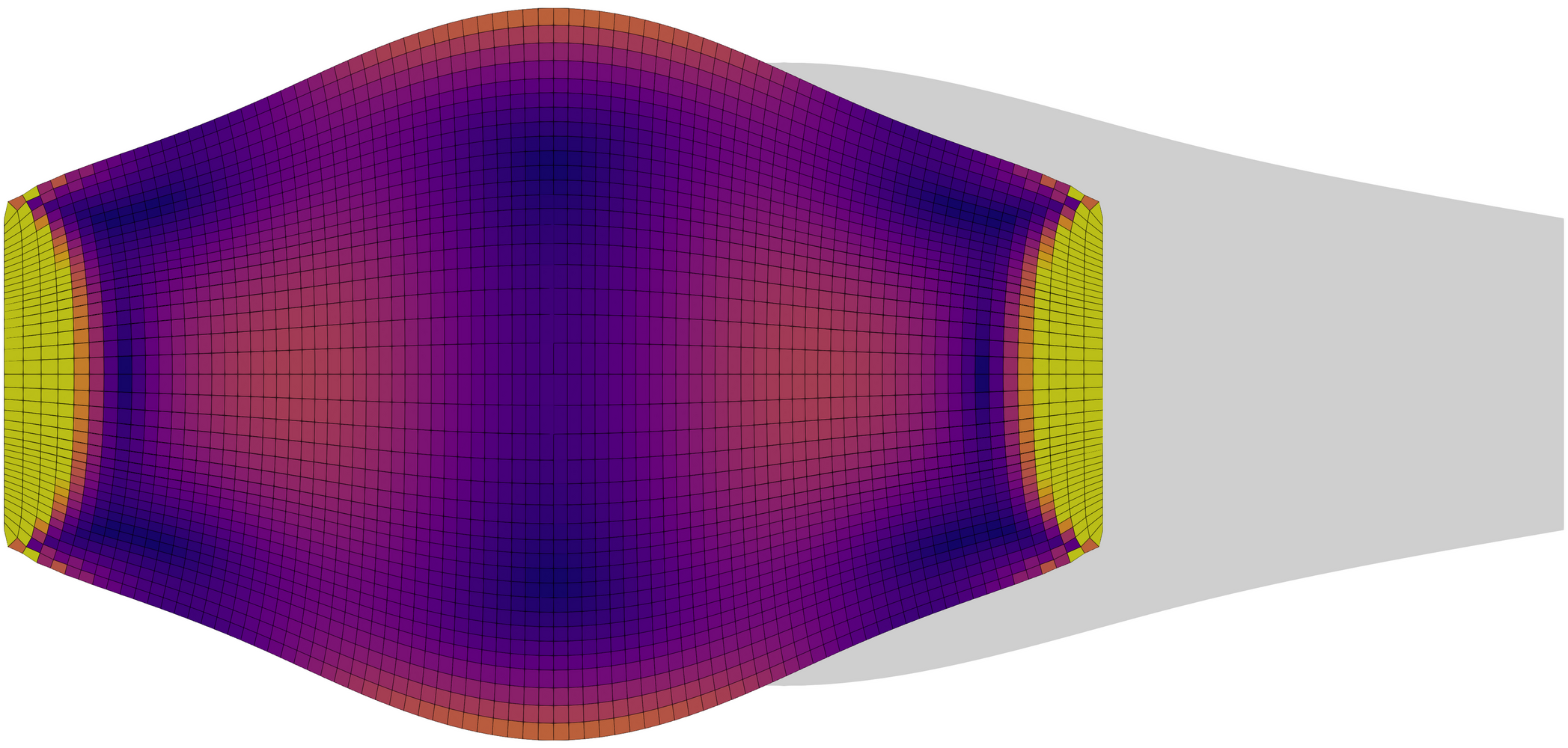}}
    \includegraphics[clip,trim=0 0 0 {.5\ht0}, width=\linewidth,left]{figures/fig_24_slice_zx_von_mises_weickenmeier_n4_fc_element_0150.png}
    \endgroup}\\
    \vspace{6pt}
    \parbox{\LWCap}{\subcaption{\raggedright \COMBI}} \hspace{-10pt}
    \parbox{\LW}{ \begingroup
    \sbox0{\includegraphics{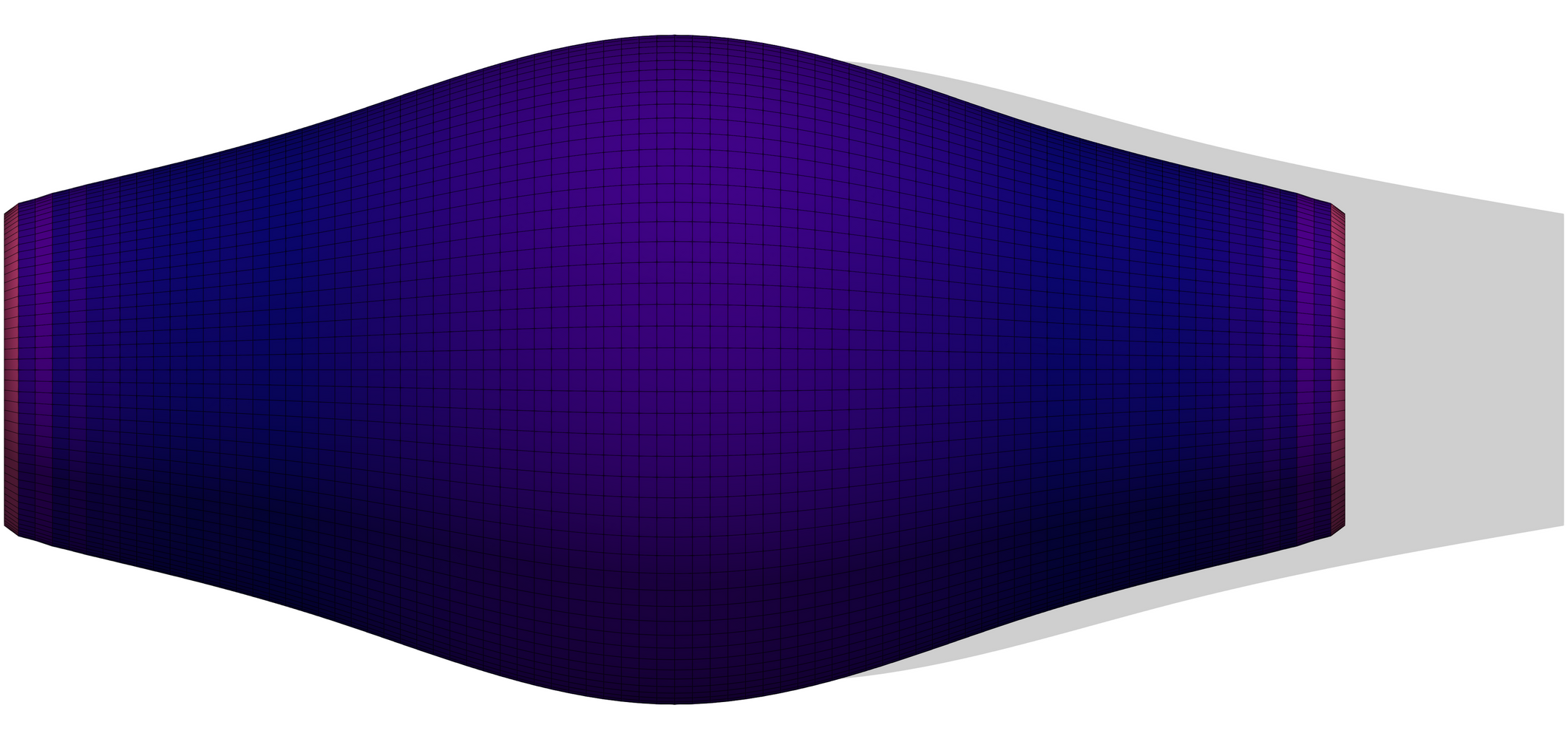}}
    \includegraphics[clip,trim=0 {.5\ht0} 0 0, width=\linewidth,left]{figures/fig_24_solid_von_mises_combi_n4_fc_element_0005.png}
    \endgroup 
    \\
    \begingroup
    \sbox0{\includegraphics{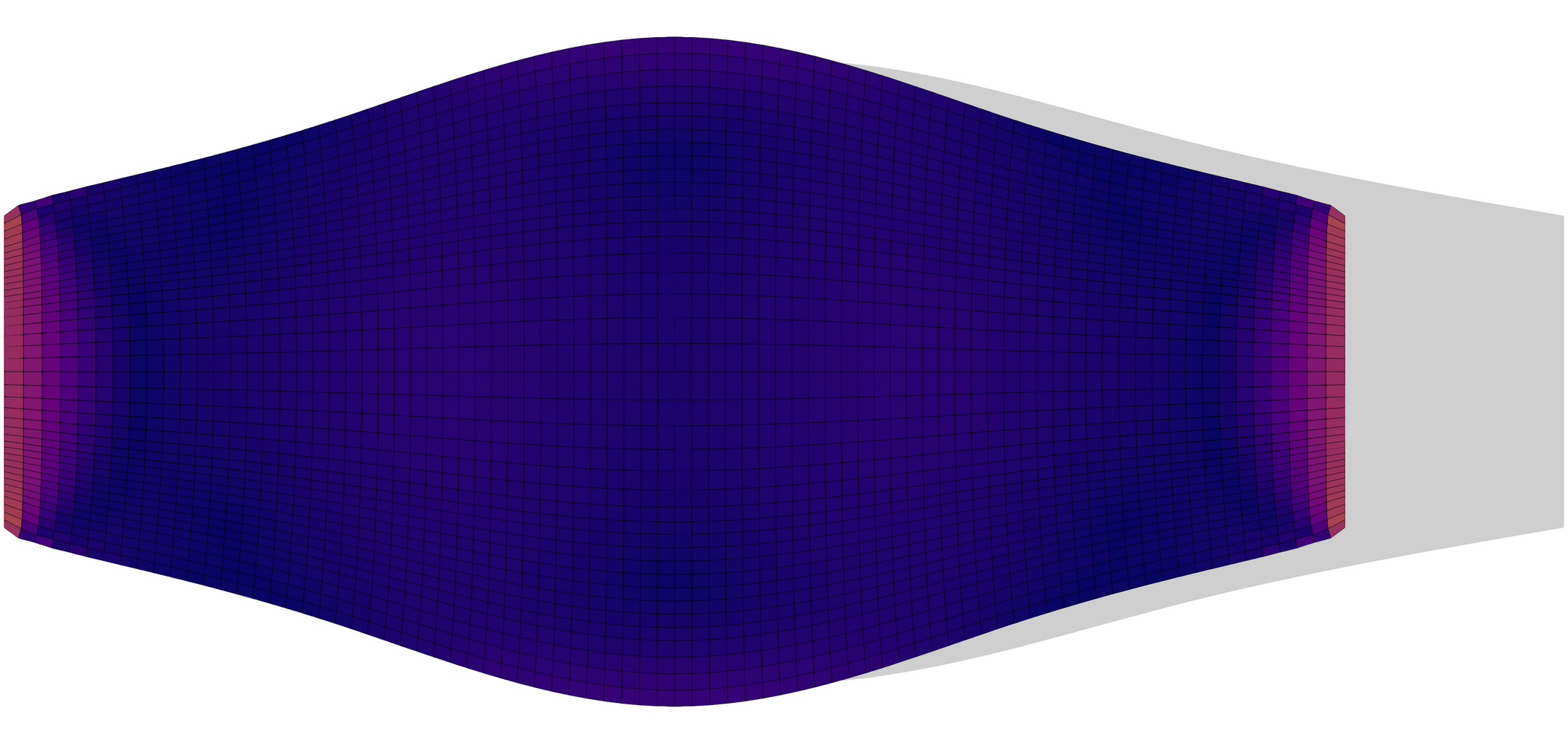}}
    \includegraphics[clip,trim=0 0 0 {.5\ht0}, width=\linewidth,left]{figures/fig_24_slice_zx_von_mises_combi_n4_fc_element_0005.png}
    \endgroup} \hspace{6pt}
    \parbox{\LW}{ \begingroup
    \sbox0{\includegraphics{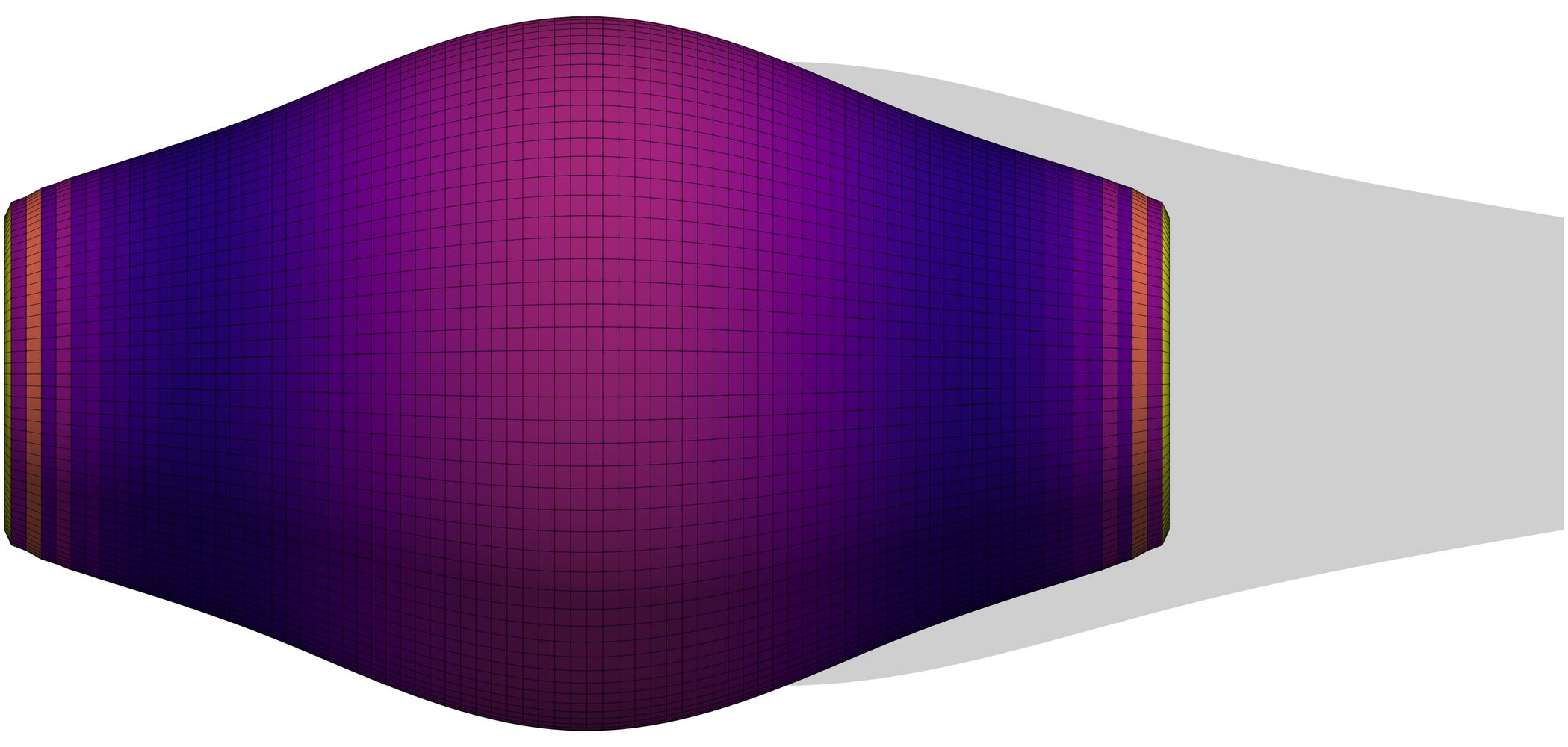}}
    \includegraphics[clip,trim=0 {.5\ht0} 0 0, width=\linewidth,left]{figures/fig_24_solid_von_mises_combi_n4_fc_element_0020.png}
    \endgroup
    \\
    \begingroup
    \sbox0{\includegraphics{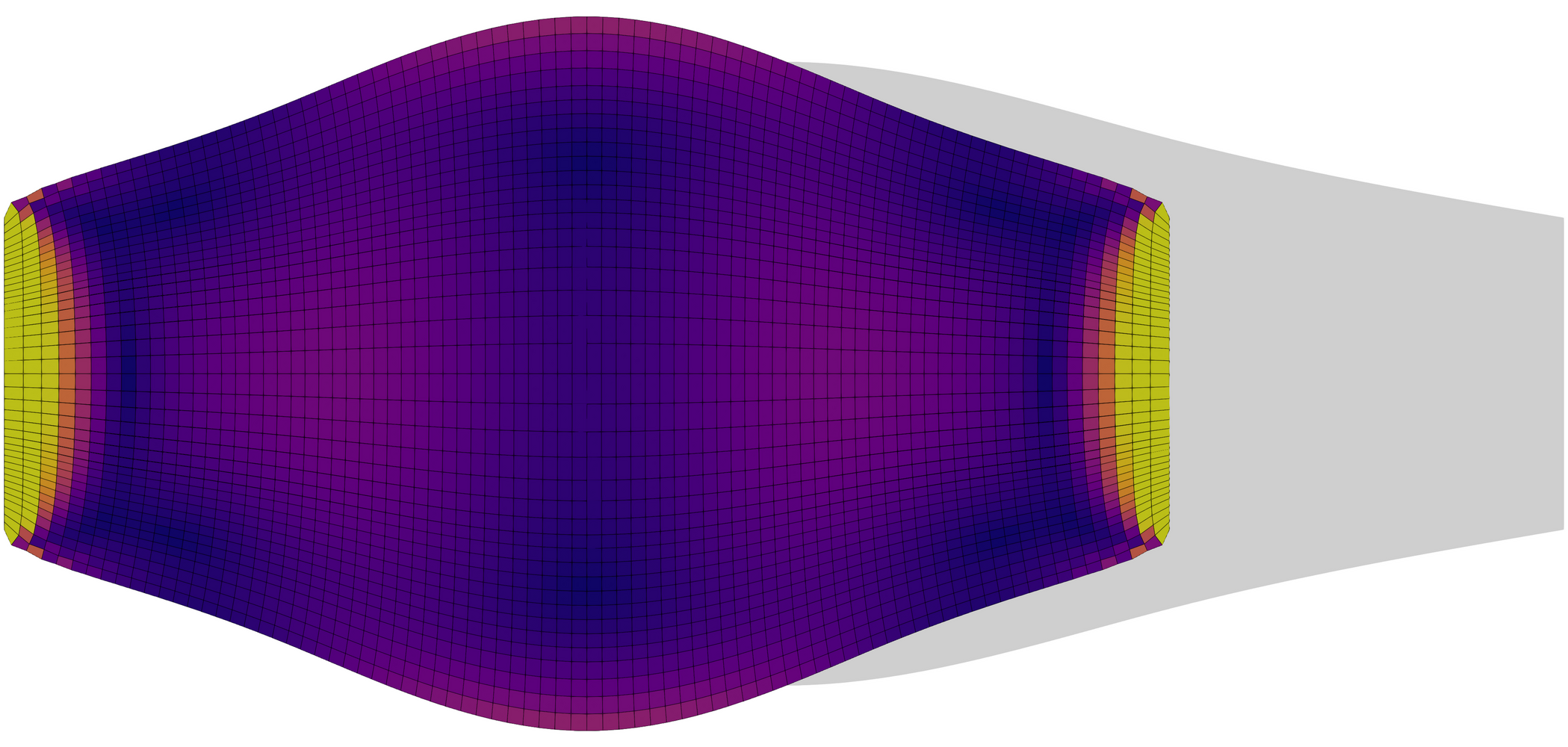}}
    \includegraphics[clip,trim=0 0 0 {.5\ht0}, width=\linewidth,left]{figures/fig_24_slice_zx_von_mises_combi_n4_fc_element_0020.png}
    \endgroup} \hspace{6pt}
    \parbox{\LW}{ \begingroup
    \sbox0{\includegraphics{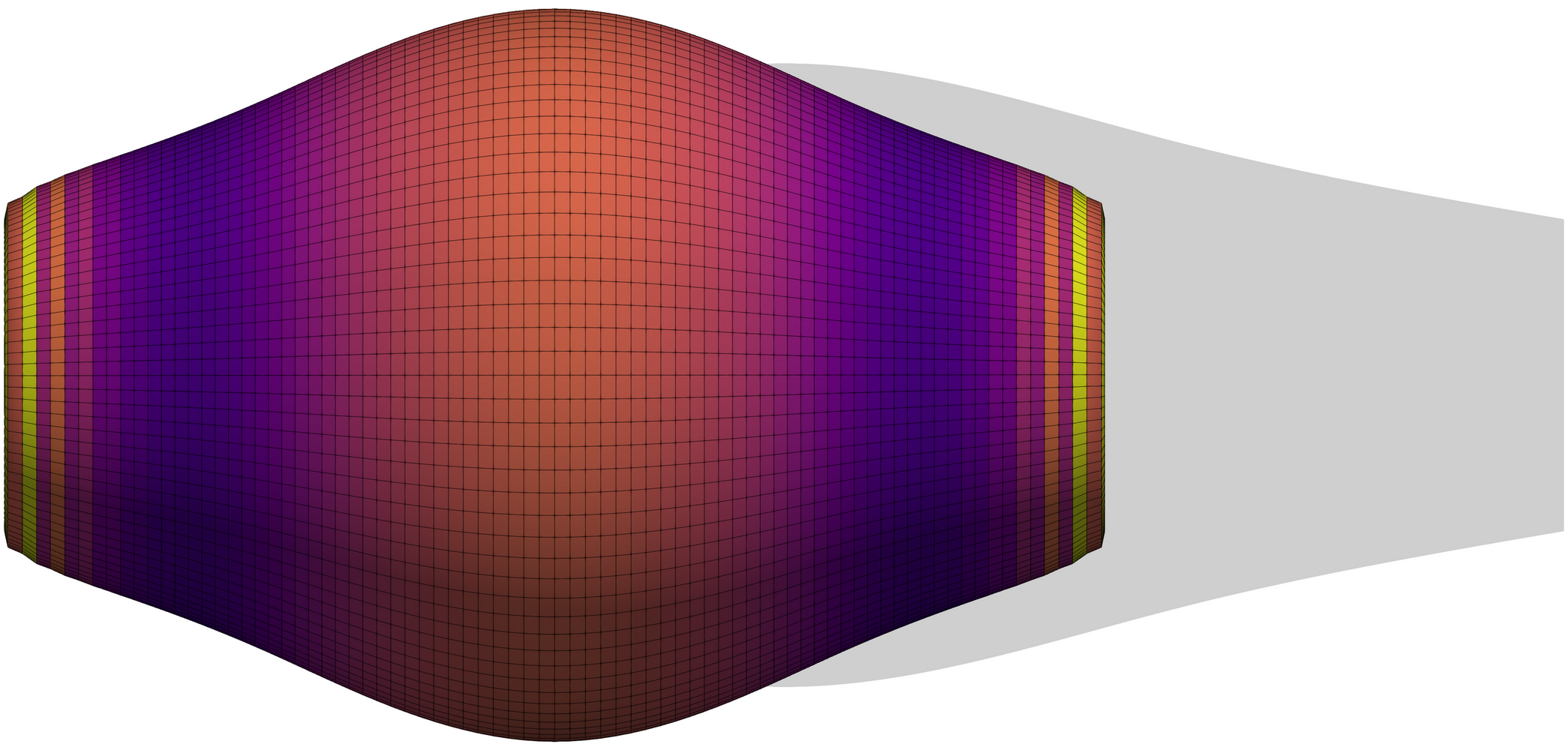}}
    \includegraphics[clip,trim=0 {.5\ht0} 0 0, width=\linewidth,left]{figures/fig_24_solid_von_mises_combi_n4_fc_element_0150.png}
    \endgroup
    \\
    \begingroup
    \sbox0{\includegraphics{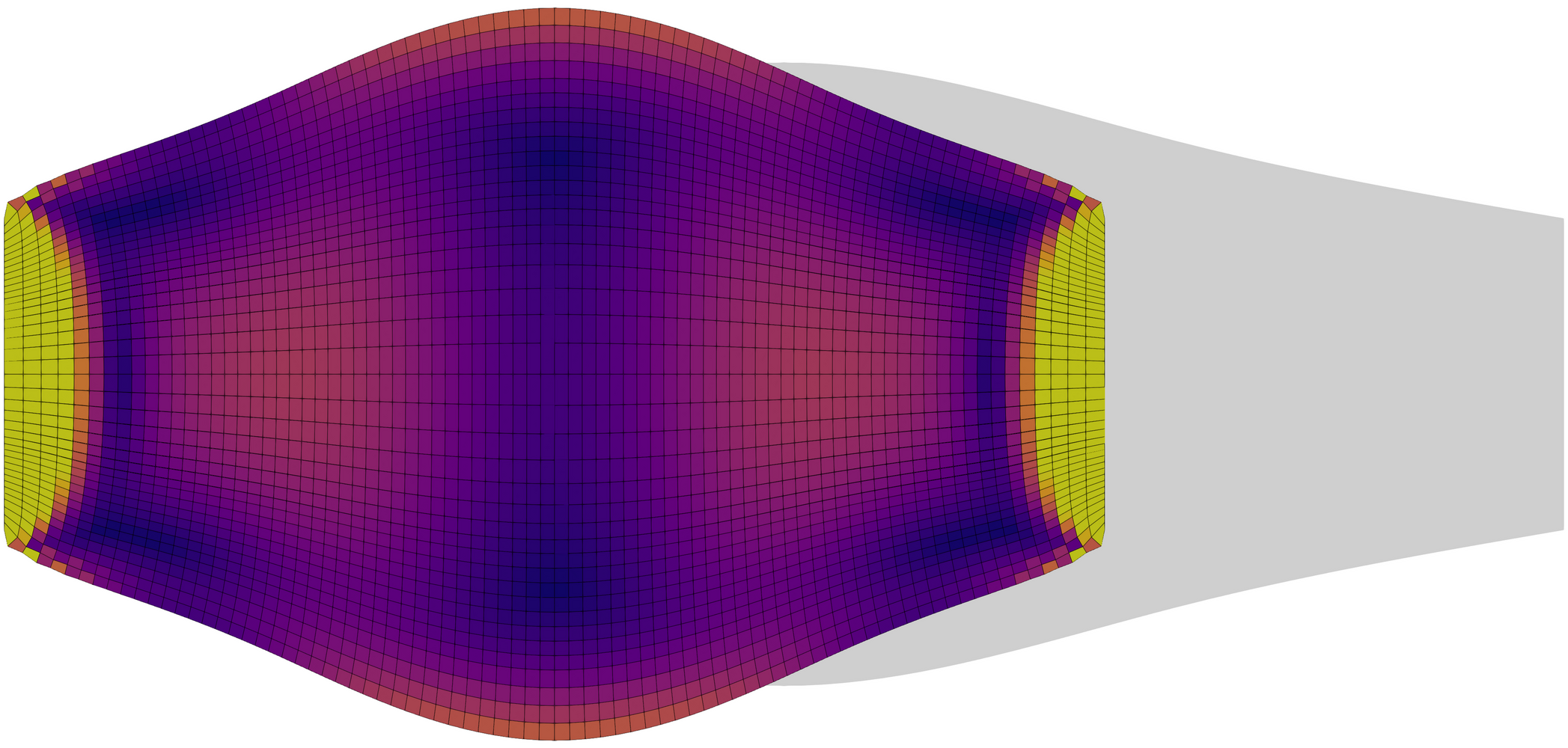}}
    \includegraphics[clip,trim=0 0 0 {.5\ht0}, width=\linewidth,left]{figures/fig_24_slice_zx_von_mises_combi_n4_fc_element_0150.png}
    \endgroup}
\end{minipage}\hfill
\begin{minipage}[c]{1.29cm}
\raggedleft
    \includegraphics[height=3cm]{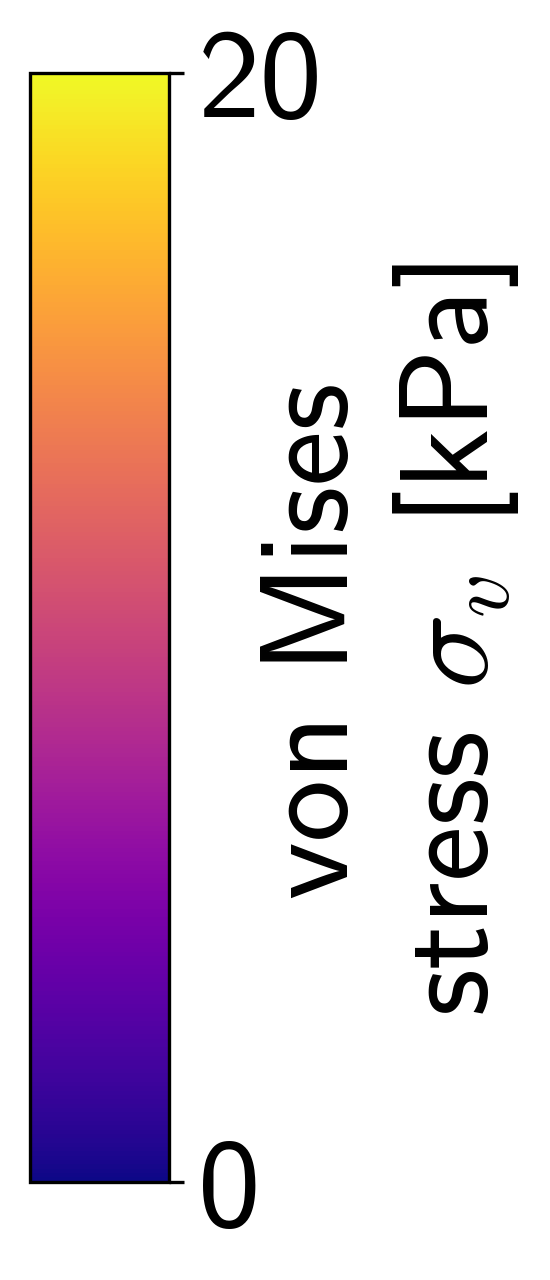}
\end{minipage}
    \caption{Von Mises stress $\sigma_\mathrm{v}$ for a free contraction of the fusiform muscle ($n=4$) at selected times. \mbox{Results are} visualized on the surface (top) and in the axial cross-section (bottom) in comparison to the initial \mbox{configuration (grey).}}
	\label{fig:fusi_free_mises_stress_vis}
\end{figure}

\FloatBarrier

\subsection{Glenohumeral joint concavity compression simulation}
\subsubsection*{Computation of humerus mass density} \label{sec:appendix_bone_density}
In our shoulder model, the approximate humerus length is $\SI{33}{\cm}$. According to \cite{haque_correlation_2024}, humerus length is positively correlated to its weight. Considering a mean length of $\SI{31}{\cm}$ and a mean dry weight of $\SI{108}{\g}$ for males \cite{ahmed_sex_2018}, we assume the model humerus dry weight to be $\SI{115}{\g}$. With an approximate dry-wet weight ratio of $0.65$ \cite{huang_effects_2003, chang_study_2018}, the model humerus wet weight (the bone weight including all organic and inorganic material) is $\SI{177}{\g}$. For a measured volume of $\SI{232}{\cm^3}$, we thus compute a mass density of $\SI{0.76}{\g\per\cm^3}$.

\subsubsection*{Definition of boundary conditions}
\bgroup
\setlength\tabcolsep{0.5em}
\begin{table}[htb]
\centering
\caption{Definition of surface and volume nodesets ids for the assignment of boundary conditions.}
\begin{tabular}[b]{@{}cccc@{}}
\toprule
\begin{tabular}[t]{lc}
\multicolumn{2}{c}{Humerus}\\
\midrule
Humeral head & 8 \\
Deltoid ins. & 5 \\
Teres minor ins. & 1 \\
Infraspinatus ins. & 4 \\
Supraspinatus ins. & 3 \\
Subscapularis ins. & 2 \\
Deltoid contact & 6 \\
RC contact & 7 \\
\end{tabular} &
\begin{tabular}[t]{L{3.05cm}c}
\multicolumn{2}{c}{Scapula with labrum}\\
\midrule
Inner volume & 34 \\
Glenoid fossa & 14 \\
Deltoid orig. (spinal, acromial) & 13 \\
Teres minor orig. & 9 \\
Infraspinatus orig. & 12 \\
Supraspinatus orig. & 11 \\
Subscapularis orig. & 10 \\
Deltoid contact & 16 \\
RC contact & 15 \\
\end{tabular} &
\begin{tabular}[t]{L{2.8cm}c}
\multicolumn{2}{c}{Deltoid}\\
\midrule
Deltoid orig. (spinal, acromial) & 18 \\
Deltoid orig. (clavicular) & 17 \\
Deltoid ins. & 19 \\
Humerus contact & 20 \\
Scapula contact & 21 \\
RC contact & 22 \\
\end{tabular} &
\begin{tabular}[t]{lc}
\multicolumn{2}{c}{Rotator cuff (RC)}\\
\midrule
Teres minor ins. & 23 \\
Infraspinatus ins. & 26 \\
Supraspinatus ins. & 25 \\
Subscapularis ins. & 24 \\
Teres minor orig. & 27 \\
Infraspinatus orig. & 30 \\
Supraspinatus orig. & 29 \\
Subscapularis orig. & 28 \\
Humerus contact & 31 \\
Scapula contact & 32 \\    
Deltoid contact & 33 \\
\end{tabular}\\
\bottomrule
\end{tabular}
\label{tab:surface_volume_nds_ids_shoulder}
\end{table}
\egroup

\begin{figure}[htb]
	\centering
    \includegraphics[width=0.9\textwidth]{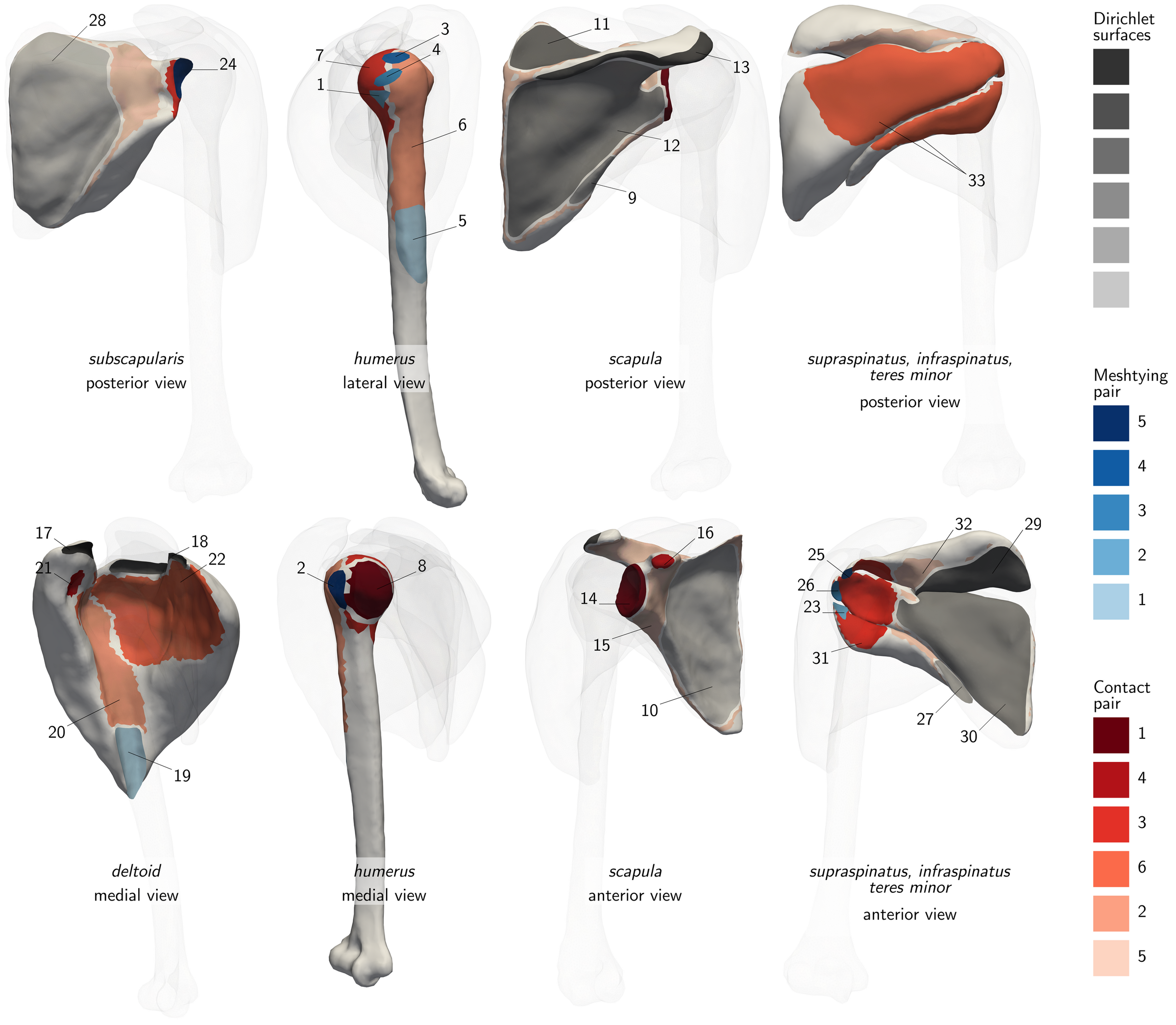}
    \caption{Surfaces defined for the shoulder model. Colors represent the assigned type of boundary condition. The surfaces are labelled according to the ids in Table \ref{tab:surface_volume_nds_ids_shoulder}.}
    \label{fig:surface_volume_shoulder}
\end{figure}

\begin{table}[htb]
\centering \caption{Definition of dirichlet boundary conditions due to fixation of the scapula and the muscle origins.}
\begin{tabular}{lc}
\toprule
    Surface nodesets (muscle origins) & 9, 10, 11, 12, 13, 17, 18, 27, 28, 29, 30 \\
    Volume nodesets (inner nodes of scapula) & 34 \\
\bottomrule
\end{tabular}
\label{tab:dirichlet_bcs_shoulder}
\end{table}

\begin{table}[htb]
\begin{minipage}[t]{0.5\linewidth}
\centering \caption{Meshtying boundary conditions.}
\begin{tabular}{lcR{0.5cm}L{0.5cm}}
\toprule
	& Pair & \multicolumn{2}{c}{Surfaces}\\ \midrule
    Deltoid ins. & 1 & 5 & 19\\
    Teres minor ins. & 2 & 1 & 23\\
    Infraspinatus ins. & 3 & 4 & 26\\
    Supraspinatus ins. & 4 & 3 & 25\\
    Subscapularis ins. & 5 & 2 & 24\\
\bottomrule
\end{tabular}
\label{tab:meshtying_bcs_shoulder}
\end{minipage}
\hfill
\begin{minipage}[t]{0.5\linewidth}
\centering 
\caption{Contact boundary conditions.}
\begin{tabular}{lccc}
\toprule
	& Pair & Master & Slave\\ \midrule
    Glenohumeral joint & 1 & 8 & 14\\
    Humerus-deltoid & 2 & 6 & 20\\
    Humerus-RC & 3 & 7 & 31\\
    Deltoid-scapula & 4 & 21 & 16\\
    RC-scapula & 5 & 32 & 15\\
    Deltoid-RC & 6 & 22 & 33\\
\bottomrule
\end{tabular}
\label{tab:contact_bcs_shoulder}
\end{minipage}
\end{table}

\end{document}